\documentclass{article}
\usepackage{url} 
\usepackage{amsmath,amssymb,amsfonts}
\usepackage{graphicx}
\usepackage{xcolor}

\begin{document}
\title{Machine Intelligence in Africa: a survey}
\author{Allahsera Auguste Tapo, Ali Traor\'e, Sidy Danioko,  Hamidou Tembine \footnote{Corresponding author: 
H. Tembine, Learning \& Game Theory Laboratory, TIMADIE. Email: tembine@landglab.com }}
 \date{}
 

\maketitle  

\begin{abstract}  In the last 5 years, the availability of large audio datasets in African countries has opened unlimited opportunities to build machine intelligence (MI)  technologies that are closer to the people and speak, learn, understand, and do businesses in  local languages, including for those who cannot read and write. Unfortunately, these audio datasets are not fully exploited by current MI tools, leaving several Africans out of MI business opportunities. Additionally, many state-of-the-art MI models are not culture-aware, and the ethics of their adoption indexes are questionable. The lack thereof is a major drawback in many applications in Africa. This paper summarizes recent developments in machine intelligence in Africa from a multi-layer multiscale and culture-aware ethics perspective, showcasing MI use cases in 54 African countries through 400 articles on MI research, industry, government actions, as well as uses in art, music, the informal economy, and small businesses in Africa. The survey also opens discussions on the reliability of MI rankings and indexes in the African continent as well as algorithmic definitions of unclear terms used in MI.

\end{abstract}


\section{Introduction}

Machine   intelligence (MI)  focuses on the creation of models, evolutionary dynamics, and algorithms that enable machines or software to co-learn from data and improve their performance over time \cite{Epstein1946,Epstein1946v2}. MI is therefore an advanced computer \& information science that allows a machine, device, software, program, code, or algorithm to interact intelligently with its environment, which means it can take measures, make decisions, perform actions, and develop strategies to maximize its chances of successfully achieving its preferences and objectives \cite{Bouare2023,Bouare2023ty,trefty1,trefty2,trefty3,trefty4,trefty5}.

\subsection*{Demystifying MI for the General Public} 

Addressing the global gap in public awareness surrounding MI, including its uses, benefits, risks, and limitations, is crucial not only in Africa but worldwide. The confusion arising from associating the term  ``intelligence" with few illustrative machine learning applications, influenced by science fiction and MI business narratives, has led to widespread uncertainty and, in some cases, fear. To unlock the potential benefits that MI holds for Africa, it is imperative to commence an educational journey for the general public, creating a foundation of informed and knowledgeable users of MI systems. This broader understanding will, in turn, contribute to nurturing the technical professionals and highly skilled specialists necessary to propel the countries' ambitious MI plans forward. 
Initiating general public awareness efforts must begin at the grassroots levels of society, ensuring that programs and content are accessible even to those with limited or no formal education. Recognizing the prevalence of informal education in many African countries, a variety of courses and training programs should be tailored to assist recipients in ascending the capacity-building pyramid.
Designing general public awareness programs that are easily consumable is key, primarily through short videos, brief audios in local languages, or interactive tradition-inspired games. These formats aim to help the audience grasp fundamental MI concepts and distinguish between myth and reality. 

{\it Regrettably, up to now, the educational journey for the general African public remains inaccessible due to the absence of machine intelligence that supports the local language - an audio-rich language in many areas. }

Here the objective is not to rank countries by their MI strategy. Some very small MI projects had very big social impact in the local population and some significant MI Innovation had almost no impact for the local population so far. What we learn from basic game theory and Pareto optimality is that  when multiple interdependent objectives are involved as it is the case in MI, the scalarization technique which maps the vector of objectives into a single  number, is not necessarily a good idea. This can be observed from the fact that 
the vector $  (1, 0) $  is not better than $ (0, 1)$ and vice-versa. These vector elements are often replaced by the well-being elements of the local population, which include monetary, non-monetary, technological, non-technological, technical, non-technical, empathetic, etc, which are not captured by a single number.
\subsection*{Literature  Review}
In \cite{refgen2}, the authors   examine MI in Africa with a special focus on challenges and opportunities. The developments in MI  have the potential to disrupt and transform socio-economic activities across industries. The countries in the Global South such as those in Africa need to tackle governance issues and lack of institutional capacity to establish the building blocks to allow MI  to flourish for them by them. It is important to also examine the roles of international communities in bridging the technological gaps in Africa by adopting a problem-driven approach where local needs and problems are contextualized into MI policy formulation rather than a blanket copy-and-paste practice that has limited the advancement of development policies in Africa. A problem-driven approach would help African countries to formulate robust MI policies that are relevant to their unique circumstances.

The deployment of basic MI technologies is proliferating on the African continent, but policy responses are still at their early stages.
The work in \cite{refgen3}  provides an overview of the main elements of MI deployment in Africa, MI's core benefits and challenges in African settings, and MI's core policy dimensions for the continent. The authors argued that for MI to build, rather than undermine, socio-economic inclusion in African settings, policymakers need to be cognisant of the following key dimensions: gender equity, cultural and linguistic diversity, and labour market shifts.

The work in \cite{refgen4} proposes a decolonized appropriation of MI in Africa.  Technocoloniality occurs when the use of technology reinforces a colonial mindset, aiming to assert power, control, and domination, often replicating historical patterns of oppression. Take digital technologies like the internet and mobile phones, for instance, which are deeply intertwined with the legacy of colonialism. This integration, originating from the West, has been imposed on other regions, notably the African continent. An illustration is the mobile app 'Free Basics' by Facebook, allowing users limited access to certain websites without data charges. However, many of these accessible sites promote services of private US companies and are not available in major African languages. This app has faced significant criticism as a manifestation of digital colonialism.

Despite dominant narratives discouraging anthropomorphizing MI, Shoko Suzuki argues against universal models for our attitudes towards MI \cite{refgen5}. We argue here that the same applies to MI ethics. We should not expect to have a one-size-fits-all in ethics. Within Africa, we should not expect to have a one-size-fits-all ethics either.

From a regional perspective, several studies have shown that machine learning technology can help address some of Africa's most pervasive problems, such as poverty alleviation, improving education, delivering quality healthcare services, and addressing sustainability challenges like food security and climate change.  In \cite{refgen6} , a critical bibliometric analysis study is conducted, coupled with an extensive literature survey on recent developments and associated applications in machine learning research with a perspective on Africa. The presented bibliometric analysis study consists of 2761 machine learning-related documents, of which 89\% were articles with at least 482 citations published in 903 journals during the past three decades. Furthermore, the collated documents were retrieved from the Science Citation Index Expanded , comprising research publications from 54 African countries between 1993 and 2021. The bibliometric study shows the visualization of the current landscape and future trends in machine learning research and its application to facilitate future collaborative research and knowledge exchange among authors from different research institutions scattered across the African continent.

Agriculture is considered as the main source of food, employment and economic development in most African countries and beyond. In agricultural production, increasing quality and quantity of yield while reducing operating costs is key. To safeguard sustainability of the agricultural sector in Africa and globally, farmers need to overcome different challenges faced and efficiently use the available limited resources. Use of technology has proved to help farmers find solutions for different challenges and make maximum use of the available limited resources. Blockchains, internet of things and machine learning innovations are benefiting farmers to overcome different challenges and make good use of resources. In \cite{refgen7} , a wide-ranging review of recent studies devoted to applications of internet of things and machine learning in agricultural production in Africa is presented. The studies reviewed focus on precision farming, animal and environmental condition monitoring, pests and crop disease detection and prediction, weather forecasting and classification, and prediction and estimation of soil properties. The work in \cite{refgenA1} explores the realization of MI potential in Africa, emphasizing the pivotal role of trust. The work in \cite{refgenA2} discusses the failure of mass-mediated feminist scholarship in Africa, highlighting normalized body-objectification as a consequence of MI. The work in \cite{refgenA3} focuses on using artificial intelligence for diabetic retinopathy screening in Africa.
The work in \cite{refgenA4} delves into ethical considerations surrounding the implementation of artificial intelligence in Africa's healthcare.
The work in \cite{refgenA5} investigates the use and impact of MI on climate change adaptation in Africa.
The work in \cite{refgenA6} maps policy and capacity for MI development in Africa.
The authors in \cite{refgenA7} explore the role of information and communication technologies, including MI, in the fight against money laundering in Africa.
The work in \cite{refgenA8} addresses the malicious use of MI in Sub-Saharan Africa, posing challenges for Pan-African cybersecurity.
The work in \cite{refgenA9} conducts a needs assessment survey for artificial intelligence in Africa.
The work in \cite{refgenA10} discusses the integration of MI in medical imaging practice from the perspectives of African radiographers.
The   authors in \cite{refgenA11} outline an agenda for journalism research on artificial intelligence in Africa.
The work in \cite{refgenA12} explores the ethics of MI in Africa, focusing on the rule of education.
The work in \cite{refgenA13} examines artificial intelligence policies in Africa over the next five years.
The work in \cite{refgenA14} discusses machine ethics and African identities, offering perspectives on MI in Africa.
In \cite{refgenA15} the authors  advocates scaling up MI to curb infectious diseases in Africa.
The work in \cite{refgenA16} addresses emerging challenges of artificial intelligence in Africa, presented in the context of responsible MI.
The work in \cite{refgenA17} scrutinizes China's role as a 'digital colonizer' in Africa, focusing on MI's impact.

The paper in \cite{refgenA18} assesses the effect of information and communication technologies  on the informal economy. The authors applied the Generalized Method of Moments  on a sample of 45 African countries from 2000 to 2017. According to the findings, the use of ICTs (mobile phone and internet) decreases the spread of the informal economy in Africa. These results are robust to a battery of robustness checks. Furthermore, the results of the mediation analysis show that the effect of ICTs on the informal economy is mediated by financial development, human capital and control of corruption. From a policy perspective, the authors suggested a quantitative and qualitative consolidation of technological infrastructures, for a sustainable mitigation of the rise of the informal sector in Africa.  

In the context of the African continent, machine learning has been used in ecology \cite{aiafricatty1}, gold mining  \cite{aiafricatty2},  education \cite{aiafricatty3}, construction \cite{aiafricatty4}, electricity \cite{aiafricatty5}, mineral resources \cite{aiafricatty6}, wheat \cite{aiafricatty7}, optical network \cite{aiafricatty8}, Teaching \cite{aiafricatty9}, fact-checking \cite{aiafricatty10}, news production \cite{aiafricatty11}, digital humanism \cite{aiafricatty12},  music \cite{aiafricatty13},  farming \cite{aiafricatty14}, low-resource languages \cite{aiafricatty15}, architecture \cite{aiafricatty16}, clinical prediction \cite{aiafricatty17}, text-based emotion \cite{aiafricatty18}, audio-based emotion \cite{aiafricatty19}, Computer Vision Community for Africans and by Africans \cite{aiafricatty20,aiafricatty21}.

\subsection*{Contribution}
The objective of this article is to present some advances in MI research, MI use cases in small businesses, and government actions in Africa. In countries, whenever available, we also highlight MI in industry, art, music, and the informal economy. As we will see, each country has its intrinsic path and cultural adaptation to MI and other emerging technologies such as Graphchains and the Internet of People.  


\begin{table}[h]
\caption{Some countries with a national MI strategy as of December 2023}
\label{tableaistratrgy}

\centering
\begin{tabular}{|l|l|}
\hline
\textbf{Country} & \textbf{Explicit national MI Strategy reported} \\
\hline
Algeria & $\checkmark$ 2021  \\
Senegal & $\checkmark$ 2023 \\
Benin & $\checkmark$ 2023 \\
Rwanda & $\checkmark$ 2023 \\
Mauritius &$\checkmark$ 2019 \\
Nigeria & $\checkmark$ 2023 \\
Egypt & $\checkmark$ 2021 \\
Tunisia & $\checkmark$ 2022 \\
Seychelles & $\checkmark$ 2019 \\
\hline
\end{tabular}
\end{table}

\begin{figure} \label{figafricakeywords} \caption{Keywords of some MI research topics in Africa} 
\includegraphics[scale=0.4]{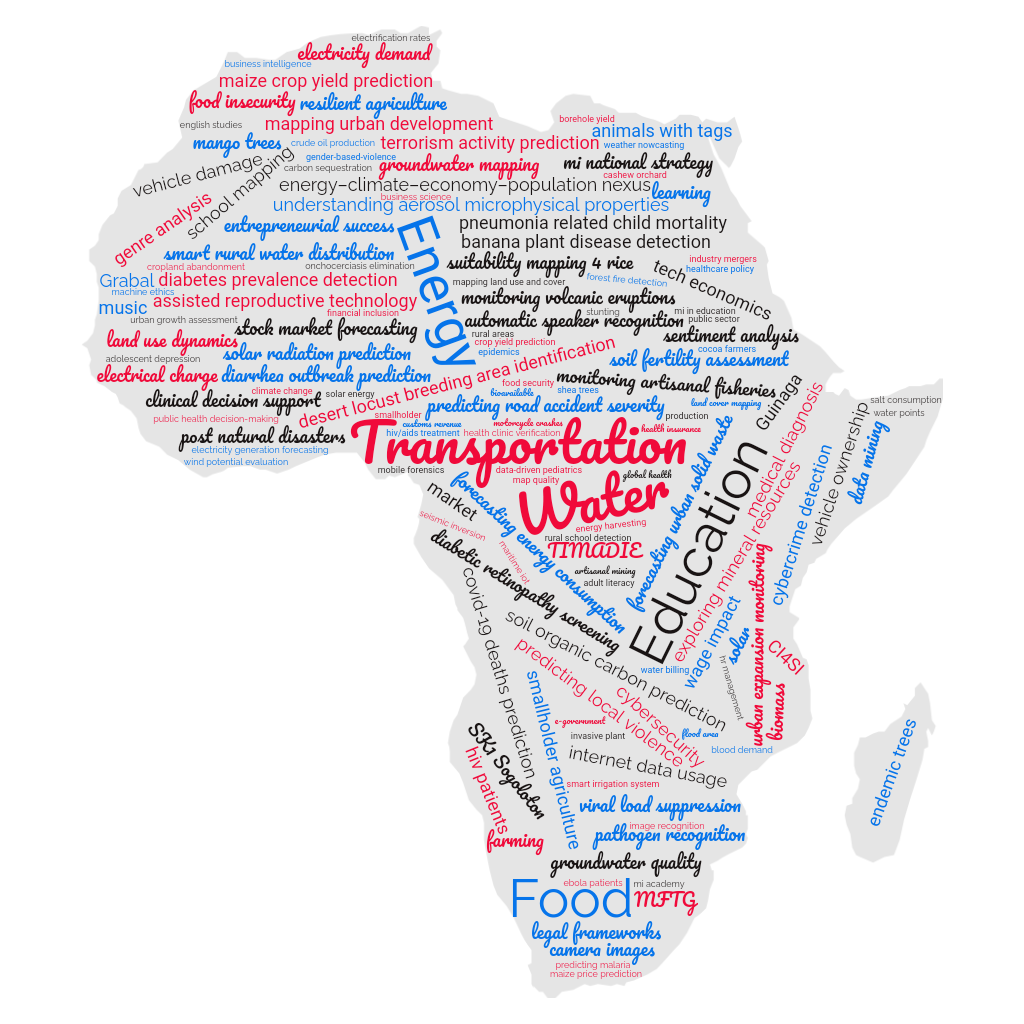}
\end{figure}

Some key findings are below: 
\begin{itemize} 
\item 
Some MI strategies (see  Table \ref{tableaistratrgy}) are dictated from outside the African continent, and some meetings are organized in very expensive hotels where the local population cannot afford a single night in a lifetime. Some reports and recommendations are made after those meetings. Unfortunately, most of these efforts have zero impact in the field and are almost useless to the local population, as they are not aware of them, except for a few selected people surrounding the funding on MI.
\item  We have carefully reviewed over 400 research articles on the use cases of MI in Africa (see Fig. \ref{figafricakeywords}). Most of the applications are at the testing level on a small dataset, with manipulated targets, and some are clearly overselling their findings. A random forest does not provide a magical output as claimed by many of these papers. As the MI field has made progress in terms of algorithms, designs, robustness, convergence, and learnability, these studies need to be updated to more practical MI algorithms than these random searches.
\item  Ranking African countries by a single uniform weighting average index is clearly unethical, as it does not take into account data transparency and concrete actions taken by governments beyond the speeches at expensive hotels and palaces.
\end{itemize}

Below we list some schools, research institutes, centers and journals  on MI established within the continent.

\begin{itemize}
\item Benin: Atlantics AI Labs: Artificial Intelligence Research Centre 
\item 
Burkina Faso: Interdisciplinary Center of Excellence in AI for Development

\item   The Artificial Intelligence \& Robotics center of excellence (AI\&R CoEs) in Addis Ababa Science and Technology University, in Ethiopia.
\item
The Ethiopian Artificial Intelligence Institute (EAII) which is now  the African Artificial Intelligence Center of Excellence.
\item 
Ghana: National Artificial Intelligence Center 
\item 
Ghana: Responsible Artificial Intelligence Lab
\item Ivory Coast:  AI and Robotics Center in Yamoussoukro
\item 
Nigeria: National Centre for Artificial Intelligence and Robotics
\item 
AIISA:  Artificial Intelligence Institute of South Africa
\item 
Algeria: National School for Artificial Intelligence 
\item 
Algeria:  House of Artificial Intelligence  
\item 
Egypt: Faculty of Artificial Intelligence
\item 
Egyptian Journal of Artificial Intelligence
\item Moroccan International Center for Artificial Intelligence
\item 
Congo: African Research Centre on Artificial Intelligence
\item 
Rwanda: Africa's Centre of Excellence in Artificial Intelligence
\item
Uganda: Artificial Intelligence and Data Science Lab
\item Artificial Intelligence Centre of Excellence  Africa in Kenya 
\item Malawi : Centre for Artificial Intelligence and STEAM - Science, Technology, Engineering, Arts and Mathematics-. 

\end{itemize}

We now list some MI companies in Africa:

African Foods Nutrition, DataBusiness-AI, CyberLabs Tech, Kumakan , SOSEB, Saintypay, Kalabaash, Dunia , Qotto, Toto Riibo, Futurafric AI , DatawareTech, Khalmax Robotics, mNotify, GreenMatics, DigiExt, CYST, CRI, Huggle.care, QualiTrace, Guinaga, Grabal, Timadie,  SK1 ART,  WETE Women-in-Drones, Tuteria, Kudi AI , Curacel , Codar Tech Africa, Afrikamart ,  NeuralSight, AIfluence, Amini, Halkin, Freshee , M-Shule, AI Connect, Qubitica, Cash Radix, AgCelerant, Arie Finance, 4Sight, Agrix Tech, Comparoshop,  eFarm, KMER MI, DASTUDY,
Teranga Capital,  Lengo AI, Semoa, Eazy Chain,  SocialGIS, Dobbee Pay, Solimi Fintech ,  Artybe, Genoskul, WenakLabs, DaTchad, ZereSoft, KivuGreen,
 BasaliTech, Plano-OneTree, eFarmers, LibraChat, Mphalane, Keti, Nalane, Qoloqo, Tincup, Loop, Hyperlink, AfriFeel, Lelapa AI, Lesan,  Xineoh, Clevva, Aerobotics, The Gearsh, Credo, Akiba Digital, Bridgement, AfricAi , Neurozone, DataProphet, Vulavula, AkilliCon, Dalil, TransformaTek, NileCode, FastAutomate, Synapse Analytics, Intixel, DevisionX, Dileny Technology, MerQ, WideBot, Aphrie, PasHakeem, MoroccoAI, Annarabic, Sigma.AI, AgriEdge, ATLAN, SudanAI, KatYos, WARM, ScorSell, LWATN, AD'VANTAGE, Business \& AI, Deepera.AI,  AQUA SAFI, Tabiri Analytics, Congretype, Openbanking, Chil AI Lab, Global Auto Systems, Wekebere, Diagnosify, Xpendly, Kwanso, Tribal Credit, Convertedin, DXwand, Sky.Garden, Save-Your-Wardrobe, Wattnow, Ubenwa Health, DataPathology.

Each country is doing its own MI path at the research level targeting concretes solutions to local problems.

Northern Africa:
\begin{itemize}
\item 
Algeria: MI applications in AgriTech, water resource management, startup incubators, strategic approach to MI.
\item 
Egypt: Diverse MI landscape from automotive cybersecurity to smart agriculture, tech innovation hubs, advancements in autonomous instruments and FastAutomate technologies.
\item 
Libya: MI applications in e-commerce, stock market prediction, smart cities, economic and societal aspects.
\item 
Morocco: MI in diverse research areas, MI-enabled startups, societal benefits from water management to challenges in the insurance sector.
\item 
Sudan: MI focus on groundwater quality assessment, disease forecasting.
\item 
Tunisia: Strong emphasis on MI in research, applications in textile industry, water management, urban solid waste forecasting, dialectal speech recognition.
\end{itemize}

Southern Africa:
\begin{itemize}
\item 
Botswana: Water billing, diabetic retinopathy screening, solar radiation prediction, gravel loss condition prediction, HIV/AIDS treatment, data mining, ART program success, clinical decision support.
\item 
Eswatini: Financial inclusion, diarrhea outbreak prediction, renewable energy, COVID-19 case prediction, invasive plant study, MI in education, maize crop yield prediction, economic development through technology.
\item 
Lesotho: Land cover mapping, carbon sequestration, soil organic carbon prediction, weather nowcasting, health insurance enrollment, education, electricity demand forecasting, legal frameworks, wage impact, industry mergers, healthcare policy.

\item 
Namibia: MI in education, predicting Gender-Based Violence, cybersecurity practices in rural areas, discriminating individual animals with tags in camera trap images, smart irrigation system for efficient farming.
\item 
South Africa: Accenture collaboration with Gordon Institute of Business Science, MI in Public Sector HR Management, MI-based medical diagnosis. 
\end{itemize}

Middle Africa :
\begin{itemize}
\item Angola: Seismic inversion, forest fire detection, urban expansion monitoring, bioavailable isoscapes. 
\item 
Cameroon: MI applications in healthcare, agriculture, small businesses, government initiatives. 
\item 
Central African Republic: IoT-based smart agriculture, MI predicting electricity mix, machine learning aiding in primate vocalization classification, smart city development.
\item 
Chad: MI for fertility rate forecasting, conflict risk projections, hate speech detection, entrepreneurial landscape.
\item 
Democratic Republic of the Congo: MI in gully erosion assessment, property tax roll creation, small businesses using MI for climate resilience, forest resource monitoring.
\item 
Equatorial Guinea: MI in researching sea level variability, economic diversification.
\item 
Gabon: MI applications in mapping land cover, monitoring coastal erosion, forest height estimation.
\item 
Republic of the Congo: African Research Centre on MI establishment, focus on research and digital technology.
\item 
Sao Tom\'e and Principe: MI applications in social protection targeting, value chain analysis in agriculture
\end{itemize}

Western Africa:

\begin{itemize} \item  
Benin: Soil fertility assessment, banana plant disease detection, electricity generation forecasting, public health decision-making, suitability mapping for rice production.
\item 
Burkina Faso: Predicting malaria epidemics, mapping urban development, forecasting energy consumption, exploring mineral resources.
\item 
Cabo Verde: Understanding aerosol properties, studying climate change impact, estimating salt consumption, monitoring volcanic eruptions.
\item 
C\^ote d'Ivoire: Machine learning for cocoa farmers, progress towards onchocerciasis elimination.
\item 
Gambia: Machine learning models for pneumonia-related child mortality, smart rural water distribution systems.
\item 
Ghana: Urban growth assessment, vehicle ownership modeling, sentiment analysis, blood demand forecasting, internet data usage analysis, severity prediction of motorcycle crashes, effects of artisanal mining, customs revenue modeling.
\item 
Guinea: Predicting viral load suppression among HIV patients, prognosis models for Ebola patients. 
\item 
Guinea-Bissau: Biomass relationships, cashew orchard mapping, learning and innovation in smallholder agriculture, automatic speaker recognition.
\item 
Liberia: Cloud computing and machine learning for land cover mapping, predicting local violence, scalable approaches for rural school detection.
\item 
Mali: Groundwater potential mapping, cropland abandonment analysis, MI in addressing global health challenges, improved recurrent neural networks for pathogen recognition, market liberalization policy analysis.
\item 
Mauritania: MI-driven insights into English studies, desert locust breeding area identification, business intelligence models for e-Government, remote monitoring of water points.
\item 
Niger: Electrical charge modeling, land use mapping using satellite time series, adult literacy and cooperative training program analysis.
\item 
Nigeria: Extensive MI applications including diabetes prevalence detection, crude oil production modeling, flood area prediction, food insecurity prediction, entrepreneurial success prediction, mobile forensics for cybercrime detection, genre analysis of Nigerian music, terrorism activity prediction, stock market forecasting, poverty prediction using satellite imagery.
\item 
Senegal: Crop yield prediction, resilient agriculture, machine learning for rice detection, monitoring artisanal fisheries, predicting road accident severity, estimating electrification rates, analyzing the energy - climate - economy - population nexus.
\item 
Sierra Leone: Initiative for rapid school mapping using MI and satellite imagery.
\item 
Togo: Novissi program expansion, machine ethics, wind potential evaluation, maize price prediction, solar energy harvesting assessment, land use dynamics forecasting, solar energy harvesting evaluation. 
\end{itemize}

Eastern Africa:
\begin{itemize}
\item 
Burundi: MI research on malaria case prediction, automated image recognition for banana plant diseases, industry focus on optimizing LPG usage.

\item Comoros: MI research in Education.
\item 
Djibouti: Research on sky temperature forecasting, deep learning for fracture-fault detection, industry applications in LPG challenges, air travel experiences.
\item 
Eritrea: Predictive lithologic mapping using remote sensing data.
\item 
Ethiopia: Machine learning to predict drought, interpretable models for evaporation in reservoirs. 
\item 
Kenya: MI Made in Africa supporting startups.
\item 
Mauritius: MI in education, maritime IoT potential, strategic approach to MI with national strategy, Mauritius MI Council, MI Academy.
\item 
Mozambique: Research areas including assessing OpenStreetMap quality, mapping land use and cover, food security, smallholder irrigated agriculture mapping, deep learning and Twitter for mapping built-up areas post-natural disasters.
\item Rwanda: National MI Policy approval.
\item Tanzania: MI applications in healthcare, Resilience Academy students using machine learning for tree-cover mapping.
\item 
Uganda: Creating high-quality datasets for East African languages.
\item 
Somalia: Sentiment analysis applied to Somali text.  
\item 
South Sudan: Machine learning to analyze fragility-related data.

\item Zambia: Machine learning in predicting stunting among children, enhancing health clinic verification efficiency.

\item 
Zimbabwe: Data-driven pediatrics, vehicle damage classification using deep learning algorithms, technology to predict and address adolescent depression.
\end{itemize}

\begin{figure} \label{figtopicsmi} \caption{Keywords of some MI research topics in Africa} 
\includegraphics[scale=0.7]{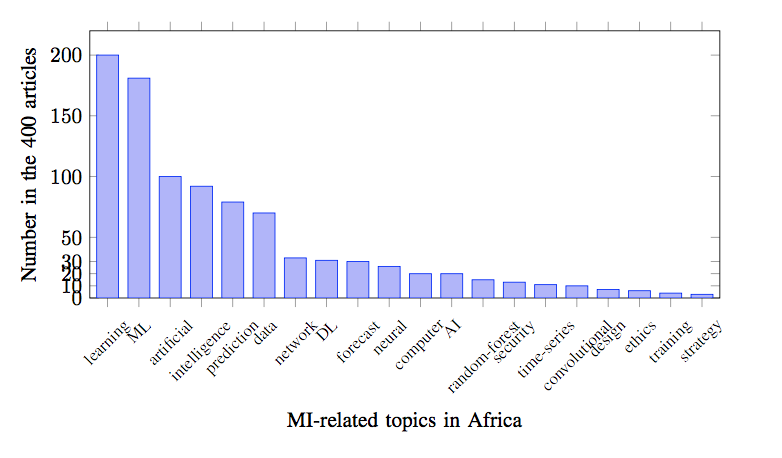}
\end{figure}

\begin{figure} \label{figtopicsmiapp} \caption{Application of MI in Africa} 
\includegraphics[scale=0.7]{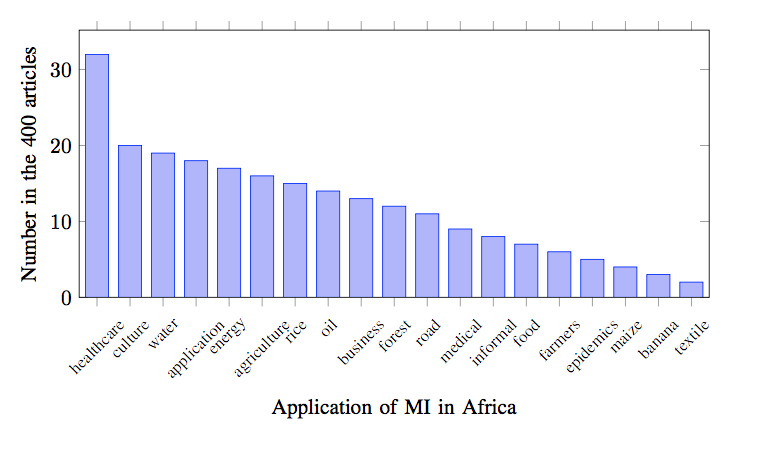}
\end{figure}

Figures \ref{figtopicsmi} and \ref{figtopicsmiapp} display the number of occurrences vs MI-related topics and their applications in Africa that appears in the title of the 400 references used in this survey.

 Examining closely 400 articles on MI in Africa, beyond the headlines, it emerges that
\begin{itemize}
\item human learning, the learning of men and women, whether children or adults, takes a much more central place in discussions than machine learning.

\item At the core is human learning, utilizing various tools, including machine assistance, as well as inspiration from nature.
\end{itemize}

 \subsection*{Content of the article}
 
The rest of article is structured as follows.  Section \ref{sec:northern} is dedicated to Northern Africa. Section \ref{sec:southern} presents it in Southern Africa. Section \ref{sec:central} presents it in Central Africa. Section \ref{sec:western} presents MI research and use cases in Western Africa. Section \ref{sec:eastern} is dedicated to Eastern Africa.    Section  \ref{sec:discussion}  examines MI indexes and limitation of  MI adoptions  in Africa. 
Section \ref{sec:ethics} discusses technical and non-technical ethics of MI in Africa.  Multi-scale multi-layer multi-modal culture-aware ethics are presented in Section \ref{sec:beyond}.
Section \ref{sec:data} presents data issues and some terms borrowed from psychology that are not undefined in the context of computer science.
 Section \ref{sec:conclusion} concludes the article.

 
\section{ MI in North Africa} \label{sec:northern}  
This section presents some of the remarkable advancements and strategic initiatives undertaken by {\it Algeria, Egypt, Libya, Morocco, Sudan, and Tunisia } in harnessing the power of MI.
In recent years, these nations have emerged as key players in the ever-evolving realm of MI, propelling the region toward technological innovation and sustainable growth. From the  economic intelligence mechanisms supporting small and medium enterprises in Algeria to the pioneering National MI strategy of Tunisia, each country contributes a unique perspective to the broader narrative of MI adoption.
We observe diverse facets of MI implementation, from educational and healthcare strategies to water management solutions and economic reforms. Noteworthy research endeavors, such as modeling groundwater quality in Sudan and forecasting urban solid waste in Tunisia, showcase the region's commitment to addressing complex challenges through cutting-edge technologies. As we navigate through the state of the art in MI across North Africa, we gain insights into the pivotal role MI plays  in shaping the future of these nations.

Algeria exhibits a growing focus on MI applications in AgriTech, emphasizing water resource management and the emergence of startup incubators. With a particular emphasis on groundwater quality and smart agriculture, Algeria's MI initiatives aim to address environmental challenges and enhance agricultural practices. The country's key achievements include the development of AgriTech solutions and the integration of MI for efficient water resource management.

Egypt stands out with a diverse MI landscape, ranging from automotive cybersecurity to smart agriculture. The country's tech innovation hubs foster research in autonomous instruments and FastAutomate technologies. In the automotive sector, the focus on cybersecurity aligns with global trends, while the application of MI in smart agriculture indicates a commitment to leveraging technology for sustainable practices. Egypt's MI achievements include advancements in autonomous instruments and a robust presence in the realm of FastAutomate technologies.

 Libya's MI research landscape spans diverse domains, including the adoption of e-commerce in SMEs, stock market prediction, and the enhancement of quality of life through smart cities. Notable studies explore MI applications in Libyan SMEs, predicting daily stock market movements with high accuracy, and leveraging MI for smart city development. Libya's MI initiatives showcase a comprehensive approach, addressing economic, financial, and societal aspects with the aim of fostering technological integration and improving living standards.

 Morocco demonstrates a multifaceted approach to MI, covering diverse research areas such as data use challenges, MI impact on human rights, and applications like automating water meter data collection. Notable achievements include the development of MI-enabled startups like Annarabic and SYGMA.AI, addressing challenges in insurance claims settlements and revolutionizing the insurance industry. Morocco's research efforts highlight a commitment to utilizing MI for societal benefits, from enhancing water management to addressing challenges in the insurance sector.

Sudan's MI focus is evident in groundwater quality assessment and disease forecasting. Research endeavors employ MI algorithms like multilayer perceptron neural networks and support vector regression to evaluate groundwater suitability for drinking. Additionally, the application of time series forecasting methodologies aids in predicting diseases like malaria and pneumonia. Sudan's MI efforts underscore a commitment to addressing critical issues in public health and environmental sustainability.

Tunisia showcases a strong emphasis on MI in research, spanning applications in the textile industry, water management, urban solid waste forecasting, and dialectal speech recognition. The country's acquisition of Instadeep, a deep tech startup, signifies global recognition of Tunisian innovations. Government initiatives include the integration of MI in public finance management, aligning with broader strategies for economic growth and accountability. Tunisia's MI landscape reflects a dynamic and innovative approach, positioning itself as a hub for entrepreneurship and technological advancements in the North African region.

Throughout the period of 2000-2023, Northern African countries have demonstrated a growing embrace of MI technologies, leveraging them across diverse sectors to address challenges and foster sustainable development in their unique socio-economic contexts.

\subsection{ Algeria}

Algeria has made significant strides in deploying MI  across various sectors from 2000 to 2023. Research initiatives highlight the crucial role of economic intelligence in supporting the growth of small and medium enterprises , emphasizing the synergy between economic intelligence and MI for SME development. Algiers serves as a hub for diverse MI-enabled platforms contributing to the rise of small and medium businesses, ranging from payroll and HR management (RAWATIB) to innovative solutions for the visually impaired (Dalil). In the informal economy, leveraging MI technologies is proposed to redefine tax policies, assess incentives for agriculture, and enhance electronic transactions, contributing to a regulated formal economy. At the governmental level, Algeria has adopted a national strategy on research and innovation in MI, inaugurated the National School for Artificial Intelligence (ENSIA), and declared 2023 as  {\it 'The Year of Artificial Intelligence'}, showcasing a comprehensive approach to advancing skills and leveraging MI in key socio-economic sectors.

\begin{table}[htb]
  \begin{center}
    \begin{tabular}{|p{0.6in}p{0.6in}p{1.4in}|}
 \hline
      Algeria & 2020-2023 & Concrete Actions \\
      \hline
      Research & $\checkmark$ & ENSIA, RAWATIB \\
      SMB & $\checkmark$ &  AkilliCon, Dalil, TransformaTek  \\
      Informal Economy & $\checkmark$  & Data 1980-2017 \\
      Government & $\checkmark$ & 2021: 100-page white paper on national MI strategy \\
      \hline
    \end{tabular}\label{tab:mytablealg}
  \end{center}
  \caption{MI in Algeria}
\end{table}

\subsubsection{ Research}  The study in  \cite{refAlgeria1} underscores the pivotal role of economic intelligence as a crucial mechanism in fostering the growth and development of small and medium enterprises in Algeria. For these enterprises, success hinges on their ability to access high-quality and timely information, enabling them to navigate a rapidly changing environment, anticipate shifts, and make optimal decisions for their survival. The study argues that integrating economic intelligence mechanisms into SMEs is not just beneficial but an imperative necessity for their development, support, and enhanced competitiveness. Through a descriptive approach that emphasizes the need for information transfer, analysis, and drawing from various sources, the results affirm the significant contribution of economic intelligence to the competitiveness, innovation, and strategic development of small and medium enterprises, empowering them to effectively face risks and bolster decision-making. In the context of Algeria's economic landscape, the study emphasizes the symbiotic relationship between economic intelligence and the MI-driven economy, particularly within the realm of small and medium businesses. The integration of these intelligence mechanisms becomes a strategic imperative, aligning with the broader national strategy for economic development. As Algeria looks towards the future, the study suggests that the synergy between economic intelligence and MI holds substantial potential for propelling the growth and resilience of SMEs, thereby contributing to the overall economic vitality of the country. 
\subsubsection{ Small Businesses} Algiers serves as a focal point for MI-powered platforms like RAWATIB, AkilliCon, Dalil, and TransformaTek, contributing to the emergence of MI-enabled small and medium businesses in Algeria. RAWATIB is an MI-powered SaaS platform that streamlines payroll and human resources  management for businesses of all sizes. The platform automates many administrative tasks involved in payroll and HR management, reducing the risk of errors and saving businesses time and resources. AkilliCon is a low-cost, low-profile Ambient Energy Harvester Terminal that can be used alone with AkilliCon specified Battery as a Power Bank. Dalil is a company focused on object recognition and navigation systems for visually impaired people. Imagine this: your friend invited you to dinner at a new restaurant downtown, but to get there, you need to go to the bus stop, take bus number 128 to the train station, and take the train into the city. Easy enough, right? Now, imagine making that trip without being able to see. It is a challenge that 314 million visually impaired people face every day. Even with expert mobility skills and the use of a cane or a guiding dog, navigation, environment detection, and recognition can be stressful. TransformaTek is a startup working toward the widespread adoption of location intelligence technologies by small businesses. Our mission is to develop community-driven platforms to democratize access to open geospatial datasets and build useful use cases for businesses.

\subsubsection{ Informal Economy} This research work  \cite{refAlgeria5}  examines Algeria's informal economy  through a comprehensive analysis employing the Multiple Indicator Multiple Causes  approach. The primary objectives include investigating key determinants, estimating size and development, examining short and long-term relationships with the formal economy, and unraveling causality directions from 1980 to 2017. Findings reveal that the tax burden, the agricultural sector, the quality of institutions, and GDP per capita emerge as pivotal determinants shaping the contours of the IE in Algeria. Understanding these factors becomes imperative for crafting effective strategies to address the informal sector. The study indicates that the informal economy constitutes an average of 33.48\% of the official GDP, witnessing a steady increase over the past 15 years. Interestingly, the research uncovers a nuanced relationship between the informal and formal economies. In the short run, the IE exerts a positive impact on the formal economy, while in the long run, this effect undergoes a reversal. Understanding these dynamics is crucial for policymakers in devising interventions that promote sustainable economic growth.To mitigate the size of the informal economy in Algeria, a multifaceted approach is proposed. Leveraging MI technologies can play a pivotal role in redefining tax policies, reassessing incentives for the agricultural sector, and fostering the widespread adoption of electronic transactions. MI-powered solutions can enhance efficiency, transparency, and compliance in economic activities, contributing to a more regulated and inclusive formal economy.

\subsubsection{ Government } In 2019, Algeria organized a workshop on MI. Participants in a workshop for the preparation of the national strategic plan for  MI 2020-2030 recommended, in Constantine, the development of a white paper on this technology to establish economic intelligence in the country. The creation of a white paper on MI serves as a roadmap to determine the appropriate mechanisms for the introduction of this technology into various socio-economic sectors, aiming to achieve economic intelligence, according to the consensus of the 150+ Algerian MI experts, including 30 researchers working abroad, gathered at the National School of Biotechnology at Salah Boubnider Constantine University. Subsequently, a 100-page white paper was presented to the public in 2021. Algeria has adopted a national strategy on research and innovation in MI, dedicated to improving Algerians' skills in MI through education, training, and research, and exploiting the potential of MI as a development tool in key socio-economic sectors (e.g. education, health, transport, energy). 
In the 2021-2022 academic year, Algeria inaugurated the National School for Artificial Intelligence  (ENSIA)  \cite{refAlgeria4}  to admit high school graduates interested in this field. The school is tailored to educate engineers in the theories and applications of MI and data science. Students will learn to develop and publish practical and innovative solutions for challenges in sectors such as health, energy, agriculture, and transportation, thereby contributing to the country's scientific and economic advancement. The House of Artificial Intelligence  \cite{refAlgeria3} comprises 13 universities nationwide and supervises 40 MI research projects. The Ministry of Higher Education and Scientific Research,  announced 2023 as "The Year of MI", on January 10th at The National School of Artificial Intelligence (ENSIA).

\subsection{Egypt}
Egypt has undergone a transformative journey in deploying MI  from 2000 to 2023, marked by the innovative endeavors of companies like NileCode, AT-Instruments, and FastAutomate. NileCode positions itself as a comprehensive tech solutions provider, committed to delivering lasting results. In the context of automotive cybersecurity, AT-Instruments introduces an anomaly detection system with MI algorithms, addressing threats in the expanding attack surface. FastAutomate revolutionizes SMB hiring with "digimates," MI candidates trained through job shadowing. Agriculture sees a paradigm shift through the "Hudhud" smart assistant, utilizing MI for disease management, land identification, and insights into informal markets. Government initiatives, such as the National Council for MI and the National MI Strategy, underscore Egypt's strategic focus on MI for sustainable development. Collaborations between the Ministry of Tourism and Antiquities and the Atomic Energy Authority highlight the pioneering use of MI in cultural preservation, reflecting Egypt's commitment to technological innovation across diverse sectors.

\begin{table}[htb]
  \begin{center}
    \begin{tabular}{|p{0.6in}p{0.6in}p{1.4in}|}
 \hline
      Egypt& 2020-2023 & Concrete Actions \\
      \hline
      Research & $\checkmark$ & E.J. AI \\
      SMB & $\checkmark$ & NileCode, FastAutomate, Synapse Analytics, Intixel, DevisionX, Dileny Technology, MerQ, WideBot, Aphrie, PasHakeem\\
      Informal Economy & $\checkmark$  & Tourism, Hudhud\\
      Government & $\checkmark$ & 2019: National MI strategy \\
      \hline
    \end{tabular}\label{tab:mytableegypt}
  \end{center}
  \caption{MI in Egypt}
\end{table}

\subsubsection{Research} The Egyptian Journal of Artificial Intelligence (E.J.AI) is a biannual refereed journal issued by the Faculty of Artificial Intelligence - Kafrelsheikh University, which publishes original and state of the art research and developments in the field of MI and related sciences. Areas of interest may include, but not limited to, machine learning, deep learning, information retrieval, computer vision, intelligent machines, robotics, Internet of Things, wireless sensor networks, cloud computing, computer science, hardware implementation, image processing and video processing (see  \cite{refegypt2}).
In the rapidly evolving landscape of education, the integration of MI has emerged as a transformative force. The scientific research work  \cite{refegypt3} presents  MI applications in educational settings, specifically focusing on the acceptance and adoption of these cutting-edge tools. As classrooms evolve into technologically advanced hubs of learning, understanding the dynamics that influence educators and students' willingness to embrace MI becomes imperative. The study employs rigorous methodologies, including the Unified Theory of Acceptance and Use of Technology, to navigate the interplay between factors such as performance expectancy, effort expectancy, and social influence. By shedding light on the nuanced relationship between MI technology and its users in education, their research aims to pave the way for informed strategies that harness the full potential of MI in shaping the future of learning in Egypt. The work in  \cite{refegypt3} investigates the acceptance of applying chat-bot technology and related MI technologies, among higher education students in Egypt. Chat-bot, as an MI technology, has garnered significant attention, especially in the education sector. Before implementing such new technology, it is vital to understand the determinants that affect students' behavioral intentions to accept or reject this technology in higher education. To comprehend this behavioral intention, the current research applied the Unified Theory of Acceptance and Use of Technology, excluding two moderators from the original model - experience and voluntariness of use. Additionally, their  research work excluded facilitating conditions and behavior use, focusing solely on the intention behavior of students. The research study also  examined the role of demographic factors (gender and age) in influencing the independent variables of the research model and the behavioral intention variable. The research outlined the study's objectives, including developing a framework for the acceptance of chat-bot technology on the behavioral intention of students in higher education in Egypt. To achieve these goals, the researcher collected data on the required variables through a questionnaire targeting students at the Arab Academy for Science and Technology and Maritime Transport (AASTMT). AASTMT was chosen due to being one of the oldest private universities in Egypt that implements MI technology in its educational system. The final sample comprised 385 responses, and data analysis involved testing, descriptive analysis, correlations, regression, and structural equation modeling. Their Results indicated a significant impact of performance expectancy, effort expectancy, and social influence on students' behavioral intention to accept chat-bot technology in their higher education in Egypt. Moreover, the results revealed no moderating role of demographic factors (gender and age) in the relationship between performance expectancy, effort expectancy, social influence, and behavioral intention.
The study  \cite{refegypt4} explores the impact of MI on the tourism industry in Egypt. MI tools, such as chatbots and personalized service recommendations, play a significant role in travel agencies, affecting various sectors within the tourism industry. The research aims to investigate the implementation of MI techniques in Egyptian tourism companies and assess employees' perceptions of using MI tools in tourism operations. Utilizing a quantitative approach, the researchers distributed an online questionnaire to tourism companies, with 320 valid responses subjected to statistical analysis. The results highlight notable differences among tourism companies offering full services when it comes to applying MI tools in their operations. Additionally, the size of tourism companies plays a role, with larger and medium-scale enterprises employing MI techniques more than their smaller counterparts. The study identifies two primary employee perspectives on MI: enthusiasm for its advantages and suspicion regarding its disadvantages. From a managerial perspective, the research sheds light on applied MI techniques in tourism and underscores the importance of implementation. This insight can assist managers in formulating policies and strategies to enhance technological infrastructure, skills, and the application of beneficial MI tools, ultimately improving performance and saving both time and money.
 

\subsubsection{ Small Businesses} In the dynamic landscape of small and medium businesses  in Egypt, NileCode emerges as a seasoned Technology House, boasting a remarkable 15-year journey in crafting seamless digital experiences. Their extensive team of tech experts, well-versed in diverse fields and grounded in technology consulting, positions NileCode as a comprehensive 360° provider of authentic tech solutions. Unlike temporary fixes and short-lived prototypes, NileCode is committed to delivering tangible and lasting results for their clients. Drawing inspiration from the iconic river Nile, NileCode is on a mission to cultivate rich and flowing experiences that precisely meet the needs of their business partners. With a blend of innovation, agility, and adaptability, the NileCode team stands out for its creative prowess and dedicated focus on addressing the persistent challenges encountered by clients. In the context of automotive cybersecurity which is very important in these geographical areas, Autonomous Instruments called AT-Instruments  introduces  an innovative anomaly detection system that runs MI algorithms on GPU. As the automotive attack surface expands rapidly, AT-instruments provides a cutting-edge solution. Its anomaly detection capabilities, coupled with MI algorithms, enable real-time monitoring and swift identification of anomalies, empowering businesses to counteract potential threats, even on the day of the attack. FastAutomate, another standout player, introduces a revolutionary approach to MI in the hiring process for SMBs. With a pool of virtual MI candidates, known as "digimates," FastAutomate offers a solution that goes beyond conventional hiring practices. These Machine employees possess diverse computer skills and are trained through job shadowing. A single video demonstration equips them to perform tasks in uncertain environments faster and with higher accuracy than their human counterparts. FastAutomate's innovative approach streamlines the hiring process for SMBs, providing them with efficient and skilled Machine employees. 

Synapse Analytics is a B2B IT company in Egypt, assisting organizations in optimizing workflows through data, machine learning, and MI models. They offer a suite of solutions to drive MI adoption across entire organizations.
Intixel, a health tech company in Egypt, utilizes MI for medical image analysis, providing modules like Cardiac MRI Segmentation and Skin Cancer Detection. Their solutions empower radiologists with efficient, automated second-eye options.
DevisionX in Egypt helps businesses integrate MI and computer vision without coding. Their platform enables organizations to label, train, and deploy custom MI-vision applications seamlessly.
Dileny Technology, based in Egypt, focuses on futuristic healthcare solutions for Africans, developing MI and medical imaging systems to enhance diagnoses and treatment administration.
MerQ, a tech company in Egypt, specializes in advanced communications and customer relationship management systems for financial organizations, deploying MI-powered interactive programs.
WideBot, a B2B CRM solutions provider, assists organizations in creating personalized customer experiences through MI-powered chatbots and offers services like data training and optimization.
Aphrie, an IT company, helps businesses leverage MI for operational efficiency, offering services such as web development, mobile apps, quality control, and cloud operations.
PasHakeem, a health tech company in Egypt, leverages MI to provide integrated healthcare services for Africans, supporting telemedicine and efficient medical record-keeping.

Tribal Credit powers startup growth in emerging markets by providing corporate cards and financial solutions. Convertedin is an automation ads platform for eCommerce and online sellers. DXwand is an AI venture founded by technology \& AI experts that are passionate about building intelligent conversational AI.

Together, these forward-thinking entities exemplify the transformative impact of MI on small and medium businesses in Egypt, showcasing how technology can be harnessed to enhance experiences, address challenges, and propel businesses into a future of innovation and efficiency.


\subsubsection{ Informal Economy} In the heart of Egypt's agricultural landscape, the "Hudhud" smart assistant project is reshaping the way farmers engage with their crops. This Arabic mobile application utilizes cutting-edge MI techniques, ushering in a qualitative development in the agricultural extension system. Tailored to individual farmers' needs, crops, and potential pests, "Hudhud" delivers instantaneous and accurate guidance, propelling Egypt towards smart, modern agriculture - a cornerstone in the construction of a digital Egypt. A notable feature of "Hudhud" is its ability to empower farmers in identifying and combating crop infections swiftly. By capturing a photo of an infected plant and submitting it through the smartphone application, farmers tap into the power of MI. The app analyzes the image, identifies the disease, and provides farmers with precise instructions to halt the infection, offering a real-time solution to a pressing problem. Beyond disease management, "Hudhud" contributes to the agricultural sector's evolution by leveraging drones  images and MI to identify reclaimable lands and recommend optimal crops for each season and region. This strategic approach enhances productivity and resource allocation, aligning with the broader vision of a digitally transformed Egypt. The initiative extends to the distribution of 2 million smart farmer cards, a transformative step towards eliminating administrative corruption and ensuring the equitable distribution of subsidies. Each card, personalized with the farmer's name, identity information, and landholding numbers, contributes to the creation of an extensive database encompassing farmers and cultivated land. This digital infrastructure not only streamlines administrative processes but also paves the way for a more transparent and accountable agricultural sector. "Hudhud" stands as a beacon for smart agricultural practices, guiding farmers through every stage - from cultivation to harvest. By addressing plant diseases, pests, and offering tailored treatments, this MI-driven initiative is propelling Egypt towards a sustainable and technologically advanced agricultural future. Furthermore, MI goes beyond the agricultural fields, playing a crucial role in shedding light on the dynamics of Egypt's informal agricultural markets. The "Hudhud" project serves as more than just a smart assistant; it acts as a conduit to comprehend the nuances of the informal market. Through the application of MI techniques, it analyzes patterns, market trends, and farmer interactions. This not only enhances the efficiency of resource allocation but also provides valuable insights into the informal market, fostering a deeper understanding of the challenges and opportunities within. As Egypt advances towards a digital agricultural landscape, the role of MI in deciphering the complexities of informal markets becomes increasingly indispensable, offering a data-driven foundation for informed decision-making and sustainable growth.
\subsubsection{ Government } In November 2019, the Egyptian government formed the National Council for MI as a partnership between governmental institutions, prominent academics and practitioners from leading businesses in the field of MI. The National Council for MI is chaired by the Minister of Communications and Information Technology. The Council is in charge of outlining, implementing and governing the MI strategy in close coordination with the concerned experts and entities. See  \cite{refegypt1}.
The National MI Strategy 2020-2025 is a key priority to help Egypt achieve its sustainable development goals. It identifies the country's plans to deepen the use of MI technologies and transform the economy. It goes beyond just adopting technology, to fundamentally rethinking business models and making deep changes to reap the benefits of productivity growth and create new areas of growth.
In 2023 In Egypt, the Ministry of Tourism and Antiquities collaborates with the Atomic Energy Authority to pioneer a workshop utilizing nuclear, radiological, and MI techniques for the restoration and documentation of ancient mummies and human remains. The workshop, organized by the Department of Restoration of Mummies and Human Remains, involves experts from various institutions. The Atomic Energy Authority's advanced laboratories, equipped with cutting-edge technologies, allow non-invasive examination and analysis of human remains. The MI Division of the authority supports restorers in their tasks. This marks the first use of MI technology on mummies in Egypt, aiming not only for video preparation and facial reconstruction but also for enhancing restoration processes. The program enables restoration workers to evaluate the restoration process before commencement, using special software on laptops or iPads to photograph and plan the restoration of available bones.
\subsection{ Libya}
In Libya, various research works have explored MI applications across different domains. Ahmed Naji \& Ahmed Abu Aeshah's research delves into the adoption of e-commerce in small and medium enterprises, emphasizing the benefits of MI in marketing, finance, data capture, and employee relationships. Najeb Masoud's work focuses on predicting movements in the Libyan Stock Market using a machine neural network  model, showcasing its accuracy in forecasting daily stock market prices. Ali AA Alarbi, Dani Strickland, and Richard Blanchard explore MI-enabled demand side management to intelligently address load shedding in a segment of Libya's grid system, aiming to make it imperceptible to consumers. Ibrahem Alsharif, Hamza Emhemed Hebrisha, and Abdussalam Ali Ahmed contribute to the enhancement of quality of life through MI, emphasizing the role of Smart Cities in addressing urbanization challenges and promoting sustainable development in the southern region of Libya. These research endeavors highlight the diverse applications of MI in Libya, spanning e-commerce, stock market prediction, energy management, and the creation of more livable urban environments. Despite the challenges, these initiatives exemplify Libya's commitment to leveraging MI for technological advancement and societal well-being.

\begin{table}[htb]
  \begin{center}
    \begin{tabular}{|p{0.6in}p{0.6in}p{1.4in}|}
 \hline
      Libya & 2020-2023 & Concrete Actions \\
      \hline
      Research & $\checkmark$ &  stock market\\
      \hline
    \end{tabular}\label{tab:mytablelibya}
  \end{center}
  \caption{MI in Libya}
\end{table}

\subsubsection{ Research}  Several research works have explored MI applications in Libya.
 \cite{refLibya1} Ahmed Naji \& Ahmed Abu Aeshah investigates MI applications in the adoption of e-commerce in small and medium enterprises (SMEs) in Libya, as presented in the Journal of University Studies for Inclusive Research. The focus is on MI in Libyan SMEs amid globalization and recent technological revolutions. The research emphasizes the integration of technology into various businesses, particularly small and medium-sized enterprises, providing them with advanced technical means for operations and digital transformation. MI proves beneficial for SMEs in marketing, finance, data capture, employee relationships, and other business domains. The study aims to examine how MI expands business operations in SMEs, exploring variables influencing its acceptance and the reality of MI and e-commerce in developing countries, with a specific focus on Libya.
 \cite{refLibya2} Najeb Masoud's research in the British Journal of Economics, Management \& Trade presents techniques and indicators of an Machine neural network model for predicting the movements of the daily Libyan Stock Market (LSM) index. The study covers the period from {\it January 2, 2007, to March 28, 2013}, using data from the emerging market of the Libyan Stock Market as a case study. Twelve technical indicators serve as inputs for the proposed models, and the forecasting ability of the ANN model is assessed using metrics like MAE, RMSE, MAPE, and R2. The results indicate that the ANN model accurately predicts movement direction with an average prediction rate of 91\%, showcasing its effectiveness in forecasting daily stock market prices. The study concludes that Machine neural networks can serve as a superior alternative technique for predicting daily stock market prices based on the strong relationship observed between parameter combinations and forecast accuracy measures.
In  \cite{refLibya3} Ali AA Alarbi, Dani Strickland, Richard Blanchard delve into MI concepts for Demand Side Shedding Management in Libya in their research presented at the 2019 8th International Conference on Renewable Energy Research and Applications (ICRERA). Their study focuses on MI-enabled demand side management, traditionally concerned with adjusting loads to meet generation and address stability or other constraints. In situations where load surpasses generation, shedding the load becomes more common. The paper specifically examines a segment of the Libyan grid system experiencing daily load shedding and explores the application of MI concepts to intelligently manage this shedding, aiming to make it imperceptible to consumers.
In  \cite{refLibya4} Ibrahem Alsharif, Hamza Emhemed Hebrisha, Abdussalam Ali Ahmed present research on the enhancement of the quality of life using MI in their article published in the African Journal of Advanced Pure and Applied Sciences (AJAPAS). Focusing on the improvement of living standards, the research emphasizes Smart Cities - urban regions leveraging advanced technology to enhance resident quality of life and optimize city operations for sustainability. Smart Cities integrate technologies such as the Internet of Things, big data analysis, and MI for effective urban planning, transportation, energy management, and public services. The authors underline the significance of smart cities in addressing the challenges of urbanization and promoting sustainable development in the southern region of Libya, aiming to create more livable and resilient environments by enhancing healthcare services, promoting energy efficiency, reducing traffic congestion, and encouraging citizen participation in civic decision-making processes.


\subsection{Morocco}

In Morocco, MI has been explored through various research works, highlighting its potential to drive progress across sectors like agriculture, healthcare, financial services, and public services. Challenges and opportunities in utilizing data and MI have been extensively studied, emphasizing the need for decisive policy responses to address issues like network limitations, educational readiness, and data availability. The impact of MI on human rights is considered, advocating for ethical frameworks guided by principles such as transparency, equity, safety, accountability, and inclusiveness. Practical applications of MI include automating water meter data collection for sustainable water use and enhancing decision-making, as well as forecasting regional tourism demand using hybrid models that outperform traditional and MI-based methods. Educational sovereignty and challenges arising from MI are scrutinized, emphasizing the importance of a national cloud computing structure for safeguarding digital sovereignty. In industries, startups like Annarabic, SYGMA.AI, Virtual Box, and ATLAN Space showcase Morocco's innovative strides in speech recognition, insurance claim settlements, VR development, and environmental monitoring. The government actively supports MI development through initiatives like UNESCO's ethics recommendations, the establishment of MoroccoAI, and hosting the MI Summer School to nurture future MI professionals. Morocco's commitment to MI education and research is further exemplified by the Moroccan International Center for MI, fostering expertise in MI and data sciences. These endeavors collectively highlight Morocco's evolving landscape in embracing and leveraging MI for societal advancement and technological innovation.
.

\begin{table}[htb]
  \begin{center}
    \begin{tabular}{|p{0.6in}p{0.6in}p{1.4in}|}
 \hline
      Morocco & 2020-2023 & Concrete Actions \\
      \hline
      Research & $\checkmark$ & AI Movement  \\
      SMB & $\checkmark$ &   MoroccoAI, Annarabic, Sigma.AI, AgriEdge\\
      Informal Economy & $\checkmark$  &  \\
      Government & $\checkmark$ & Maison de l'intelligence artificielle Oujda , Moroccan International Center for Artificial Intelligence  \\
      \hline
    \end{tabular}\label{tab:mytablemorocco}
  \end{center}
  \caption{MI in Morocco}
\end{table}

\subsubsection{Research}  Several research works have explored MI applications in Morocco.

The authors of  \cite{refMorocco1} present the challenges and opportunities for developing the use of data and MI in Morocco. MI has the potential to drive progress, development, and democratization if governments adeptly handle the challenges. It can significantly enhance productivity growth by extending opportunities in crucial African development sectors such as agriculture, healthcare, financial services, and public services. MI holds the promise of enabling employees, entrepreneurs, and enterprises to compete globally and spearhead economic development by providing access to high-quality digital tools. However, addressing the accompanying roadblocks requires decisive policy responses. The implementation of MI will necessitate significant adjustments for workers, employers, and businesses, opening new ethical questions that demand thoughtful answers. Specific constraints in Africa, including network limitations, educational institution readiness, and the availability of digital data, further compound the ethical issues. Aggressive efforts are essential for Africa to overcome these challenges, and success will position the continent to catch up with nations that have already taken strides in MI development. Although the path ahead is intricate, the government's effectiveness will be measured by its capacity to facilitate collaboration among all stakeholders, including state and civil society, academics, industry, and national and international entities.

The work  \cite{refMorocco2} introduces the topic of MI's impact on the enjoyment of human rights and proposes initial considerations for a framework at the national level. This paper primarily comprises a literature review on the subject, accompanied by concrete recommendations for overseeing MI in Morocco. It advocates for extensive consultations among stakeholders, aiming for co-regulation that culminates in the development of an ethical code. This code, guided by a human rights-based approach, is designed to address key principles, including Transparency \& Trust, Equity, Safety, Human freedom \& autonomy, Accountability \& Justice, Dignity \& Integrity, Sustainability, and Solidarity \& Inclusiveness.

The research presented in  \cite{refMorocco3} centers on the development of a fully MI-based system for automating water meter data collection in Morocco. As the demand for water resources continues to rise, monitoring becomes crucial for the rational and sustainable use of this vital resource. Currently, water meter data collection in Morocco is predominantly performed manually once a month due to cost and time constraints. This manual approach often leads to inaccurate estimations and calculations, resulting in customer disputes over inflated invoices. The paper proposes a comprehensive MI-based system for automating water meter data collection, comprising a Recognition System (RS) and a web services platform. This framework offers a range of services for both customers and water service providers, including consumption monitoring, leak detection, visualization of water consumption, and potable water coverage on a geographic map. Additionally, it serves as a robust tool for facilitating accurate decision-making through multiple reporting services. The primary component of the RS is a Convolutional Neural Network model trained on a proposed MR-AMR (Moroccan Automatic Meter Reading) dataset, achieving an impressive accuracy of 98.70\% during the model test phase. The system underwent thorough testing and validation through experiments.
The research outlined in  \cite{refMorocco4} explores the forecasting of regional tourism demand in Morocco, comparing traditional and MI-based methods to ensemble modeling. Tourism stands as a key economic contributor to Moroccan regions, accounting for 7.1\% of the total GDP in 2019. However, this sector remains highly susceptible to external shocks such as political and social instability, currency fluctuations, natural disasters, and pandemics. To mitigate these challenges, policymakers employ various techniques to forecast tourism demand for informed decision-making. The study specifically forecasts tourist arrivals to the Marrakech-Safi region using annual data from 1999 to 2018. Three conventional approaches (ARIMA, AR, and linear regression) are contrasted with three MI-based techniques (SVR, XGBoost, and LSTM). Hybrid models, combining both conventional and MI-based methods through ensemble learning, are then developed. The results reveal that these hybrid models outperform both conventional and MI-based techniques, showcasing their ability to overcome individual method limitations.

 \cite{refMorocco5} examines Educational Sovereignty and challenges posed by MI in Morocco. The study offers a concise and focused analysis of the primary threats to educational sovereignty in Morocco in the era of MI. It sheds light on the concept of educational sovereignty within the Moroccan media and political discourse surrounding "ministries of sovereignty." The article outlines key initiatives and projects in Morocco aimed at overcoming challenges to the educational system and sovereignty in the age of MI, recognizing these systems as indispensable tools in learning and classroom practices. The study primarily delves into the impacts of using foreign languages, the proliferation of foreign schools and transnational universities, the consequences of Moroccan educators and learners extensively employing foreign EdTech, and the influence of platforms and programs by major tech companies like GAFAM (Google, Apple, Facebook, Amazon, and Microsoft). It also addresses the threats arising from the absence of a national cloud computing structure, essential for safeguarding digital sovereignty and protecting the personal data of learners and the educational system. The article contends that despite Morocco's aspirations to establish a digital ecosystem capable of preserving educational sovereignty, numerous subjective and objective obstacles still impede these plans.

The study presented in  \cite{refMorocco6} employed factor analysis, a confirmatory method designed to identify latent factors from observable variables. This method involves assigning a set of observable characteristics to each latent variable, and by setting parameters (loadings) to 0 in confirmatory factor analysis, further analysis becomes possible, allowing correlations between latent factors and, if needed, additional correlations between residual errors. This process offers a comprehensive description of hidden variables. The central question posed by the authors was: To what extent does digitalization contribute to reducing inequalities in technological acceptance in Morocco? Confirmatory factor analysis, a specialized form of structural equation modeling, was used. In this approach, a model is predefined, specifying the number of factors, potential relationships between these factors, connections between these factors and the observed variables, error terms associated with each observed variable, and possible correlations between them. The sample focused on stakeholders of the Digital Development Agency, with a selected group of 60 stakeholders, comprising 20 companies, 20 associations, and 20 cooperatives. Utilizing the Unified Theory of Acceptance and Use of Technology (UTAUT) model, the study provides estimates indicating that ease of use, quality of service, anticipated performance, and the influence of the professional body are all concepts contributing to psychological and motivational acceptance for the use of MI.

The research presented in  \cite{refMorocco7} explores the opportunities and challenges associated with integrating MI  into International Financial Reporting Standards (IFRS) within accounting systems. Focused on the context of Morocco, the paper provides insights into various aspects of finance and accounting in the country. It addresses challenges related to adoption, education, and technical expertise in Morocco, alongside examining the ethical and legal implications of MI-based accounting systems. Their article further presents into the importance of maintaining compliance with IFRS requirements and ensuring the integrity and transparency of financial data. The authors conclude that a comprehensive understanding of IFRS and the adoption of MI in accounting systems are crucial for effectively navigating the challenges and opportunities presented by these developments.

\subsubsection{ Small Businesses}  Annarabic is an MI startup in Morocco that aims to empower every Arab voice. Their focus areas include MI  Software API, Customer Satisfaction, Automatic Speech Recognition (ASR), Customer Feedback, and Natural Language Processing. The startup is actively involved in developing speech recognition systems for Arabic dialects, offering services such as audio transcription for call centers, audio intelligence for call centers, retail, and social media companies (covering sentiment analysis, keyword flagging, etc.), as well as Chatbot/Voicebot solutions to facilitate customer navigation and order processing, and subtitling and captioning for videos.
SYGMA.AI revolutionizes the insurance industry by significantly reducing the time required to settle car accident claims. With an average settlement time of less than 5 minutes, the fully automated end-to-end SAAS solution utilizes MI visual inspection to detect car damages, estimate repair costs, and manage claims based solely on mobile photos. SYGMA.AI's capabilities extend to eliminating fraud, thereby saving valuable time and money for insurance companies and their clients.
Virtual Box, located in the heart of Casablanca, is an independent multidisciplinary studio dedicated to creating and developing Virtual Reality applications, interactive maps, and immersive experiences utilizing 3D, 360 Video, and Photography/VR.
AgriEdge offers a data-driven decision support platform for crop production management. Leveraging data from satellites, drones, field sensors, weather, and market prices, it provides farmers with decision proposals related to strategic processes like irrigation, fertilization, plant disease management, and yield prediction. The platform delivers these proposals through user-friendly mobile and web applications, specifying the right place, time, and quantity.
Moroccan startup ATLAN Space has secured 1.1 million USD in Series A funding to further develop its MI system guiding unmanned aircraft on data collection and tracking missions over large geographical areas. The investment, led by Maroc Numeric Fund II, builds on the startup's initial seed funding in 2019. ATLAN Space's unique technology, recognized globally, supports governments and institutions in combating environmental crimes like illegal fishing or deforestation. Founded in 2016 by Badr Idrissi, ATLAN Space has received acclaim, including recognition from Nvidia as one of its top ten MI startups in 2018. DataPathology is a Moroccan medtech startup that provides remote pathology consultation.
 
\subsubsection{ Government } In terms of development and adoption of MI, Morocco has made some progress, mainly through the establishment of academic institutions, hosting international conferences, implementing effective training and educational programs, and setting up large-scale data centers.

In December 2018, UNESCO and the Polytechnic University Mohamed VI  organized their very first important International forum on Artificial Intelligence in Africa,  in  Bengu\'erir, Morocco. The Forum's objective was to enrich the global reflection on artificial intelligence by drawing up a complete inventory of the situation and towards an assessment of an African scale, by taking into account challenges, opportunities and issues specific to local contexts. About 150 participants representing the Member States as well as high-level public and private sector partners participated to this event.

Morocco stands out as one of the earliest countries to implement UNESCO's recommendations on MI ethics, adopted during the 41st session of UNESCO's General Conference in November 2021 in Paris.
Established in 2019, MoroccoAI is a prominent Non-Governmental Organization , spearheaded by esteemed MI scientists and researchers in Morocco and abroad. MoroccoAI core focus is on promoting MI education and fostering excellence in research and innovation, in Morocco and across the broader African landscape and beyond.
The MI Summer School offers an immersive and transformative experience, empowering participants to excel in these technologies. The  organizers, Al Akhawayn University and MoroccoAI, come together in a strategic partnership through this distinguished summer school to foster a highly proficient workforce and cultivate the next generation of MI professionals and leaders. The first MI Summer School of Morocco was held in Ifrane  Morocco from  July 17th-21st, 2023. 
MI movement, the Moroccan International Center for Artificial Intelligence is a center of excellence in MI that aims to foster the emergence of Moroccan expertise in MI and Data Sciences.

On the Knowledge Campus, Oujda, {\it Maison de l'intelligence artificielle} MIA-UMPO-Maroc is designed to better understand MI and its transformations. MI opens up new avenues for the dissemination of expert and ethical knowledge to the general public while providing opportunities to develop innovative and collaborative projects by bringing together various stakeholders in the MI ecosystem. Data, artificial intelligence, and the Internet of Things are present in our lives, reshaping our relationship with knowledge and the economy, potentially reshaping our future and transforming our societies. That's why the Mohammed Premier University, like the MIA in Sophia Antipolis, wanted to create a space where reflections and experiments related to AI focus on uniting all stakeholders around a future shaped by MI.
MIA-UMPO-Maroc aims to invigorate institutional, academic, and industrial collaboration around new technologies and their challenges. The space will be regularly animated by events on innovation and AI, allowing the public to learn, exchange, and develop knowledge. MIA-UMPO-Maroc aims to create a real dynamic of AI acculturation through public experiences and the promotion of applied research with a significant societal and economic impact.
MIA-UMPO-Maroc consists of an Ethical Watch Observatory, Exhibition Hall, Coworking Space, and Experimentation Laboratory.

\subsection{Sudan}
In Sudan, the deployment of MI  from 2000 to 2023 has been marked by impactful research initiatives and innovative applications. MI played a crucial role in assessing groundwater quality in northern Khartoum State, employing multilayer perceptron neural networks and support vector regression to model suitability for drinking. The models demonstrated efficiency, showcasing their potential in groundwater quality evaluation. Additionally, MI addressed health challenges in Gedaref State, forecasting malaria and pneumonia cases using ARIMA and Prophet models. The industry sector witnessed the emergence of ATLAN Space's MI-enabled drones for environmental conservation, contributing to efforts like combatting desertification. The SudanAI model, a collaborative venture between local researchers and global MI organizations, exemplifies Sudan's commitment to creating a cutting-edge MI solution tailored to the nuances of the Arabic language spoken in Sudan. These endeavors collectively highlight Sudan's stride towards leveraging MI for research, environmental conservation, and linguistic adaptation, showcasing a growing engagement with technological advancements to address societal challenges. 

\begin{table}[htb]
  \begin{center}
    \begin{tabular}{|p{0.6in}p{0.6in}p{1.4in}|}
 \hline
      Sudan & 2020-2023 & Concrete Actions \\
      \hline
      Research & $\checkmark$ & SudanAI \\
      SMB & $\checkmark$ &  ATLAN  \\
      Government & $\checkmark$ &  \\
      \hline
    \end{tabular}\label{tab:mytablesudan}
  \end{center}
  \caption{MI in Sudan}
\end{table}

\subsubsection{Research} 
The work in  \cite{refSudan1} investigated groundwater quality in northern Khartoum State, Sudan, utilizing MI algorithms. The authors employed multilayer perceptron  neural network and support vector regression  to assess groundwater suitability for drinking. The groundwater quality was evaluated through the groundwater quality index (GWQI), a statistical model using sub-indices and accumulation functions to reduce data dimensionality. In the first stage, GWQI was calculated based on 11 physiochemical parameters from 20 groundwater wells, indicating that most parameters exceeded World Health Organization standards, except EC and NO3-. The GWQI ranged from 21 to 396, classifying samples into excellent (75\%), good (20\%), and unsuitable (5\%) water categories. 

To overcome computational challenges, the study applied MI techniques, specifically MLP neural network and SVR models. The models were trained and validated on a dataset divided into 80\% for training and 20\% for validation. Comparison of predicted  and actual (calculated GWQI) models using MSE, RMSE, MAE, and R2 criteria demonstrated the robustness and efficiency of MLP and SVR models. Overall, groundwater quality in north Khartoum is deemed suitable for human consumption, except for BH 18, which exhibits highly mineralized water. The developed approach proves advantageous for groundwater quality evaluation and is recommended for incorporation into groundwater quality modeling.

The work in  \cite{refSudan2} centers on Endemic Diseases: A Case Study of Gedaref State in Sudan, employing MI technologies.
Smart Health, a crucial component, enhances healthcare through services like disease forecasting and early diagnosis. Although numerous machine learning algorithms support S-Health services, the optimal choice for disease forecasting remains uncertain. Gedaref State faces persistent challenges with malaria and pneumonia due to heavy rainfall. Predicting future trends in these diseases is vital for effective prevention and control. This study utilizes a time series dataset from the state's ministry of health to estimate malaria and pneumonia cases in Gedaref State, Sudan, five months later. Two forecasting methodologies, ARIMA and Prophet models, are applied, comparing their performance in predicting diseases. Data collected from January 2017 to December 2021 reveals that the ARIMA technique outperforms FB-Prophet in forecasting both malaria and pneumonia cases in Gedaref State.

\subsubsection{  Small Businesses } MI-enabled Drones for good.  ATLAN Space is one of many startups innovating with drone technology on the continent. Across Southern Africa, drones are used to protect elephants and rhinos from poaching. In Sudan, a startup wants to drop Acacia tree seeds from the sky to tackle desertification, and in South Africa, drones are used in agriculture to monitor crop health and detect disease.  
The Sudanese MI model, also known as the SudanAI, is an MI model that is specifically designed to understand and process the Arabic language spoken in Sudan, as well as capture the nuances of Sudanese dialects and cultural references. The development of the Sudanese MI model is a collaborative effort between local Sudanese researchers, engineers, and data scientists, who are working in partnership with global MI organizations to create a cutting-edge MI solution that addresses the unique needs and challenges of Sudan. 



\subsection{Tunisia}

In Tunisia, the integration of MI  has been instrumental in diverse research endeavors and strategic government initiatives from 2015 to 2023. MI empowers the Tunisian textile industry by providing a competitive edge through informed decision-making, as outlined in \cite{refTunisia3}. Additionally, the application of low-power blockchain and MI in water management exemplifies Tunisia's Industry 4.0 endeavors, enhancing services and reinforcing trust among stakeholders \cite{refTunisia4}. Research efforts address environmental challenges, such as forecasting urban solid waste using sequential MI models, emphasizing the effectiveness of LSTM and bidirectional LSTM in predicting optimal waste bin numbers \cite{refTunisia5}. On the linguistic front, Tunisia advances MI through the creation of an end-to-end speech recognition system for the Tunisian dialect, showcasing innovative applications in under-resourced languages \cite{refTunisia6}. The technology industry is booming, marked by Instadeep's acquisition, highlighting Tunisia's emergence as an entrepreneurial and innovative hub in deep tech, attracting global attention \cite{refTunisia3}. In the governmental sphere, Tunisia's commitment to reform includes incorporating MI in Public Finance Management Information Systems, enhancing transparency and accountability to rebuild public trust \cite{refTunisia1}, \cite{refTunisia2}. These collective efforts underscore Tunisia's strategic use of MI across sectors for innovation, environmental sustainability, and governance. 

\begin{table}[htb]
  \begin{center}
    \begin{tabular}{|p{0.6in}p{0.6in}p{1.4in}|}
 \hline
      Tunisia & 2020-2023 & Concrete Actions \\
      \hline
      Research & $\checkmark$ &  TunSpeech, textile, Afro-Mediterranean meeting in the field of Artificial Intelligence \\
    Industry &  & Instadeep\\
    %
SMB & $\checkmark$ &  KatYos, WARM, ScorSell, LWATN, AD'VANTAGE, Business \& AI  
   \\
      Informal Economy & $\checkmark$  & Deepera.AI  \\
      Government & $\checkmark$ & PFMIS, National MI Strategy  \\
      \hline
    \end{tabular}\label{tab:mytabletunisia}
  \end{center}
  \caption{MI in Tunisia}
\end{table}

\subsubsection{ Research}
 \cite{refTunisia3} aims to empower the Tunisian textile industry with MI. Textile holds paramount importance in Tunisia, being the leading industry in terms of employment and the number of companies. It is crucial for the country to stay abreast of the latest technologies to maintain global competitiveness. MI is identified as a technology providing a competitive advantage, aiding sector leaders in making informed business and strategic decisions. The article introduces MI in the context of the Tunisian textile industry, outlines use cases, and provides recommendations to industry stakeholders. The article maintains a deliberately high level to be accessible to a diverse audience.

 \cite{refTunisia4} focuses on low-power blockchain to study water management in Tunisia within the context of Industry 4.0. The industrial sector is evolving towards Industrial IoT (IIoT) and Industry 4.0, where blockchain technology can address limitations related to security and data reliability in IoT. The article presents a new platform integrating MI and smart contracts to monitor and track water consumption in Tunisia. The proposed solution enhances water management services, offering benefits such as consumption monitoring, traceability, security, water leak detection, and visualization of water consumption and drinking water coverage. This approach aims to strengthen trust and security among various stakeholders.
 \cite{refTunisia5} utilizes sequential MI models to forecast urban solid waste in the city of Sousse, Tunisia. Urban solid waste poses a significant environmental challenge, with waste generation linked to economic growth, industrialization, urbanization, and population growth. The article focuses on predicting solid waste generation based on monthly recorded waste amounts to determine the optimal number of waste bins. Various MI regression and classification approaches are evaluated, highlighting the effectiveness of sequential models—specifically, long short-term memory and bidirectional LSTM  - in predicting the number of waste bins compared to other methods.
 \cite{refTunisia6} focuses on Tunisian dialectal end-to-end speech recognition using deepspeech. Automatically recognizing spontaneous human speech and transcribing it into text is an increasingly important task. However, freely available models are scarce, especially for under-resourced languages and dialects, as they require extensive data to achieve high performance. The paper outlines an approach to constructing an end-to-end Tunisian dialect speech system based on deep learning. To achieve this, a Tunisian dialect paired text-speech dataset named "TunSpeech" was created. Existing Modern Standard Arabic (MSA) speech data was combined with dialectal Tunisian data, reducing the Out-Of-Vocabulary rate and improving perplexity. Conversely, the introduction of synthetic dialectal data through text-to-speech increased the Word Error Rate.
\subsubsection{ Industry} Tunisia, a  North African country, is making significant strides in the technology industry. With a rapidly evolving entrepreneurship and innovation ecosystem, Tunisia is gradually establishing itself as a leading destination for entrepreneurship and innovation, especially in deep tech. The acquisition of Instadeep, a deep tech startup founded in 2014 by Tunisians Karim Beguir and Zohra Slim, has sent shockwaves through the Tunisian entrepreneurship and innovation ecosystem. BioNTech acquired Instadeep for 362 million  euros upfront, with an additional 200 million euros contingent on future performance. Such acquisitions are pivotal milestones for developing ecosystems. Instadeep's success underscores the potential of Tunisian startups to innovate, compete globally, and attract attention from major players in the tech industry. 

IEEE AMCAI 2023 was organized by ATIA (Association Tunisienne pour l'Intelligence Artificielle) with the Financial Co-Sponsorship of IEEE Africa Council, and the Technical Co-Sponsorship of IEEE Tunisia Section, IEEE SPS Tunisia Chapter, IEEE CIS Tunisia Chapter, IEEE SMC Tunisia Chapter in December 2023, at Hammamet, Tunisia. IEEE AMCAI’2023 is the premier Afro-Mediterranean meeting in the field of Artificial Intelligence. Its purpose is to bring together researchers, engineers, and practitioners from both Mediterranean and African countries to discuss and present their research results around Artificial Intelligence and its applications.
The conference emphasizes applied and theoretical researches in MI to solve real problems in all fields, including engineering, science, education, agriculture, industry, automation and robotics, transportation, business and finance, design, medicine and biomedicine, bioinformatics, human-computer interactions, cyberspace, security, Image and Video Recognition, agriculture, etc. 

\subsubsection{ Small Businesses} Founded in 2019, ScorSell, Mobile classifieds that Inspires everyone in the world to start selling. It provides a highly scalable e-commerce platform based on new technologies such as machine intelligence , machine learning , image recognition , data science . With a very friendly user experience and attractive design, ScorSell aims to create a secure payment system based on smart contract technology with low fees. 

LWATN leads the legal tech frontier in Tunisia with its state-of-the-art MI-powered chatbot. This groundbreaking tool not only facilitates seamless conversations and delivers precise legal guidance with a clear voiceover in various languages, but also prioritizes accessibility.

AD'VANTAGE is a startup that comes up with an innovative idea which would rethink classical loyalty programs used by companies to ensure their costumers faithfulness.

WARM is an online marketplace that helps connect buyers and sellers to find the right piece of furniture and decor for their homes and offices. WARM is conceived as a smart platform that answers our current furniture consumption problems that might be a frustrating experience. Lack of online presence for firsthand brands or dealing with strangers and organizing heavy pickups for secondhand products.

KatYos connects opticians and eyewear buyers by offering a 100\% digital experience, from online fitting to corrective lens processing.
It is an optometry marketplace that allows opticians to digitize their business easily and without investment by offering an immersive, unique, safe and secure shopping experience to their customers with the help of an enhanced digital mirror, where customers can evaluate a potentially unlimited number of frames anywhere and anytime.

In Tunisia, Business \& AI innovates with MI-powered decision-making solutions to help companies improve operational processes and make informed decisions.  Save Your Wardrobe is a FashionTech startup using AI to make Fashion more digital and sustainable. Wattnow help compagnies gain actionable insights on their overall energy usage. 


\subsubsection{ Informal Economy } Deepera.AI in Tunisia develops smart solutions for investors and stock traders, incorporating MI into financial tasks and offering products like stock exchange management tools.

\subsubsection{ Government}  In 2015, the Tunisian government initiated a series of public sector reforms to enhance government operations and address citizen needs. Recognizing the necessity for continued reforms, it is advocating for the implementation of an accountable Public Finance Management Information System (PFMIS) and proposes introducing MI into the current financial system, identified as a high-risk area for corruption.
In 2016, the government outlined its vision for MI, along with other priorities, in a strategy document detailing a five-year development plan for Tunisia (2016-2020). This strategy was later complemented by the government's economic and social roadmap for 2018-2020.
The economic and social roadmap for 2018-2020 aims to expedite the reforms initiated under the earlier five-year development plan. The overarching goal of the development plan is to ensure human rights, social well-being, and economic growth in Tunisia. To formulate its National MI Strategy, the Secretary of State for Research established a Task Force in 2018 to oversee the project and a Steering Committee to develop a methodology and action plan for the strategy. To kick off this national initiative, the UNESCO Chair on Science, Technology, and Innovation Policy, in collaboration with the National Agency for Scientific Research Promotion-ANPR, will host a workshop titled “National MI Strategy: Unlocking Tunisia's capabilities potential.” The primary goal of the gathering was to share and discuss the proposed framework, methodology, and action plan put forth by the Steering Committee for designing the strategy. The event took place at ENIT, specifically in the Mokhtar Latiri Amphitheater.
Concrete MI applications in the Public Finance Management Information System (PFMIS) has conducted in Tunisia. The PFMIS comprises core subsystems that furnish the government with essential information for planning, executing, and monitoring public finances. Its scope and functionality encompass fraud detection, budget efficiency, and financial analytics.
By employing a combination of machine learning, big data, and natural language processing techniques, MI aids auditors and finance officials at the Ministry of Finance in managing the vast amounts of data essential for meeting transparency and accountability requirements in fulfilling fiduciary responsibilities to Tunisian taxpayers. This, in turn, contributes to rebuilding public trust in the government  \cite{refTunisia1},  \cite{refTunisia2}.


\section{ Southern Africa} \label{sec:southern} 

In Southern Africa, MI activities are diverse.


Botswana explores water billing, diabetic retinopathy screening, solar radiation prediction, gravel loss condition prediction, HIV/AIDS treatment, data mining, ART program success, and clinical decision support. Eswatini engages in financial inclusion, diarrhoea outbreak prediction, renewable energy, COVID-19 case prediction, invasive plant study, MI in education, maize crop yield prediction, and economic development through technology. Lesotho integrates MI in land cover mapping, carbon sequestration, soil organic carbon prediction, weather nowcasting, health insurance enrollment, education, electricity demand forecasting, legal frameworks, wage impact, industry mergers, and healthcare policy.

Namibia's MI landscape is equally dynamic, exploring MI in education, predicting Gender-Based Violence through machine learning, and examining cybersecurity practices in rural areas. Other endeavors include using MI to discriminate individual animals with tags in camera trap images and developing a smart irrigation system for efficient farming. The UNESCO Southern Africa sub-Regional Forum on Artificial Intelligence in Windhoek further emphasizes the commitment to sustainable and ethical MI use in the region.

In South Africa, Accenture collaborates with the Gordon Institute of Business Science to guide businesses through digital transformation, emphasizing the integration of MI  in Public Sector Human Resource Management and MI-based medical diagnosis for improved healthcare.

The pervasive use of MI across these regions reflects its transformative impact in various domains.

\subsection{Botswana}

\begin{table}[htb]
  \begin{center}
    \begin{tabular}{|p{0.6in}p{0.6in}p{1.4in}|}
 \hline
      Botswana & 2020-2023 & Concrete Actions \\
      \hline
      Research & $\checkmark$ &  water\\
      SMB & $\checkmark$ &    \\
      Informal Economy & $\checkmark$  &  \\
      Government & $\checkmark$ &  \\
      \hline
    \end{tabular}\label{tab:mytablebotswana}
  \end{center}
  \caption{MI in Botswana}
\end{table}

\subsubsection{Research}

\cite{refBotswana1} Gaboalapswe's research explores an MI and data analytics model to address domestic water billing crises in Botswana. The study integrates MI and Big Data Analytics for efficient household water meter reading and proposes an enterprise integration system for water utility corporations.
\cite{refBotswana2} Kgame, Wu, and Geng investigate eye care physicians' knowledge and perceptions of using MI in screening for diabetic retinopathy in Botswana. The study reveals limited awareness of MI technology among clinicians but positive attitudes toward its potential in diabetic retinopathy screening.
\cite{refBotswana3} Sampath Kumar, Prasad, Narasimhan, and Ravi discuss the application of machine neural networks for predicting solar radiation in Botswana. The study utilizes data from multiple locations and encourages the development of a significant model for estimating solar potential in the country.
\cite{refBotswana4} Oladele, Vokolkova, and Egwurube apply a Feed Forward Neural Network  to predict gravel loss condition for low volume road networks in Botswana. The FFNN model proves accurate in predicting gravel loss, providing valuable information for future maintenance interventions.
\cite{refBotswana5} Dr. Chandrasekaran focuses on clustering as an MI technique in drug resistance of HIV/AIDS patients in Botswana. The study aims to address challenges in the Messiah ARV program and improve the effectiveness of HIV/AIDS treatment.
\cite{refBotswana6} Anderson, Masizana-Katongo, and Mpoeleng explore the potential of data mining in Botswana. The paper discusses how organizations in Botswana can benefit from data mining technologies to transform raw data into valuable information for strategic decision-making.
\cite{refBotswana7} Smartson and Thabani Nyoni analyze Botswana's ART program success using machine neural networks. The study employs an ANN model to forecast upward trends in annual ART coverage from 2019 to 2023, emphasizing the importance of improving ART access.
\cite{refBotswana8} Ndlovu et al. evaluate the feasibility and acceptance of a mobile clinical decision support system (VisualDx) in Botswana. Healthcare workers express interest in using mHealth technology, highlighting the potential benefits of VisualDx in resource-constrained environments.







\subsection{Eswatini}

\begin{table}[htb]
  \begin{center}
    \begin{tabular}{|p{0.6in}p{0.6in}p{1.4in}|}
 \hline
      Eswatini & 2020-2023 & Concrete Actions \\
      \hline
      Research & $\checkmark$ &  eSwatini \\
      SMB & $\checkmark$ &    \\
      Informal Economy & $\checkmark$  &  \\
      Government & $\checkmark$ &  \\
      \hline
    \end{tabular}\label{tab:mytableeswatini}
  \end{center}
  \caption{MI in Eswatini}
\end{table}

\subsubsection{Research}

\cite{refEswatini1}) explores the application of machine learning on financial inclusion data in Eswatini, focusing on small-scale businesses. Contributes to understanding the extent of financial inclusion and the challenges faced by entrepreneurs.
\cite{refEswatini2} develops a machine learning-based surveillance system for predicting diarrhoea outbreaks in Eswatini. The study employs supervised learning models and highlights the effectiveness of Support Vector Machine (SVM) in data classification.
\cite{refEswatini3}) focuses on classifying financial inclusion datasets in Eswatini using SVM and logistic regression. Recommends attention to enhancing financial inclusion, particularly in Hhohho, Shiselweni, and Lubombo regions, with mobile money as a key indicator.
\cite{refEswatini4} presents research on renewable energy sources in Eswatini, emphasizing the need for alternative energy to supplement existing hydro power. Aims to contribute to the country's self-sufficiency in power generation.
\cite{refEswatini5} utilizes Machine neural networks to predict daily COVID-19 cases in Eswatini, providing insights into disease forecasting. The study emphasizes the importance of adhering to WHO guidelines for disease prevention and control.
\cite{refEswatini6} applies machine-learned Bayesian networks to study the invasion dynamics of the highly invasive plant, Chromolaena odorata, in Eswatini. Contributes to understanding factors influencing the species' geographic distribution patterns.
\cite{refEswatini7} proposes a theoretical framework for integrating MI into Library and Information Science  curricula at the University of Eswatini. Contributes to the ongoing discourse on incorporating MI in education.
\cite{refEswatini8} addresses the critical issue of predicting maize crop yields in Eswatini using machine learning, providing insights into how this technology can contribute to alleviating poverty and ensuring food security.
\cite{refEswatini9} highlights the work of DataNet in eSwatini, showcasing how the organization contributed to the country's economic development through innovative strategies and technology implementations, such as the National Fire and Emergency Response System.







\subsection{Lesotho}

\begin{table}[htb]
  \begin{center}
    \begin{tabular}{|p{0.6in}p{0.6in}p{1.4in}|}
 \hline
      Lesotho & 2020-2023 & Concrete Actions \\
      \hline
      Research & $\checkmark$ & Trade \\
      SMB & $\checkmark$ &  BasaliTech, Plano-OneTree, eFarmers, LibraChat, Mphalane, Keti, Nalane, Qoloqo
   \\
      Informal Economy & $\checkmark$  &  \\
      Government & $\checkmark$ &  \\
      \hline
    \end{tabular}\label{tab:mytablelesotho}
  \end{center}
  \caption{MI in Lesotho}
\end{table}

\subsubsection{Research}
The work in \cite{refLesotho1} focuses on the integration of machine learning and open-access geospatial data for land cover mapping. The authors developed a framework using 10-m resolution satellite images and machine learning techniques for land cover mapping in Lesotho. This work highlights the use of MI for precise mapping in developing countries.

In \cite{refLesotho2}, the study addresses carbon sequestration potential in Lesotho's croplands. Through the application of a quantile random forest model, the research emphasizes the role of sustainable soil management practices in replenishing carbon stocks. The AI-driven model aids in determining soil organic carbon content, showcasing the integration of MI for environmental assessment.

The paper \cite{refLesotho3} presents an innovative methodology for operational land cover mapping in Lesotho using Earth Observation data. The study utilizes machine learning techniques, specifically a Random Forest Classifier, for generating standardized annual land cover maps. This demonstrates the effective use of MI in overcoming challenges related to limited in-situ data, a common issue in developing countries.

In \cite{refLesotho4}, the focus is on predicting soil organic carbon content using hyperspectral remote sensing in a mountainous region of Lesotho. The study employs machine learning algorithms, specifically random forest regression, to accurately estimate soil organic carbon. This work showcases the utility of MI in predicting soil properties for agricultural and environmental purposes.

The work by Senekane et al. in \cite{refLesotho5} explores weather nowcasting using deep learning techniques. The study employs multilayer perceptron, Elman recurrent neural networks, and Jordan recurrent neural networks for short-term weather forecasting. The high accuracies achieved highlight the effectiveness of MI in predicting weather variations.

In \cite{refLesotho6}, the authors utilize machine learning techniques to predict health insurance enrollment and analyze the take-up of health services in sub-Saharan Africa. The study demonstrates that applying different machine learning approaches improves the identification of excluded population groups, emphasizing the role of MI in enhancing health policy research.

The work in \cite{refLesotho7} investigates students' intention to learn MI in Lesotho's secondary schools. The study identifies factors influencing students' intentions, emphasizing the impact of attitudes, confidence, self-efficacy, and subjective norms. This work underscores the importance of creating a supportive environment for MI education and highlights AI's role in shaping educational intentions.

In \cite{refLesotho8}, the study focuses on short-term electricity demand forecasting for Lesotho. Utilizing ABB Nostradamus software, the research achieves accurate day-ahead, week-ahead, and hour-ahead electricity demand forecasting results. The study recommends bilateral contracts and the use of AI-based forecasting for optimal power procurement, showcasing the significance of MI in energy planning.

The work by Ramokanate in \cite{refLesotho9} critically examines the approach under the Lesotho Electronic Transactions and Electronic Commerce Bill 2013, specifically regarding the use of electronic agents in trade and commerce. The analysis recommends considering the use of the law of agency for instances involving electronic agents. This emphasizes the need for legal frameworks to adapt to advancements in MI.

In \cite{refLesotho10}, the study explores the impact of MI on average wages in Southern Africa. Using a panel VAR approach, the research reveals a significant negative relationship between MI and average wages. The study suggests policy directions focusing on wage stabilization, income redistribution, and skill development to address the effects of MI on wages.

The article in \cite{refLesotho11} investigates knowledge retention in the cross-border mergers of Lesotho's telecommunications industry. The study identifies challenges and successes in knowledge retention during mergers, emphasizing the need for formal policies. The research underscores the role of MI in optimizing knowledge management processes during industry transformations.

The research by Katende et al. in \cite{refLesotho12} evaluates the impact of a multi-disease integrated screening and diagnostic model for COVID-19, TB, and HIV in Lesotho. The study utilizes AI-driven screening and diagnostic services, maintaining integrated testing despite disruptions. The findings highlight the positive effects of MI in ensuring timely diagnoses and linkage to healthcare services.

In \cite{refLesotho13}, the study predicts health insurance enrollment and analyzes the take-up of health services in 38 sub-Saharan African countries. Using machine learning techniques, the research identifies excluded population groups, improving targeting for health policy. This work emphasizes the role of MI in enhancing the accuracy of predictions and informing healthcare decisions.

Across these works, the common theme is the pervasive use of MI in addressing diverse challenges in Lesotho. From educational intentions and energy forecasting to legal frameworks, labor markets, industry mergers, and healthcare policy, these studies collectively demonstrate the versatile and transformative impact of MI across various domains.


\subsubsection{Small Businesses}

 BasaliTech, a Lesotho non-profit that aims to address gender parity in STEM through training such as computing, programming( web and mobile), and electronics and robotics for young girls and children.
 Plant One Tree, a  youth-empowered environmental organization that seeks to plant all trees.
eFarmers Lesotho, Agro mobile App: e-extension and virtual marketing.
Mphalane provides automated chat assistants for your business. Mphalane is a company with a product called LibraChat. LibraChat is a communication channel for every entity. The product is a 24-hour chatbot focused on increasing communication, enquiries, shopping, ordering and many other forms of services that may require communication.
Founded in 2018, KETI aims to use MI to improve healthcare outcomes in Lesotho and beyond. The startup has developed a platform that uses machine learning algorithms to analyze medical data and identify patterns that can help doctors make more accurate diagnoses and treatment decisions.
Nalane is a social venture that engages in viable social enterprising initiatives to raise resources to sustain its mission of ensuring that public education works for all in Lesotho and South Africa. Through inclusive and comprehensive after-school programs at primary school level, they especially target  Orphans and Vulnerable Children to increase education attainment rates. 
Qoloqo focuses on improving access to financial services for underserved communities. Qoloqo’s platform uses MI to analyze financial data and provide personalized recommendations to users, such as which savings accounts or loans would be the best fit for their needs. The startup has already partnered with several banks and microfinance institutions in Lesotho, and is expanding into other countries in the region






\subsection{Namibia}

\begin{table}[htb]
  \begin{center}
    \begin{tabular}{|p{0.6in}p{0.6in}p{1.4in}|}
 \hline
      Namibia & 2020-2023 & Concrete Actions \\
      \hline
      Research & $\checkmark$ & Gender-Based Violence \\
      SMB & $\checkmark$ &  Tincup, Loop, Hyperlink, AfriFeel  \\
      Government & $\checkmark$ &  Windhoek Statement  \\
      \hline
    \end{tabular}\label{tab:mytablenamibia}
  \end{center}
  \caption{MI in Namibia}
\end{table}

\subsubsection{Research}

\cite{refNamibia1} reports some experiences of teaching MI in a Namibian university in collaboration with a Finnish university and a few companies. Within the Computing Education Community, only a minority of research reports have experience teaching MI, and very little research has been conducted on teaching and learning MI in Africa. Given the high importance and impact of MI, this is alarming. Learning and teaching MI in an African higher education setting provides unique challenges compared to the standardized approach in the Global North. Our undergraduate course in MI was carried out in a novel way that emphasized the creative application of MI to meet the requirements of the Fourth Industrial Revolution (4IR). We chose an approach that helps Computer science graduates to explore and get inspired by the opportunities of MI at the ground.

Gender-Based Violence (GBV) is a pervasive issue that poses a significant threat to individuals and communities worldwide. Many countries, faces the pressing challenge of combating GBV as an impediment towards ensuring the safety and well-being of its citizens. Despite the coordinated efforts taken by several governments and its people, there remains a cause for concern regarding curbing GBV globally. Women, children, and other vulnerable members of society continue to be exposed to this significant threat. Current measures are primarily reactive, with focus on punishing perpetrators rather than foreseeing and preventing future occurrences. \cite{refNamibia2} provides a machine leaning-based analytical process for predicting the occurrence of GBV to aid in early prevention strategies. A preliminary case study of the analytical process was recommended using Namibia as one of the South African countries. External dataset of GBV were extracted from kaggle platform containing some Namibia records for preliminary study and model testing. It was recommended in this study that Namibia should employ various machine learning  algorithms as a test case. These algorithms should be compared and evaluated for their predictive accuracy in forecasting GBV events in Namibia. Additionally, the case study should conduct exploratory data analysis to identify key trends and drivers of GBV in Namibia, providing valuable insights for policymakers and intervention programs. By leveraging advanced ML techniques, the case study would contribute to evidence-based decision-making, policymaking, planning, and resource allocation aimed at reducing GBV incidents in Namibia. If the analytical process is duly followed and applied in Namibia, the research outcome would have the potentials to inform policymakers, law enforcement agencies, and social organizations about the underlying causes and risk factors associated with GBV, enabling them to implement effective intervention strategies

Globally, Information Communication Technology  device usage has seen a steep rise over the last few years. This also holds in developing countries, which have embarked on connecting the unconnected or previously disadvantaged parts of their populations. This connectivity enables people to interact with cyberspace, which brings opportunities and challenges. Opportunities such as the ability to conduct business online, attend online education, and perform online banking activities. Challenges experienced are the cost of Internet access and more worrying cyber-risks and potential for exploitation. There remain pockets of communities that experience sporadic connectivity to cyberspace, these communities tend to be more susceptible to cyber-attacks due to issues of lack/limited awareness of cyber secure practices, an existent culture that might be exploited by cybercriminals, and overall, a lackluster approach to their cyber-hygiene. The work in \cite{refNamibia3} presents a qualitative study conducted in rural Northern Namibia. Our findings indicate that both secure and insecure cybersecurity practices exist. However, through the Ubuntu and Uushiindaism Afrocentric lenses, practices such as sharing mobile devices without passwords among the community mirror community unity. Practices such as this in mainstream research can be considered insecure. We also propose interrogating “common” secure cybersecurity practices in their universality of applicability

The use of technology in ecology and conservation offers unprecedented opportunities to survey and monitor wildlife remotely, for example by using camera traps. However, such solutions typically cause challenges stemming from the big datasets gathered, such as millions of camera trap images. MI is a proven, powerful tool to automate camera trap image analyses, but this is so far largely been restricted to species identification from images. The work in \cite{refNamibia4} develops and tests an MI algorithm that allows discrimination of individual animals carrying a tag (in this case a patagial yellow tag on vultures) from a large array of camera trap images. Such a tool could assist scientists and practitioners using similar patagial tags on vultures, condors and other large birds worldwide. We show that the overall performance of such an algorithm is relatively good, with 88.9\% of all testing images (i.e. those not used for training or validation) correctly classified using a cut‐off discrimination of 0.4. Specifically, performance was high for correctly classifying images with a tag (95.2\% of all positive images correctly classified), but less so for images without a tag (87.0\% of all negative images). The correct classification of images with a tag was, however, significantly higher when the tag code was at least partly readable compared with the other cases. Overall, this study underscores the potential of MI for assisting scientists and practitioners in analysing big datasets from camera traps.

Farmers in Namibia currently operate their irrigation systems manually, and this seems to increase labor and regular attention, especially for large farms. With technological advancements, the use of automated irrigation could allow farmers to manage irrigation based on a certain crops' water requirements. The work in  \cite{refNamibia5}  looks at the design and development of a smart irrigation system using IoT. The conceptual design of the system contains monitoring stations placed across the field, equipped with soil moisture sensors and water pumps to maintain the adequate moisture level in the soil for the particular crop being farmed. The design is implemented using an Arduino microcontroller connected to a soil moisture sensor, a relay to control the water pump, as well as a GSM module to send data to a remote server. The remote server is used to represent data on the level of moisture in the soil to the farmers, based on the readings from the monitoring station


\subsubsection{Small Businesses}

AfriFeel Digital Innovations:  In today’s fast-paced world, finding the best travel deals is crucial for both leisure and business travelers. Afrifeel travel app makes it effortless to discover the most up-to-date and exclusive travel bargains.

 Hyperlink InfoSystem is well known to craft the most innovative \& eye catchy mobile apps \& websites. It offers a wide range of customized services in mobile apps, website development, AR-VR Development, Game Development, Blockchain Development and much more. 

Loop Technologies deliver exceptional experiences that drive your business forward. Floor standing machines, tablets, signage, selfservice data collection and more.

TinCup Digital: TinCup was established in order to facilitate the delivery of integrated digital and marketing services to clients in Namibia, clients who until now have had to be content with existing above-the-line strategies dictating the direction of integrated campaigns. We aim to rewrite the advertising and marketing handbook in Namibia. With decades of combined experience in the Namibian digital, marketing and advertising landscape, we offer clients a holistic view of their brands in real life, from the first point of contact through to campaign conceptualisation, development, integration and alignment with current strategies



\subsubsection{Government}

UNESCO, in collaboration with the Government of Namibia, is organized the UNESCO Southern Africa sub-Regional Forum on Artificial Intelligence, under the theme 'Towards a sustainable development-oriented and ethical use of Artificial Intelligence' in Windhoek, Namibia, on 7-9 September 2022.
This Forum was the first sub-regional Forum on MI in Africa, following the first UNESCO Forum on Artificial Intelligence in Africa in Bengu\'erir, Morocco, in December 2018 , which called for the organization of sub-regional Forums in Africa to facilitate exchanges, elaboration of strategic frameworks and action plans in line with unique sub-regional and national contexts, in view of an MI Strategy for Africa.

The Ministers and senior experts on MI from Southern African countries convened and deliberated on the development of a sustainable development-oriented and ethic use of MI in Southern Africa.

The Windhoek Statement on Artificial Intelligence in Southern Africa made several recommendations on MI in Africa.


\subsection{South Africa}

\begin{table}[htb]
  \begin{center}
    \begin{tabular}{|p{0.6in}p{0.6in}p{1.4in}|}
 \hline
      SA & 2020-2023 & Concrete Actions \\
      \hline
      Research & $\checkmark$ &  Vulavula\\
     Industry & $\checkmark $ &  DataProphet \\
      SMB & $\checkmark$ &   Lelapa AI, Lesan,  Xineoh, Clevva, Aerobotics, The Gearsh, Credo, Akiba Digital, Bridgement, AfricAi , Neurozone, Congretype, Openbanking
\\
      Government & $\checkmark$ & AIISA:  Artificial Intelligence Institute Of South Africa  \\
      \hline
    \end{tabular}\label{tab:mytablesa}
  \end{center}
  \caption{MI in South Africa}
\end{table}

\subsubsection{Research}

The world around us is changing at a furious pace. It's left many established businesses shaken, with executives questioning their organisations' longevity. To participate in a digital future, business transformation is critical and increasingly urgent. A strategic approach is essential. To help businesses chart the way forward, Accenture has partnered with the Gordon Institute of Business Science (GIBS) to provide insight into digital technologies and the future of business \cite{refSA1} .

 The Fourth Industrial Revolution has transformed modern society by ushering in the fusion of advances in robotics, the Internet of Things, genetic engineering, quantum computing, and MI among others. MI brings a range of different technologies and applications to interact with environments that comprise both the relevant objects and the interaction rules and have the capacity to process information in a way that resembles intelligent behaviour. Similarly, artificial intelligence is also being used in the human resources management processes and functions in the public sector to map sequences to actions. The study explores the opportunities, challenges, and future prospects of integrating MI and Public Sector Human Resource Management in South Africa's public sector. The study in \cite{refSA2}  examines the dynamics surrounding the adoption, implementation and operationalisation of the 4IR in the management of human resources in the SA public sector in this unfolding dispensation. Data was collected using the extensive review of written records such as books, journal articles, book chapters among others which were selected for use in this study. Data was analysed using content and thematic analysis techniques. The study established that Artificial Intelligence is beneficial in the sense that it can improve public service delivery in South Africa as the HRM personnel is enabled to focus more on the strategic areas of management by taking over routine tasks, and that it helps minimize bias in public service recruitment and selection. In contrast, research on potential challenges has revealed that combining MI and Public Sector Human Resource Management may pose a threat to white-collar jobs. This study may lead to practical applications of MI to support the HR functions of public sector entities in SA. The public managers are better informed about the impediments, gaps and opportunities that may arise from using MI in managing human resources in SA's public sector. This study contributes to the body of knowledge as it unpacks and informs the dynamics associated with the implementation of MI in managing human resources in public sector entities.

Historically, South Africa has battled numerous chronic diseases such as Cancer, Diabetes, and Tuberculosis. Even though significant efforts are made in the medical diagnostic industry to detect and treat these chronic diseases, these efforts have fallen short due to their higher diagnostic costs, shortage in infrastructure, equipment, and highly skilled technicians at the required time, resulting in reduced access to healthcare for patients. In recent years, the field of MI-based medical diagnosis has gained prominence because of its low cost, less infrastructure, equipment, and technician requirements. In addition, MI-based medical diagnosis reduces diagnostic time with a significantly high level of accuracy. The work in \cite{refSA3}  conducts a systematic literature review of 32 collated MI articles. We present our findings, and scope the tools, techniques, and algorithms from a South African Context. The scope of this literature review involves (1) conducting an attribute analysis of literature that includes studying disease, temporal, and spatial aspects of literature as well as stages in developing MI-based medical diagnosis tool; (2) conducting a conceptual analysis of literature that includes studying applications, algorithms, techniques, and performance measures related to different MI stages; and (3) scoping the insights from the literature from a South African context that involves proposing a framework for developing MI-based medical diagnostic tool, hardware and software requirements, and deployment strategies into underdeveloped medical diagnostic provinces of South Africa.

The work in \cite{refSAllm1} focuses on adapting pretrained ASR models to low-resource clinical speech using epistemic uncertainty-based data selection. In \cite{refSAllm2}, the authors propose FonMTL, a multitask learning approach for the Fon language. \cite{refSAllm3} introduces pretrained vision models for predicting high-risk breast cancer stage. AfriNames, discussed in \cite{refSAllm4}, addresses challenges in ASR models for African names. \cite{refSAllm5} presents GFlowOut, a model incorporating dropout with generative flow networks. AfriQA, outlined in \cite{refSAllm6}, focuses on cross-lingual open-retrieval question answering for African languages. AfriSpeech-200, as described in \cite{refSAllm7}, introduces a pan-African accented speech dataset for ASR. \cite{refSAllm8} investigates low-compute methods for low-resource African languages. MasakhaNEWS, discussed in \cite{refSAllm9}, involves news topic classification for African languages.
The work in \cite{refSAllm10} focuses on MasakhaPOS, a part-of-speech tagging system for typologically diverse African languages. \cite{refSAllm11} investigates language representation in multilingual models. PuoBERTa, introduced in \cite{refSAllm12}, involves the training and evaluation of a curated language model for Setswana. MphayaNER, as discussed in \cite{refSAllm13}, addresses named entity recognition for Tshivenda. \cite{refSAllm14} explores fine-tuning multilingual pretrained African language models. Unsupervised cross-lingual word embedding representation for English-isiZulu is presented in \cite{refSAllm15}. Consultative engagement of stakeholders toward a roadmap for African language technologies is detailed in \cite{refSAllm16}. The study in \cite{refSAllm17} evaluates the performance of large language models on African languages. 

\subsubsection{Industry}

Based in South Africa, DataProphet accelerates your smart factory journey with an integrated suite of AI-powered technology. These tools seamlessly integrate your machine data with existing infrastructure, streaming high-quality data for consistent, real-time access. The intuitive and secure platform for data visualization and KPI tracking across your plant facilitates a smooth transition to high-impact data-to-value decision-making. Prescriptive AI guides you in optimizing control plans and benchmarking manufacturing performance, enhancing efficiency and returns. DataProphet's state-of-the-art MI tools seamlessly combine data orchestration, insightful contextualization, and digital-era continuous improvement. Deep learning algorithms interpret complex process interactions, prescribing optimal recipes and precise steps for implementation. Plants witness significant improvements in quality, production, and sustainability KPIs, achieving peak production efficiency.
\subsubsection{Small Businesses}

Lelapa AI: At the research level, A new venture called Lelapa AI, is trying to use machine learning to create tools that specifically work for Africans. Vulavula, a new MI tool that Lelapa released, converts voice to text and detects names of people and places in written text (which could be useful for summarizing a document or searching for someone online). It can currently identify four languages spoken in South Africa - isiZulu, Afrikaans, Sesotho, and English - and the team is working to include other languages from across Africa.  The tool can be used on its own or integrated into existing MI tools like online conversational chatbots. 
Lesan: Lesan, a Berlin-based AI startup that is developing translation tools for Ethiopian languages. 
AJALA is a  London-based startup that provides voice automation for African languages.  
 Lelapa AI and Lesan are just two of the startups developing speech recognition tools for African languages. In February, Lelapa AI raised 2.5 million USD in seed funding, and the company plans for the next funding round in 2025. But African entrepreneurs say they face major hurdles, including lack of funding, limited access to investors, and difficulties in training MI to learn diverse African languages. 
 
 In South Africa, The Gearsh is a mobile app connecting artists with fans, providing a platform for bookings and learning. Credo is a Peer 2 Peer Lending Platform, while Akiba Digital focuses on data and technology SAAS solutions.
Bridgement, based in South Africa, helps small businesses optimize cash flow through its online platform offering instant advances on outstanding invoices.
The AfricAi Project introduces DanAi, an MI-powered chatbot designed to revolutionize daily tasks with contextual knowledge for each African country. Neurozone, a neuroscience company in South Africa, optimizes brain/body systems for wellness and high performance, assembling expertise across various fields for a comprehensive Model of Brain Performance.
Congretype Pty. Ltd primarily focuses on providing societal-based solutions in renewable energy, ICT for development, and Climate-Smart Agriculture.
Trusource: A new secure, safe, and revolutionary way for consumers to share financial information, allowing them to move, manage, and make more money with real-time payments and credit approvals. All done on one platform powered by Trusource.

Xineoh is a company based in Sandton, South Africa, that has created a platform capable of predicting consumer behavior using machine intelligence. For instance, VideoLlama is an implementation of Xineoh's consumer behavior prediction platform. It utilizes the platform to recommend movies and TV shows to users. The algorithm suggests films or TV shows to viewers one at a time. Based on user feedback on each suggestion being good, average, or bad, it learns the personal tastes of each viewer and recommends movies and shows they are in the mood for at that moment.
Clevva, a South African technology company, specializes in decision navigation. It assists staff, customers, and digital workers through desired decision journeys, ensuring they consistently achieve the right outcomes in a contextually appropriate and compliant manner.
Aerobotics, another South African company, uses machine intelligence and drones to help farmers manage their farms, trees, and fruits. Its technology tracks and assesses the health of these crops, identifying when trees are sick, monitoring pests and diseases, and conducting analyses for better yield management. The company has advanced its technology, providing farmers with independent and reliable yield estimates and harvest schedules by collecting and processing images of trees and fruits from citrus producers early in the season. In turn, farmers can prepare their inventory, forecast demand, and ensure the best product quality for their customers.



\subsubsection{Government}

The Machine Intelligence Institute of Africa (MIIA) is an African non-profit organization founded in 2016. MIIA aims to transform and help build an MI-powered Africa through a strong, innovative and collaborative Machine Intelligence, MI and Data Science community, consisting of individuals and key players in the African Artificial Intelligence Ecosystem. MIIA's growing network consists of stakeholders in the African MI Ecosystem, including thousands of members as well as key decision-makers in NGOs, NPOs, academia, businesses, and the public sector. 
On 30 November 2022, The Department Of Communication \& Digital Technologies launched the Artificial Intelligence Institute Of South Africa (AIISA). On The Same Day, The University Of Johannesburg launched the First AI Hub Of AIISA. AIISA will focus on three sectors of the economy: the fourth industrial revolution in manufacturing, healthcare, agriculture and food processing.

The Centre for Artificial Intelligence Research (CAIR) is a South African national research network that conducts foundational, directed and applied research into various aspects of Artificial Intelligence. CAIR has nodes at five South African universities: the University of Cape Town, University of KwaZulu-Natal, North-West University, University of Pretoria and Stellenbosch University. CAIR was founded  in November 2011 as a joint research centre between the University of KwaZulu-Natal and the Council for Scientific and Industrial Research (CSIR). In 2015 CAIR expanded to four other South African universities with the CSIR playing a coordinating role. CAIR is primarily funded by the Department of Science and Technology (DST), as part of the implementation of South Africa's ICT Research, Development and Innovation (RDI) Roadmap. Advances in ICT supported by the RDI Roadmap aim to guide South Africa to a state of digital advantage that will strengthen economic competitiveness and enable an enhanced quality of life for all South Africans.



\section{ Middle Africa} \label{sec:central} 

In Central Africa, In Angola, there's a focus on seismic inversion, forest fire detection, urban expansion monitoring, and bioavailable isoscapes. In Cameroon, MI is making significant strides in various sectors. In healthcare, MI is employed for infant mortality rate analysis, tuberculosis detection through chest radiography interpretation, HIV management using medical imaging, and smartphone-based MI classifiers for cervical cancer screening. The country also sees agricultural innovation with MI-driven Agri-FinTech solutions and crop disease diagnosis systems. Small businesses leverage MI for e-commerce enhancement, crop and soil monitoring, and skill development platforms. Additionally, there's a government initiative establishing a MI training center to digitize the economy.
The Central African Republic focuses on IoT-based smart agriculture, MI predicting electricity mix, and machine learning aiding in primate vocalization classification. The country is spearheading the development of a smart city, Renaissance City, with MI consulting playing a pivotal role. The government leads a crypto project, the Sango Initiative, introducing a digital-first economy backed by blockchain. The Sango platform further facilitates e-visas, digital company registration, and tokenization of natural resources.
In Chad, MI is employed for fertility rate forecasting, conflict risk projections, hate speech detection in social media, and predicting hotspots of food insecurity. The country's entrepreneurial landscape sees the emergence of an edtech solution, Genoskul, providing online training courses. Tech hub WenakLabs supports startups, offering incubation, mentorship, and financing opportunities. Chad is also actively working on a National Cybersecurity Strategy to combat cyber threats.
In the Democratic Republic of the Congo, MI is employed for gully erosion assessment using machine learning methods like Random Forest and MaxEnt. Additionally, machine learning and computer vision aid in property tax roll creation, addressing challenges in resource-constrained environments. In the realm of small businesses, KivuGreen leverages MI to enhance climate resilience for farmers, showcasing the positive impact on agriculture and livelihoods.
Equatorial Guinea utilizes MI in researching sea level variability, demonstrating a critical application in climate change mitigation. The country explores MI for economic diversification, recognizing its potential in reducing vulnerability to oil and gas price fluctuations.
Gabon's MI applications range from mapping land cover using cloud computing and machine learning to monitoring coastal erosion through convolutional neural networks. Deep learning frameworks are also employed for forest height estimation, contributing to forest resource monitoring.
The Republic of the Congo stands out with the establishment of the African Research Centre on MI. This pioneering center focuses on advancing research and digital technology, offering education and skills development to promote Africa's integration and inclusive economic growth.
Sao Tom\'e and Principe showcases innovative MI applications in social protection targeting through satellite data and value chain analysis in the agricultural sector. These applications address challenges related to poverty mapping, program eligibility, and sustainable agricultural practices.
 Central Africa demonstrates a growing embrace of MI across various sectors, showcasing both research advancements and practical applications that contribute to addressing local challenges and fostering economic development. Overall, MI is playing a transformative role across research, industry, small businesses, and government initiatives in Central Africa.

\subsection{Angola}
% ...

\begin{table}[htb]
  \begin{center}
    \begin{tabular}{|p{0.6in}p{0.6in}p{1.4in}|}
 \hline
      Angola & 2020-2023 & Concrete Actions \\
      \hline
      Research & $\checkmark$ &  cancer\\
      \hline
    \end{tabular}\label{tab:mytablean}
  \end{center}
  \caption{MI in Angola}
\end{table}

\subsubsection{Research}

\cite{refAngola1} discusses Deep Learning Seismic Inversion as a case study from offshore Angola. The deep learning seismic inversion results showed improved accuracy in identifying reservoir properties, surpassing traditional seismic inversion approaches.
\cite{refAngola2} explores the use of GNSS reflectometry and machine learning to detect fire disturbances in forests in Angola. The research successfully applies different machine learning techniques to identify burned areas, providing valuable insights for monitoring forest disturbances.
\cite{refAngola3} focuses on Luanda, Angola. Their paper integrates geographical information systems, remote sensing, and machine learning to monitor urban expansion. The study evaluates the effectiveness of index-based classification and employs unsupervised machine learning algorithms, revealing insights into the urban growth of Luanda.

\cite{refAngola4} presents a bioavailable strontium isoscape of Angola, crucial for understanding the transatlantic slave trade. The study uses a machine learning framework to create the first bioavailable 87Sr/86Sr map for Angola, aiding in identifying the geographic origin of enslaved individuals and contributing to archaeological and forensic studies.
In \cite{refAngola5} Zolana R Joao's research focuses on a "Road Construction Assessment Model (RC-AM)" designed to prevent contract overbilling in Angola. The study explores the use of satellite imagery and machine learning to identify road layers and measure road lengths, providing triggers for road inspections and addressing the issue of poor road quality.
\cite{refAngola6} Investigates stored maize in Angola, this study identifies insects and fungi associated with maize under various storage conditions. The findings contribute to improving food security in Angola by understanding the pests affecting stored maize in different provinces, offering insights into pest management strategies.

In \cite{refAngola7} Geraldo AR Ramos and Lateef Akanji apply MI for technical screening of enhanced oil recovery (EOR) methods. The study utilizes a five-layered feedforward backpropagation algorithm, incorporating fuzzy logic reasoning and neural networks to screen and predict suitable EOR techniques for oilfields.
\cite{refAngola8}. In this work, Geraldo AR Ramos and Lateef Akanji employ a neuro-fuzzy simulation study to screen candidate reservoirs for enhanced oil recovery  projects in Angolan oilfields. The study combines fuzzy logic and neural network techniques to assess the suitability of EOR techniques based on rock and fluid data.
\cite{refAngola9} presents a case study on "3D Water Saturation Estimation Using AVO Inversion Output and Machine Neural Network" in offshore Angola. The study utilizes high-resolution seismic data and neural network techniques to estimate water saturation in the reservoir, providing insights for hydrocarbon prospection and development.

%
%
%
%
%
%
%

\subsection{Cameroon}

\begin{table}[htb]
  \begin{center}
    \begin{tabular}{|p{0.6in}p{0.6in}p{1.4in}|}
 \hline
      Cameroon & 2020-2023 & Concrete Actions \\
      \hline
      Research & $\checkmark$ &  Healthcare\\
      SMB & $\checkmark$ & Agrix Tech, Comparoshop,  eFarm, KMER MI, DASTUDY\\
      Government & $\checkmark$ &  AI training center\\
      \hline
    \end{tabular}\label{tab:mytableyyy}
  \end{center}
  \caption{MI in Cameroon}
\end{table}

\subsubsection{Research}

In a recent study \cite{RefCameroon1}, researchers utilized artificial Neural Networks to analyze the infant mortality rate in Cameroon. The data, spanning {\it from 1960 to 2020}, with projections reaching 2030, demonstrated the model's stability in forecasting. According to the model (12, 12, 1), the predicted Infant Mortality Rate (IMR) is approximately 48 per 1000 live births annually. This suggests a need for government action, including strengthening primary healthcare, enhancing immunization coverage, and providing training for health workers to reduce infant mortality. 

In a study \cite{RefCameroon2}, Deep Learning  neural networks were employed to interpret chest radiography (CXR) for screening and triaging pulmonary tuberculosis (TB). This retrospective evaluation assessed three DL systems (CAD4TB, Lunit INSIGHT, and qXR) for detecting TB-associated abnormalities in chest radiographs from outpatients in Nepal and Cameroon. All 1196 individuals underwent Xpert MTB/RIF assay and CXR readings by radiologists and DL systems, with Xpert as the reference standard. DL systems showed higher specificities compared to radiologists, potentially reducing Xpert MTB/RIF tests by 66\% while maintaining sensitivity at 95\% or better. However, performance variations across sites emphasize the importance of selecting scores based on the screened population. DL systems are valuable for TB programs with limited human resources and available automated technology

 In a study \cite{RefCameroon3}, the focus is on combating the spread of HIV using Medical Imaging. With the global challenge posed by HIV, there's a hot debate on applying MI to manage the disease. Over 6,000 people are newly diagnosed with HIV annually in the United States, where an estimated 1.2 million have the disease. Ongoing medical care for HIV-positive individuals is crucial, and recent studies suggest that MI can enhance the speed and accuracy of HIV detection. For example, a Northwestern University study used a deep learning algorithm to identify HIV-positive individuals with 94.2\% accuracy, outperforming manual diagnosis. MI, especially MI, is also improving patient-specific treatment strategies, as demonstrated by a University of California, San Francisco study. Tailored treatment regimens based on machine learning algorithms proved more accurate than conventional methods. MI is further utilized to monitor the virus's development in HIV-positive individuals, as seen in a Stanford University study. With faster and more accurate diagnoses, personalized treatment plans, and enhanced virus monitoring, MI has the potential to revolutionize HIV control, offering more effective and efficient treatment for patients.

In a study \cite{RefCameroon4}, the focus is on a protocol for a two-site clinical trial validating a smartphone-based MI classifier for identifying cervical precancer and cancer in HPV-positive women in Cameroon. Cervical cancer remains a significant public health challenge in low- and middle-income countries (LMICs) due to financial and logistical issues. The WHO recommends HPV testing as the primary screening method in LMICs, followed by visual inspection with acetic acid (VIA) and treatment. However, VIA is subjective and dependent on the healthcare provider's experience. To enhance accuracy, our study aims to assess the performance of a smartphone-based Automated VIA Classifier (AVC) using MI to distinguish precancerous and cancerous lesions from normal cervical tissue. The AVC study is nested in the ongoing cervical cancer screening program "3T-study" (Test, Triage, and Treat), involving HPV self-sampling, VIA triage, and treatment if needed. After applying acetic acid on the cervix, precancerous and cancerous cells whiten more rapidly than non-cancerous ones, and their whiteness persists stronger over time.


\subsubsection{Small Businesses}

Agrix Tech, headquartered in Yaound\'e, Cameroon, is an Agri-FinTech company dedicated to assisting small-scale farmers in transitioning from subsistence farming to commercial ventures, aiming to maximize their profits. The company provides farmers with a comprehensive package, including financing, farm inputs, expert advice, insurance, and, whenever possible, access to markets. Leveraging machine learning and satellite data, Agrix Tech facilitates improved credit decisions, while automated operations ensure cost-effectiveness and scalability.

AllGreen startup introduces ROSALIE, an MI system designed to diagnose crop diseases in real-time with accuracy. Beyond diagnosis, ROSALIE also offers notifications to guide owners on adopting effective practices for eradicating these diseases, contributing to improved crop health and yields.

Comparoshop operates as a SaaS toolkit utilizing MI to enhance e-commerce in Africa. The platform equips web merchants with solutions, enabling them to set up a sales site in just 5 minutes, automate product catalog updates, enhance user experience, increase shopping cart conversions, and stay updated on real-time market trends and competition.

eFarm, a web and mobile platform for agricultural products, incorporates MI to optimize crop and soil monitoring, enhancing agriculture's efficiency and efficacy. This innovative approach, developed during TechStars Start-up Weekend, aims to significantly increase productivity, playing a crucial role in the global fight against hunger and facilitating access to markets for farmers worldwide.

DASTUDY, an MI-type platform, is a local innovation developed by Cameroonian entrepreneur David Kenfack. This platform enables users to acquire new skills, share content, receive answers based on local socio-cultural realities, and seek assistance with daily tasks such as administrative writing, analysis, and translation. DASTUDY serves as a knowledge-sharing platform between learners and professionals, facilitating the exchange of documents, exercises, and software in academic, professional, and cultural fields. Additionally, it features a virtual assistant named TSAF, powered by Generative MI, designed by this  Cameroonian startup. The platform addresses the need to enhance local skills and provide visibility to content producers in the local community.

KMER AI is a platform dedicated to promoting MI and showcasing innovative technologies and research in Cameroon. Our goal is to educate students on the fundamentals of MI and provide a space for Cameroonian researchers, students, and industries to share their work and interests in the field. We also aim to create an opportunity for the relatively few Cameroonians involved in the field to collaborate, network, and communicate, motivating and encouraging others to join this journey and address challenges in our own context. By uniting theoretical and applied machine learning techniques in cutting-edge areas such as health, agriculture, climate change, marketing, education, transportation, linguistics, and art, we aspire to generate impactful ideas. These ideas will inspire students and companies to harness the power of MI, contributing to the acceleration of Cameroon's emergence.



\subsubsection{Government}

In 2019, Cameroon's first MI training center has been established through a partnership between the state-owned telecommunications operator  Camtel and the university of Yaounde I.
The training facility will be hosted by the University's National Advanced School of Engineering in Yaounde, which already houses a high-tech 3D printing center. The project, valued at FCFA 1.3 billion, covers the construction of learning infrastructure, procurement of training equipment, training of trainers, curriculum design, and learner scholarships, aiming to be fully operational within two and a half years.
As part of the country's digitization strategy, there's a target to multiply the number of direct and indirect jobs in ICT by 50. The establishment of the training resource aligns with the government's vision to digitize the economy and develop relevant skill sets. The center plans to run research and development postgraduate programs with market-related portfolios, intending to improve the employability of job-seekers through transformative education and training systems. The goal is to meet the knowledge, competencies, skills, research, innovation, and creativity needs required to nurture the future of the MI sector. The unnamed manager stated that they envision training one hundred individuals initially, with twenty-five percent of pioneer students benefiting from fully-funded scholarships offered by the telecommunications operator and its partners, as well as the university through Polytech. The project aims to position Cameroon among the leaders in MI training on the continent.


\subsection{Central African Republic}
% ...

\begin{table}[htb]
  \begin{center}
    \begin{tabular}{|p{0.6in}p{0.6in}p{1.4in}|}
 \hline
    CAR & 2020-2023 & Concrete Actions \\
      \hline
      Research & $\checkmark$ &  electricity \\
      SMB & $\checkmark$ &    \\
      Informal Economy & $\checkmark$  &  \\
      Government & $\checkmark$ & Web3 Sango, National Bitcoin Treasury  \\
      \hline
    \end{tabular}\label{tab:mytablecar}
  \end{center}
  \caption{MI in Central African Republic}
\end{table}

\subsubsection{Research}

Desertification poses a threat to the Central African Republic's economy heavily reliant on agriculture, especially in arid zones with low rainfall. To address this, \cite{RefCAR1} introduces a smart agriculture solution using the Internet of Things. The system, based on a Raspberry Pi 3 B+, employs temperature and soil moisture sensors, a submersible water pump, and a relay. TV Whitespace (TVWS) is the inactive or unused space found between channels actively used in UHF and VHF spectrum. TVWS frequency spans from 470 MHz - 790 MHz.
Using a TVWS network, farmers access a web interface on their smartphones to monitor field temperature and soil moisture, remotely activating or deactivating irrigation. This solution aims to enhance economic resilience and promote sustainable agriculture in arid regions.
In \cite{RefCAR2}, a machine-learning model is developed to predict Africa's electricity mix based on planned power plants and their success likelihood. This addresses the limitations of energy scenarios, offering insights into the risks associated with planned power-generation projects. By utilizing a large dataset and country-level characteristics, the model accurately predicts project outcomes. Contrary to some rapid transition scenarios, the study suggests that non-hydro renewables in Africa's electricity generation may remain below 10\% in 2030, highlighting potential carbon lock-in risks.
\cite{RefCAR3} explores the automatic classification of primate vocalizations in Central Africa using deep learning. The study compares various neural network architectures and employs data augmentation techniques to address the small training dataset. The best model, a standard 10-layer CNN, achieves high accuracy in classifying primate species. The results showcase the effectiveness of augmentations and training tricks, providing insights into improving the classification of primate vocalizations.
In the context of 12 years of politico-military crises in the Central African Republic, \cite{RefCAR4} proposes algorithms for remediating lost school and university years. These algorithms are implemented using an API Framework Rasa chatbot, considering the social imbalance in education caused by conflicts and worsened by the Covid-19 pandemic. The chatbot offers guidance on normal, professional, and specific training, considering the learner's province and the Covid-19-induced social distancing measures.
\cite{RefCAR5} focuses on the application of machine learning for diagnosing 18 common pediatric diseases in the Central African Republic. The study addresses the healthcare challenges in the region, developing a diagnostic system based on decision tree, random forest, and neural network models. Results indicate high diagnostic accuracy, showcasing the potential of MI in improving healthcare access and quality in developing countries.


\subsubsection{Small Businesses}

Renaissance City is set to become the pioneering smart city in the heart of Africa, situated in the Central African Republic, ushering in new business and community prospects. WebRobot, the official MI and Big-Data consulting company for this ambitious endeavor, will play a key role in shaping a sustainable, innovative, technological, and economic hub. In collaboration with the Mistahou Financial Group, this project aims to propel Renaissance City to the forefront of smart city development in Africa.
 


\subsubsection{Government}

The Sango Initiative, initiated by the Central African Republic National Assembly and endorsed by the President, is a crypto project aimed at fostering a digital-first, blockchain-based economy. Project Sango, launched on July 15, 2022, introduces the Sango coin, representing the country's move towards a digital future. This national digital currency, backed by Bitcoin, forms the core of a new digital monetary system, including a "Digital National Bank" and "National Bitcoin Treasury." The government's collaboration with the private sector mirrors a public-private partnership. Expanding the Sango blockchain project, the Central African Republic has introduced tokenization of its land and natural resources as of July 2023.
The E-visa, facilitated through the Web3 Sango platform, is a digitally issued visa streamlining entry into the Central African Republic for specific purposes. Unlike traditional visas, the e-visa application, payment, and approval occur digitally with the Sango platform's integrated wallet feature.
Digital company registration involves the official incorporation of businesses through the Web3 Sango platform powered by blockchain. This process, payable in cryptocurrency, allows submitting necessary documents to government authorities for establishing a legal entity for business in the Central African Republic.
In blockchain technology, tokenization represents the digital representation of ownership and rights over the Central African Republic's natural resources on the Sango blockchain. This approach enhances management, monitoring, and international investment opportunities in the country's resources.
The Sango tokenization platform offers various agricultural and forest plots across the country, enabling investors to concession and engage in agriculture or forest exploitation. Investors can further tokenize their concessions to attract additional investment.


\subsection{Chad}
% ...
\subsubsection{Research}

\begin{table}[htb]
  \begin{center}
    \begin{tabular}{|p{0.6in}p{0.6in}p{1.4in}|}
 \hline
      Chad & 2020-2023 & Concrete Actions \\
      \hline
      Research & $\checkmark$ &  lingua franca \\
      SMB & $\checkmark$ & Genoskul, WenakLabs, DaTchad, ZereSoft   \\
      Informal Economy & $\checkmark$  &  \\
      Government & $\checkmark$ & National Cybersecurity Strategy  \\
      \hline
    \end{tabular}\label{tab:mytablechad}
  \end{center}
  \caption{MI in Chad}
\end{table}

\cite{RefChad2} employs machine learning algorithms to analyze remote sensing and ground-truth Lake Chad's level data. The study addresses environmental challenges faced by Lake Chad due to climate change and anthropogenic activities. Results reveal associations between climate variables, remote sensing, and ground-truth lake levels. Random Forest Regression outperforms other models, emphasizing soil temperature's role in remote sensing lake level fluctuations. The study provides insights for integrated water management in the Lake Chad basin, considering climate change, vulnerability, human activities, and water balance. \cite{RefChad1} forecasts the Total Fertility Rate  in Chad using a machine learning approach. The study applies an Artificial Neural Network  to analyze TFR data from 1960 to 2018, with out-of-sample forecasting for 2019-2030. Model evaluation criteria indicate its stability, predicting an increase in annual total fertility rates in Chad. Recommendations include boosting demand for family planning services, enhancing accessibility to sexual and reproductive health services, and empowering women through education, labor participation, and promoting women's rights.

\cite{RefChad3} projects armed conflict risk in Africa towards 2050 using a machine learning approach. Utilizing the CoPro ML framework, the study explores sub-national conflict risk for different socio-economic and climate scenarios. Results align with socio-economic storylines, projecting conflict intensification in severe scenarios. The study identifies hydro-climatic indicators' limited but contextual role in conflict drivers. Challenges include inconsistent data availability, but ML models present a viable approach for armed conflict risk projections, informing climate security policy-making.
\cite{RefChad4} Forecasts conflict in Africa using automated machine learning systems. The ViEWS competition focuses on predicting changes in the level of state-based violence for the next six months at the PRIO-GRID and country level. The study explores combinations of autoML systems and limited datasets, emphasizing the endogenous nature of conflict. Key findings include the improved predictive performance of autoML and the superiority of the Dynamics model. The Dynamics model, utilizing limited data related to state-based violence and its spatial-temporal structure, won the ViEWS competition for "predictive accuracy" at the PGM level.
\cite{RefChad5} Predicts Hotspots of Food Insecurity in Chad. The paper addresses the persistent issue of food insecurity, examining factors such as poverty, conflicts, and climate contributing to the lack of access to nutritious food. By forecasting hotspot regions for food insecurity, the study aims to provide program managers with lead time to organize and coordinate efforts. Chad, with high hunger rates and poverty exacerbated by environmental degradation, desertification, and conflict, is the focus. The paper classifies food insecurity based on the Cadre Harmonis\'e phases, offering policy recommendations.
\cite{RefChad6} Examines Measurements and determinants of extreme multidimensional energy poverty using machine learning. The study calculates the depth, intensity, and degrees of energy poverty in developing countries, revealing widespread 'severe' energy poverty across multiple dimensions. Machine learning identifies the most influential socioeconomic determinants of extreme multidimensional energy poverty, including household wealth, house size and ownership, marital status of the main breadwinner, and residence of the main breadwinner. The findings have policy significance for addressing severe energy poverty through incentives, resource allocation, and special assistance.
\cite{RefChad7} Studies Delineation of Groundwater Potential Zones in the Eastern Lake Chad Basin using ensemble tree supervised classification methods. The paper employs machine learning to map groundwater potential in crystalline domains. Twenty classifiers are trained on 488 boreholes and excavated wells, with random forest and extra trees classifiers outperforming others. Relevant explanatory variables include fracture density, slope, SAR coherence, topographic wetness index, basement depth, distance to channels, and slope aspect. The study emphasizes the importance of using a large number of classification algorithms, the impact of performance metrics on variable relevance, and the contribution of seasonal variations in satellite images to groundwater potential mapping.
The work in \cite{RefChad8,RefChad8v2} examines the Detection of Hate Speech Texts Using Machine Learning Algorithm. The article focuses on identifying hate speech in social media, particularly in "lingua franca," a mix of local Chadian and French languages. The dataset consists of 14,000 comments from popular Facebook pages, categorized as hate, offense, insult, or neutral. Natural Language Processing techniques clean the data, and three word embedding methods (Word2Vec, Doc2Vec, Fasttext) are applied. Four machine learning methods (LR, SVM, RF, KNN) classify different categories, with the FastText-SVM combination achieving 95.4\% accuracy in predicting comments containing insults.
\cite{RefChad9} examines a Machine Learning-Based Model of Boko Haram. The book presents a computational modeling effort to understand Boko Haram's behavior, gathering data from 2009 to 2016. Predictive models generate forecasts of Boko Haram attacks every month, allowing real-world testing of accuracy. The book introduces Temporal Probabilistic (TP) rules to explain predictions and enhance understanding of Boko Haram's behaviors for counter-terrorism analysts, law enforcement, policymakers, and diplomats.


\subsubsection{Small Businesses}

Genoskul is an edtech solution developed by a Chadian start-up, providing access to online training courses and tutors. It features a smart assistant to deliver relevant answers to users' queries. Through its Android app, users can register using an email or phone number, gaining access to services like virtual classrooms for discussions with fellow learners. These virtual rooms connect learners from diverse backgrounds for intellectual exchange, supervised by qualified teachers to prepare effectively for national and international secondary and higher education exams and competitions. Genoskul offers courses spanning various professions, from loincloth shoe making and shea butter processing to rabbit breeding, public management, sustainable development, and civic action. In support of its growth, Genoskul secured CFAF 5 million (approximately \$8,149) in funding and received backing from Chad Innovation, an incubator that provided the start-up with a stand at Gitex Africa 2023 in Marrakech, Morocco.

In the emerging entrepreneurial landscape of Chad, tech hub WenakLabs stands at the forefront of innovation and creativity. Located in N'Djamena, the hub offers incubation, mentorship, and financing opportunities to startups, facilitating the transformation of ideas into thriving companies.
Founded in 2014 through collaboration between Chadian bloggers, the French platform Mondoblog, local network JerryClan Tchad, and tech entrepreneur Abdelsalam Safi, who currently serves as the CEO of WenakLabs. Over the years, the hub has closely partnered with local and international entities such as AfriLabs, Oxfam, the French Institute, Moov Africa, Sahel Innov, and UNICEF. This collaboration aims to provide resources and tools supporting Chadian entrepreneurs in launching and growing their businesses.
WenakLabs encourages youth entrepreneurship by offering a fab lab and a media lab. The fab lab serves as an open-access digital production space, providing digitally controlled machines to the public for designing and creating physical objects collaboratively. It is open to young individuals seeking to acquire skills for quickly transforming their ideas into physical prototypes.
The media lab provides an innovative communication environment for incubatees and project leaders, utilizing new technologies to process, visualize, and share information. It serves as a platform to disseminate reliable and engaging information, inspiring communities to take action.
With approximately 120 startups incubated and 70 projects developed in Chad, WenakLabs boasts achievements such as ZereSoft, a platform modernizing agriculture and the rural world with 2.0 tools, DaTchad, a data journalism agency project, and Nomad Learning, an SMS-based learning platform.
The incubator actively organizes programs and events, including the Startup Weekend N'Djamena and DENE MAGIC, a 2022 initiative aiming to provide digital skills to women for better access to decent jobs. WenakLabs also provides advice, training, and coaching to socioeconomic and digital development actors in the country.



\subsubsection{Government}

Chad is on the path to establishing a National Cybersecurity Strategy. The Ministry of Telecommunications and Digital Economy, along with the National Agency for Computer Security and Electronic Certification, launched the development of this strategy on December 14, with the Telecommunications Minister in attendance. Developed in partnership with the International Telecommunication Union, the future National Cybersecurity Strategy aims to find means to better combat all forms of cyber attacks.
"It is important to assess the stakes related to cybersecurity to define and prioritize responses to establish a strategy capable of providing greater digital security to all structures. To strengthen the regulations, the government has decided to make significant progress in implementing the National Cybersecurity Strategy, which has lagged behind for some years. In 2019, a gathering involving participants from 32 national and regional institutions took place in the country. One of the resolutions from the discussions was to accelerate the process of developing the national cybersecurity strategy in Chad.
In February 2022, Chad also hosted cybersecurity experts from various countries and the sub-region to discuss issues related to evaluation methodology, strategic cybersecurity policy, online commerce, banking, legal and regulatory frameworks, and technology standards.
In December 2022, Chad accelerated its efforts to strengthen cybersecurity. On December 5, two bills were adopted to enhance the country's cybersecurity: the first ratifies Ordinance No. 007/PCMT/2022 of August 31, 2022, regarding cybercrime and cyber defense, and the second ratifies Ordinance No. 008/PCMT/2022 of August 31, 2022, regarding cybersecurity.


\subsection{Democratic Republic of the Congo}
% ...

\begin{table}[htb]
  \begin{center}
    \begin{tabular}{|p{0.6in}p{0.6in}p{1.4in}|}
 \hline
    DRC & 2020-2023 & Concrete Actions \\
      \hline
      Research & $\checkmark$ &  \\
      SMB & $\checkmark$ &  KivuGreen   \\
      Informal Economy & $\checkmark$  &  \\
      Government & $\checkmark$ &  e-health, e-learning\\
      \hline
    \end{tabular}\label{tab:mytabledrc}
  \end{center}
  \caption{MI in Democratic Republic of the Congo}
\end{table}

\subsubsection{Research}

\cite{refDRC1} considers four machine learning methods to examine gully erosion in Democratic Republic of the Congo.
Soil erosion by gullying causes severe soil degradation, resulting in profound socio-economic and environmental damages in tropical and subtropical regions. To mitigate these adverse effects and ensure sustainable natural resource management, preventing gullies is imperative. Effective gully management strategies begin with devising appropriate assessment tools and identifying driving factors and control measures. Machine learning methods play a crucial role in identifying these driving factors for implementing site-specific control measures. Their study aimed to assess the effectiveness of four machine learning methods (Random Forest (RF), Maximum Entropy (MaxEnt), Artificial Neural Network, and Boosted Regression Tree (BRT)) in identifying gully-driving factors and predicting gully erosion susceptibility in the Luzinzi watershed, Walungu territory, eastern Democratic Republic of the Congo. Gullies were initially identified through field surveys and digitized using a high-resolution image from Google Earth. Of the 270 identified gullies, 70\% (189) were randomly selected for training the machine learning methods with topographical, hydrological, and environmental factors. The remaining 30\% (81 gullies) were used for testing the methods. They have used the area under the receiver operating characteristic (AUROC) method.
Results showed that RF and MaxEnt algorithms outperformed other methods, with RF (AUROC = 0.82) and MaxEnt (AUROC = 0.804) exhibiting higher prediction accuracies than BRT (AUROC = 0.69) and ANN (AUROC = 0.55). TSS results indicated that RF and MaxEnt were the best methods in predicting gully susceptibility in Luzinzi watershed. Factors like Digital Elevation Model, Normalized Difference Water Index, Normalized Difference Vegetation Index, slope, distance to roads, distance to rivers, and Stream Power Index played key roles in gully occurrence. Considering these factors is crucial for policymakers to develop strategies to reduce the risk of gully occurrence and related consequences at the watershed scale in eastern DRC.

Developing countries often face financial constraints in providing public goods. Property taxation is seen as a promising local revenue source due to its efficiency, ability to capture real estate value growth, and potential for progressivity. However, ineffective property tax collection is common in many low-income countries, often due to missing or incomplete property tax rolls.
In a large Congolese city, the work in \cite{refDRC2} employs machine learning and computer vision models to build a property tax roll. Training the algorithm on 1,654 randomly selected properties assessed during government land surveyors' in-person visits, along with property characteristics from administrative data or extracted from photographs, we achieve promising results. The best machine learning algorithm, trained on administrative data, achieves a cross-validated R2 of 60\%, with 22\% of predicted values within 20\% of the target value. Computer vision algorithms, relying on property picture features, perform less effectively, with only 9\% of predicted values within 20\% of the target value for the best algorithm. These findings suggest that even in contexts with limited property information, simple machine learning methods can assist in constructing a property tax roll, particularly when the government can only collect a small number of property values through in-person visits.


\subsubsection{Small Businesses}

In North Kivu, a province of the Democratic Republic of the Congo, many people earn their livelihood as small-scale farmers. However, climate change is making farming increasingly challenging due to unpredictable weather patterns. This not only threatens their income but also their food security. KivuGreen, a youth-led enterprise, is enhancing the climate resilience of small-scale farmers in the Democratic Republic of the Congo through a mobile service that provides real-time forecasts and climate-smart agricultural advice. Building the adaptive capacity of these farmers faced challenges, such as limited internet connectivity in rural areas and the prevalence of older mobile phones. To overcome these hurdles,  KivuGreen's service is made accessible via SMS, eliminating the need for expensive technology or an internet connection. The impact has been significant, with small-scale farmers increasing their agricultural yield by 40\% and their income by 30\%.
In addition to enhancing food security,  the increased income allows farmers to invest in their children's education, healthcare, sanitary facilities, and non-polluting energy sources. By boosting the agricultural yield and income of small farmers, KivuGreen contributes to creating more job opportunities for young people in North Kivu.



\subsubsection{Government}

In 2022, the government of Democratic Republic of Congo organized a forum  under the theme "Artificial intelligence (AI), myth or reality". According to the authorities of this Central African country, the capital Kinshasa populated by 17 million souls  will soon be a smart city thanks to the implementation of the “Smart City” project launched in 2019.

In low-income countries (such as the Democratic Republic of the Congo) a specific challenge for public health professionals is how to accurately distinguish malaria symptoms from other febrile illnesses with similar characteristics. In remote rural areas, microscopic diagnosis is slow, inaccurate and done in non-specialized laboratories; lab technicians have difficulty reading smear results to rapidly determine the species and stage of plasmodium, putting healthcare professionals and their patients in these areas at a particular disadvantage.
Through a web connection the system will support physicians and healthcare professionals in any location [in the country] rapidly identify malaria type and severity for individual patients and so prescribe the optimal treatment

A Prototype web-based platform and tools in the Democratic Republic of the Congo to connect rural health care professionals with malaria experts.
The e-health platform will also have a growing library of e-learning resources on malaria – including articles, reports, videos, documentation and quizzes – that healthcare professionals can consult to deepen their knowledge of diagnosis and treatment.

\subsection{Equatorial Guinea}
% ...

\begin{table}[htb]
  \begin{center}
    \begin{tabular}{|p{0.6in}p{0.6in}p{1.4in}|}
 \hline
      Equatorial Guinea & 2020-2023 & Concrete Actions \\
      \hline
      Research & $\checkmark$ &  \\
      SMB & $\checkmark$ &    \\
      Informal Economy & $\checkmark$  &  \\
      Government & $\checkmark$ &  \\
      \hline
    \end{tabular}\label{tab:mytableeg}
  \end{center}
  \caption{MI in Equatorial Guinea}
\end{table}

\subsubsection{Research}

\cite{refEG1} investigates sea level variability and predictions using artificial neural networks and other machine learning techniques in the Gulf of Guinea. The rising sea level due to climate change poses a critical threat, particularly affecting vulnerable low-lying coastal areas such as the Gulf of Guinea (GoG). This impact necessitates precise sea level prediction models to guide planning and mitigation efforts for safeguarding coastal communities and ecosystems. The study presents a comprehensive analysis of mean sea level anomaly (MSLA) trends in the GoG between 1993 and 2020, covering three distinct periods (1993–2002, 2003–2012, and 2013–2020). It investigates connections between interannual sea level variability and large-scale oceanic and atmospheric forcings. Additionally, the performance of artificial neural networks (LSTM and MLPR) and other machine learning techniques (MLR, GBM, and RFR) is evaluated to optimize sea level predictions. The findings reveal a consistent rise in MSLA linear trends across the basin, particularly pronounced in the north, with a total linear trend of 88 mm/year over the entire period. The highest decadal trend (38.7 mm/year) emerged during 2013–2020, and the most substantial percentage increment (100\%) occurred in 2003–2012. Spatial variation in decadal sea-level trends was influenced by subbasin physical forcings. Strong interannual signals in the spatial sea level distribution were identified, linked to large-scale oceanic and atmospheric phenomena. Seasonal variations in sea level trends are attributed to seasonal changes in the forcing factors. Model evaluation indicates RFR and GBR as accurate methods, reproducing interannual sea level patterns with 97\% and 96\% accuracy, respectively. These findings contribute essential insights for effective coastal management and climate adaptation strategies in the GoG.
\cite{refEG2} explores the spatial ecology and conservation of leatherback turtles (Dermochelys coriacea) nesting in Bioko, Equatorial Guinea. Bioko Island (Equatorial Guinea) hosts essential nesting habitat for leatherback sea turtles, with the main nesting beaches found on the island's southern end. Nest monitoring and protection have been ongoing for more than two decades, although distribution and habitat range at sea remain to be determined. This study uses satellite telemetry to describe the movements of female leatherback turtles (n = 10) during and following the breeding season, tracking them to presumed offshore foraging habitats in the south Atlantic Ocean. Leatherback turtles spent 100\% of their time during the breeding period within the Exclusive Economic Zone  of Equatorial Guinea, with a core distribution focused on the south of Bioko Island extending up to 10 km from the coast. During this period, turtles spent less than 10\% of time within the existing protected area. Extending the border of this area by 3 km offshore would lead to a greater than threefold increase in coverage of turtle distribution (29.8 ± 19.0\% of time), while an expansion to 15 km offshore would provide spatial coverage for more than 50\% of tracking time. Post-nesting movements traversed the territorial waters of Sao Tom\'e and Principe (6.4\% of tracking time), Brazil (0.85\%), Ascension (1.8\%), and Saint Helena (0.75\%). The majority (70\%) of tracking time was spent in areas beyond national jurisdiction (i.e. the High Seas). This study reveals that conservation benefits could be achieved by expanding existing protected areas stretching from the Bioko coastal zone, and suggests shared migratory routes and foraging space between the Bioko population and other leatherback turtle rookeries in this region.

\cite{refEG3} explores knowledge about Fang Traditional Medicine: an informal health-seeking behavior for medical or cultural afflictions in Equatorial Guinea. This study delves into a range of informal health-seeking behaviors, including the use of Fang Traditional Medicine (FTM) for medical or cultural afflictions in Equatorial Guinea (EQ). The research covers therapeutic methods, health problems addressed, the learning process, traditional medicine user profiles, and the social images of Fang Traditional Healers (FTHs). Ethnography was employed as a qualitative strategy using emic-etic approaches. Semi-structured interviews were conducted with 45 individuals, including 6 community leaders, 19 tribal elders, 7 healthcare professionals, 11 FTHs, and 2 relatives of traditional healers in 5 districts of EQ. FTM offers a cure for malaria and treatments for reproductive health issues, bone fractures, and cultural illnesses. Several methods used to learn FTM are based on empirical observation, without the need for traditional schooling. For example, watching a family member or the spirits/ancestors can reveal healing knowledge. Materials from forests, including tree barks and plants, and rituals are used to keep Fang populations healthy. In addition, two rituals known as 'osuin' and 'etoak' (infusions of tree barks with the blood of sacrificed animals) are the most commonly used treatments. Elders and women are the most active consumers of FTM, playing a relevant role in curing medical and cultural afflictions in Fang communities. The informal health-seeking behavior among the Fang community is conditioned by the explanation model of illness.





\subsubsection{Government}

Equatorial Guinea's economic landscape, predominantly reliant on the oil and gas sector, presents a vulnerability due to its susceptibility to fluctuating global prices and the finite nature of these resources. Embracing MI technologies offers a strategic avenue for economic diversification. By investing in research and development in MI, Equatorial Guinea can foster innovation across various sectors.

\subsection{Gabon}
% ...

\begin{table}[htb]
  \begin{center}
    \begin{tabular}{|p{0.6in}p{0.6in}p{1.4in}|}
 \hline
      Gabon & 2020-2023 & Concrete Actions \\
      \hline
      Research & $\checkmark$ &  Grand Libreville\\
   Industry &  $\checkmark$ & Oil and Gas \\
      \hline
    \end{tabular}\label{tab:mytablegabon}
  \end{center}
  \caption{MI in Gabon}
\end{table}

\subsubsection{Research}
\cite{refGabon1} examines cloud computing and machine learning in support of country-level land cover and ecosystem extent mapping in Liberia and Gabon. Liberia and Gabon joined the Gaborone Declaration for Sustainability in Africa, established in 2012, with the goal of incorporating the value of nature into national decision-making by estimating the multiple services obtained from ecosystems using the natural capital accounting framework. In this study, we produced 30-m resolution, 10-class land cover maps for the 2015 epoch for Liberia and Gabon using the Google Earth Engine cloud platform to support the ongoing natural capital accounting efforts in these nations. We propose an integrated method of pixel-based classification using Landsat 8 data, the Random Forest classifier, and ancillary data to produce high-quality land cover products for a broad range of applications, including natural capital accounting. Our approach focuses on a pre-classification filtering (Masking Phase) based on spectral signature and ancillary data to reduce the number of pixels prone to be misclassified, thereby increasing the quality of the final product. The proposed approach yields an overall accuracy of 83\% and 81\% for Liberia and Gabon, respectively, outperforming prior land cover products for these countries in both thematic content and accuracy. Their approach is   replicable and was able to produce high-quality land cover products to fill an observational gap in up-to-date land cover data at the national scale for Liberia and Gabon.
\cite{refGabon2} studies the Grand Libreville, Gabon coastline using machine learning and convolutional neural network detection and automatic extraction methods. Coastal erosion, worsened by climate change and natural occurrences like droughts and marine flooding, is a major problem. Countries situated along coastlines face significant challenges in preserving their land and protecting their people and assets. To mitigate the damage caused by sea encroachment on land, effective monitoring tools and methods are required. This study uses Object-oriented Analysis (OBIA), Pixel-Oriented Analysis (PBIA), and Convolutional Neural Network  methods to automatically detect and extract the Greater Libreville coastline based on Pl\'eiades very high-resolution satellite images dating from 2022. Three test areas were chosen and extracted, showing competitive Overall Accuracy values with the OBIA method and the CNN model. However, the OBIA method using the Random Forest algorithm achieved the highest accuracy rates, reaching 95\%, 90\%, and 80\% for the three test areas, respectively.
\cite{refGabon3} utilizes a deep learning framework for the estimation of forest height from bistatic TanDEM-X data. Up-to-date canopy height model  estimates are crucial for forest resource monitoring and disturbance analysis. This study presents a deep learning approach for the regression of forest height from TanDEM-X bistatic interferometric synthetic aperture radar (InSAR) data. The proposed fully convolutional neural network framework is trained using reference CHM measurements derived from the LiDAR LVIS airborne sensor from NASA, acquired during the joint NASA-ESA 2016 AfriSAR campaign over five sites in Gabon, Africa. The DL model achieves an overall performance of 1.46-m mean error, 4.2-m mean absolute error, and 15.06\% mean absolute percentage error when tested on all considered sites. Additionally, a spatial transfer analysis provides insights into the generalization capability of the network when trained and tested on datasets acquired over different locations and types of tropical vegetation. The results are promising and align with state-of-the-art methods based on both physical-based modeling and data-driven approaches, with the advantage of requiring only one single TanDEM-X acquisition at inference time.

\subsubsection{Industry} Neural networks have been used in the oil and gas industry in  \cite{refGabonty1,refGabonty2,refGabonty3,refGabonty4,refGabonty5,refGabonty6}.






\subsection{Republic of the Congo}
% ...

\begin{table}[htb]
  \begin{center}
    \begin{tabular}{|p{0.6in}p{0.6in}p{1.4in}|}
 \hline
     Congo & 2020-2023 & Concrete Actions \\
      \hline
      Research & $\checkmark$ &  \\
      SMB & $\checkmark$ &    \\
      Informal Economy & $\checkmark$  &  \\
      Government & $\checkmark$ &  African Research Centre on artificial intelligence\\
      \hline
    \end{tabular}\label{tab:mytablecongo}
  \end{center}
  \caption{MI in Congo}
\end{table}






\subsubsection{Government}

On February 24, 2022, the Economic Commission for Africa (ECA) and the Government of the Republic of Congo inaugurated a groundbreaking center devoted exclusively to advancing research through MI, aiming to propel digital technology in Africa across areas such as digital policy, infrastructure, finance, skills, digital platforms, and entrepreneurship. The African Research Centre on artificial intelligence, funded through the ECA and other partners, will provide the necessary technology education and skills to promote Africa's integration, contributing to generating inclusive economic growth, stimulating job creation, breaking the digital divide, and eradicating poverty for the continent's socio-economic development, ensuring Africa's ownership of modern tools of digital management.
The center was officially launched by the UN Under-Secretary-General and Executive Secretary of the ECA, and the Prime Minister of Congo under the auspices of President Denis Sassou Nguesso. The event was attended by African ministers responsible for ICT and the digital economy.
With the full support of the Government of the Republic of Congo, the Center, the first of its kind in Africa, will serve as a regional hub for the development of emerging technologies in the region. A partnership agreement to develop the project was signed in March 2021 by the Republic of the Congo and ECA during the official opening ceremony of the 7th session of the African Regional Forum on Sustainable Development. UN partners include the United Nations Industrial Development Organization (UNIDO), UNESCO, the International Telecommunications Union, Alibaba Jack Ma Foundation, and other key ECA partners.
Congo will function as a regional MI hub across the continent, providing access to the deepest and highest quality pool of MI talent. Aligned with Agenda 2063, the MI Centre introduces a new dynamic to Africa's participation in the global value chain. Global companies choosing to locate in regional hubs can benefit from strong government support, low business costs, and access to world-class MI clusters.
The MI Centre envisions working collaboratively with multiple stakeholders to establish linkages for a collaborative environment between industry, institutions, government, public and private sectors. The strategic pillars of the center include the provision of state-of-the-art MI research facilities, collaboration with top-ranked universities in Africa, building a network of highly skilled researchers, and providing support and training to citizens to become scholars, researchers, leaders, and engaged individuals required to deliver digital transformation in society.
The African Research Centre on MI is now established in the DENIS SASSOU-N'GUESSO University of Kint\'el\'e, recognized as a platform for business analysis on the continent. The MI Centre offers hybrid modes of training in MI and robotics for researchers, youths, and interested citizens. It also provides basic MI and Robotics skill-oriented training for talented elementary and senior school students. The pursuit of a Master's of Science Degree in MI and data science in collaboration with the University of Denis Sassou Nguesso is available at the MI Centre.
With the aim of increasing the number of STEAM (Science, Technology, Engineering, and Mathematics) graduates and joining the ranks of the world's highest educated workforce, the depth and quality of the MI Centre's program content are remarkable. As part of the growing global digital and knowledge economy, the ECA works towards enabling countries to promote, adopt new and emerging technologies, and promote digital skills to deliver the transformation of their economies.

\subsection{Sao Tom\'e and Principe}
% ...
\subsubsection{Research}

\begin{table}[htb]
  \begin{center}
    \begin{tabular}{|p{0.6in}p{0.6in}p{1.4in}|}
 \hline
     Sao Tom\'e and Principe & 2020-2023 & Concrete Actions \\
      \hline
      Research & $\checkmark$ & pepper value chain   \\
      SMB & $\checkmark$ &    \\
      Informal Economy & $\checkmark$  &  \\
      Government & $\checkmark$ &  \\
      \hline
    \end{tabular}\label{tab:mytablestp}
  \end{center}
  \caption{MI in Sao Tom\'e and Principe}
\end{table}

\cite{RefSTP1} develops Guiding mechanisms for Social Protection Targeting Through Satellite Data in Sao Tom\'e and Principe. Social safety net programs often target the poorest and most vulnerable populations. However, in many developing countries, there is a lack of administrative data on the relative wealth of the population to support the selection process for potential beneficiaries of these programs. Therefore, the selection process often involves a multi-methodological approach, starting with geographical targeting for the selection of program implementation areas. 

To facilitate this stage of the targeting process in Sao Tom\'e and Principe, their article develops High-Resolution Satellite Imagery (HRSI) poverty maps, providing estimates of poverty incidence and program eligibility at a highly detailed resolution (110 m x 110 m). The analysis combines poverty incidence and population density to facilitate the geographical targeting process. 

Their work  demonstrates that HRSI poverty maps can serve as key operational tools to aid decision-making in geographical targeting and efficiently identify entry points for rapidly expanding social safety net programs. Unlike HRSI poverty maps based on census data, poverty maps based on satellite data and machine learning can be updated frequently at a low cost, supporting more adaptive social protection programs.
\cite{RefSTP2} examines the role of value chains analysis in the agricultural sector, focusing on the case of pepper value chains in Sao Tom\'e e Principe. Despite significant progress in yield and productive capacity, global inequality persists, with hundreds of millions suffering from hunger and undernourishment. Climate change poses a significant threat to agriculture, emphasizing the need for sustainable agricultural value chains that promote and protect natural resources while being inclusive and supportive of smallholders. Value chain analysis, often used in policy development by NGOs, international organizations, and governments, plays a crucial role in promoting economic progress for less favored smallholders in the global market. In Sao Tom\'e and Principe, where conditions such as dimension, insularity, and poverty are specific, the promotion of value chains is particularly dependent on foreign aid. The case of the pepper value chain reveals similarities between formal and informal producers in terms of production systems, but the formal sector tends to be more stable and financially protected than the informal sector.







\section{Western Africa} \label{sec:western}

Western Africa  has a total area of 5,112,903 squared kilometers, with a population of 418,544,337. The principal activities include agriculture, live stock management, fishing, trade. 
In Western Africa, MI activities are diverse, with notable research initiatives in Benin focusing on soil fertility assessment, banana plant disease detection, electricity generation forecasting, public health decision-making for breast cancer, and suitability mapping for rice production. Burkina Faso is engaged in predicting malaria epidemics, mapping urban development, forecasting energy consumption, exploring mineral resources through airborne geophysical data, and modeling monthly energy consumption. Cabo Verde is involved in understanding aerosol microphysical properties, studying climate change's impact on endemic trees, estimating salt consumption using machine learning, and monitoring volcanic eruptions. C\^ote d'Ivoire explores machine learning for cocoa farmers and progress towards onchocerciasis elimination. Gambia investigates machine learning models for pneumonia-related child mortality and smart rural water distribution systems. Ghana's MI research spans urban growth assessment, vehicle ownership modeling, public sentiment analysis, blood demand forecasting, internet data usage analysis, severity prediction of motorcycle crashes, effects of artisanal mining, and customs revenue modeling. Guinea focuses on predicting viral load suppression among HIV patients and prognosis models for Ebola patients. Guinea-Bissau delves into biomass relationships, cashew orchard mapping, learning and innovation in smallholder agriculture, and automatic speaker recognition for monitoring PLHIV. Liberia engages in cloud computing and machine learning for land cover mapping, predicting local violence, remote sensing for land cover studies, scalable approaches for rural school detection, and open challenges for mapping urban development. Mali's research covers groundwater potential mapping, borehole yield predictions, cropland abandonment analysis, image recognition with deep convolutional neural networks, MI's role in addressing global health challenges, improved recurrent neural networks for pathogen recognition, and market liberalization policy analysis. Mauritania explores MI-driven insights into English studies, desert locust breeding area identification, business intelligence models for e-Government, and remote monitoring of water points. Niger's research includes electrical charge modeling, land use mapping using satellite time series, and adult literacy and cooperative training program analysis. Nigeria showcases an extensive array of MI applications, including diabetes prevalence detection, crude oil production modeling, flood area prediction, food insecurity prediction, entrepreneurial success prediction, mobile forensics for cybercrime detection, genre analysis of Nigerian music, terrorism activity prediction, stock market forecasting, and poverty prediction using satellite imagery. Senegal's research spans crop yield prediction, resilient agriculture, machine learning for rice detection, monitoring artisanal fisheries, predicting road accident severity, estimating electrification rates, and analyzing the energy-climate-economy-population nexus. Sierra Leone collaborates with UNICEF's Giga Initiative for rapid school mapping using MI and satellite imagery. Togo contributes to the Novissi program expansion, machine ethics, wind potential evaluation, maize price prediction, solar energy harvesting assessment, land use dynamics forecasting, and solar energy harvesting evaluation. The government of Togo hosts the Artificial Intelligence Week in 2024, emphasizing MI's role in the country's development. The humanitarian sector in Togo leverages machine learning algorithms and mobile phone data for effective aid distribution.
Despite numerous startups in Western Africa, \cite{plantinga2023responsible} found that  most of these countries are missing from the MI ecosystem.
\cite{jaldi2023artificial} reports the following number of companies that specialize in MI in Western Africa: Nigeria: 456, Ghana: 115, Ivory Coast: 29, Senegal: 23.
\cite{gwagwa2021road} reports on the opportunities in agriculture. \cite{adeshina2023role} investigate the role of MI in SDGs from an African perspective.

\subsection{Benin}

\begin{table}[htb]
  \begin{center}
    \begin{tabular}{|p{0.6in}p{0.3in}p{1.3in}|}
      \hline
      Benin & 2020-2023 & Concrete Actions \\
      \hline
      Research & {\checkmark} &   soil, banana, rice, electricity, public health\\
      Government & {\checkmark}   & SNIAM, SENIA, national MI strategy\\
      \hline
    \end{tabular}\label{tab:mytablebenin}
  \end{center}
  \caption{MI in Benin}
\end{table}

\subsubsection{Research}

The work in \cite{refBenin1} assesses soil fertility status in Benin using digital soil mapping and machine learning techniques. Published in Geoderma Regional in 2022.
The work in \cite{refBenin2} detects banana plants and their major diseases through aerial images and machine learning methods, focusing on a case study in DR Congo and the Republic of Benin.  The work in \cite{refBenin3} involves short-term electricity generation forecasting using machine learning algorithms, with a case study of the Benin Electricity Community (CEB). 
The work in \cite{refBenin4} utilizes mathematical modeling and machine learning for public health decision-making, with a focus on the case of breast cancer in Benin. 
The work in \cite{refBenin5} maps suitability for rice production in inland valley landscapes in Benin and Togo using environmental niche modeling.

%
%
%

\subsubsection{Government}
Benin has an institutional framework composed of several entities, including the Ministry of Digitalization, the Agency for Information Systems and Digital, the Authority for Personal Data Protection (APDP), and the S\`em\`e City Development Agency (ADSC). In addition, some private actors, primarily startups and citizen associations, are already exploring avenues and organizing training on application development and MI usage. Benin also has training institutes such as the Institute of Training and Research in Computer Science  and the Institute of Mathematics and Physical Sciences  equipped with a supercomputer, providing significant computing power for MI development in Benin and the entire sub-region. The Benin Government, through its action program, has made digitalization a cornerstone of economic and social progress. Significant investments in this sector since 2016 demonstrate a strong political will to develop the digital economy and transform the country into a regional platform for digital services sustainably. Major digital sector projects (data centers, interoperability, e-Administration, e-Services, Open Data, etc.) will generate massive data, requiring proper management and utilization to ensure the creation of value remains within the Beninese economy.
Machine intelligence emerges as a tool to effectively address this issue and support Benin's influence in strategic sectors such as education, health, agriculture, environment, and tourism. However, to harness its potential and strengthen existing initiatives in artificial intelligence, Benin needs to identify its strengths and define objectives to take a leadership position in the sub-region. To achieve this, the Ministry of Digitalization has initiated the development of the National Strategy for Machine Intelligence and Big Data and its action plan, collaboratively designed with stakeholders and submitted for approval by the Council of Ministers. In its session on January 18, 2023, the Council of Ministers approved the National  Artificial Intelligence and Big Data Strategy (SNIAM) 2023-2027. Driven by the vision of making Benin a country that shines by leveraging its massive data through MI systems and technologies, it consists of four programs implemented in three phases over five years, with a portfolio containing one hundred twenty-three actions impacting the public and private sectors. Its adoption positions Benin as a country capable of seizing current and future opportunities related to artificial intelligence and massive data processing, making it more attractive for investments from the private sector and development partners.
With a projected amount of four billion six hundred eighty million (4,680,000,000) CFA francs over a five-year period, the implementation of this strategy provides an opportunity to leverage MI in targeted development areas to position the country as a major player in MI in West Africa.
For the second consecutive year, the Digital Entrepreneurship and Machine Intelligence Fair (SENIA) took place in Cotonou on May 12 and 13, 2023. Bringing together nearly 1,000 participants, the event reflects Benin's ambition to become a major player in MI in West Africa. The debates aim to generate impactful initiatives benefiting the government, private sector, and Beninese society as a whole. Organized by the Ministry of Digitalization, SENIA brings together Beninese talents and international experts specializing in artificial intelligence and data science. Through numerous conferences and demonstrations, SENIA is not only a platform for exchange but also a business and networking environment for industry professionals.

With the goal of fostering research, education, and implementation of MI in Africa, an Artificial Intelligence Research Centre has launched in Cotonou, Benin Republic.
Founded in March 2018, Atlantic AI Labs is attempting to unleash the potential of MI for sustainable development in healthcare, precision agriculture, education, unmanned aerial vehicles, clean energy, environmental protection, and wildlife conservation.

\subsection{Burkina Faso}

\begin{table}[htb]
  \begin{center}
    \begin{tabular}{|p{0.6in}p{0.3in}p{1.2in}|}
      \hline
      Burkina Faso& 2020-2023 & Concrete Actions \\
      \hline
      Research & {\checkmark} &  CITADEL, urban systems, energy, mining  \\
      SMB & {\checkmark} & African Foods Nutrition, DataBusiness-AI, CyberLabs Tech, Kumakan , SOSEB, Saintypay, Kalabaash, Dunia , Qotto, Toto Riibo \\
      Informal Economy &    & \\
      Government & {\checkmark}   & Interdisciplinary Center of Excellence in Artificial Intelligence for Development \\
      \hline
    \end{tabular}\label{tab:mytableburkina}
  \end{center}
  \caption{MI in Burkina Faso}
\end{table}

\subsubsection{Research}

The work in \cite{refBurkina1} focuses on predicting malaria epidemics in Burkina Faso using machine learning. 
The work in \cite{refBurkina2} maps patterns of urban development in Ouagadougou, Burkina Faso, utilizing machine learning regression modeling with bi-seasonal Landsat time series. 
The work in \cite{refBurkina3} explores machine learning models to predict town-scale energy consumption in Burkina Faso. 
The work in \cite{refBurkina4} employs machine learning techniques on airborne geophysical data for mineral resources exploration in Burkina Faso. 
The work in \cite{refBurkina5} focuses on forecasting models for monthly energy consumption using machine learning in Burkina Faso.


\subsubsection{Small Businesses}
Qotto sells solar home systems to rural households in West Africa with a pay-as-you-go model. At Qotto, we transform the daily life of our customers by installing solar kits. To help them achieve their dreams, we offer electric autonomy with great flexibility and quality customer service.
Toto Riibo is a Burkina Faso-based online food ordering and delivery service.
Dunia Payment is an Ouagadougou-based mobile wallet startup that lets its users send and receive money, pay in stores with a simple QR Code.
African Foods Nutrition is an agri-food processing unit with an industrial focus. Their solutions involve the innovative production of standard nutritional flours based on three types of cereal and two legumes, nutritional flours enriched with Moringa, nutritional flours enriched with baobab powder, nutritional flours enriched with dried fruits, nutritional flours enriched with other vegetables, nutritional flours for breastfeeding women, nutritional flours for pregnant women and anemic individuals, baby compotes, and natural wellness products.
DataBusiness-AI is a consulting company specializing in MI solutions.
CyberLabs Tech develops MI-enabled robots for agriculture.
Kumakan Studio creates games that promote African culture, myths, legends, and folklore to entertain and showcase African treasures to the world. Additionally, they create games for social change to help solve local problems.
SOSEB, also known as SOS Energie Burkina, is a commercial enterprise with a social and environmental mission that aims to protect the environment and promote sustainable development in Burkina Faso. In this regard, it distributes solar electrification kits, solar irrigation kits, and ecological cooking solutions (such as solar ovens, biomass cookers, ecological coal, etc.). Saintypay is assisting businesses and users in facilitating money transfers between Africa and Dubai.
Kalabaash operates in the digital field and consists of a team of dynamic experts with various skills. Kalabaash aims to be a leader in the field of big data and artificial intelligence in Burkina Faso.



\subsubsection{Government}

After inaugurating significant electronic communication infrastructure in Bobo-Dioulasso, on September 4,  the country hosted the 16th edition of the Digital Week from September 8 to 10, 2020, in Ouagadougou, under the theme: "Artificial Intelligence: Opportunities and Challenges." The Interdisciplinary Center of Excellence in Artificial Intelligence for Development (CITADEL) will host researchers from Burkina Faso seeking a conducive environment for conducting high-quality, globally competitive, interdisciplinary research relevant to the African context. CITADEL will also train new talents with versatile skills to meet the needs of local industry and research. It is located at the Virtual University of Burkina Faso and is endowed with funding of one million Canadian dollars.

\subsection{Cabo Verde}

\begin{table}[htb]
  \begin{center}
    \begin{tabular}{|p{0.6in}p{0.3in}p{1.2in}|}
      \hline
      Cabo Verde& 2020-2023 & Concrete Actions \\
      \hline
      Research & {\checkmark} & salt, climate change, volcanic eruptions monitoring  \\
      \hline
    \end{tabular}\label{tab:mytablecap}
  \end{center}
  \caption{MI in Cabo Verde}
\end{table}

\subsubsection{Research}

The work in \cite{refCaboVerde1} aims to understand aerosol microphysical properties based on 10 years of data collected at Cabo Verde, utilizing an unsupervised machine learning classification.
The work in \cite{refCaboVerde2} explores the implications of climate change on the distribution and conservation of Cabo Verde endemic trees. 
The work in \cite{refCaboVerde3} involves the development, validation, and application of a machine learning model to estimate salt consumption in 54 countries, including Cabo Verde. 

The work in \cite{refCaboVerde4} utilizes Sentinel-1 GRD SAR data for volcanic eruptions monitoring, focusing on the case-study of Fogo Volcano in Cabo Verde during 2014/2015. 

%
%
%
%
%
%

\subsection{C\^ote d'Ivoire}
\begin{table}[htb]
  \begin{center}
    \begin{tabular}{|p{0.6in}p{0.6in}p{1.4in}|}
 \hline
      Ivory Coast & 2020-2023 & Concrete Actions \\
      \hline
      Research & $\checkmark$ &  Cacao\\
      SMB & $\checkmark$ &   Futurafric AI \\
      Government & $\checkmark$ &  AI and Robotics Center in Yamoussoukro\\
      \hline
    \end{tabular}\label{tab:mytableivory}
  \end{center}
  \caption{MI in Ivory Coast}
\end{table}

\subsubsection{Research}
The work in \cite{refIC1} investigates Machine Learning as a Strategic Tool for Helping Cocoa Farmers in C\^ote D'Ivoire.
The work in  \cite{refIC2} examines progress towards onchocerciasis elimination in C\^ote d'Ivoire.

%
%
%

\subsubsection{Government}

In 2018, MainOne launched Abidjan data center, which offers capacity for 100 racks. MainOne  has launched a second Cote d’Ivoire data center in 2023. This new Tier III-quality data center is in the Village of ICT \& Biotechnology of Cote d'Ivoire (VITIB) in Grand Bassam, on the outskirts of Abidjan.

In April 2023, the Ministry of Communication and Digital Economy of C\^ote d'Ivoire, in collaboration with Smart Africa, inaugurated the Cybersecurity Innovation Center. Located at the African Higher School of Information and Communication Technologies (ESATIC) in Abidjan, this center focuses on combating cybercrime. Aligned with the agreement signed in September 2022 with the Republic of C\^ote d'Ivoire, the center aims to enhance digital skills, including the implementation of the Smart Africa Digital Academy (SADA). Supported by ESATIC and Hitachi Systems Security Inc., this innovation center serves as a tool to improve national cybersecurity culture through awareness and skill development for the target population.

There are several ongoing initiatives in C\^ote d'Ivoire to promote the development of MI. These include the Digital Transformation Initiative of C\^ote d'Ivoire, creating an environment for MI development, and the Africa MI Initiative, a partnership between the Ivorian government and the World Bank to foster MI development. The country hosts universities like the University of Abidjan and the MI and Robotics Center in Yamoussoukro actively involved in MI research and application development, particularly in areas like health, finance, and agriculture. In 2022, a Franco-Ivorian collaboration was established through the Franco-Ivorian Hub for Education, launching the Master of Science BIHAR to support joint diploma programs between Ivorian and French institutions, emphasizing the crucial role of data in Africa's digital future. The ESTIA aims to globally disseminate its Master of Science BIHAR through Digital Associate Campuses in partner universities for local tutoring of remote learners.

In August 2023, the issue of MI  in the context of Customer Experience was discussed  during the second edition of LONACI Online mornings at the Ivory Coast National Lottery (LONACI), at the lagoon hall of Ivoire Trade Center in Abidjan-Cocody. LONACI Online mornings serve as a platform for discussions on digital topics by Lonaci, aiming to "better connect" with its clients. During this event, various speakers from Yadec Consulting, Futurafric Artificial Intelligence, and Willis Towers Watson shared their insights on the topic of MI.

In September 2023, accompanied by the UNESCO Assistant Director for Social and Human Sciences, the Minister of Good Governance and Anti-Corruption chaired, in Abidjan-Plateau, the launch of the implementation of the recommendation on the ethics of Machine Intelligence  in C\^ote d'Ivoire. The event was attended by representatives from various organizations, including the High Authority for Audiovisual Communication, the Commission for Access to Public Information and Public Documents, the Virtual University, the Association of Bloggers of C\^ote d'Ivoire, the National Union of Journalists of C\^ote d'Ivoire, and civil society.
Given the scope of MI use and the main theme focusing on digital issues, the Minister of Good Governance mentioned having enlisted the support of the Minister of Digital Economy to serve as Co-lead. In the implementation of this recommendation, the ministerial department will focus on the governance and ethics of MI, while the Ministry of Communication and Digital Economy will address the technical aspects of MI use in various sectors.

\subsection{Gambia}

\begin{table}[htb]
  \begin{center}
    \begin{tabular}{|p{0.6in}p{0.6in}p{1.4in}|}
 \hline
     Gambia & 2020-2023 & Concrete Actions \\
      \hline
      Research & $\checkmark$ &  water\\
      SMB & $\checkmark$ &    \\
      Informal Economy & $\checkmark$  &  \\
      Government & $\checkmark$ &  \\
      \hline
    \end{tabular}\label{tab:mytablegambia}
  \end{center}
  \caption{MI in Gambia}
\end{table}
\subsubsection{Research}

The work in \cite{refGambia1} examines how to deploy Machine Learning Models Using Progressive Web Applications: Implementation Using a Neural Network Prediction Model for Pneumonia Related Child Mortality in The Gambia. The work in \cite{refGambia2} investigates smart rural water distribution systems in the Gambia.

%
%
%
%

\subsection{Ghana}

\begin{table}[htb]
  \begin{center}
    \begin{tabular}{|p{0.6in}p{0.6in}p{1.4in}|}
 \hline
      Ghana & 2020-2023 & Concrete Actions \\
      \hline
      Research & $\checkmark$ &  gold mining\\
    Industry &  $\checkmark$ &  Mining\\
      SMB & $\checkmark$ & Diagnosify, Xpendly, Kwanso, DatawareTech, Khalmax Robotics, mNotify, GreenMatics, DigiExt, CYST, CRI, Huggle.care,QualiTrace    \\
      Informal Economy & $\checkmark$  & energy \\
      Government & $\checkmark$ & National Artificial Intelligence Center, Responsible Artificial Intelligence Lab \\
      \hline
    \end{tabular}\label{tab:mytableghana}
  \end{center}
  \caption{MI in Ghana}
\end{table}

\subsubsection{Research}

The work in \cite{refGhana1} assesses urban growth in Ghana using machine learning and intensity analysis, with a focus on the New Juaben Municipality. 

The work in \cite{refGhana2} models vehicle ownership in the Greater Tamale Area, Ghana, utilizing machine learning techniques. 

The work in \cite{refGhana3} employs statistical analysis and machine learning to study public sentiment on the Ghanaian government. 

The work in \cite{refGhana4} utilizes machine learning algorithms for forecasting and backcasting blood demand data at Tema General Hospital in Ghana. 

The work in \cite{refGhana5} conducts a historical analysis and time series forecasting of internet data usage and revenues in Ghana using a machine learning-based Facebook Prophet model.

The work in \cite{refGhana6} focuses on severity prediction of motorcycle crashes in Ghana using machine learning methods. Published in the International Journal of Crashworthiness in 2020.

The work in \cite{refGhana7} explores the local effects of artisanal mining in Ghana, providing empirical evidence. 

The work in \cite{refGhana8} introduces GC3558, an open-source annotated dataset of Ghana currency images for classification modeling. 

The work in \cite{refGhana9} applies Ito calculus and machine learning for the projection of forward US dollar-Ghana cedi rates. 

The work in \cite{refGhana10} uses machine learning and Google Earth Engine to quantify the spatial distribution of artisanal goldmining in Ghana, focusing on the conversion of vegetation to gold mines. 

The work in \cite{refGhana11} models customs revenue in Ghana using novel time series methods. 

The work in \cite{refGhana12} applies machine learning to analyze jump dynamics in US Dollar-Ghana Cedi exchange returns.

\subsubsection{Industry} 
Neural networks have been used in the mining industry in  \cite{refghananewty1, refghananewty2, refghananewty3, refghananewty4, refghananewty5}.


\subsubsection{Small Businesses}

DatawareTech is a data analytics company with a mission to empower organizations to gain insights from data for strategic decision-making.

Khalmax Robotics is an EdTech company that provides robotics and MI products in education.

mNotify is an MI-powered Customer Engagement tool for SMEs, facilitating exponential growth.

GreenMatics develops affordable autonomous solar-powered agricultural robots.

DigiExt provides technical services to rice farmers for optimal crop growth, resulting in cost savings.

CYST is a software innovation company founded in 2013 and based in Ghana. It specializes in artificial intelligence to create simple and easy-to-use technology solutions relevant to local markets while adhering to international standards. CYST also has a research arm called CRI (CYST Research Institute).

Huggle.care uses Machine Intelligence to enhance how people find the best care for the symptoms they experience.

The QualiTrace concept is built on the idea of traceability, allowing consumers to trace food produce back to the farm gates. QualiTrace is an Agri-Tech startup that utilizes track and trace technology to authenticate agricultural inputs (such as seeds and fertilizers) and outputs. QualiTrace not only authenticates but also provides a clear, simple means by which players in any given supply chain can trace products along the chain to the final consumer.

Xpendly  is an MI-powered startup that uses artificial intelligence to diagnose skin diseases, predict the name of the disease, determine its severity level, and assign patients to pharmaceutical services or dermatologists. Xpendly is a personal finance management app that enables young African Millennials to manage their finances in one place, build alternative credit profiles with their financial activity, and access tailored financial products that help them save and invest.
A solution to prevent, track, and monitor road traffic accidents by offering passengers, drivers, and regulators an app for identifying accident locations and monitoring speed.


\subsubsection{Informal Economy}

Informal enterprises learn how to produce goods and services through cumulative and diverse ways. However, there is limited empirical evidence on how learning processes influence the innovation of informal enterprises in Africa. The paper \cite{refghanan1} examines the effects of two learning processes (apprenticeship and formal interactions) on the product innovativeness of informal enterprises in Ghana. Employing unique survey data on 513 enterprises and the Type II Tobit model, our analyses revealed that apprenticeship, on the one hand, enhances the technological capability of enterprises leading to product innovativeness, while competitive formal interactions, on the other hand, provide important market feedback that enhances the innovativeness of enterprises. In addition, financially constrained informal enterprises that compete with formal enterprises in product markets performed poorly with their new products, compared with their counterparts who were not financially constrained. The work in \cite{refghanan2} focuses on informal energy consumption in Ghana.
The work in 
\cite{refghanan3} studies the drivers of undeclared works using machine learning.
\subsubsection{Government}
In 2019, the government launched the National Artificial Intelligence Center to promote the development and adoption of MI in the country. The center is tasked with developing policies, strategies, and frameworks that will guide the development and deployment of MI in Ghana. 

In 2022, the Kwame Nkrumah University of Science and Technology  has been awarded a grant to fund the establishment of a Responsible Artificial Intelligence Lab  under the AI4D Africa Multidisciplinary Labs project initiated by International Development Research Centre. The Responsible Artificial Intelligence Lab  is hosted at the Kwame Nkrumah University of Science and Technology in Ghana. RAIL seeks to be a first step in establishing a sustainable approach to nurturing local talent to engage in multidisciplinary, responsible MI for development research and innovation with a focus on women and that that responds to capacity requirements of the public and private sector.

The Responsible AI Network - Africa was founded through a partnership between the Faculty of Electrical and Computer Engineering at Kwame Nkrumah University of Science and Technology  in Ghana and the Institute for Ethics in Artificial Intelligence  at the Technical University of Munich in Germany: The aim is to build a network of scholars working on the responsible development and use of AI in Africa.

\subsection{Guinea}

\begin{table}[htb]
  \begin{center}
    \begin{tabular}{|p{0.6in}p{0.6in}p{1.4in}|}
 \hline
      Guinea & 2020-2023 & Concrete Actions \\
      \hline
      Research & $\checkmark$ &  water\\
      SMB & $\checkmark$ &    \\
      Informal Economy & $\checkmark$  &  \\
      Government & $\checkmark$ &  \\
      \hline
    \end{tabular}\label{tab:mytableguinea}
  \end{center}
  \caption{MI in Guinea}
\end{table}
\subsubsection{Research}

The work in \cite{refConakry1} focuses on the development of machine learning algorithms to predict viral load suppression among HIV patients in Conakry, Guinea.

The work in \cite{refConakry2} transforms clinical data into actionable prognosis models, utilizing a machine-learning framework and a field-deployable app to predict the outcome of Ebola patients. 

The work in \cite{refConakry3}  uses machine learning in epidemiology. It characterizes of risk factors related to the occurrence of pulmonary and extra pulmonary tuberculosis in the province of Settat.

%
%
%
%
%
%

\subsection{Guinea-Bissau}

\begin{table}[htb]
  \begin{center}
    \begin{tabular}{|p{0.6in}p{0.6in}p{1.4in}|}
 \hline
      Guinea-Bissau & 2020-2023 & Concrete Actions \\
      \hline
      Research & $\checkmark$ &  cashew, speaker recognition\\
      SMB & $\checkmark$ &    \\
      Informal Economy & $\checkmark$  &  \\
      Government & $\checkmark$ &  \\
      \hline
    \end{tabular}\label{tab:mytablebissau}
  \end{center}
  \caption{MI in Guinea-Bissau}
\end{table}
\subsubsection{Research}

The work in \cite{refGuineaBissau1} explores the relationship between above ground biomass and ALOS PALSAR data in the forests of Guinea-Bissau. 
The work in \cite{refGuineaBissau2} focuses on mapping cashew orchards in Cantanhez National Park, Guinea-Bissau. 
The work in \cite{refGuineaBissau3} delves into endogenous learning and innovation in African smallholder agriculture, drawing lessons from Guinea-Bissau.
The work in \cite{refGuineaBissau4} introduces an Automatic Speaker Recognition  application for monitoring PLHIV in the cross-border area between the Gambia, Guinea-Bissau, and Senegal. 

%
%
%
%
%
%

\subsection{Liberia}

\begin{table}[htb]
  \begin{center}
    \begin{tabular}{|p{0.6in}p{0.6in}p{1.4in}|}
 \hline
      Liberia & 2020-2023 & Concrete Actions \\
      \hline
      Research & $\checkmark$ &  anti-violence,  land cover change \\
      SMB & $\checkmark$ &    \\
      Informal Economy & $\checkmark$  &  \\
      Government & $\checkmark$ &  \\
      \hline
    \end{tabular}\label{tab:mytableliberia}
  \end{center}
  \caption{MI in Liberia}
\end{table}
\subsubsection{Research}

The work in \cite{refLiberia1} employs cloud computing and machine learning to support country-level land cover and ecosystem extent mapping in Liberia and Gabon. 

The work in \cite{refLiberia2} investigates predicting local violence in Liberia using evidence from a panel survey. 

The work in \cite{refLiberia3} focuses on remote sensing, machine learning, and change detection applications for land cover studies in Liberia. 

The work in \cite{refLiberia4} presents a scalable approach using convolutional neural networks and satellite imagery for detecting rural schools in Africa.

The work in \cite{refLiberia5} introduces an open machine learning challenge to map urban development and resilience in diverse African cities from aerial imagery.

%
%
%
%
%
%

\subsection{Mali}

\begin{table}[htb]
  \begin{center}
    \begin{tabular}{|p{0.6in}p{0.6in}p{1.4in}|}
 \hline
      Mali & 2020-2023 & Concrete Actions \\
      \hline
      Research & $\checkmark$ &  Predominantly Oral Languages, MI in Africa in 20 Questions, image recognition, farmers insurance,\\
      SMB & $\checkmark$ &   Guinaga, Grabal, Timadie \\
      Art \& Music &  & SK1 ART,  Dronegraphy \\
      Informal Economy & $\checkmark$  &  WETE\\
      Government & $\checkmark$ &  CIAR-Mali\\
      \hline
    \end{tabular}\label{tab:mytablemali}
  \end{center}
  \caption{MI in Mali}
\end{table}

\subsubsection{Research}

A recent book \cite{Bouare2023}  titled "MI in Africa in 20 Questions"  with 11 co-authors from Africa addresses key questions raised by MI in Africa was published in June 2023:
\begin{itemize}
\item Question 1: The first question addressed is that of the  definition of intelligence, natural intelligence, human intelligence, machine intelligence, artificial intelligence, and collective intelligence.
\item Question 2: In an African context, the issue of generative machine intelligence emerges with particular importance, especially on online platforms. The authors of this book prompt us to reflect on the opportunities and challenges this approach may present for Africa, addressing issues related to the utopia and opportunity of generative machine intelligence in various sectors such as the economy, finance, and agriculture.
\item Question 3: Conversational agents generate keen interest. The book examines potential errors of these conversational agents and the need to correct them to ensure better interaction with African users.
\item Question 4: The book also tackles intriguing questions about delusions, illusions, confabulations, and hallucinations in generative machine intelligence. Do these phenomena actually exist or are they just analogies?
\item Question 5: Machine intelligence also finds applications in the financial sector in Africa. The book explores opportunities and challenges related to using machine intelligence for banking without traditional banks in the informal sector.
\item Question 6: In the agricultural sector, can machine intelligence be considered a delusion or an opportunity for Africa? This book guides us through a thorough reflection on the use of machine intelligence in African agriculture, examining potential advantages such as cost reduction and optimization of agricultural practices.
\item Question 7: Manure is becoming increasingly interesting for mass agriculture in Africa. This book invites you to explore this alternative with machine intelligence from plants and animal waste, particularly cattle urine.
\item Question 8: In the livestock sector, can machine intelligence be an ally in herd monitoring and improving animal welfare? This book encourages us to ponder the effects of machine intelligence in livestock.
\item Question 9: A crucial question arises: can the use of machine intelligence contribute to reducing the cost of animal feed in Africa? The authors of this book explore the possibilities offered by machine intelligence to optimize animal feeding processes, highlighting the economic and environmental implications of this approach. What is the optimal strategy between producing livestock feed oneself, arranging with local producers, and buying imported concentrates from other continents?
\item Question 10: The book also raises questions about access to machine intelligence in African contexts where internet access may be limited. How can the integration of machine intelligence be envisioned in environments where connectivity is a major challenge?
\item Question 11: Another essential question addressed in this book concerns empowering women through machine intelligence. This book urges us to reflect on the socio-economic implications of this technology, evaluating whether its adoption can strengthen or, conversely, amplify existing inequalities.
\item Question 12: The book also examines the risks of inequalities and stereotypes arising from the massive use of machine intelligence. How can we ensure that machine intelligence systems do not reproduce biases and discriminations present in African society? This book pushes us to question practices and responsibilities related to machine intelligence in the fight against inequalities and stereotypes.
\item Question 13: The spread of misinformation is a major challenge in the digital world. The book explores the possibility that machine intelligence may accentuate this phenomenon and highlights measures to mitigate this risk in Africa.
\item Question 14: In the fight against crime, can machine intelligence be an effective tool, or does it represent a double-edged sword? The book invites us to reflect on the implications for security, confidentiality, and human rights in using machine intelligence to combat crime.
\item Question 15: How can machine intelligence contribute to civil protection in Africa? The book examines the potential applications of this technology in disaster prevention, crisis management, and improving the resilience of African communities.
\item Question 16: The health sector is also explored in the book. How can machine intelligence be used to improve healthcare in Africa? This book raises crucial questions about potential illegal practices and ethical issues related to the use of machine intelligence in the healthcare field.
\item Question 17: Can the massive integration of machine intelligence pose energy supply problems in certain African countries? This book leads us to reflect on the implications of this technology on electricity demand and to identify sustainable solutions to address this challenge.
\item Question 18: In the field of road safety, can the widespread use of machine intelligence lead to greater inattention, distraction, and insecurity? The book raises important questions about the potential consequences of integrating machine intelligence into vehicles and the need to ensure increased safety on African roads.

\item Question 19: The nineteenth question in this book discusses the potential offered by intelligence for auditing, but its reliability depends on several factors, including the quality of data, the accuracy of algorithms, and the ability to adapt to changes.
\item 
Question 20: The integration of machine intelligence into education sparks many discussions about the role of machines in teaching and learning. This twentieth and final question, far from the least, explores the question of who should be responsible for the use of machine intelligence in education and what the proposed content will be. The use of machine intelligence in education can have several advantages, such as personalized learning, providing instant feedback to students, and access to online learning resources. However, it is important to consider who will be responsible for designing and implementing these technologies.
\end{itemize}

Most languages of the world are predominantly oral, with little to no writing tradition. In recent years, the African continent was completely missing on the NLP map, but due to efforts of grassroots communities such as Masakhane, Africa is now present on the NLP map focused on African languages. Despite all the efforts made thus far, more progress is needed.  
As of August 2023, Mali's population stands at approximately 22 million individuals. The country boasts 13 national languages and an illiteracy rate of 65\%. Interestingly, 80\% of the population communicates in Bambara. In the past, French held the position of the official language; however, merely 35\% of the people were adept in functional literacy in French. Notably, Mali has recently embraced a new constitution that relegates French from its former status while elevating all 13 national languages to the rank of official languages. 

DONIYA-SO a  legal non-governmental platform  on Data Science and MI in Bamako, Mali is studying B.A.M.B.A.R.A: Breaking Audio Multilingual Barriers Advancing Research Across Africa. The MI project  aims to build high-quality audio/sound/speech datasets for Predominantly Oral Languages  (POLs). The work seeks to narrow this language gap by gathering top-notch audio/sound/speech datasets for Bambara. This effort will not only facilitate research outcomes for the remaining 12 languages of Mali but also extend its benefits beyond. The state-of-the-art machine learning including deep learning  models requires a significant amount of data, which is not readily available in low-resource settings, and this is the case for POLs. If funded, this project has the potential to increase the amount of data usable by researchers, and developers in the field of ML. This, when successfully implemented, has the potential to increase literacy rate, open our people to the world, and the world to our people.

The work in \cite{refMali1} focuses on preprocessing approaches in machine-learning-based groundwater potential mapping in Mali, specifically the Koulikoro and Bamako regions. 
The work in \cite{refMali2} presents multiclass spatial predictions of borehole yield in southern Mali using machine learning classifiers. 
The work in \cite{refMali3} assesses cropland abandonment from violent conflict in central Mali using SENTINEL-2 and Google Earth Engine. 
The work in \cite{refMali4} introduces a deep convolutional neural network for image recognition in Mali. 
The work in \cite{refMali5} explores the potential of artificial intelligence in addressing global health challenges, focusing on antimicrobial resistance and the impact of climate change on disease epidemiology. 
The work in \cite{refMali6} utilizes an improved recurrent neural network with LSTM for the recognition of pathogens through image classification. 
The work in \cite{refMali7} analyzes three decades of market liberalization policy in Mali, specifically focusing on grain markets. 
The work in \cite{refMali8} introduces SatDash, an interactive dashboard for assessing land damage in Nigeria and Mali. 
The work in \cite{refMali9} reflects on the role of digital technology in peace processes, emphasizing the need to make peace with uncertainty. 
The work in \cite{refMali10} evaluates machine learning and deep learning classifiers for offensive language detection in a code-mixed Bambara-French corpus. 
The work in \cite{refMali11} presents a typology of Malian farmers and their credit repayment performance using an unsupervised machine learning approach. 
The work in \cite{refMali12} applies principal component analysis to hydrochemical data from groundwater resources in Bamako, Mali.


\subsubsection{Small Businesses}

There are also opportunities for private sector investment in MI-based solutions for agriculture in Mali. Startups and tech companies are developing innovative solutions to address challenges faced by farmers. For instance, Timadie is an innovative platform of platforms that is based on graphchain technologies adapted for less connected environments. 

In the agriculture sector, the Guinaga platform has physically engaged with over 400 farmers and is developing a graphchain as a traceability tool for food products. Graphchains such as MangueChain, Karit\'eton, Yuton, and Rizton have contributed to the valorization of local products in Senegal, Mali, and Burkina Faso.  It utilizes MI, Blockchain, and Graphchain technologies to optimize the supply chain, air pollution, and understand production, consumption, sales, purchases, transportation, storage, and more in Mali, Burkina Faso, and Senegal. The data contributes to building a Kariteton, incorporating shea butter production. MI techniques aid in estimating production 8 weeks in advance to reduce demand-supply mismatch. Guinaga also employs deep learning in its Manguechain for tracking mangoes from tree to consumer tables. Guinaga's knitting and crochet club is a dedicated platform for cotton and indigo plant cultivation for natural dyes, semi-automatic machine knitting, providing a space for knowledge sharing, creativity, and support for knitting enthusiasts.

Since 2019, SK1 Sogoloton uses deep learning for videos analytics  for information dissemination in  25 languages across Africa. Users can access reliable and verifiable information, participate in information production, and combat misinformation in Africa. The platform has over 10 million video views and 200 million interactions as of November  2023. CI4SI, from the learning and game theory laboratory, emphasizes collective intelligence and encourages collaboration to solve  societal problems.

The Grabal platform is a catalyst for connecting traditional livestock breeders  in Africa, promoting the preservation of genetic diversity. MoutonChain, in particular, has ensured the traceability of sheep, providing consumers with the necessary confidence during the Tabaski period from 2019 to 2023. Grabal links breeders directly to buyers from the field, utilizing MI-enabled drones to estimate animal food, headcount, and safe paths in Mali, Burkina Faso, and Senegal. 

WETE has opened new perspectives for the economic empowerment of women in Africa. To date, 150 African women CEOs are present on the platform. Women-in-Drones promotes the participation of women in emerging sectors such as drone technology and dronegraphy. To date, 23 women have been trained in drone piloting and the uses of MI in video-processing in Mali. The particularity here is that these trained women uses MI-enabled drones in their professional works such as mineral extraction, geo-information, and crisis management.

By fostering interconnectivity and encouraging co-opetition, Timadie provides an environment conducive to the emergence of new ideas, fruitful collaborations, and significant social innovations in multiple African countries.
 
Timadie, hosting various platforms, has trained students at several schools on MI applications in geo-information systems, economy, transportation, energy, healthcare, and agriculture.

\subsubsection{Art \& Music}  SK1'ART tests the accuracy and reliability of several music and Dogon mask generated by MI. Together with Malian artists the platform aims to be create an online Malian Got Talent combined with MI and blockchain technologies.


\subsubsection{Government}
The groundbreaking ceremony on June 7, 2023, marked a crucial moment for Mali's recent technological ambitions. Colonel Assimi Goita, leading the transitional government, initiated the construction of the Center for Artificial Intelligence and Robotics (CIAR-Mali) in Kati,  Koulikoro region in Mali. This visionary project, designated as a Public Scientific, Technical, and Cultural Establishment, is set to become a hub for cutting-edge research, development, and educational activities in artificial intelligence and robotics.
With an estimated construction cost of 3.3 billion F CFA (4.5 million euros), the CIAR-Mali underscores the government's firm belief in the transformative power of artificial intelligence. This strategic investment not only signifies a commitment to technological advancement but also presents an unparalleled opportunity for the youth of Mali and Africa. Aspiring individuals keen on contributing to the continent's technological evolution now have a potential avenue through the CIAR-Mali.
Fast forward to October 19, 2023, the ratification bill for the creation of the CIAR-Mali received the unanimous approval of the National Transitional Council. This legislative approval solidified the center's role in fostering innovation and knowledge transfer within the domains of artificial intelligence and robotics. However, despite the legislative green light, the center has not yet commenced its operational phase as of the current date.
The CIAR-Mali underscores the nation's forward-looking approach, emphasizing the pivotal role that artificial intelligence and robotics play in shaping the future of Mali and contributing to the broader technological landscape of the African continent.

\subsection{Mauritania}
\subsubsection{Research}

The work in \cite{refMauritania1} explores MI-driven insights into the factors influencing students' choice of English studies as a major at the University of Nouakchott Al Aasriya, Mauritania. Published in the International Journal of Technology, Innovation, and Management in 2022.

The work in \cite{refMauritania2} leverages MI methodologies for identifying desert locust breeding areas in Mauritania through Earth Observation.

The work in \cite{refMauritania3} employs MI techniques to survey business intelligence models for e-Government in Mauritania. The authors, El Arby Chrif et al., showcase the MI applications at The International Conference on Artificial Intelligence and Smart Environment in 2022.

The work in \cite{refMauritania4} utilizes MI and IoT (LoRa) technology for the remote monitoring of water points in Mauritania. This AI-infused study is published in the Indonesian Journal of Electrical Engineering and Informatics  in 2022.

%
%
%
%
%
%

\subsection{Niger}

\begin{table}[htb]
  \begin{center}
    \begin{tabular}{|p{0.6in}p{0.6in}p{1.4in}|}
 \hline
      Niger & 2020-2023 & Concrete Actions \\
      \hline
      Research & $\checkmark$ &  electricity, training\\
      SMB & $\checkmark$ &    \\
      Informal Economy & $\checkmark$  &  \\
      Government & $\checkmark$ &  \\
      \hline
    \end{tabular}\label{tab:mytableniger}
  \end{center}
  \caption{MI in Niger}
\end{table}

\subsubsection{Research}

\cite{refNiger1} examines Electrical Charge of Niamey City using a  neural network model.
\cite{refNiger2}  studies land use mapping using Sentinel-1 and Sentinel-2 time series in a heterogeneous landscape in Niger, Sahel.

\cite{refNiger3} examines adult literacy and cooperative training programs in  Niamey, Niger.

%
%
%
%
%
%

\subsection{Nigeria}

\begin{table}[htb]
  \begin{center}
    \begin{tabular}{|p{0.6in}p{0.6in}p{1.4in}|}
 \hline
      Nigeria& 2020-2023 & Concrete Actions \\
      \hline
      Research & $\checkmark$ &  Patato\\
      SMB & $\checkmark$ &  Tuteria, Kudi AI , Curacel , Codar Tech Africa\\
     Art \& Music & $\checkmark $ & Afrobeats, Infinite Echoes\\
      Informal Economy & $\checkmark$  &  \\
      Government & $\checkmark$ & NCAIR \\
      \hline
    \end{tabular}\label{tab:mytablenigeria}
  \end{center}
  \caption{MI in  Nigeria}
\end{table}

\subsubsection{Research}

The work in \cite{refNigeria1} by Oyebode and Orji focuses on detecting factors responsible for diabetes prevalence in Nigeria using social media and machine learning. Obite et al. contribute to the modeling of crude oil production in Nigeria, identifying an eminent model for application \cite{refNigeria2}. Ighile et al. apply GIS and machine learning to predict flood areas in Nigeria \cite{refNigeria3}. Villacis et al. explore the role of recall periods in predicting food insecurity in Nigeria using machine learning \cite{refNigeria4}. McKenzie and Sansone analyze the competition between man and machine in predicting successful entrepreneurs in Nigeria \cite{refNigeria5}. Goni and Mohammad present a machine learning approach to a mobile forensics framework for cybercrime detection in Nigeria \cite{refNigeria6}. Folorunso et al. dissect the genre of Nigerian music with machine learning models \cite{refNigeria7}. Lawal et al. predict floods in Kebbi state, Nigeria, using machine learning models \cite{refNigeria8}. Nwankwo et al. focus on predicting house prices in Lagos, Nigeria, using machine learning models \cite{refNigeria9}. Panjala et al. identify suitable watersheds across Nigeria using biophysical parameters and machine learning algorithms for agri-planning \cite{refNigeria10}.

McKenzie and Sansone, in \cite{refNigeria11}, presents  the challenges of predicting entrepreneurial success, drawing evidence from a business plan competition in Nigeria. Gladys and Olalekan present a machine learning model for predicting color trends in the textile fashion industry in southwest Nigeria \cite{refNigeria12}. Odeniyi et al. predict terrorist activities in Nigeria using machine learning models \cite{refNigeria13}. Ogundunmade and Adepoju model liquefied petroleum gas prices in Nigeria using time series machine learning models \cite{refNigeria14}. Ekubo and Esiefarienrhe utilize machine learning to predict low academic performance at a Nigerian university \cite{refNigeria15}. Muhammad and Varol propose a symptom-based machine learning model for malaria diagnosis in Nigeria \cite{refNigeria16}. Salele et al. model run-off in pervious and impervious areas using SWAT and a novel machine learning model in Cross River State, Nigeria \cite{refNigeria17}. Adeeyo and Osinaike model the oil viscosity of Nigerian crudes using machine learning \cite{refNigeria18}.
Jean et al., in \cite{refNigeria19}, combine satellite imagery and machine learning to predict poverty in Nigeria. Ibrahim et al. predict potato diseases in smallholder agricultural areas of Nigeria using machine learning and remote sensing-based climate data \cite{refNigeria20}. Oyebode and Orji, in \cite{refNigeria21}, explore social media and sentiment analysis in the context of the Nigeria presidential election 2019. Eneanya et al. examine environmental suitability for lymphatic filariasis in Nigeria \cite{refNigeria22}.
Obulezi et al. predict transportation costs inflated by fuel subsidy removal policy in Nigeria using machine learning \cite{refNigeria23}. Mbaoma et al. use geospatial and machine learning-driven air pollution modeling in Agbarho, Delta State, Nigeria \cite{refNigeria24}. Achara et al. investigate financial institution readiness and adoption of machine learning algorithms and performance of select banks in Rivers State, Nigeria \cite{refNigeria25}. Oyewola et al. propose a new auditory algorithm for stock market prediction in the Nigerian stock exchange, focusing on the oil and gas sector \cite{refNigeria26}.

The work in \cite{refNigeria22} by Oyebode and Orji explores social media and sentiment analysis, specifically examining the Nigeria presidential election in 2019. Eneanya et al. in \cite{refNigeria28} investigate the environmental suitability for lymphatic filariasis in Nigeria.
Obulezi et al. \cite{refNigeria29} focus on machine learning models for predicting transportation costs inflated by fuel subsidy removal policy in Nigeria. Mbaoma et al. \cite{refNigeria30} utilize geospatial and machine learning-driven air pollution modeling in Agbarho, Delta State, Nigeria. Achara et al. \cite{refNigeria31} study financial institution readiness and the adoption of machine learning algorithms, examining the performance of select banks in Rivers State, Nigeria. Oyewola et al. \cite{refNigeria27} introduce a new auditory algorithm for stock market prediction in the Nigerian stock exchange, particularly emphasizing the oil and gas sector


\subsubsection{Small Businesses}

Tuteria is an online platform that connects people who are seeking to learn anything with those who live near them and are available to teach them. It provides an environment that offers safety, accountability and quality. Globally, conventional methods of education and learning are being challenged. They are moving from centralized to distributed, from standardized to personalized. These trends have repeatedly demonstrated their ability to deliver better learning outcomes and Tuteria fits in well with this trend.

Kudi AI has a familiar origin as {\it Kudi} means money in the Hausa language, reflecting its simplicity in service options. Kudi is a conversational agent powered by machine intelligence, based in Nigeria, that assists with your finances simply by asking. Kudi uses a conversational machine intelligence system to interact with you on a daily basis. It helps you transfer money, track your account details, purchase airtime, pay recurring bills, and also reminds you when some of these bills are due.

Codar Tech Africa in Lagos, Nigeria was created in December 2021 realizing that tech training was inadequate in Africa’s biggest economy. With the swiftly-changing job market because of advancements in generative AI, there is  an importance and urgency of teaching tech skills that would set people apart in an AI-influenced landscape. To date, Codar Africa  provides hands-on training experience in various aspects of tech including; data analysis, web development and design, Search Engine Optimization, and Cybersecurity.

Nigerian-based Curacel is an MI platform that aims to drive insurance penetration in emerging markets via APIs enabling insurers to connect with digital distribution channels and administer their claims. Founded in 2019, the MI platform recently raised \$3 million in seed funding. Initially the platform was intended to be an electronic health information management platform for healthcare providers, enabling clinics to digitize and manage paper records, appointments, patient communications, billing and reporting through a web app. 

 Ubenwa Health is a MedTech startup in Nigeria building the future of automated sound-based medical diagnostics.

\subsubsection{Art \& Music} 

The Afrobeats music genre from Lagos has recently kept millions on their feet and challenged preconceptions of African music. Its popularity is growing globally, with an increasing number of players in the entertainment industry trying to get a slice of it.
When machine intelligence  apps began spreading in Nigeria's music industry, Eclipse Nkasi thought his days as a producer were numbered. However, he took a step back, identified opportunities as well as threats, and used the technology to generate a whole new Afrobeats album in his studio on the outskirts of Lagos. MI doesn't have to replace what we have. It gives people a new experience... and that's how I believe MI is really going to shake things. In the past, it would have taken him thousands of dollars and up to three months to compose the tracks, recruit musicians, record performances, refine them in a traditional studio, and release them to fans.
Nigerian artists activated MI algorithms and set them to work, assisting in creating the nine-track album "Infinite Echoes." They instructed it to auto-generate song lyrics and titles, including "God Whispers," "Love Tempo," and "Dream Chaser." Then, they modified the words themselves to fit their chosen theme - a struggling artist not giving up on their passion for creating music. Next, they used another MI tool to generate the tunes. Nkasi recorded some vocals and fed them into yet another app, transforming his vocals into the voice of the album's generated singer - a virtual "singer" named Mya Blue, who appears online as a computer animation in front of her audience. Certain things may become obsolete due to MI, but it should also create opportunities for artists to reinvent themselves and improve their work more efficiently. The technology is already transforming the industry and could have a positive impact on production values and other technical aspects of the recording process. However, there are still uncertainties and areas, including copyright, that need consideration and development.

Initially launched in February 2023 in the US and Canada, followed by the UK and Ireland in May, DJ is a personalized MI guide that offers users a carefully curated lineup of music, accompanied by commentary on the tracks and artists, all delivered in a realistic voice. The feature aims to foster a stronger connection between users and their music, enabling discovery through tailored recommendations.
The initial voice model for DJ was based on Spotify’s Head of Cultural Partnerships, Xavier 'X' Jernigan. However, the latest rollout extends the offering to include commentary in English for listeners in various international markets.
African countries where the DJ feature is now accessible include Botswana, Burundi, eSwatini, the Gambia, Ghana, Kenya, Lesotho, Liberia, Malawi, Namibia, Nigeria, Rwanda, Sierra Leone, South Africa, Tanzania, Uganda, Zambia, and Zimbabwe. Never before has listening felt so completely personal to each and every user, thanks to the powerful combination of Spotify’s personalization technology, generative AI, and a dynamic, expressive voice.
Users can access the AI DJ by opening the Spotify mobile app on their iOS or Android devices. After launching the app, they can navigate to the Music Feed on the homepage and tap play on the DJ feature.


\subsubsection{Government}
The National Centre for Artificial Intelligence and Robotics (NCAIR) is one of NITDA's special purpose vehicles created to promote research and development on emerging technologies and their practical application in areas of Nigerian national interest. The center, a state-of-the-art facility, along with its modern digital fabrication laboratory, is co-located in the same building complex with the Office for Nigerian Digital Innovation, at No. 790 Cadastral Zone, Wuye District, Abuja. NCAIR as a digital innovation and research facility is focused on Artificial Intelligence, Robotics and Drones, Internet of Things, and other emerging technologies, aimed at transforming the Nigerian digital economy, in line with the National Digital Economy Policy and Strategy. NCAIR is also focused on creating a thriving ecosystem for innovation-driven entrepreneurship, job creation and national development. In 2023, Nigeria is extending an invitation to scientists of Nigerian heritage, as well as globally renowned experts who have worked within the Nigerian market, to collaborate in the formulation of its National Artificial Intelligence Strategy. According to the Minister of Communications, Innovation and Digital Economy, the National Information Technology Development Agency has initiated the development of a National MI Strategy. The action will impact the way the government formulates new technological solutions for its critical national challenges. As a result, the government is broadening its co-creation strategy by assembling a selection of leading MI researchers with Nigerian heritage from around the world. The Nigerian government recognizes that MI has developed into a versatile technology, reshaping production and services and holds immense potential for influencing societal progress and economic expansion.  According to a white paper titled Co-creating a National Artificial Intelligence Strategy for Nigeria, a sophisticated method was used to pinpoint accomplished MI researchers with Nigerian roots, using global MI publication data and advanced machine learning models. Research index was created to locate influential machine intelligence researchers of Nigerian heritage. As the preliminary research phase concludes, the Nigerian government seeks public involvement, acknowledging the potential for errors and aiming to tap into collective knowledge and insights. The government of Nigeria via the National Centre for Artificial Intelligence and Robotics aims to  establish communities of MI developers nationwide to influence the country's technological future. The initiative  started in three states in 2023, followed by strategic planning for its extension to additional states and, eventually, all local government areas.

\subsection{Senegal}

\begin{table}[htb]
  \begin{center}
    \begin{tabular}{|p{0.6in}p{0.6in}p{1.4in}|}
 \hline
      Senegal & 2020-2023 & Concrete Actions \\
      \hline
      Research & $\checkmark$ &  mango\\
      SMB & $\checkmark$ &  Afrikamart , Teranga Capital,    \\
      Informal Economy & $\checkmark$  & Lengo AI  \\
      Government & $\checkmark$ & Data Center, National MI strategy \\
      \hline
    \end{tabular}\label{tab:mytablesene}
  \end{center}
  \caption{MI in Senegal}
\end{table}

\subsubsection{Research}

The work in \cite{refSenegal1} by Sarr and Sultan predicts crop yields in Senegal using machine learning methods, focusing on climatology. In \cite{refSenegal2}, Nyasulu et al. contribute to resilient agriculture in the Sahel region, employing machine learning for weather prediction. MMbengue et al. evaluate machine learning classification methods for rice detection using Earth observation data in Senegal \cite{refSenegal3}. Bayet et al. apply a machine learning approach to enhance the monitoring of Sustainable Development Goals, concentrating on Senegalese artisanal fisheries \cite{refSenegal4}. Dia et al. present a hybrid model for predicting road accident severity in Senegal, utilizing artificial intelligence and empirical studies \cite{refSenegal5}.
The work in \cite{refSenegal6} by State et al. explores explainability in practice by estimating electrification rates in Senegal using mobile phone data. In \cite{refSenegal7}, Sarkodie et al. conduct an empirical analysis of the energy-climate-economy-population nexus in Senegal and other countries. Kebe et al. share their experience with detecting and classifying Quality-Of-Service problems in MV/LV distribution substations in Senegal using artificial intelligence \cite{refSenegal8}. Seck and Diakit\'e develop supervised machine learning models for predicting renal failure in Senegal \cite{refSenegal9}. Lee et al. investigate the intersection of colonial legacy and environmental issues in Senegal through language use \cite{refSenegal10}.
The work in \cite{refSenegal11} by Dione et al. focuses on designing Part-of-Speech-Tagging resources for Wolof in Senegal. Alla et al. leverage LSTM to translate French to Senegalese local languages, with Wolof as a case study \cite{refSenegal12}. Dia et al. present an empirical study on predicting the severity of road accidents in Senegal using a hybrid model \cite{refSenegal13}. Dione et al. propose an IoT-based e-health model for developing countries, with Senegal as a case study \cite{refSenegal14}. \"{O}zdogan and Govind examine three decades of forest cover change in Senegal using remote sensing \cite{refSenegal15}. Traor\'e et al. analyze nonlinear price transmission in the rice market in Senegal, employing a model-based recursive partitioning approach \cite{refSenegal16}. Drame et al. conduct an analysis and forecast of energy demand in Senegal using SARIMA and LSTM models \cite{refSenegal17}. Sarron et al. investigate the efficiency of machine learning for mango yield estimation in Senegal under heterogeneous field conditions \cite{refSenegal18}. Diop et al. apply a machine learning approach to the classification of Okra \cite{refSenegal19}. Moustapha Mbaye et al. propose a new machine learning workflow for creating an optimal waiting list in hospitals \cite{refSenegal20}.


\subsubsection{Small Businesses}

Afrikamart is an agritech that facilitates the pick-up, shipping, and trading of fresh fruits and vegetables between small producers and urban retailers via a digital platform. Afrikamart was founded to address agricultural loss issues present throughout the production chain, from producer to retailer.  Afrikamart is supported by Acceleration Technologies, a 2.5 million euros program that aims to finance and support fifteen digital start-ups in Sub-Saharan Africa, supported by AFD through the Digital Africa initiative. In Senegal, Teranga Capital is in charge of implementing this program. 


\subsubsection{Informal Economy}

Lengo AI   is the first AI-Driven Intelligence Platform for for Fast Moving Consumer Goods companies in Africa for the Informal Sector.

\subsubsection{Government}
 
In December 2016, during the State visit to France by President Macky Sall, Ministers of Economy, Finance, Planning, and Higher Education and Research were instructed to sign a 15 million euros financing deal with the Public Investment Bank (BPI). This funding aimed at establishing the National Center for Scientific Computing (CNCS) in Diamniadio, equipped with a parallel calculator, the most powerful south of the Sahara excluding South Africa. The funding covered mobility, maintenance, and training, benefiting research in artificial intelligence, Big Data, cybersecurity, robotics, and scientific computing. The CNCS, operational today, contributes to sectors like agriculture, health, genomics, biotechnology, ICT, mining, gas, oil, energy, security, meteorology, climate change, coastal erosion, water management, navigation, and environmental data exploitation. The supercomputer, with a computing power of around 320 Tflops and integrated storage capacity of 21 Terabytes, aligns with the Senegal Emergent Plan in agrohydrology and the mining sector, facilitating numerical simulations for innovation in meteorology, climatology, image processing, vegetation growth, and mineral exploration.

In May 2021 was the inauguration of the Data Center of Diamniadio, Senegal. This infrastructure is a major revolution for the digital sovereignty of our country, which for the first time will have its own digital data storage structure. Among other features, the Data Center will make it possible to generalize very high speed across the national territory, to satisfy, at affordable costs, government and private sector requests for hosting and operation of computer platforms and data.
This Data Center will also facilitate the dematerialization of procedures. 

In May 2022, An experts consultative meeting on developing a continental strategy for Machine Intelligence in Africa was successfully held in Dakar, Senegal. This was held on the margins of the 6th Calestous Juma Executive Dialogue (CJED), organized by the African Union High-Level Panel on Emerging Technologies (APET).  The CJED convenes policy and decision makers, executives, youth, and relevant stakeholders to deliberate on harnessing appropriate innovations and emerging technologies for Africa's socio-economic development.
APET has prioritized and recommended MI as an emerging technology worth harnessing for Africa's socio-economic development. In the APET "AI for Africa"  report  launched in December 2022, the panel provides guidelines for African countries on how best to exploit AI-based technologies for the continent's advancement. The high-level panel further recommended developing a continental Machine Intelligence strategy for Africa, necessitating this expert consultative meeting.

In June 2023, Senegal developed its National MI Strategy. This strategy, developed through extensive stakeholder consultations, aimed to position Senegal as an MI leader in West Africa.
A delegation composed of members from the Ministry of Communications, Telecommunications and Digital Economy  of Senegal, academia and the private sector embarked on a study tour to Rwanda, an African pioneer in digital innovation. This tour, a strategic move in operationalizing Senegal's National MI strategy, was not just about technology transfer but a deeper dive into the complexities of MI in an African context. The initiative, led by Enabel, one of the implementing partners of the AU-EU D4D Hub Project, aimed to gather insights and best practices to operationalize Senegal's National MI Strategy.

\subsection{Sierra Leone}

\begin{table}[htb]
  \begin{center}
    \begin{tabular}{|p{0.6in}p{0.6in}p{1.4in}|}
 \hline
      Sierra Leone& 2020-2023 & Concrete Actions \\
      \hline
      Research & $\checkmark$ &  \\
      SMB & $\checkmark$ &    \\
      Informal Economy & $\checkmark$  &  \\
      Government & $\checkmark$ &  \\
      \hline
    \end{tabular}\label{tab:mytablesierra}
  \end{center}
  \caption{MI in Sierra Leone}
\end{table}

\subsubsection{Research}

UNICEF's Giga Initiative endeavors to map every school on the Planet. Knowing the location of schools is the first step to accelerate connectivity, online learning, and initiatives for children and their communities, and drive economic stimulus, particularly in lower-income countries. Development Seed \cite{refSierraLeone1} is working with the UNICEF Office of Innovation to enable rapid school mapping from space across Asia, Africa, and South America with AI. In seven months of development and implementation, we added 23,100 unmapped schools to the map in Kenya, Rwanda, Sierra Leone, Niger, Honduras, Ghana, Kazakhstan, and Uzbekistan.
To accomplish this we built an end-to-end scalable MI model pipeline that scans high-resolution satellite imagery from Maxar, applies our highly refined algorithm for identifying buildings that are likely to be schools, and flags those schools for human review by  a Data Team. 

%
%
%
%
%
%

\subsection{Togo}

\begin{table}[htb]
  \begin{center}
    \begin{tabular}{|p{0.6in}p{0.6in}p{1.4in}|}
 \hline
      Togo & 2020-2023 & Concrete Actions \\
      \hline
      Research & $\checkmark$ &  maize,  MI ethics\\
      SMB & $\checkmark$ & Semoa, Eazy Chain,  SocialGIS, Dobbee Pay, Solimi Fintech ,  Artybe    \\
      Government & $\checkmark$ & GiveDirectly, African Cybersecurity Resource Center,  Artificial Intelligence Week \\
      \hline
    \end{tabular}\label{tab:mytabletogo}
  \end{center}
  \caption{MI in Togo}
\end{table}

\subsubsection{Research}
The research efforts in \cite{refTogo1} focused on helping the government expand the Novissi programme from informal workers in Greater Lom\'e to poorer individuals in rural regions of the country, and were designed to meet the government's two stated policy objectives: first, to direct benefits to the poorest geographic regions of the country; and second, to prioritize benefits to the poorest mobile subscribers in those regions. Individuals without access to a mobile phone could not receive Novissi payments, which were delivered digitally using mobile money. The approach they developed,  uses machine learning to analyze non-traditional data from satellites and mobile phone networks.   The work in \cite{refTogo2} by Kohnert (2022) explores machine ethics and African identities, offering perspectives on the role of artificial intelligence in Africa.  The work in \cite{refTogo3} assessed wind potential in Togo's Kara region using artificial neural networks, offering both static and dynamic evaluations.
The work in \cite{refTogo4} focused on predicting maize prices in Lome, Togo, utilizing a Hidden Markov Chain Model.
The work in \cite{refTogo5} compared survey-based impact estimation with digital traces in the context of randomized cash transfers in Togo.
The work in \cite{refTogo6} forecasted land use and cover dynamics in the Ago\`enyiv\'e Plateau, Togo, using a combination of remote sensing, machine learning, and local perceptions.
The work in \cite{refTogo7} employed artificial neural networks to evaluate solar energy harvesting in Togo, providing insights into sustainable development.


\subsubsection{Small Businesses} 

Sema-Kiosk and Cashpay are Semoa's main products. Semoa-Kiosk which support Cashpay and permit to make different virtual transactions and mainly deposit cash on a mobile account. An innovation who permit to customer to make transactions any time without depend of a physical place.

Eazy Chain is a digital logistics dashboard that enables small businesses to track, manage and monitor all their shipping operations by air, ocean and road. With Eazy Chain, small businesses can also pay their suppliers abroad in foreign currencies and track their consolidated shipments from all over the world to their destinations.

SocialGIS is the African Geospatial Intelligence Agency is a startup that works on Free Geomatics Technologies and uses big data and open data.
SocialGIS provides GIS services and solutions, and tools for data collection and visualization, good governance, landscape restoration, nature preservation and sanitation, agriculture, urbanization and more.

Dobbee Pay strives to establish a platform facilitating users in receiving and transferring funds through mobile money, banks, and cryptocurrency platforms. This digital tool stands out by creating interoperability among existing payment methods, allowing users to make group payments to a maximum of one million people with just one click.

Solimi Fintech is on a mission to reduce cash usage by 40\% within five years by democratizing access to financial services. The long-term vision involves leveraging AI to create a world where all financial and commercial transactions can be managed online. The app will feature an integrated chatbot.

Artybe is a platform that seamlessly blends MI with African culture to showcase Togo as an exceptional tourist destination, emphasizing environmental preservation for future generations. Beyond promoting tourism, the app will function as a versatile learning platform covering IT, fitness, agriculture, swimming, and more. Users can tailor their courses based on their preferences and the country's wealth, considering their availability and financial capacity.



\subsubsection{Government}

In March 2021, the African Development Bank granted \$2 million to the African Cybersecurity Resource Center (ACRC) for Financial Inclusion, aiming to combat cybercrime and enhance the resilience of digital financial ecosystems. Located in Lomé, the African Centre for Coordination and Research in Cybersecurity, established through a partnership between the government and the United Nations Economic Commission for Africa (UNECA), will monitor, detect, and share cybersecurity intelligence with African governments, policymakers, law enforcement, and security agencies. Cybercrime, costing Africa an estimated \$4 billion annually, remains a significant concern. The center will also spearhead internet security research, particularly crucial as hacking groups increasingly deploy sophisticated deep learning software to infiltrate African government websites, banks, hospitals, power companies, and telecommunication firms.

Fast forward to March 2022, the inaugural Cybersecurity Summit, co-organized by Togo and UNECA, convened Heads of State and Government, private sector leaders, and civil society representatives to discuss Africa's pressing cybersecurity challenges. During the summit, member states endorsed the Lom\'e Declaration on Cybersecurity and the Fight Against Cybercrime, commonly referred to as the Lomé Declaration. This commitment signifies member states' pledge to sign and ratify the African Union's Malabo Convention, one of the world's most comprehensive cybersecurity conventions, aiming to strengthen African cooperation in combating cyber threats. 

 Togo will host Artificial Intelligence Week (AIS) in March 2024. The event will be organized by CONIIA-Togo, the Togolese branch of the Conseil International pour l'Intelligence Artificielle, and Human-AI, a structure specialized in the development of new technologies. Scheduled from March 19-24, 2024, the SIA will gather Togo's MI stakeholders. They will take stock of the current state of MI and explore its opportunities. The event will focus on raising awareness of advances in artificial intelligence among the general public, students, decision-makers, and institutions.
 
 Governments and humanitarian groups can use machine learning algorithms and mobile phone data to get aid to those who need it most during a humanitarian crisis.
  Researchers team helped Togo's Ministry of Digital Economy and GiveDirectly, a nonprofit that sends cash to people living in poverty, turn this insight into a new type of aid program.The simple idea behind this approach, is that wealthy people use phones differently from poor people. Their phone calls and text messages follow different patterns, and they use different data plans, for example. Machine learning algorithms - which are fancy tools for pattern recognition - can be trained to recognize those differences and infer whether a given mobile subscriber is wealthy or poor \cite{refTogo1}.

\section{Eastern Africa} \label{sec:eastern} 

Burundi's MI research landscape encompasses malaria case prediction using deep learning models and automated image recognition for diagnosing banana plant diseases. In industry, there's a focus on optimizing LPG usage with Gaslink. In small businesses, Neural Labs Africa employs MI for medical image diagnosis, while Wiggles Technologies provides custom software solutions. The government plays a role in supporting MI adoption through initiatives like the National MI Strategy in Seychelles. Shinzwani dictionary construction and orthographic choice is built in the Comoro Islands.

In Djibouti, research spans from improving sky temperature forecasting to using deep learning for fracture-fault detection in groundwater models. In industry, Farnbec addresses LPG challenges with Gaslink, and MI Connect enhances air travel experiences. Eritrea's research involves predictive lithologic mapping using remote sensing data. Ethiopia applies machine learning to predict drought and uses interpretable models for evaporation in reservoirs.

In Kenya, AI Made in Africa supports startups in finding diverse tech talents.
The educational app, targeting school students, offers an engaging approach to learning about fruits and vegetables, presenting names in French and Mauritian Creole. Backed by a dataset of 1600 images, machine learning classifiers were tested, revealing TensorFlow's outstanding accuracy of 98.1\%. 
In the broader context, Mauritius demonstrates a strategic approach to MI, with a national strategy and established entities like the Mauritius MI Council and MI Academy. The government's focus on the ocean economy aligns with the strategy, emphasizing the potential of maritime IoT. Additionally, various companies, including Qubitica, Cash Radix, AgCelerant, Arie Finance, and 4Sight Holdings, contribute to the integration of MI across industries, small businesses, and governmental initiatives.

In Mozambique, MI activities encompass a range of research areas. From assessing OpenStreetMap quality using unsupervised machine learning to mapping land use and cover, the studies reveal insights into data contributors, LULC changes, and MI practices in education. Additionally, initiatives focus on food security, smallholder irrigated agriculture mapping, and leveraging deep learning and Twitter for mapping built-up areas post-natural disasters like Cyclone IdMI and Kenneth in 2019.

Rwanda has approved the National MI Policy to harness MI's benefits. In Tanzania, MI applications in healthcare are explored, and Resilience Academy students use machine learning for tree-cover mapping. Uganda focuses on creating high-quality datasets for East African languages. In South Sudan, machine learning is used to analyze fragility-related data, and in Somalia, sentiment analysis is applied to Somali text.

Meanwhile, in Zambia, machine learning aids in predicting stunting among children and enhances the efficiency of health clinic verification through algorithms like Random Forest. The country also faces air quality challenges in mining towns, necessitating improved environmental monitoring. In Zimbabwe, the focus is on data-driven pediatrics to enhance pediatric care, effective vehicle damage classification using deep learning algorithms, and the use of technology to predict and address adolescent depression. These initiatives underscore the transformative potential of technology in diverse sectors across these African nations.

Each country's MI landscape reflects a unique blend of research, industry applications, small businesses, and government initiatives, showcasing the diverse ways MI is contributing to development across Eastern Africa.

\subsection{Burundi}

\begin{table}[htb]
  \begin{center}
    \begin{tabular}{|p{0.6in}p{0.6in}p{1.4in}|}
 \hline
  Burundi & 2020-2023 & Concrete Actions \\
      \hline
      Research & $\checkmark$ &  water\\
      SMB & $\checkmark$ &    \\
      Informal Economy & $\checkmark$  &  \\
      Government & $\checkmark$ &  \\
      \hline
    \end{tabular}\label{tab:mytableBurundi}
  \end{center}
  \caption{MI in Burundi}
\end{table}

\subsubsection{Research}

Malaria continues to be a major public health problem on the African continent, particularly in Sub-Saharan Africa. Nonetheless, efforts are ongoing, and significant progress has been made. In Burundi, malaria is among the main public health concerns. The work in \cite{refBurundi1} built machine-learning based models to estimates malaria cases in Burundi. The forecast of malaria cases was carried out at province level and national scale as well. Long short term memory  model, a type of deep learning model has been used to achieve best results using climate-change related factors such as temperature, rainfall, and relative humidity, together with malaria historical data and human population. With this model, the results showed that at country level different tuning of parameters can be used in order to determine the minimum and maximum expected malaria cases. The univariate version of that model  which learns from previous dynamics of malaria cases give more precise estimates at province-level, but both models have same trends overall at province-level and country-level.

 Bananas are  the dominant crop in Burundi. The surface area under cultivation is estimated at 200,000 to 300,000 ha, representing 20 to 30\% of the agricultural land. Data from Burundi's Ministry of Agriculture and Livestock indicate food security and nutrition continue to worsen, with 21 percent of the population food insecure. This could be exacerbated by various plant diseases such as the Banana Bunchy Top Disease. The disease has been reported in Angola, Benin, Burundi, Cameroon, Central African Republic, Republic of Congo, DRC, Equatorial Guinea, Gabon, Malawi, Mozambique, Nigeria, Rwanda, South Africa, and Zambia.   The East African Highlands is the zone of secondary diversity of a type of bananas called the AAA-EA types. These bananas are genetically close to the dessert banana types but have been selected for use as beer, cooking, and dessert bananas.
 
Banana cultivation in Burundi is grouped into three different categories. Banana for beer/wine in which juice is extracted and fermented accounts for around 77 percent of the national production by volume. Fourteen percent of bananas are grown for cooking, and finally, about five percent are dessert bananas which are ripened and directly consumed. With recent advances in machine learning, researchers were convinced that new disease diagnosis based on automated image recognition was technically feasible. Minimizing the effects of disease threats and keeping a matrix mixed landscaped of banana and non-banana canopy is a key step in managing a large number of diseases and pests \cite{refBurundi2}.

%
%
%
%
%
%

\subsection{Comoros}

\begin{table}[htb]
  \begin{center}
    \begin{tabular}{|p{0.6in}p{0.6in}p{1.4in}|}
 \hline
      Comoros & 2020-2023 & Concrete Actions \\
      \hline
      Research & $\checkmark$ &  water\\
      SMB & $\checkmark$ &    \\
      Informal Economy & $\checkmark$  &  \\
      Government & $\checkmark$ &  \\
      \hline
    \end{tabular}\label{tab:mytablec}
  \end{center}
  \caption{MI in Comoros}
\end{table}

\subsubsection{Research}

The work in \cite{refcomoros1} discusses information and communication uses in education the Comoros.

In  \cite{refcomoros2}   a  Shinzwani dictionary construction and orthographic choice in the Comoro Islands is presented.

%
%
%
%
%

\subsubsection{Government}
 In 2019, the government launched the Comoros Digital Plan, which aims to promote the use of digital technologies, including AI, to drive economic growth and improve public services.

In August 2023, The Banque Centrale des Comores (BCC) officially began work on the country’s first National Financial Inclusion Strategy (NFIS) as part of an AFI-led training workshop.
The event that occurred on 10th August, aimed to guide key stakeholders and BCC staff in formulating and implementing an NFIS to drive forward the country’s broader financial inclusion ambitions.
With this move, the BCC hopes to increase access to financial services and raise awareness among stakeholders of the pivotal role financial inclusion could play in reinforcing the country’s economic stability and the financial well-being of its people using machine intelligence technologies 

\subsection{Djibouti}

\begin{table}[htb]
  \begin{center}
    \begin{tabular}{|p{0.6in}p{0.6in}p{1.4in}|}
 \hline
      Djibouti & 2020-2023 & Concrete Actions \\
      \hline
      Research & $\checkmark$ &  water\\
      SMB & $\checkmark$ &    \\
      Informal Economy & $\checkmark$  &  \\
      Government & $\checkmark$ &  \\
      \hline
    \end{tabular}\label{tab:mytabled}
  \end{center}
  \caption{MI in Djibouti}
\end{table}

\subsubsection{Research}

The building exchanges heat with different environmental elements: the sun, the outside air, the sky, and the outside surfaces. To correctly account for building energy performance, radiative cooling potential, and other technical considerations, it is essential to evaluate sky temperature. It is an important parameter for the weather files used by energy building simulation software for calculating the longwave radiation heat exchange between the exterior surface and the sky. In the literature, there are several models to estimate sky temperature. However, these models have not been completely satisfactory as far as the hot and humid climate is concerned. In this case, the sky temperature remains overestimated. The work in \cite{refDjibouti1}  is to provide a comprehensive analysis of the sky temperature measurement conducted, for the first time in Djibouti, with a pyrgeometer, a tool designed to measure longwave radiation as a component of thermal radiation, and an artificial neural network  model for improved sky temperature forecasting.A systematic comparison of known correlations for sky temperature estimation under various climatic conditions revealed their limited accuracy in the region, as indicated by low R2 values and high root mean square errors (RMSEs). To address these limitations, we introduced an ANN model, trained, validated and tested on the collected data, to capture complex patterns and relationships in the data. The ANN model demonstrated superior performance over existing empirical correlations, providing more accurate and reliable sky temperature predictions for Djibouti's hot and humid climate. This study showcases the effectiveness of an integrated approach using pyrgeometer-based sky temperature measurements and ANNs for sky temperature forecasting in Djibouti. Our findings support the use of advanced machine learning techniques to overcome the limitations of existing correlations and improve the accuracy of sky temperature predictions, particularly in hot and humid climates.

The work in \cite{refDjibouti2} examines fracture-fault detection using deep learning.  Accurate estimation of groundwater flow is crucial in arid regions where permanent surface water is absent. In several groundwater simulation models, an important parameter for identifying areas with high potential for groundwater resources is the accurate fracture-fault detection. In the present study we propose a deep learning approach to detect fracture-fault structures in the Ali Faren sub-catchment of Ambouli Wadi in Djibouti. Our deep convolutional neural network (Deep-CNN) model is trained on high-spatial resolution multispectral satellite images using wadi streamline as labels. Fracture-fault structures are extracted using stepwise elimination based on geological characteristics observed in relief images derived from PALSAR-1/2 data. Their results demonstrate that the proposed Deep-CNN model accurately detects fracture-fault lines, achieving a validation accuracy of 0.9684, precision of 0.9124, recall of 0.9701, and F1 of 0.8997. The proposed model has the potential to identify potential areas for groundwater resources across the country, contributing to sustainable water management and improving Djibouti's water security. 
 
%
%
%
%
%
%
%

\subsection{Eritrea}

\begin{table}[htb]
  \begin{center}
    \begin{tabular}{|p{0.6in}p{0.6in}p{1.4in}|}
 \hline
    Eritrea & 2020-2023 & Concrete Actions \\
      \hline
      Research & $\checkmark$ &  water\\
      SMB & $\checkmark$ &    \\
      Informal Economy & $\checkmark$  &  \\
      Government & $\checkmark$ &  \\
      \hline
    \end{tabular}\label{tab:mytableeri}
  \end{center}
  \caption{MI in Eritrea}
\end{table}

\subsubsection{Research}

 A regional bedrock map provides a foundation from which to build geological interpretations. However, rapid and accurate bedrock mapping in an area that lacks outcrop is a common problem, especially in regions with sparse data. A historic bedrock map from an Au and base metal project in the Kerkasha district, Eritrea, is significantly improved by predicting bedrock distribution in areas previously mapped as transported overburden. Publicly-available remote sensing data (DTM and ASTER) were combined with airborne geophysical data (magnetics and radiometrics) to provide features for bedrock prediction \cite{refEritrea1} . Remote sensing data were pre-processed using Principal Components Analysis  to yield an equal number of principal components  as input features. Four iterations were trialled, using different combinations of remote sensing PC features. The two initial trials used all available remote sensing data but compared results when feature ranking and selection is applied to reduce the number of PCs used for training and classification. The subsequent two trials used subsets of available remote-sensing data, selected based on domain expertise (i.e., the domain-specific knowledge of a geologist), with all respective PCs were retained. Five-fold cross-validation scores were highest when a DTM, magnetics, and radiometrics data were included as input features. However, qualitative visual appraisal of predicted results across trials, complemented by maps of class membership uncertainty (using a measure of entropy), indicate that geologically-meaningful results are also produced when radiometrics are omitted and only the DTM and magnetics are used. The study concludes with a generalized workflow to assist geologists who are seeking to improve the bedrock interpretation of areas under cover in a single area of interest. Domain expertise is shown to be critical for the selection of appropriate input features and validation of results during predictive lithologic mapping.

%
%
%
%
%

\subsection{Ethiopia}

\begin{table}[htb]
  \begin{center}
    \begin{tabular}{|p{0.6in}p{0.6in}p{1.4in}|}
 \hline
      Ethiopia & 2020-2023 & Concrete Actions \\
      \hline
      Research & $\checkmark$ &  water\\
      SMB & $\checkmark$ &  iCog Labs  \\
      Government & $\checkmark$ &  EAII\\
      \hline
    \end{tabular}\label{tab:mytableeth}
  \end{center}
  \caption{MI in Ethiopia}
\end{table}

\subsubsection{Research}

This study \cite{refEthiopia1}  applies machine learning to the rapidly growing societal problem of drought. Severe drought exists in Ethiopia with crop failures affecting about 90 million people. The Ethiopian famine of 1983-85 caused a loss of  400,000 - 1,000,000 lives. The present drought was triggered by low precipitation associated with the current El Ni$\tilde{n}$o and long-term warming, enhancing the potential for a catastrophe. In this study, the roles of temperature, precipitation and El Ni$\tilde{n}$o are examined to characterize both the current and previous droughts. Variable selection, using genetic algorithms with 10-fold cross-validation, was used to reduce a large number of potential predictors (27) to a manageable set (7). Variables present in $ 70\%$ of the folds were retained to classify drought (no drought). Logistic regression and Primal Estimated sub-GrAdient SOlver for SVM (Pegasos) using both hinge and log cost functions, were used to classify drought. Logistic regression (Pegasos) produced correct classifications for 81.14\% (83.44\%) of the years tested. The variable weights suggest that El Ni$\tilde{n}$o plays an important role but, since the region has undergone a steady warming trend of  1.6 Celsius since the 1950s, the larger weights associated with positive temperature anomalies are critical for correct classification.

The work \cite{refEthiopia2} develops an Interpretable machine learning for predicting evaporation from Awash reservoirs in Ethiopia. An in-depth understanding of a key element such as lake evaporation is particularly beneficial in developing the optimal management approach for reservoirs. In this study, we first aim to evaluate the applicability of regressors Random Forest, Gradient Booting, and Decision Tree, K-Nearest Neighbor, and XGBoost architectures to predict daily lake evaporation of five reservoirs in the Awash River basin, Ethiopia. The best performing models, Gradient Boosting and XGBoost, are then explained through an explanatory framework using daily climate datasets. The interpretability of the models was evaluated using the Shapley Additive explanations (SHAP). The factors with the greatest overall impact on the daily evaporation for GB and XGboost Architecture were the SH, month, Tmax, and Tmin for Metehara and Melkasa, and Tmax, Tmin, and month had the greatest impact on the daily evaporation for Dubti. Furthermore, the interpretability of the models showed good agreement between the  simulations and the actual hydro-climatic evaporation process. This result allows decision makers to not only rely on the results of an algorithm, but to make more informed decisions by using interpretable results for better control of the basin reservoir operating rules.


\subsubsection{Small Businesses}

iCog Labs is a team of software professionals dedicated to advancing the frontier of research and applications in machine intelligence and delivering quality products to clients. It is based in Addis Ababa, Ethiopia.

%
%
%
%
%
\subsubsection{Government}

The Artificial Intelligence and Robotics is one of the centers of excellence which is identified by the ministry of science and technology to be established in Addis Ababa Science and Technology University.The Artificial Intelligence \& Robotics center of excellence (AI\&R CoEs) is established with the aim to create a close collaboration between the academia and industries in the fields of Artificial intelligence and robotics.

The Ethiopian Artificial Intelligence Institute (EAII) has become African Artificial Intelligence Center of Excellence, the Ministry of Innovation and Technology of Ethiopia (MInT) confirmed in November 2023.
The Ethiopian AI institute was promoted as the continent’s center of excellence during (5th Ordinary Session of the African Union Specialized Technical Committee on Communication and ICT (STC-CICT-5) which is being held at the African Union conference hall, in Addis Ababa.
The Ethiopian Artificial Intelligence Institute was proposed to become the “African Artificial Intelligence Center of Excellence”. The proposal was  accepted and approved by the members of the ICT and Communication Ministers of African countries. 
 
\subsection{Kenya}

\begin{table}[htb]
  \begin{center}
    \begin{tabular}{|p{0.6in}p{0.6in}p{1.4in}|}
 \hline
      Kenya & 2020-2023 & Concrete Actions \\
      \hline
      SMB & $\checkmark$ &  NeuralSight, AIfluence, Amini, Halkin, Freshee , M-Shule, AI Connect \\
      Informal Economy & $\checkmark$  &  \\
      Government & $\checkmark$ &  2019 Kenya's Distributed Ledger Technology and Artificial Intelligence Taskforce , AICEA  \\
      \hline
    \end{tabular}\label{tab:mytableken}
  \end{center}
  \caption{MI in Kenya}
\end{table}

\subsubsection{Research}

Yego et al. (2021) conducted a comparative analysis of machine learning models for predicting insurance uptake in Kenya, emphasizing the role of insurance in financial inclusion and economic growth \cite{refKenya1}.
Mulungu et al. present a machine learning approach to assess the economic impact of integrated pest management practices for mango fruit flies in Kenya \cite{refKenya2}.
Alharahsheh and Abdullah (2021) predict individuals' mental health status in Kenya using machine learning methods \cite{refKenya3}.
Yego et al. (2023) optimize pension participation in Kenya through a comparative analysis of tree-based machine learning algorithms and logistic regression classifier \cite{refKenya4}.
Pius et al. (2021) employ supervised machine learning to model the demand for outpatient health-care services in Kenya using artificial neural networks and regression decision trees \cite{refKenya5}.
Shah et al. (2023) predict postpartum hemorrhage (PPH) in a Kenyan population using machine learning algorithms \cite{refKenya6}.
Ondiek et al. (2023) develop a recommender system for STEM enrollment in Kenyan universities using machine learning algorithms \cite{refKenya7}.
Wilson et al. (2017) demonstrate that ensemble machine learning and forecasting can achieve 99\% uptime for rural handpumps \cite{refKenya8}.
Lees et al. (2022) apply deep learning for vegetation health forecasting in Kenya \cite{refKenya9}.
Gram-Hansen et al. (2019) map informal settlements in developing countries using machine learning and low-resolution multi-spectral data \cite{refKenya10}.
Orare (2019) develops a travel time prediction model for Nairobi city using machine learning algorithms \cite{refKenya11}.
Kochulem et al. (2023) conduct a mass valuation of unimproved land value in Nairobi County \cite{refKenya12}.
Pius Kamando (2023) proposes a tree-based neural network for forecasting outpatient health-care services demand in Nairobi County, Kenya \cite{refKenya13}.
Kuria (2014) utilizes machine learning for flood forecasting in the Nzoia river basin, western Kenya \cite{refKenya14}.
Onyango (2021) develops a Twitter sentiment analysis tool for detecting crime hotspots in Nairobi, Kenya \cite{refKenya15}.
Magiya (2020) predicts package delivery time for motorcycles in Nairobi \cite{refKenya16}.
Muthoka et al. (2021) map Opuntia stricta in the arid and semi-arid environment of Kenya using Sentinel-2 imagery and ensemble machine learning classifiers \cite{refKenya17}.
Omolo (2016) creates a mobile and web-based application for security intelligence gathering in Nairobi County \cite{refKenya18}.
Mbani et al.(2020)  employ artificial intelligent agents for crime mapping in Nairobi City County, Kenya \cite{refKenya19}.
Omondi and Boitt (2020) model the spatial distribution of soil heavy metals using a random forest model- a case study of Nairobi and Thirirka Rivers’ confluence \cite{refKenya20}.
Muchuku (2023) assesses recurrent neural networks as a prediction tool for quoted stock prices on the Nairobi Securities Exchange \cite{refKenya21}.


\subsubsection{Small Businesses}

Halkin designs, manufactures and operates Unmanned Aerial Systems (UAS). Halkin is   able to implement embedded systems through our software engineers for additional capability such as faster processing; incorporation to various sensors and systems, providing our own failsafe procedures; Incorporation of image processing, MI and Machine Learning. 

Freshee is a mobile marketplace for deals at every food, drink and entertainment venue to help users Save More and Explore while getting rewarded for loyalty at their favorite places. The venue discovery industry is broken. Venues face several hours of low/no footfall daily, struggle to advertise promotional offers, and effectively subscribe customers to loyalty programs. Customers miss out on deals and venues they would love to visit.

Neural Labs Africa is an innovative medical technology Company using Artificial Intelligence  to transform medical imaging diagnosis.
We have developed (NeuralSight) a technology that screens medical images for Radiologists and Hospitals in real-time. NeuralSight can identify over 20 respiratory, heart and breast diseases which include: Pneumonia, Tuberculosis, COVID-19, Pneumothorax, Cardiomegaly, Benign breast Tumor, Malignant breast Cancer, Atelectasis, Infiltration, etc. 

AI connect is on Conversational MI and Omni-channel customer engagement platform that connects the air traveler to the airline ecosystem using artificial intelligence, machine learning, and customer engagement excellence. 
Farnbec adresses  firsthand the challenges and inconveniences of cooking with LPG. That's why Farnbec developed Gaslink as a solution. Gaslink, is being developed with the goal of revolutionizing the way that Households and Restaurant Chains manage their LPG usage for clean cooking. Using advanced technology including NB-IoT, cloud computing, AI, and APIs, our solution provides real-time tracking and monitoring of LPG cylinder usage, as well as valuable insights and recommendations. 

Wiggles Technologies is a custom software development company that provides dedicated groups of highly-skilled and creative programmers. We deliver custom software applications and mobile solutions, run software testing, perform in-depth product analyses, and provide technology management, support and expertise.

AI Made in Africa helps Startups and  SMEs find talented, diverse tech talents by matching them with candidates of the best culture fit while providing practical levels of flexibility.

Founded in 2017, M-Shule is the first personalized knowledge-building platform in Africa, connecting learners to tailored learning, evaluation, activation, and data tools through SMS and chatbot. Meaning "mobile school" in Swahili, M-Shule combines SMS with artificial intelligence to reach offline or marginalized communities, offering self-paced, interactive, and personalized resources. Initially focusing on academic courses, M-Shule has expanded to include professional courses, life skills, data collection, and behavior change. To date, the platform has reached over twenty thousand households, not only in Kenya but also across East Africa. M-Shule has demonstrated success in over 30 Kenyan counties, Uganda, and Tanzania, covering more than 6 skill development domains and 7 languages, including Dholuo, English, Kamba, Kikuyu, Kiswahili, Ng’aturkana, and Somali.

Nairobi-based climate tech startup Amini is focused on solving Africa's environmental data gap through artificial intelligence and satellite technology and has raised \$2 million in a pre-seed funding round. The Kenyan startup was founded in 2022 was designed to address Africa's data scarcity, facilitate capital investment, promote climate resilience, and accelerate economic development opportunities in the region. Furthermore, the platform also provides access to valuable environmental data analytics, including drought, flood, soil and crop health. This data can be processed to forecast crop yields for smallholder farmers in seconds and to measure the impact of natural disasters across the region. Before the funding, the company initially focused on the insurance industry, however, it is now experiencing rapid expansion into supply chain monitoring, specifically at the "last mile", or the initial stages of the global supply chain.

  Founded  in 2019, AIfluence uses advanced machine learning algorithms to match influencers with a target demographic through its audience-first strategy. The Kenyan startup in 2021 raised a \$1 million seed funding round to accelerate the expansion of its MI-powered marketing platform. The MI-powered marketing platform allows advertisers to onboard and coordinate hundreds to thousands of micro and nano influencers per campaign, generating authentic peer-to-peer conversations and superior conversion.
Sky.Garden is a Kenyan mobile SaaS eCommerce Platform for African retailers.

%
%

\subsubsection{Government}

In 2019, Kenya's Distributed Ledger Technology and Artificial Intelligence Taskforce report provided the government with a strategic direction on developing a roadmap to uphold human rights when adopting emerging technologies like MI. The report recommends leveraging blockchain and MI to combat corruption and enhance state transparency.
The report assesses emerging technologies and their deployment globally, recommending that the government utilize Blockchain Technology and machine intelligence to combat and eliminate corruption, safeguarding the interests of citizens. It advises the government to leverage Blockchain and MI technology solutions to fight corruption and enhance transparency due to their record immutability. 

AICE: Founded in 2020, the AI Centre of Excellence is passionate about creating value and sustainable impact within the African Intelligence and Machine Learning space by
Transforming Data Scientists \& Software engineers into MI and ML Engineers,
Creating sustainable impact through Research and Development,
Providing custom MI as a Service and building MI solutions.
From challenges to solutions, the AI Centre of excellence aims to develop impact within the MI space that allows for growth, innovation and creativity.

\subsection{Madagascar}

\begin{table}[htb]
  \begin{center}
    \begin{tabular}{|p{0.6in}p{0.6in}p{1.4in}|}
 \hline
      Madagascar & 2020-2023 & Concrete Actions \\
      \hline
      Research & $\checkmark$ &  water\\
      SMB & $\checkmark$ &    \\
      Informal Economy & $\checkmark$  &  \\
      Government & $\checkmark$ &  \\
      \hline
    \end{tabular}\label{tab:mytableMadagascar}
  \end{center}
  \caption{MI in  Madagascar}
\end{table}

\subsubsection{Research}

The work in \cite{refMadagascar1} by Clément Le Ludec, Maxime Cornet, Antonio A Casilli explores the impact of MI on labor, focusing on France outsourcing tasks to workers in Madagascar. The study unveils the intricate production chain of MI, revealing the reliance on data workers in low-income countries.
The work in \cite{refMadagascar2} by Fahafahantsoa Rapelanoro Rabenja discusses the PASSION Project in Madagascar and Guinea, using MI for dermatological data collection. The study aims to address the scarcity of dermatologists, emphasizing the potential of MI in enhancing data collection on skin conditions.
The work in \cite{refMadagascar3} by Paola Tubaro, Antonio A Casilli, Marion Coville delves into the role of digital platform labor in MI development. The study highlights micro-workers' functions in MI preparation, verification, and impersonation, emphasizing the enduring significance of micro-work in contemporary MI production processes.
The work in \cite{refMadagascar4} by Daniele Silvestro, Stefano Goria, Thomas Sterner, Alexandre Antonelli introduces a framework, CAPTAIN, for spatial conservation prioritization using reinforcement learning. The study demonstrates the efficacy of MI in maximizing biodiversity protection under limited budgets, presenting a promising approach for conservation in a resource-limited world.
The work in \cite{refMadagascar5} by Sandro Valerio Silva, Tobias Andermann, Alexander Zizka, Gregor Kozlowski, Daniele Silvestro addresses the global conservation crisis for tree species. The study employs MI to estimate and map the conservation status of over 21,000 tree species, revealing insights into threatened species distribution and providing efficient approximations of extinction risk assessments.

 The work in \cite{refMadagascar6} by Harimino Andriamalala Rajaonarisoa et al. characterizes the evolution of precipitation in Southern Madagascar using High Order Fuzzy Time Series. The study models annual precipitation data, determining hyperparameters and fuzzy sets to interpret the characteristic evolution of precipitation.
The work in \cite{refMadagascar7} by RABENIAINA Anjara Davio Ulrick and RAKOTOVAO Niry Arinavalona presents a method for modeling the onset and end dates of the monsoon season in Northern Madagascar. It employs a Machine Neural Fuzzy Inference System (ANFIS) based on MI and zonal wind data, providing estimates for the monsoon season.
The work in \cite{refMadagascar8} by JB Koto, TR Ramahefy, S Randrianja focuses on the extraction of knowledge from civil status data (surname and first name) using MI. The study demonstrates the application of MI and Python tools to analyze and visualize patterns in the formulation of names in Madagascar.
The work in \cite{refMadagascar9} by Paola Tubaro and Antonio A Casilli explores the role of micro-work in the "back-office" of MI, particularly in the automotive industry. The study highlights the labor-intensive process of MI production, emphasizing the structural need for micro-workers in data annotation, tagging, and labeling for smart solutions in the industry.

The work in \cite{refMadagascar10} by TR Rasamoela and J Szpytko explores the implementation of telematics in the transport system in Antananarivo, Madagascar. The paper emphasizes the significance of reliable transportation for economic growth and poverty reduction, proposing the use of telematics as a solution to enhance the transport sector in Madagascar. 
The work in \cite{refMadagascar11} by Manuel Dominguez-Rodrigo  et al. introduces a breakthrough method that utilizes MI  and computer vision techniques to achieve high accuracy in the classification of modern and ancient bone surface modifications. The study demonstrates the potential of MI in objectively identifying hominin butchery traces in the archaeological record.
The work in \cite{refMadagascar12} by Matteo Giuliani et al. presents the Climate State Intelligence framework, employing MI to detect the state of multiple global climate signals. The framework enhances seasonal forecasts, particularly in the Lake Como basin, providing valuable information for water system operations and improving system performance.
The work in \cite{refMadagascar13} by Dominique Badariotti et al. introduces SIMPEST, an agent-based model designed to simulate plague epidemics in Madagascar. The research focuses on understanding the behavior and spread of plague in the environment, aiming for better control and management of this epidemiological case.
The work in \cite{refMadagascar14} by Ala Saleh Alluhaidan investigates public perception of drones as a tool for telecommunication technologies. The study explores how the public views drones, particularly in healthcare applications, and identifies concerns related to safety, security, and privacy. The results highlight the need for increased public awareness and education about drone technology.

The work in \cite{refMadagascar15} examines the use of fuzzy inference modeling to predict the beginning and ending dates of rain in the coastal areas of South East Madagascar. The model, based on MI and fuzzy logic, covers the period from 1980 to 2017 and demonstrates excellent performance with a calculated MAPE of less than 10\%.
The work in \cite{refMadagascar16} presents an abstractive text summarization approach for the Malagasy language. Utilizing the Scheduled Sampling model and deep learning, the study focuses on summarizing content in a more natural and harmonious manner. The results indicate the applicability of deep learning to the Malagasy language.
The work in \cite{refMadagascar17} focuses on estimating deforestation in tropical humid and dry forests in Madagascar from 2000 to 2010. Using multi-date Landsat satellite images and a random forests classifier, the study provides high-resolution deforestation maps with reliable uncertainty estimates, crucial for forest conservation and management.
The work in \cite{refMadagascar18} explores the enhancement of a budget simulation model for decentralized territorial authorities in Madagascar using MI. The study emphasizes the importance of predictive analyses in better managing budget implementation by considering various factors such as economic, political, and performance indicators.
The work in \cite{refMadagascar19} investigates climatic factors affecting monthly rainfall variability in a remote region of Madagascar. Machine learning models, analyzing past weather conditions and relevant climate indices, contribute to the development of short-to-medium-range rainfall outlook models.
The work in \cite{refMadagascar20} applies an Machine neural network approach to forecast infant mortality rate in Madagascar. Covering the period 1960-2020, the study's stable model predicts that the infant mortality rate will be around 35/1000 live births per year in the out-of-sample period, emphasizing the need for maternal and child care programs.
The work in \cite{refMadagascar21} introduces a one-dimensional convolutional neural network for visible and near-infrared spectroscopy to improve soil phosphorus prediction in Madagascar. The study demonstrates the model's superior predictive accuracy compared to traditional regression methods, contributing to effective fertilizer management and ecosystem sustainability.

%
%
%
%
%
%


\subsection{Malawi}

\begin{table}[htb]
  \begin{center}
    \begin{tabular}{|p{0.6in}p{0.6in}p{1.4in}|}
 \hline
      Malawi & 2020-2023 & Concrete Actions \\
      \hline
      Research & $\checkmark$ &  water\\
      SMB & $\checkmark$ &    \\
      Informal Economy & $\checkmark$  &  \\
      Government & $\checkmark$ & Centre for Artificial Intelligence and STEAM - Science, Technology, Engineering, Arts and Mathematics   \\
      \hline
    \end{tabular}\label{tab:mytablemal}
  \end{center}
  \caption{MI in Malawi}
\end{table}

\subsubsection{Research}

In \cite{refMalawi1}, 
Poverty alleviation in Malawi is explored through machine learning models utilizing existing survey data to predict poor and non-poor households. Open-source algorithms such as Logistic Regression, Extra Gradient Boosting Machine, and Light Gradient Boosting Machine demonstrate accuracy comparable to full feature sets, suggesting the potential for shorter, lower-cost surveys.

In \cite{refMalawi2} machine learning is used, specifically a random forest model, on high-frequency household survey data in southern Malawi to infer predictors of food insecurity. The model outperforms others, emphasizing the significance of location and self-reported welfare as predictors. Various models are evaluated for forecasting food security outcomes.

\cite{refMalawi3} introduces an energy-climate-water framework, combining satellite data and machine learning, to assess the impact of hydro-climatic variability on hydropower reliability in Malawi. The approach, validated for the period 2000 - 2018, mitigates data scarcity and enhances understanding of vulnerabilities in the power sector.

The work in \cite{refMalawi4} examines legislation in Malawi.  Legal research in Malawi faces challenges with limited resources. This interdisciplinary research builds tools for annotating Malawi criminal law decisions with legal meta-data using machine learning tools, spaCy, and Gensim LDA. The study sets the foundation for classifying Malawi criminal case law according to the International Classification of Crime for Statistical Purposes.

In \cite{refMalawi5}, the authors compare Machine Learning  methods with hedonic pricing using household survey data from Uganda, Tanzania, and Malawi. ML methods such as Boosting, Bagging, Forest, Ridge, and LASSO outperform OLS models, providing superior prediction of rental values in housing surveys.

The work in \cite{refMalawi6} explores the adoption of conservation agriculture  in Malawi, finding that peer effects, particularly adoption by neighbors, play a crucial role. The study highlights the significance of considering social dynamics and peer influence in promoting CA interventions.

%
%
%
%
%
%
\subsubsection{Government}

In October 2023, Malawi launched its first-ever Centre for Artificial Intelligence and STEAM - Science, Technology, Engineering, Arts and Mathematics -  at the Malawi University of Science and Technology. Established with support from various U.S.-based universities, the center aims to provide solutions to the country's innovation and technology needs. 
%

\subsection{Mauritius}

\begin{table}[htb]
  \begin{center}
    \begin{tabular}{|p{0.6in}p{0.6in}p{1.4in}|}
 \hline
      Mauritius & 2020-2023 & Concrete Actions \\
      \hline
      Research & $\checkmark$ &  FruVegy\\
      SMB & $\checkmark$ &  Qubitica, Cash Radix, AgCelerant, Arie Finance, 4Sight  \\
      \hline
    \end{tabular}\label{tab:mytableMauritius}
  \end{center}
  \caption{MI in Mauritius}
\end{table}

\subsubsection{Research}

The study \cite{refMauritius1} looks at plants called invasive flora alien species (IAS), which can harm the variety of life in tropical forests. They focused on one specific plant, strawberry guava, and used pictures from satellites that anyone can access to learn more about how these plants affect tropical forests. This might be the first time someone used these free satellite pictures to create a map of strawberry guava and the first time they used this method to map invasive species in Mauritius.

In a park in Mauritius called Black River Gorges National Park (BRGNP), the researchers did some on-the-ground observations and collected 4670 samples to understand how much strawberry guava covered different areas. They used 70\% of this information to teach their computer models and make them better, and the other 30\% they kept to test how accurate their models were. They used special satellite images and a tool called Google Earth Engine for this. They also used some calculations to help them understand the colors and textures of the strawberry guava plants in the pictures.

Their computer models, called Random Forest and Support Vector Machine, did a really good job. RF was recommended for future studies because it was very accurate (97.60\%  $\pm$ 0.20\% with 95\% confidence) and made predictions in more consistent areas. They also found that strawberry guava was most common in the central parts of BRGNP and on steeper slopes. Surprisingly, the amount of strawberry guava didn't change much from 2016 to 2020.

In \cite{refMauritius2}, the authors explore new ways, like using computers to analyze lots of data, to predict how bad accidents might be.

They tried different computer methods, like Support Vector Machine, Gradient Boosting, Logistic Regression, Random Forest, and Naive Bayes, all using a programming language called Python. The method called Gradient Boosting did the best job in figuring out how severe accidents could be. It was right about 83.2\% of the time, which is pretty good, and it had an AUC of 83.9\%, showing it's effective in making these predictions.

\cite{refMauritius3} examines proper identification of plant species has major benefits for a wide range of stakeholders ranging from forestry services, botanists, taxonomists, physicians, pharmaceutical laboratories, organisations fighting for endangered species, government and the public at large. Consequently, this has fueled an interest in developing automated systems for the recognition of different plant species. A fully automated method for the recognition of medicinal plants using computer vision and machine learning techniques has been presented. Leaves from 24 different medicinal plant species were collected and photographed using a smartphone in a laboratory setting. A large number of features were extracted from each leaf such as its length, width, perimeter, area, number of vertices, colour, perimeter and area of hull. Several derived features were then computed from these attributes. The best results were obtained from a random forest classifier using a 10-fold crossvalidation technique. With an accuracy of 90.1\%, the random forest classifier performed better than other machine learning approaches such as the k-nearest neighbour, naive Bayes, support vector machines and neural networks. These results are very encouraging and future work will be geared towards using a larger dataset and high-performance computing facilities to investigate the performance of deep learning neural networks to identify medicinal plants used in primary health care. To the best of our knowledge, this work is the first of its kind to have created a unique image dataset for medicinal plants that are available on the island of Mauritius. It is anticipated that a web-based or mobile computer system for the automatic recognition of medicinal plants will help the local population to improve their knowledge on medicinal plants, help taxonomists to develop more efficient species identification techniques and will also contribute significantly in the protection of endangered species.

The research work 
\cite{refMauritius4} studies flood prediction using Machine neural networks in Mauritius.
The average temperature of the earth is increasing at an alarming rate and it has been envisaged to increase by a factor of about 1.4 to 5.8 degree Celsius by the year 2100. An increase in the atmospheric temperature entails the occurrence of many extreme events such as stronger heat waves, formation of intense cyclones, unprecedented flash floods and severe drought events  which are set to impact greatly on both the global economy and society. Among the various natural disasters, which affect mankind, flash floods have been reported to cause more casualties in terms of economic loss, death tolls and infrastructural damages. Flooding has become a recurrent phenomenon in the recent decade accounting for about 73\% of damages caused by natural disasters which in turn results in an overall loss of about \$30 billions. Flash floods are thus a global phenomenon affecting major parts of the world  as indicated for the year 2018, which marked the occurrence of several deadly flash floods in Kerala, France and Vietnam. In \cite{refMauritius4}  the focus is  on Mauritius, which is a small island located in the Indian Ocean, off the east coast of Africa and Madagascar. The morphological landscape of Mauritius consists of highlands and coastal regions in a relatively small geographical area of 1865 km² such that it is typical for the island to experience several microclimates on the same day in different regions. Their study is especially motivated by the occurrence of a series of flash floods in Mauritius.

Receiving and managing complaints effectively are important for organisations which aim to provide excellent customer service. In order for this to happen, organisations should make it quick and easy for users to report issues. In \cite{refMauritius5} , a smart mobile application for complaints management in Mauritius is described. Users of this mobile application can report issues for different organisations using a single application on their smartphones. They can register complaints using text, images or videos, and they do not have to specify which authority the complaint is directed to. Instead, the application uses text and image analysis alongside a Convolutional Neural Network in order to direct complaints to the correct utility organisations. The classifiers have been trained to identify different categories of complaints for each local utility organisation. Users are notified regarding the status of their complaints and can use the application to directly communicate with the personnel.

Small Island Developing States (SIDS), like Mauritius, share similar sustainable development challenges inherent to their characteristics. Growth in the global energy demand and fears of energy supply disruptions, have triggered much debate geared towards the necessity for sustainable energy planning. Accurate forecasting of future electricity demand is an essential input to this process. Such forecasts are also important in regional or national power system strategy Management. Non linearity of the factors adds complexity to the electricity load forecasting process. Statistical learning theory, in the form of Support Vector Machines, have been used successfully to tackle nonlinear regression and time series problems. However application to the electricity demand forecasting problem with focus on SIDS'characteristics is lacking. The article \cite{refMauritius6}  focuses on the application of SVMs to forecast electricity demand of a SIDS member, Mauritius. A two years ahead forecast, for 2008 and 2009, was derived using monthly time series data from years 1996 to 2007. The inputs considered were historical electricity demand and prices, temperature, humidity, population and GDP.

To facilitate the recognition and classification of medicinal plants that are commonly used by Mauritians, a mobile application which can recognize seventy different medicinal plants has been developed in \cite{refMauritius7}. A convolutional neural network  based on the TensorFlow framework has been used to create the classification model. The system has a recognition accuracy of more than 90\%. Once the plant is recognized, a number of useful information is displayed to the user. Such information includes the common name of the plant, its English name and also its scientific name. The plant is also classified as either exotic or endemic followed by its medicinal applications and a short description. Contrary to similar systems, the application does not require an internet connection to work. Also, there are no pre-processing steps, and the images can be taken in broad daylight. Furthermore, any part of the plant can be photographed. It is a fast and non-intrusive method to identify medicinal plants. This mobile application will help the Mauritian population to increase their familiarity of medicinal plants, help taxonomists to experiment with new ways of identifying plant species, and will also contribute to the protection of endangered plant species.

The research article in \cite{refMauritius8} investigates the application of supervised machine learning techniques to predict the price of used cars in Mauritius. The predictions are based on historical data collected from daily newspapers. Different techniques like multiple linear regression analysis, k-nearest neighbors, naive bayes and decision trees have been used to make the predictions. The predictions are then evaluated and compared in order to find those which provide the best performances. A seemingly easy problem turned out to be indeed very difficult to resolve with high accuracy. All the four methods provided comparable performance.

In \cite{refMauritius9} a Machine Learning Technique called the Support Vector Machine  is adopted on the Stock Exchange of Mauritius (SEM) to determine if stock market returns are predictable based on information from past prices, allowing arbitrage opportunities for abnormal profit generation. The serial correlation test, used as benchmark, and the SVM technique show evidence that previous information on share prices as well as the indicators constructed are useful in predicting share price movements. The implications of the study are that investors have the prospect of adopting speculative strategies and profits from trading based on information and advanced techniques and models are possible.

In this era of education and technology, it is undeniable that there is a growing interaction between machine and humans. Student performance is of prime importance as education is the key to success. At the university of Mauritius, the number of students enrolled in a course does not match the number of students graduating as not every student complete their academic cycle of 3 or 4 years. Some extend their course duration as they have to repeat the whole year or several modules, while others exit with a certificate or diploma since they lack the required number of credits to obtain a degree. Unfortunately, the registration of some students with very low average marks are terminated. The research work \cite{refMauritius10}  investigates a machine learning model to predict the performance of university students on a yearly basis. The model will forecast student performance and help take necessary actions before it is too late. The classification technique is used to train the proposed model using an existing student dataset. The training phase generates a training model that can then be used to predict student performance based on parameters such as attendance, marks, study hours, health or average performance. Different algorithms are evaluated and the classification and prediction algorithms which are more accurate are recommended

The Mauritius MI Strategy 2018, established by the government, aims at making MI  a cornerstone of the next development model by recognizing the potential of technology to improve growth, productivity and quality of life. In this regard, MI has already started to shape the legal sector, for instance, by assisting law practitioners to identify and minimize bias in client intake, offer initial consultation solutions, expand the scope of information for law practitioners and predict the outcome of future legal cases, among others. Nevertheless, while the legal profession worldwide is facing pressure to innovate and transform, the emergence of MI is causing significant disruption to long-established practices in the legal world, especially since this particular sector has traditionally under-utilized technology. The work in \cite{refMauritius11} seeks to assess the influence of MI on employees from the legal profession mainly in terms of their performance, their reaction, and adaptability to change and to identify the challenges faced by these employees in Mauritius in adopting MI for their operational activities.

Sentiment analysis is becoming increasing important with the rise in the amount of content on social media. However, sentiment analysis remains challenging for under-resourced languages such as Kreol Morisien (KM), the native language of Mauritius. In fact, it has been observed that in Mauritius, social media comments often consist of more than one language among English, French and Kreol Morisien. In work in \cite{refMauritius12}  first creates an annotated dataset of 1300 sentences and then outline a framework through which sentiment analysis can be performed on social media comments. We propose a KM sentiment analyzer using two algorithms namely Support Vector Machine (SVM) and Multinomial Naive Bayes (MNB). Our results show that SVM outperforms MNB for sentiment analysis in Kreol Morisien, achieving an accuracy of 66.15\% after pre-processing techniques stopwords removal and spell checking are applied. This paper highlights the need to develop further tools in order to enable natural language processing of Kreol Morisien.

Fruit Flies impact the field of agriculture in a negative way affecting the economy of a host country. This work in \cite{refMauritius13} presents an identification system that can be deployed on a mobile application. The identification system uses Convolutional Neural Network (CNN) to learn the key visual features of fruit flies to be able to perform detections. The system comprises of two main aspects; a detection model and a classification model. The detection model uses Single Snapshot Detector (SSD) MobileNet V2 FPNLite 640x640 model which is converted to a TensorFlow Lite version and hosted locally on the mobile phone. The classification model uses Xception model which is hosted on the Google Cloud Platform where requests are made to the cloud from the mobile application. A custom image dataset of nine(9) fruit flies was created in two ways: Two fruit flies predominant in Mauritius were obtained from the Entomology Department of the Ministry of Agro Industry and Food Security and photographs were taken. The remaining fruit flies' images were obtained through web scraping. Transfer learning has been successfully used to produce the SSD MobileNetV2 FPNLite 640x640 model with a loss of 31\% and the Xception model with an accuracy of 75.5\%.

Birds communicate with their colonies through sound and inform them of potential problems like forest fires. The identification of bird sounds is therefore very important and has the potential to solve some global problems. Convolutional neural networks (CNNs) are sophisticated deep learning algorithms that have proven to be effective in image processing and in sound classification. The work in \cite{refMauritius14} describes the work done to develop a tool using a deep learning model for classifying Mauritius bird sounds from audio recordings. A dataset obtained from the Xeno-canto bird song sharing site, which hosts a vast collection of labeled and classified recordings, is used to fine-tune three pre-trained CNN models, namely InceptionV3, MobileNetV2 and RestNet50 and a custom model. The neural network's input is represented by spectrograms created from downloaded mp3 files. Time shifting and pitch stretching have been used for data augmentation. The best performing model has been integrated into a website to identify birds sounds recordings. In this work, transfer learning has been used successfully to produce a model with a weighted accuracy of 84\%. Although a custom CNN was trained, better accuracy was achieved through the use of transfer learning.

Nowadays, many people are unaware of the benefits of fruits and vegetables which has resulted in their reduced consumption. This has inevitably led to a rise in diseases such as obesity, high blood pressure and heart diseases. To this end, \cite{refMauritius15} developed FruVegy which is an android app which can automatically identify fruits and vegetables and then display its nutritional values. The app can identify forty different fruits/vegetables. The app is specially targeting school students who will find it easy and fun to use and this, we believe, will increase their interest in the consumption of fruits and vegetables. Furthermore, the names of the fruits and vegetables are also available in French and in Mauritian Creole. A dataset  of 1600 images from 40 different fruits and vegetables is proposed. There was an equal number of images for each fruit/vegetable. Features such as shape, color and texture were extracted from each image. Different machine learning classifiers were tested but random forest with 100 trees produced the best result with an accuracy of 90.6\%. However, with TensorFlow, an average accuracy of 98.1\% was obtained under different scenarios.


\subsubsection{Small Businesses}
Qubitica: Decentralized autonomous organization. It offers QBIT, an ERC-20 token that serves as a membership token and a currency. It works on blockchain and MI projects with the help of developers and entrepreneurs. It allows investors to invest in projects these projects. It focuses on projects such as mining, trading \& exchange, news platform, voting system, node services, accounting, and more.

Cash Radix: AI-based trading signals for forex market assets. It provides forex recommendations with the market analysis, Machine intelligent technical analysis software robot, indicators, and more. It features charts, fundamental analysis, broker recommendation, forex, calculator, and more. It also offers a training platform for forex market trading.

AgCelerant: Provider of phygital agriculture solutions to agribusinesses using earth observation, IoT, and MI. They offer physically based, digitally-driven solutions to secure the sustainability, transparency, and sourcing of food. The platform connects smallholders contract farming, producers with banks, insurers, input providers, and agro-industries to control risks, secure transactions, reduce frictions and sustainably improve productivity and welfare. It implements a business model in which smallholder farmers and larger-scale investors are simultaneously accompanied and protected as they empower themselves to respond to growing and changing customer needs.

Arie Finance : Software for core banking. It offers core banking solutions that enable bank accounts, cashflows, payments, investments, and more. It provides a dashboard to monitor verifications, customers, exchange rates; omnichannel communication including mobile and  web applications, and more.

4Sight Holdings Limited (4Sight) is a public company listed on the JSE AltX  incorporated on 29 June 2017 in accordance with the laws of the Republic of Mauritius. 4Sight focuses on a cross-section of established, new, and emerging technologies. These include MI solutions with machine learning, big data, cloud and business intelligence solutions, digital twin and simulation, information and operational technologies, production scheduling, horizontal and vertical integration, industrial internet of things, cloud service provider, robotic process automation and augmented and virtual reality solutions.



\subsubsection{Government} 
  In 2018, the government has developed a national MI strategy and established the Mauritius MI Council and the Mauritius MI Academy to support the development and implementation of MI technology in the country.  The strategy focuses on how MI can support the ocean economy, which comprises over 10\% of Mauritius' GDP. For example, it suggests investment into a maritime Internet of Things (IoT).
The strategy also established an MI Council that advises the government on supporting Mauritius' MI ecosystem. Both the MI Strategy and the Mauritius 2030 Strategic Plan prioritize developing local talent, such as through making programming a required university course.

\subsection{Mozambique}

\begin{table}[htb]
  \begin{center}
    \begin{tabular}{|p{0.6in}p{0.6in}p{1.4in}|}
 \hline
      Mozambique & 2020-2023 & Concrete Actions \\
      \hline
      Research & $\checkmark$ &  water\\
      SMB & $\checkmark$ &    \\
      \hline
    \end{tabular}\label{tab:mytablemo}
  \end{center}
  \caption{MI in Mozambique}
\end{table}
\subsubsection{Research}

\cite{refMozambique1} examines the quality of OpenStreetMap in Mozambique using unsupervised machine learning. Anyone can contribute geographic information to OpenStreetMap (OSM), regardless of their level of experience or skills, which has raised concerns about quality. When reference data is not available to assess the quality of OSM data, intrinsic methods that assess the data and its metadata can be used. In this study, we applied unsupervised machine learning for analysing OSM history data to get a better understanding of who contributed when and how in Mozambique. Even though no absolute statements can be made about the quality of the data, the results provide valuable insight into the quality. Most of the data in Mozambique (93\%) was contributed by a small group of active contributors (25\%). However, these were less active than the OSM Foundation's definition of active contributorship and the Humanitarian OpenStreetMap Team (HOT) definition for intermediate mappers. Compared to other contributor classifications, our results revealed a new class: contributors who were new in the area and most likely attracted by HOT mapping events during disaster relief operations in Mozambique in 2019. More studies in different parts of the world would establish whether the patterns observed here are typical for developing countries. Intrinsic methods cannot replace ground truthing or extrinsic methods, but provide alternative ways for gaining insight about quality, and they can also be used to inform efforts to further improve the quality. We provide suggestions for how contributor-focused intrinsic quality assessments could be further refined

Accurate land use and land cover (LULC) mapping is essential for scientific and decision-making purposes. \cite{refMozambique2}   maps LULC classes in the northern region of Mozambique between 2011 and 2020 based on Landsat time series processed by the Random Forest classifier in the Google Earth Engine platform. The feature selection method was used to reduce redundant data. The final maps comprised five LULC classes (non-vegetated areas, built-up areas, croplands, open evergreen and deciduous forests, and dense vegetation) with an overall accuracy ranging from 80.5\% to 88.7\%. LULC change detection between 2011 and 2020 revealed that non-vegetated areas had increased by 0.7\%, built-up by 2.0\%, and dense vegetation by 1.3\%. On the other hand, open evergreen and deciduous forests had decreased by 4.1\% and croplands by 0.01\%. The approach used in this paper improves the current systematic mapping approach in Mozambique by minimizing the methodological gaps and reducing the temporal amplitude, thus supporting regional territorial development policies.

 \cite{refMozambique3} provides an overview of the most important practices in the field of MI  used in educational contexts, with a focus on the main platforms used for teaching (LMS) to support the development of a research work at Eduardo Mondlane University  in Mozambique. To that end, definitions and descriptions of relevant terms, a brief historical overview of MI  in education and an overview of the common goals and practices of using computational methods in educational contexts are provided. The state of the art regarding the adaptation and use of MI is presented and we discuss the potential benefits and the open challenges.

\cite{refMozambique4} introduces food security and advanced imaging radiometer datasets and ML models. As part of the chapter, satellite radiometer, dairy, food security and satellite data, global vegetation - Cropland and Vegetation Index, and the Normalized Difference Vegetation Index  are also covered. Their work also covers Mozambique cashew nuts market, agriculture, and industrialization. It concludes with two machine learning models that specifically look at Mozambique cashew nuts production model and Mozambique cashew nuts and Normalized Difference Vegetation Index  model.

\cite{refMozambique5}  maps smallholder irrigated agriculture in sub-Saharan Africa using remote sensing techniques is challenging due to its small and scattered areas and heterogenous cropping practices. A study was conducted to examine the impact of sample size and composition on the accuracy of classifying irrigated agriculture in Mozambique's Manica and Gaza provinces using three algorithms: random forest (RF), support vector machine (SVM), and Machine neural network (ANN). Four scenarios were considered, and the results showed that smaller datasets can achieve high and sufficient accuracies, regardless of their composition. However, the user and producer accuracies of irrigated agriculture do increase when the algorithms are trained with larger datasets. The study also found that the composition of the training data is important, with too few or too many samples of the “irrigated agriculture” class decreasing overall accuracy. The algorithms' robustness depends on the training data's composition, with RF and SVM showing less decrease and spread in accuracies than ANN. The study concludes that the training data size and composition are more important for classification than the algorithms used. RF and SVM are more suitable for the task as they are more robust or less sensitive to outliers than the ANN. Overall, the study provides valuable insights into mapping smallholder irrigated agriculture in sub-Saharan Africa using remote sensing techniques.

\cite{refMozambique6} uses deep learning and Twitter for mapping in Mozambique. Accurate and detailed geographical information digitizing human activity patterns plays an essential role in response to natural disasters. Volunteered geographical information, in particular OpenStreetMap (OSM), shows great potential in providing the knowledge of human settlements to support humanitarian aid, while the availability and quality of OSM remains a major concern. The majority of existing works in assessing OSM data quality focus on either extrinsic or intrinsic analysis, which is insufficient to fulfill the humanitarian mapping scenario to a certain degree. This paper aims to explore OSM missing built-up areas from an integrative perspective of social sensing and remote sensing. First, applying hierarchical DBSCAN clustering algorithm, the clusters of geo-tagged tweets are generated as proxies of human active regions. Then a deep learning based model fine-tuned on existing OSM data is proposed to further map the missing built-up areas. Hit by Cyclone IdMI and Kenneth in 2019, the Republic of Mozambique is selected as the study area to evaluate the proposed method at a national scale. As a result, 13 OSM missing built-up areas are identified and mapped with an over 90\% overall accuracy, being competitive compared to state-of-the-art products, which confirms the effectiveness of the proposed method.

\cite{refMozambique7} compares seven machine learning algorithms. Logistic regression (LR) is the most common prediction model in medicine. In recent years, supervised machine learning (ML) methods have gained popularity. However, there are many concerns about ML utility for small sample sizes. In this study, we aim to compare the performance of 7 algorithms in the prediction of 1-year mortality and clinical progression to AIDS in a small cohort of infants living with HIV from South Africa and Mozambique. The data set (n = 100) was randomly split into 70\% training and 30\% validation set. Seven algorithms (LR, Random Forest (RF), Support Vector Machine (SVM), K-Nearest Neighbor (KNN), Naive Bayes (NB), Machine Neural Network , and Elastic Net) were compared. The variables included as predictors were the same across the models including sociodemographic, virologic, immunologic, and maternal status features. For each of the models, a parameter tuning was performed to select the best-performing hyperparameters using 5 times repeated 10-fold cross-validation. A confusion-matrix was built to assess their accuracy, sensitivity, and specificity. RF ranked as the best algorithm in terms of accuracy (82,8\%), sensitivity (78\%), and AUC (0,73). Regarding specificity and sensitivity, RF showed better performance than the other algorithms in the external validation and the highest AUC. LR showed lower performance compared with RF, SVM, or KNN. The outcome of children living with perinatally acquired HIV can be predicted with considerable accuracy using ML algorithms. Better models would benefit less specialized staff in limited resources countries to improve prompt referral in case of high-risk clinical progression.

\cite{refMozambique8} focuses on machine learning aspects of Bantu language Emakhuwa of Mozambique.  Major advancement in the performance of machine translation models has been made possible in part thanks to the availability of large-scale parallel corpora. But for most languages in the world, the existence of such corpora is rare. Emakhuwa, a language spoken in Mozambique, is like most African languages low-resource in NLP terms. It lacks both computational and linguistic resources. In this paper we describe the creation of the Emakhuwa-Portuguese parallel corpus, which is a collection of texts from the Jehovah's Witness website and a variety of other sources including the African Story Book website, the Universal Declaration of Human Rights and Mozambican legal documents. The dataset contains 47,415 sentence pairs, amounting to 699,976 word tokens of Emakhuwa and 877,595 word tokens in Portuguese. After normalization processes which remain to be completed, the corpus will be made freely available for research use.

%
%
%
%

%
%
%
%
%
%
%
%
%



\subsection{Rwanda}

\begin{table}[htb]
  \begin{center}
    \begin{tabular}{|p{0.6in}p{0.6in}p{1.4in}|}
 \hline
      Rwanda & 2020-2023 & Concrete Actions \\
      \hline
      Research & $\checkmark$ &  water\\
      SMB & $\checkmark$ &  AQUA SAFI, Tabiri Analytics   \\
      Informal Economy & $\checkmark$  &  \\
      Government & $\checkmark$ &  Africa's Centre of Excellence in Artificial Intelligence\\
      \hline
    \end{tabular}\label{tab:mytablerw}
  \end{center}
  \caption{MI in Rwanda}
\end{table}

\subsubsection{Research}

The work in \cite{refrwanda1} examines rainfall-induced landslide prediction in Ngororero District, Rwanda.
The study described in \cite{refrwanda2} focuses on predicting out-of-pocket health expenditures in Rwanda using machine learning techniques.
\cite{refrwanda3} presents a case study on modeling and mapping soil nutrient depletion in the humid highlands of East Africa, specifically in Rwanda, using ensemble machine learning.
The paper in \cite{refrwanda4} employs a machine learning approach to predict the demand for essential medicines in Rwanda based on consumption data.
The research outlined in \cite{refrwanda5} utilizes machine learning techniques for predicting stunting among under-5 children in Rwanda.
\cite{refrwanda6} explores the early detection of students at risk of poor performance in Rwanda's higher education using machine learning techniques.
In \cite{refrwanda7}, the authors compare different machine learning classifiers to predict hospital readmission of heart failure patients in Rwanda.
\cite{refrwanda8} focuses on predicting landslide susceptibility and risks in the upper Nyabarongo catchment of Rwanda using GIS-based machine learning simulations.
The study detailed in \cite{refrwanda9} uses machine learning and remote sensing to value property in Rwanda.
\cite{refrwanda10} introduces a machine learning-based triage tool for children with acute infection in a low-resource setting in Rwanda.
\cite{refrwanda11} employs machine learning and the Internet of Things for malaria outbreak prediction in Rwanda.
The research in \cite{refrwanda12} predicts crop yields for Irish potato and maize in Rwanda using machine learning models.
\cite{refrwanda13} compares supervised machine learning algorithms for road traffic crash prediction models in Rwanda.
\cite{refrwanda14} proposes a data-driven predictive machine learning model for efficiently storing temperature-sensitive medical products, such as vaccines, in Rwandan pharmacies.
\cite{refrwanda15} applies deep learning techniques to estimate greenhouse gases emissions from agricultural activities in Rwanda.
The work in \cite{refrwanda16} focuses on creating farmers' awareness of fall armyworms pest detection at an early stage in Rwanda using deep learning.
\cite{refrwanda17} utilizes UAV-based mapping to support decision-making for banana cultivation in Rwanda.
\cite{refrwanda18} introduces a convolutional neural network for checkbox detection on Rwandan perioperative flowsheets.
\cite{refrwanda19} provides ground truths to support remote-sensing inference of irrigation benefits and effects in Rwanda.
\cite{refrwanda20} compares tree-based models and logistic regression classifiers for predicting business success in Rwanda.
\cite{refrwanda21} introduces an ensemble mode decomposition combined with SVR-RF model for predicting groundwater levels in Eastern Rwandan aquifers.
\cite{refrwanda22} explores IoT and ML-based precision agriculture, focusing on Rwanda's coffee industry.
The research in \cite{refrwanda23} leverages convolutional neural networks and satellite images to map informal settlements in urban settings of the city of Kigali, Rwanda.


\subsubsection{Small Businesses}

Aqua Safi assists fish farmers in improving fish productivity by providing a system to check water quality, manage fish feeding, and adopt best practices for fisheries yield.
Tabiri Analytics is a cybersecurity company building the first affordable, comprehensive, and automated cybersecurity-as-a-service solution for enterprises in underserved markets, utilizing machine learning.

%
%
%

\subsubsection{Government}
In 2023, the Cabinet of the Republic of Rwanda has approved the National Artificial Intelligence Policy, which The Future Society  supported in drafting from 2020 to 2021. The Office of the Prime Minister announced the Policy's approval in a Cabinet resolution communiqu\'e in April 2023. 
The National Artificial Intelligence Policy for the Republic of Rwanda serves as a roadmap to 
enable Rwanda to harness the benefits of MI and mitigate its risks. Building on the mission of the 
Vision 2050, Smart Rwanda Master Plan and other key national plans and policies, it equips 
Rwanda to harness MI for sustainable and inclusive growth. By mobilizing local, regional, and 
international stakeholders, it positions Rwanda to become a leading African Innovation Hub and 
Africa's Centre of Excellence in Artificial Intelligence. The National AI Policy has been developed by MINICT and RURA, with support by GIZ FAIR 
Forward, the Centre for the 4th Industrial Revolution Rwanda (C4IR) and The Future Society. The National AI Policy, which promotes and fosters Rwanda's inclusive and sustainable 
socio-economic transformation, is oriented around the following vision and mission statements.  Bringing together key national stakeholders, including line ministries, regulation authorities, academia, private sector, start-ups, CSOs and development partners, the two-day workshop was held at Lemigo Hotel in Kigali, Rwanda from 26 to 27 September 2023. It was organized by the Rwandan National Commission for UNESCO in collaboration with the UNESCO Regional Office for Eastern Africa, the Ministry of ICT \& Innovation (MINICT), the German National Commission for UNESCO, Rwanda Development Board, University of Rwanda, the National Council for Science and Technology, and the Rwanda Information Society Authority. 
The UNESCO Recommendation on the Ethics of AI, adopted in November 2023 by all 193 UNESCO member states, formed the basis of the national workshop, which focused on the implementation of the Recommendation in Rwanda.

\subsection{Seychelles}

\begin{table}[htb]
  \begin{center}
    \begin{tabular}{|p{0.6in}p{0.6in}p{1.4in}|}
 \hline
      Seychelles & 2020-2023 & Concrete Actions \\
      \hline
      Research & $\checkmark$ &  D. sechellia\\
      SMB & $\checkmark$ &    \\
      Informal Economy & $\checkmark$  &  \\
      Government & $\checkmark$ & National AI Strategy \\
      \hline
    \end{tabular}\label{tab:mytableseychelles}
  \end{center}
  \caption{MI in Seychelles}
\end{table}

\subsubsection{Research}

The research in \cite{refseychelles1} investigates the geomorphological drivers influencing deeper reef habitats around Seychelles, providing insights into the underwater landscape.
\cite{refseychelles2} presents a study mapping the national seagrass extent in Seychelles using PlanetScope NICFI data, contributing to the understanding of coastal ecosystems in the region.
In \cite{refseychelles3}, supervised machine learning is employed to reveal introgressed loci in the genomes of Drosophila simulans and D. sechellia, shedding light on genetic interactions between these species.
\cite{refseychelles4} introduces Bayesian models for multiple outcomes, with an application to the Seychelles Child Development Study, offering a statistical framework for analyzing diverse factors influencing child development in Seychelles. 

%
%
%
%

\subsubsection{Government}
 In 2019, the government of Seychelles launched the National AI Strategy, which aims to promote the development and adoption of MI in Seychelles while ensuring that the benefits are shared equitably. The strategy focuses on four key areas: education and skills development, research and innovation, regulatory framework, and ethical considerations.

\subsection{Somalia}

\begin{table}[htb]
  \begin{center}
    \begin{tabular}{|p{0.6in}p{0.6in}p{1.4in}|}
 \hline
      Somalia & 2020-2023 & Concrete Actions \\
      \hline
      Research & $\checkmark$ &  \\
      SMB & $\checkmark$ &    \\
      Informal Economy & $\checkmark$  &  \\
      Government & $\checkmark$ &  \\
      \hline
    \end{tabular}\label{tab:mytablesomalia}
  \end{center}
  \caption{MI in Somalia}
\end{table}

\subsubsection{Research}

Jetson is an experimental project launched by UNHCR's Innovation Service in 2017 to better understand how data can be used to predict movements of people in Sub-Saharan Africa, particularly in the Horn of Africa. The project combines data science, statistical processes, design-thinking techniques, and qualitative research methods. It actively seeks new data sources, new narratives, and new collaborations in order to keep iterating, and improving. Jetson initially focused on understanding the catalysts that cause people to flee their homes in Somalia. Extensive field research resulted in the definition of ten key variables of forced displacement, such as commodity market prices, rainfall, and violent conflicts. Supported by machine learning, these variables inform an index that allows for short-term predictions of expected migration flows out of Somalia. To fulfil its mission, Jetson works in collaboration with partners such as the World Meteorological Organization, the Met Office in the UK, academia, and other UN institutions such as UN Global Pulse. Overall, Project Jetson demonstrates an innovative use of machine learning in the context of forced migration movements: It runs short-term predictions more efficiently, at a higher frequency, and at lower costs than traditional calculations. Potentially, the project can be replicated to other contexts that currently are regions of frequent forced migration out-flows.

Understanding and analysing sentiment in user-generated content has become crucial with the increasing use of social media and online platforms. However, sentiment analysis in less-resourced languages like Somali poses unique challenges. The work in \cite{refSomalia1} test presents the performance of three ML algorithms (DTC, RFC, XGB) and two DL models (CNN, LSTM) in accurately classifying sentiment in Somali text. The CC100-Somali dataset, comprising 78M monolingual Somali texts from the Common crawl snapshots, is utilized for training and evaluation. The study employed rigorous evaluation techniques, including train-test splits and cross-validation, to assess classification accuracy and performance metrics. The results demonstrated that DTC achieved the highest accuracy among ML algorithms, 87.94\%, while LSTM achieved the highest accuracy among DL models, 88.58\%. This study's findings contribute to sentiment analysis in less-resourced languages, specifically Somali, and provide valuable insights into the performance of ML and DL techniques. Moreover, the study highlights the potential of leveraging both ML and DL approaches to analyze sentiment in Somali text effectively. The results and evaluation metrics benchmark future research in sentiment analysis for Somali and other low-resource languages. 

%
%
%
%
%
%

\subsection{South Sudan}

\begin{table}[htb]
  \begin{center}
    \begin{tabular}{|p{0.6in}p{0.6in}p{1.4in}|}
 \hline
     South  Sudan  & 2020-2023 & Concrete Actions \\
      \hline
      Research & $\checkmark$ &  \\
      SMB & $\checkmark$ &    \\
      Informal Economy & $\checkmark$  &  \\
      Government & $\checkmark$ &  \\
      \hline
    \end{tabular}\label{tab:mytabless}
  \end{center}
  \caption{MI in South Sudan}
\end{table}

\subsubsection{Research}

\cite{refSouthSudan1} introduces and applies a set of  machine intelligence techniques to analyze multi-dimensional fragility-related data. Our analysis of the fragility data collected by the OECD for its States of Fragility index showed that the use of such techniques could provide further insights into the non-linear relationships and diverse drivers of state fragility, highlighting the importance of a nuanced and context-specific approach to understanding and addressing this multi-aspect issue. They applied the methodology  to South Sudan, one of the most fragile countries in the world to analyze the dynamics behind the different aspects of fragility over time. The results could be used to improve the Fund's country engagement strategy (CES) and efforts in the country.

%
%
%
%
%


\subsection{Tanzania}

\begin{table}[htb]
  \begin{center}
    \begin{tabular}{|p{0.6in}p{0.6in}p{1.4in}|}
 \hline
      Tanzania & 2020-2023 & Concrete Actions \\
      \hline
      Research & $\checkmark$ &  water\\
      SMB & $\checkmark$ &    \\
      Informal Economy & $\checkmark$  &  \\
      Government & $\checkmark$ &  \\
      \hline
    \end{tabular}\label{tab:mytabletanzania}
  \end{center}
  \caption{MI in Tanzania}
\end{table}

\subsubsection{Research}

The study in \cite{refTanzania1} aims to explore the current status, challenges, and opportunities for MI application in the health system in Tanzania.
A scoping review was conducted using the Preferred Reporting Items for Systematic Review and Meta-Analysis Extensions for Scoping Review (PRISMA-ScR). They searched different electronic databases such as PubMed, Embase, African Journal Online, and Google Scholar.
Eighteen (18) studies met the inclusion criteria out of 2,017 studies from different electronic databases and known MI-related project websites. Amongst MI-driven solutions, the studies mostly used machine learning and deep learning for various purposes, including prediction and diagnosis of diseases and vaccine stock optimisation. The most commonly used algorithms were conventional machine learning, including Random Forest and Neural network, Naive Bayes K-Nearest Neighbour and Logistic regression.
This review shows that MI-based innovations may have a role in improving health service delivery, including early outbreak prediction and detection, disease diagnosis and treatment, and efficient management of healthcare resources in Tanzania. Their results indicate the need for developing national MI policies and regulatory frameworks for adopting responsible and ethical AI solutions in the health sector in accordance with the World Health Organisation  guidance on ethics and governance of MI for health. 

In Dar es Salaam and many other cities, poorer neighborhoods tend to have less vegetation and tree canopy around them. Take Namanga, a densely packed informal settlement in eastern Dar es Salaam. Despite abutting some of the city's wealthiest and greenest neighborhoods, residents of Namanga must endure weather extremes with barely any green cover to provide shade or absorb run-off from sudden downpours. Understanding which districts have ample or inadequate tree canopy cover is challenging but it is becoming more feasible due to increasingly detailed satellite imagery. Machine learning algorithms are a key technology for interpreting such imagery. Still, algorithms to detect features such as tree canopy are only as good as the data they are built on. When their team applied an off-the-shelf tree detection algorithm (developed using a tree canopy dataset from California) to satellite imagery of Dar es Salaam, the results were unsatisfactory \cite{refTanzania2}. Data labeling is the process of adding meaningful information to raw data so that computers can learn to recognize patterns in it, for example, annotating recordings of human speech with the words they contain or identifying objects in photographs.  To produce detailed tree-cover maps of Dar es Salaam and Freetown, the capital of Sierra Leone, the Resilience Academy students began by developing their own large dataset of labeled satellite imagery. Using an open-source labeling tool developed by Azavea, the students loaded high-resolution satellite imagery, divided it into grid cells, and drew accurate boundaries around the tree canopy. By labeling just 1\% of the city in this way, the resulting dataset enabled a machine learning model to learn how to recognize its trees, distinguishing tree canopy from grass, buildings and other features, even in shady conditions.

%
%
%
%
%
%

\subsection{Uganda}

\begin{table}[htb]
  \begin{center}
    \begin{tabular}{|p{0.6in}p{0.6in}p{1.4in}|}
 \hline
      Uganda & 2020-2023 & Concrete Actions \\
      \hline
      Research & $\checkmark$ & speech datasets\\
      SMB & $\checkmark$ &  Chil Ai Lab, Global Auto Systems, Wekebere   \\
      Informal Economy & $\checkmark$  &  \\
      Government & $\checkmark$ & Artificial Intelligence and Data Science Lab  \\
      \hline
    \end{tabular}\label{tab:mytableug}
  \end{center}
  \caption{MI in Uganda}
\end{table}

\subsubsection{Research}

The project \cite{refUganda1} aims  deliver open, accessible, and high-quality text and speech datasets for low-resource East African languages from Uganda, Tanzania, and Kenya. Taking advantage of the advances in NLP and voice technology requires a large corpora of high quality text and speech datasets. This project will aim to provide this data for these languages: Luganda, Runyankore-Rukiga, Acholi, Swahili, and Lumasaaba. The speech data for Luganda and Swahilli will be geared towards training a speech-to-text engine for an SDG relevant use-case and general-purpose ASR models that could be used in tasks such as driving aids for people with disabilities and development of MI tutors to support early education. Monolingual and parallel text corpora will be used in several NLP applications that need NLP models, including natural language classification, topic classification, sentiment analysis, spell checking and correction, and machine translation. 


\subsubsection{Small Businesses}

Chil AI utilizes Machine Learning and Artificial Intelligence to offer Telehealth services, Electronic medical records, E-consultation, automated laboratory results interpretation, E-referral, and E-pharmacy services to African women. Global Auto Systems aims to revolutionize the healthcare system in Uganda by using AI and Cloud Computing technologies to improve patient outcomes while reducing the total cost of care. Wekebere is a health social enterprise striving to engineer innovative healthcare solutions that give expectant mothers in low-resource settings the healthy lives they deserve.

%
%

\subsubsection{Government}

 The Artificial Intelligence and Data Science lab in Uganda specializes in the application of artificial intelligence and data science - including, for example, methods from computer vision, natural language processing and predictive analytics-to problems in the developing world.
Applications include  Natural language processing for under-resourced languages, automated diagnosis of both crop and human diseases, auction design for mobile commodity markets, analysis of traffic patterns in African cities, and of telecoms and remote sensing data for anticipating the spread of infectious diseases



\subsection{Zambia}

\begin{table}[htb]
  \begin{center}
    \begin{tabular}{|p{0.6in}p{0.6in}p{1.4in}|}
 \hline
      Zambia & 2020-2023 & Concrete Actions \\
      \hline
      Research & $\checkmark$ &  Mining\\
      SMB & $\checkmark$ &    \\
      Informal Economy & $\checkmark$  &  \\
      Government & $\checkmark$ &  \\
      \hline
    \end{tabular}\label{tab:mytablez}
  \end{center}
  \caption{MI in Zambia}
\end{table}

\subsubsection{Research}

Stunting is a global public health issue. We sought to train and evaluate machine learning (ML) classification algorithms on the Zambia Demographic Health Survey (ZDHS) dataset to predict stunting among children under the age of five in Zambia. The authors of \cite{refZambia1} applied Logistic regression , Random Forest, SV classification, XG Boost (XgB) and Naive Bayes algorithms to predict the probability of stunting among children under five years of age, on the 2018 ZDHS dataset. We calibrated predicted probabilities and plotted the calibration curves to compare model performance. We computed accuracy, recall, precision and F1 for each machine learning algorithm. About 2327 (34.2\%) children were stunted. Thirteen of fifty-eight features were selected for inclusion in the model using random forest. Calibrating the predicted probabilities improved the performance of machine learning algorithms when evaluated using calibration curves. RF was the most accurate algorithm, with an accuracy score of 79\% in the testing and 61.6\% in the training data while Naive Bayesian was the worst performing algorithm for predicting stunting among children under five in Zambia using the 2018 ZDHS dataset. ML models aids quick diagnosis of stunting and the timely development of interventions aimed at preventing stunting.

\cite{refZambia2} examines audits in Zambia.  Independent verification is a critical component of performance-based financing (PBF) in health care, in which facilities are offered incentives to increase the volume of specific services but the same incentives may lead them to over-report. We examine alternative strategies for targeted sampling of health clinics for independent verification. Specifically, we empirically compare several methods of random sampling and predictive modeling on data from a Zambian PBF pilot that contains reported and verified performance for quantity indicators of 140 clinics. Our results indicate that machine learning methods, particularly Random Forest, outperform other approaches and can increase the cost-effectiveness of verification activities.

Air quality monitoring in Zambian mining towns is an important issue due to the high levels of pollution caused by mining activities. Zambia is a country rich in minerals and mining is a significant contributor to its economy. However, mining activities have also led to increased levels of air pollution in mining towns, affecting the health of local communities. According to the Ministry of Mines, the major sources of air pollution in the Copperbelt are smelters, mining, and quarrying among others. Additionally, the Ministry of Mines reports that major pollutants include sulfur dioxide (SO2), oxides of nitrogen (NOx), particulate matter, carbon monoxide (CO), dust, Carbon dioxide, etc. There are several government agencies engaged in management that can help with these environmental issues, including the Zambia Environmental Management Agency (ZEMA). A research was investigated  in \cite{refZambia3} by using a thorough review of the literature, furthermore, a qualitative study was conducted at ZEMA the primary institution for environmental monitoring, and specifically, interviews were conducted. This was done in order to gain an in-depth overview of the current state of the art for environmental pollutant monitoring in affected mining towns. According to the findings presented here, the country has not made enough investments in environmental monitoring technologies and instead relies on funded projects that render the agency responsible for preventing and controlling ambient pollution inoperable after the projects are completed, despite the fact that there are plenty of mineral resources available and more are still to be discovered. The research suggested new techniques for comparing ambient air pollutant levels to national guideline limits based on the limitations of its results. This study uses data from an ongoing obstetrical cohort in Lusaka, Zambia that uses early pregnancy ultrasound to estimate GA. Our intent was to identify the best set of parameters commonly available at delivery to correctly categorize births as either preterm (37 weeks) or term, compared to GA assigned by early ultrasound as the gold standard. Trained midwives conducted a newborn assessment (72 hours) and collected maternal and neonatal data at the time of delivery or shortly thereafter. New Ballard Score (NBS), last menstrual period (LMP), and birth weight were used individually to assign GA at delivery and categorize each birth as either preterm or term. Additionally, machine learning techniques incorporated combinations of these measures with several maternal and newborn characteristics associated with prematurity and SGA to develop GA at delivery and preterm birth prediction models. The distribution and accuracy of all models were compared to early ultrasound dating. Within our live-born cohort to date (n = 862), the median GA at delivery by early ultrasound was 39.4 weeks. Among assessed newborns with complete data included in this analysis (n = 468), the median GA by ultrasound was 39.6 weeks. Using machine learning, we identified a combination of six accessible parameters (LMP, birth weight, twin delivery, maternal height, hypertension in labor, and HIV serostatus) that can be used by machine learning to outperform current GA prediction methods. For preterm birth prediction, this combination of covariates correctly classified 94\% of newborns and achieved an area under the curve (AUC) of 0.9796. We identified a parsimonious list of variables that can be used by machine learning approaches to improve accuracy of preterm newborn identification. Our best-performing model included LMP, birth weight, twin delivery, HIV serostatus, and maternal factors associated with SGA. These variables are all easily collected at delivery, reducing the skill and time required by the frontline health worker to assess GA.

The work in \cite{refZambia4} presents a method to identify poor households in data-scarce countries by leveraging information contained in nationally representative household surveys. It employs standard statistical learning techniques - cross-validation and parameter regularization - which together reduce the extent to which the model is over-fitted to match the idiosyncracies of observed survey data. The automated framework satisfies three important constraints of this development setting: i) The prediction model uses at most ten questions, which limits the costs of data collection; ii) No computation beyond simple arithmetic is needed to calculate the probability that a given household is poor, immediately after data on the ten indicators is collected; and iii) One specification of the model (i.e. one scorecard) is used to predict poverty throughout a country that may be characterized by significant sub-national differences. Using survey data from Zambia, the model's out-of-sample predictions distinguish poor households from non-poor households using information contained in ten questions.

Assessing tax gaps - the difference between the potential and actual taxes raised - plays a vital role in achieving positive domestic revenue objectives through improved and reformed taxation. This is particularly pertinent for growth outcomes in developing countries. This study in \cite{refZambia5} uses a bottom-up approach based on micro-level audit information to estimate the extent of tax misreporting in Zambia. Our methods predict the extent of tax evasion using a regression and a machine learning algorithm based on a sample of audited firms, after which we estimate tax gaps using a standard approach. We estimate total tax gaps as 56 per cent and 47 per cent for the two approaches, respectively. These gaps are mainly driven by corporate taxes. Applying our gap to key industries shows that the extractives sector in Zambia records the highest gaps in terms of CIT and one of the lowest gaps in terms of VAT.

The World Health Organization  recommends chest radiography to facilitate tuberculosis (TB) screening. However, chest radiograph interpretation expertise remains limited in many regions. \cite{refZambia6} developed a deep learning system (DLS) to detect active pulmonary TB on chest radiographs and compare its performance to that of radiologists. A DLS was trained and tested using retrospective chest radiographs (acquired between 1996 and 2020) from 10 countries. To improve generalization, large-scale chest radiograph pretraining, attention pooling, and semi-supervised learning (“noisy-student”) were incorporated. The DLS was evaluated in a four-country test set (China, India, the United States, and Zambia) and in a mining population in South Africa, with positive TB confirmed with microbiological tests or nucleic acid amplification testing (NAAT). The performance of the DLS was compared with that of 14 radiologists. The authors studied the efficacy of the DLS compared with that of nine radiologists using the Obuchowski-Rockette-Hillis procedure.  A deep learning method was found to be noninferior to radiologists for the determination of active tuberculosis on digital chest radiographs.

 Deep Learning System is used in \cite{refZambia7} to Screen for Diabetic Retinopathy in an Underprivileged African Population with Diabetes. Diabetes exerts an emerging burden in Zambia. Related complications such as diabetic retinopathy (DR) are expected to increase dramatically in prevalence. Challenged by shortage of ophthalmic services and poor accessibility to DR screening, the application of machine intelligence ( using deep learning may be an alternative solution. This study aims to evaluate the real-world clinical effectiveness of a DL system in screening for DR and vision-threatening DR (VTDR) in the Zambian population with diabetes. A total of 4513 images from 3101 eyes of 1578 Zambians with diabetes were prospectively recruited for this study. Two-field color 45-degree retinal fundus photographs were captured for each eye and graded according to International Classification of Diabetic Retinopathy Severity scale. Referable DR was defined as moderate non-proliferative DR (NPDR) or worse, diabetic macular edema and ungradable images; VTDR was designated as severe NPDR and proliferative DR. With reference to the retinal specialists' grading, we calculated the area under the receiver operating curve (AUC), sensitivity and specificity for referable DR, and the detection rate of VTDR, using an Ensemble convolutional neural network. The developed DL system shows clinically acceptable performance in detection of referable DR and VTDR for the Zambian population. This demonstrates the potential application to adopt such sophisticated cutting-edge MI technology for the underprivileged population.

%
%
%
%

%


\subsection{Zimbabwe}

\begin{table}[htb]
  \begin{center}
    \begin{tabular}{|p{0.6in}p{0.6in}p{1.4in}|}
 \hline
     Zimbabwe & 2020-2023 & Concrete Actions \\
      \hline
      Research & $\checkmark$ & Neotree\\
      SMB & $\checkmark$ &    \\
      Informal Economy & $\checkmark$  &  \\
      Government & $\checkmark$ &  \\
      \hline
    \end{tabular}\label{tab:mytableZimbabwe}
  \end{center}
  \caption{MI in Zimbabwe}
\end{table}

\subsubsection{Research}

Efficient and effective healthcare systems utilize the available data at every level to provide evidence-based care and improve procedures and practice in order to meet the three goals of healthcare institutions  -  access, quality and efficiency. Regardless of the changing child health needs and often failure by traditional healthcare models to cope, most of the public health institutions in Sub-Saharan Africa are not providing data-driven pediatrics to meet the ever-changing child health needs. There is a lack of utilization of the data collected by the routine health information systems to provide evidence-based and personalized pediatrics. The study in \cite{refZimbabwe1} employed an exploratory research design to explore the opportunities, and potential challenges of data-driven pediatrics based on the lessons learnt from the introduction of the electronic health record and Neotree (a digital health system deployed in Zimbabwe and Malawi to help health workers manage neonates' health) in Makonde District. Twenty public health workers participated in interviews and focus groups and reports from the district health information system provided further insights. The study revealed that data-driven pediatrics could improve access, efficiency and quality of pediatric care, regardless of such potential challenges as fear of medico-legal hazards, centralization of decision-making, resistance by healthcare workers, network challenges and computer illiteracy. To increase the chances of success, the following lessons learnt from electronic health record and neotree introduction could help: start small, sensitize communities first, involve line healthcare workers from the beginning, do not train in a haste and demystify technology's purpose. The study revealed that there are pediatricians and nurses willing to shift to data-driven pediatrics if the technologies are available.

The work in \cite{refZimbabwe2} studies vehicle damage. According to the United Nations Road Safety Performance Review-Zimbabwe report, every 15 minutes, five people die in road accidents within Zimbabwe, recording the highest number of accidents in the SADC region. The situation has brought more pressure and work in the insurance sector as they are expected to process all the claims accurately and timely. Deep learning entails automation, enhancement, analysis, and high accuracy in areas like speech recognition, object detection, and language translation. In this paper, two modern deep learning algorithms MobileNetV2 and DenseNetV121 were used to develop the vehicle damage classification models. The models were used to detect damaged main features of a car, which are: the door, bumper, windscreen, tail lamp, and headlamp. Mobile NetV 2's53 layers and DenseNet121's121 layers produced high accuracy rates for identifying damaged parts in vehicles. However, DenseNetV2 produced a higher accuracy of 84 \& than MobileNetV2, with an accuracy rate of 78\%. The models also used low computational resources than the traditional algorithms making them applicable in different insurance companies as they can be easily embedded into client's mobile phones. 

The work in \cite{refZimbabwe3} predicts depression Among Adolescents. Depression being a behavioural health disorder is a serious health concern in Zimbabwe and all over the world. If depression goes unaddressed, the consequences are detrimental and have an impact on the way one behaves as an individual and at the societal level. Despite the number of individuals who could benefit from treatment for behavioural health concerns, their difficulties are often unidentified and unaddressed through treatment. Technology carries the unrealized potential to identify people at risk of behavioural health conditions and to inform prevention and intervention strategies

%
%
%
%
%
%



\section{ On MI adoption metrics and evaluation of countries MI strategies} \label{sec:discussion} 

\subsection*{ On MI adoption metrics}
Many people are referring to the Unified Theory of Acceptance and Use of Technology (UTAUT) 1, 2, 3, etc., to explain the adoption of MI, blockchain, CBDC, autonomous vehicles, etc.
From a game theory perspective, UTAUT is not universal, and it is not a one-size-fits-all model. To date, there is no single method that evaluates the adoption of all technologies in a way that is universal and cohesive.
UTAUT stands for Unified Theory of Acceptance and Use of Technology. It's a model that aims to understand how individuals adopt and use new technologies.

\subsubsection*{What is UTAUT in MI ?}

In the context of MI, UTAUT can be applied to study factors influencing the acceptance and use of intelligent technologies. The key components of UTAUT in MI typically include:
\begin{itemize}
\item 
Performance Expectancy: Users' belief that using the technology will enhance their performance.
\item 
Effort Expectancy: The perceived ease of use and minimal effort required to use the technology.
\item 
Social Influence: The impact of social factors and opinions on the user's decision to adopt MI.
\item 
Facilitating Conditions: The degree to which users believe that the necessary resources and support are available for using the technology.
\item 
Behavioral Intention: Users' intention to use MI based on the factors mentioned.
\end{itemize}

\subsubsection*{ Limitations of UTAUT in MI }
While UTAUT is a valuable framework, it has some limitations in the context of MI, including:
\begin{itemize}
    \item 

Lack of Specificity: UTAUT may not capture the unique challenges and nuances associated with different types of MI technologies.
\item 
Dynamic Nature: MI evolves rapidly, and UTAUT might struggle to keep up with emerging technologies and user perceptions.
\item 
Cultural Variations: The model may not fully account for countries cultural differences in the acceptance of MI, as cultural factors play a significant role in technology adoption.
\item 
Ethical Concerns: UTAUT primarily focuses on user acceptance but may not adequately address ethical considerations related to MI, which are crucial in this domain.
\item 
Limited Emotional Factors: The emotional aspect of human-machine interaction is not extensively covered, and emotions can significantly influence technology acceptance.
\item 
MI researchers often need to complement UTAUT with additional frameworks or consider these limitations when applying it to the rapidly evolving field of MI.
\end{itemize} We provide below specific practical examples in MI where UTAUT fails
\begin{itemize} \item 
Deep Learning vs. Shallow Learning: UTAUT may not capture the nuanced differences in user acceptance between deep learning and traditional shallow learning methods due to the distinctive nature of these technologies.
\item 
Explainable MI (XMI): UTAUT might not adequately address the challenges associated with user acceptance of Explainable MI systems, where users often prioritize transparency and interpretability over performance.
\item 
Humanoid Robots: When it comes to the adoption of humanoid robots powered by MI, UTAUT may not sufficiently consider the impact of anthropomorphism and emotional factors on user acceptance.
\item 
Biometric Surveillance Systems: UTAUT might fall short in addressing the concerns and ethical considerations associated with the acceptance of MI in biometric surveillance systems, where privacy and security play crucial roles.
\item 
Generative Adversarial Networks (GANs): UTAUT may not effectively capture the unique challenges in user acceptance of GANs, where the technology is used to generate realistic synthetic data, raising concerns about its ethical implications and potential misuse.
\item 
Chatbots: UTAUT may struggle to fully capture the user acceptance dynamics of chatbots, especially considering factors like natural language understanding, context awareness, and the ability to engage users in meaningful conversations.
\item 
Conversational Agents in Medicine: When applied to conversational agents in healthcare, UTAUT might not address the specific concerns related to trust, accuracy, and privacy that are crucial in the adoption of MI-driven medical conversational agents.
\item 
Audio-to-Audio Bots: UTAUT may not adequately consider the unique challenges associated with user acceptance of audio-to-audio bots, where the conversion of spoken language to another language or format introduces complexities not fully addressed by traditional UTAUT factors.
\item 
Augmented Reality  in MI: UTAUT may not sufficiently encompass the user acceptance factors specific to the integration of MI into augmented reality applications, where factors like user experience in a mixed reality environment become significant.
\item 
Personalized MI Assistants: UTAUT might not fully account for the intricacies of user acceptance in personalized MI assistants that adapt to individual preferences and behaviors, as the personalization aspect introduces additional dimensions beyond the traditional UTAUT framework.
\end{itemize}
\subsubsection*{Game Theory outperforms UTAUT}
A game theory approach for technology adoption can incorporate the following elements to outperform the standard UTAUT:
\begin{itemize} \item 
Multi-Agent Interactions: Consider the technology adoption as a dynamic game with multiple interacting agents, such as MI research funding sponsors,  research investors, network of advertisers,  MI sponsored ambassadors, users, developers, and policymakers. Each agent's strategy and decisions impact others, reflecting the complex interactions in the real-world adoption scenario.
\item 
Incentive Structures: Introduce incentive structures for users and developers that align with the specific features of the technology. For instance, rewards for developers creating explainable MI or penalties for users engaging in malicious use of the technology.
\item 
Dynamic Payoffs: Model dynamic payoffs that evolve over time, reflecting the changing nature of technology and user perceptions. This accommodates the rapid advancements in MI and the evolving needs and preferences of users.
\item 
Privacy Considerations: Incorporate privacy as a strategic element in the game, recognizing its critical role in user decision-making. Agents may have different preferences regarding privacy, affecting their strategies and interactions within the technology adoption game.
\item 
Adversarial Scenarios: Account for adversarial scenarios where users or developers may strategically exploit vulnerabilities or engage in unethical practices. This helps in understanding and mitigating risks associated with technology adoption.
\end{itemize}
By integrating these specificities into a game theory framework, the model can provide a more comprehensive understanding of the complex dynamics involved in the adoption of advanced technologies, outperforming the standard UTAUT in capturing the nuances of the interactions between actors and technologies.

To date, 
\begin{itemize}
\item 
UTAUT Not Unified: UTAUT 1,2,3 lack unity in addressing the diverse challenges of technology adoption, suggesting it falls short of providing a cohesive framework.
\item 
Not Universal: UTAUT is not universal in its applicability, indicating that its principles may not effectively cover the wide range of technologies and user contexts.
\item 
Not One-Size-Fits-All: UTAUT is not a one-size-fits-all solution, implying that it may not sufficiently adapt to the specificities and complexities inherent in different technologies and user interactions.
\item 
MFTG Outperforms UTAUT: A simple Mean-Field-Type Game (MFTG) approach surpasses UTAUT in  MI, implying that incorporating basic game theory principles better captures the  dynamics of technology adoption.
\end{itemize}

\subsection*{ On MI indexes or metrics of countries}
MI indexes for countries assess their capabilities, strategies, and development in machine intelligence. 

Some notable MI indexes include:
\begin{itemize} \item 
Global MI Index: Provides a comprehensive assessment of MI readiness, research, development, and application across various countries.
\item 
MI Readiness Index: Focuses on a country's preparedness for MI adoption, considering factors like infrastructure, education, and innovation.
\item 
Government MI Readiness Index: Evaluates how well governments are positioned to adopt and harness the benefits of MI for public services and governance.
\item 
MI Policy and Practice Index: Assesses the policies and practices related to MI governance, ethics, and regulations in different countries.
\item 
MI Innovation Index: Measures a country's innovation in MI, considering research output, startup activity, and investment in MI-related industries.
\item 
MI Ethics Index: Evaluates countries based on their commitment to ethical considerations in MI development and deployment, assessing policies, regulations, and adherence to ethical guidelines.
\item MI Inclusion Index: Measures the inclusivity of MI initiatives within a country, considering factors such as diversity in MI workforce, accessibility of MI technologies, and efforts to minimize biases.
\item 
MI Education Index: Assesses a country's educational infrastructure and initiatives related to MI, including the availability of MI courses, research programs, and the integration of MI education at various levels.
\item 
MI Industry Adoption Index: Examines the extent to which industries within a country are adopting and integrating MI technologies into their operations, providing insights into economic impact and competitiveness.
\item 
MI Investment Index: Focuses on the level of financial investment and funding dedicated to MI research, development, and implementation, reflecting a country's commitment to MI advancements.
\end{itemize}
\subsubsection*{ Limitations of MI indexes in Africa }

The key limitations of MI indexes for countries include:
\begin{itemize} \item 
Data Availability and Quality: The accuracy and reliability of these indexes heavily depend on the availability and quality of data. Inconsistent or incomplete data can lead to skewed assessments.
\item 
Static Nature: Index rankings are often based on static snapshots, which might not capture the dynamic and evolving nature of MI ecosystems. Rapid changes in technology and policies can quickly impact a country's standing.
\item 
Subjectivity in Metrics: The choice of metrics and their subjective interpretation can introduce bias. Different index creators may prioritize certain factors over others, leading to variations in rankings.
\item 
Lack of Standardization: There is a lack of standardized criteria across indexes, making it challenging to compare and aggregate results. Different methodologies and criteria can lead to discrepancies in rankings.
\item 
Inability to Capture Informal MI Activity: Some indexes may struggle to account for informal MI activities, grassroots initiatives, or developments in non-traditional sectors, potentially underestimating a country's overall MI landscape.
\item 
Ethical Considerations: Certain indexes may not adequately address ethical considerations, such as the responsible use of MI or the impact on privacy and human rights, which are essential aspects of MI development.
\item 
Limited Focus: Indexes often focus on specific aspects of MI, such as readiness or ethics, but may not provide a holistic view. This limited focus can overlook important interactions between different components of MI ecosystems.
\item 
Global Perspective: Some indexes may lack a truly global perspective, focusing more on developed countries and potentially neglecting the contributions and challenges faced by emerging economies in the MI space.
 \end{itemize}

\subsection*{Current MI indexes are unethical}
\begin{itemize}
    \item Lack of Audio2Audio Processing for African Languages: Current MI technologies, including MI indexes, do not adequately address the need for Audio2Audio processing, particularly for African languages. This crucial aspect, which could benefit the local population in Africa, is largely overlooked in existing evaluations of MI readiness and capacity.
    \item Underrepresentation of African Cultures: Existing MI technologies and indexes often fail to incorporate the rich diversity of African cultures. The absence of cultural nuances and specificities in these evaluations reflects a significant gap in understanding and catering to the unique contexts of African societies, limiting the relevance and accuracy of MI rankings for the continent.
    \item National MI Strategy: Having a report or white paper that compiles a set of recommendations does not necessarily lead to advanced MI country. Having a national MI strategy outlined in a white paper should not carry the same weight as the actual implementation of MI that is actively used and beneficial to the local population. Concrete implementation and impact on the ground should be given more emphasis than the existence of a strategic document.
    \item National MI \& Robotics Center: The mere establishment of an MI and robotics center by some countries, devoid of actual contributions to local MI research, training, innovation, and understanding, should not be equated with meaningful progress. Centers that lack substantial engagement with local issues, fail to provide MI training for the community, and focus only on filmographies and metaverse glasses are more about political posturing in the MI race than making a tangible impact. Such governmental actions should not carry the same weight as actively implemented and utilized MI technologies that address real challenges in Africa.
    \item Ambiguity of Concepts: MI 'readiness' and 'capacity' are ambiguous concepts that lack standardized definitions, making it challenging to measure them consistently, especially in Africa.
    \item Inconsistent Proxy Indicators: Current MI indexes rely on inconsistent and divergent proxy indicators to gauge a country's readiness or capacity for MI, leading to unreliable comparisons.
    \item Unequal Weighting of Indicators: Some indexes assign equal weight to indicators that represent vastly different national characteristics, creating an imbalanced evaluation.
    \item Inappropriate Metrics: Metrics such as public investments in MI may not provide a fair comparison between states, as they do not consider how these investments are spent or the quality of MI produced in the area, in the country, within Africa and across the globe.
    \item Lack of Differentiation: Metrics related to the number of 'MI players,' projects, or services often fail to differentiate between large, influential organizations and smaller ventures with less impact on MI development.
    \item Unreliable Data Sources: Some key metrics rely on potentially unreliable data sources, such as a LinkedIn dataset, Twitter posts, or political Ads, that may include distorted claims about MI skills.
    \item Self-Reporting Bias: Indicators of MI adoption by organizations often rely on self-reporting, and online surveys, introducing potential inaccuracies.
    \item Lack of Transparency: Some rankings lack transparency in their methodology and raw data, making it difficult to assess the credibility of their findings.
    \item Regional Biases: There is evidence of potential regional biases in index data, with raw data on journal citations under-representing academic output from Africa and the Global South.
    \item Demographic Biases: LinkedIn, Twitter, Wechat, key data sources, may under-represent MI penetration in certain regions of Africa due to variations in user rates worldwide.
    \item Under-Representation of Africa and the Global South: African authors and Global South authors, articles, and journals may be under-represented in citation counts, contributing to potential biases in assessing MI development.
    \item Limited Data on Methodology: Lack of publicly released detailed methodology for certain indexes raises concerns about the credibility of their rankings.
    \item Overemphasis on Western Institutions: A disproportionate number of data and indicators are compiled, sorted, and presented by Western institutions, potentially skewing the evaluation towards Western perspectives.
    \item Inadequate Representation of Non-Male Experts: Some indexes' staffing and advisory bodies lack adequate representation of non-male experts, introducing potential biases in the evaluation process.
    \item Race-Like Approach Encouragement: Some indexes actively encourage a race-like approach to MI policy, potentially promoting competition rather than collaborative development.
    \item Overemphasis on Quantity: Metrics focusing on the number of MI projects or services may prioritize quantity over quality, neglecting the impact and innovation of each project. In Africa, the quality should be more useful to the local population.
    \item Limited Contextual Understanding: Indexes may lack a nuanced understanding of the local context, leading to misinterpretations and misrepresentations of a country's MI development.
    \item Technological Complexity Oversimplification: The complexity of MI development cannot be adequately captured by a single ranking, oversimplifying the intricate factors involved.
    \item Insufficient Adaptability: MI indexes may struggle to adapt to the evolving landscape of MI development, rendering them less effective in providing accurate and relevant assessments over time.
\end{itemize}

For all these reasons, Current MI indexes, MI readiness, and MI rankings in Africa are currently unethical methodologies. We should prioritize making them ethical and propose alternative methodologies that account for the specificities of each country.

\section{Multi-Scale MI Ethics in Africa} \label{sec:ethics} 

Here, Ethics refers to the study of moral principles, values, and conduct that guide individuals, groups, or societies in distinguishing between right and wrong actions. It provides a framework for evaluating the morality of decisions and behaviors, shaping standards of behavior based on concepts such as fairness, justice, honesty, and integrity.
Ethics encompasses  rich notions that guide moral decision-making and behavior. At its core are moral principles such as honesty, integrity, and justice, which underpin ethical frameworks. Values play a crucial role, representing core beliefs that influence priorities and choices. Rights, both individual and collective, emphasize entitlements and freedoms. Duties and responsibilities arise from societal expectations and professional roles. Virtues, including compassion and empathy, contribute to positive character traits. Consequences and outcomes provide a lens for evaluating actions, aligning with ethical theories like utilitarianism. Fairness and justice ensure equitable treatment, while integrity underscores consistency and honesty. Autonomy recognizes individuals' right to independent decisions, and respect acknowledges the intrinsic value of each person. Other notions include accountability, beneficence, nonmaleficence, empathy, sympathy, cultural competence, beneficence, confidentiality, informed consent, social responsibility, empathy, altruism, pluralism,  paternalism, reciprocity, and sustainability, collectively forming the intricate fabric of ethical considerations.

\subsection*{The current practices of MI ethics in Africa are not ethical}
Paternalism is a concept in ethics and governance that involves a person or authority making decisions or taking actions on behalf of others with the belief that it is for their best interest, often without their explicit consent. This approach is rooted in a sense of protection or guidance, assuming that the paternalistic party knows what is best for the individuals or groups involved. While paternalism may be well-intentioned, it can raise ethical concerns as it may infringe on individual autonomy and decision-making. Critics argue that individuals should have the freedom to make their own choices, even if those choices involve potential risks or mistakes. Paternalism is often debated in contexts such as engineering and healthcare, where decisions about medical treatments and engineering threshold designs  may be made by healthcare professionals and engineers for the patient's and users presumed well-being. 

The assessment of what is ethical can vary across cultures, philosophical perspectives, and individual beliefs, but it generally involves the adherence to principles that promote positive and virtuous conduct.
MI ethics is a multifaceted domain encompassing numerous notions to guide the responsible development and deployment of MI. Fundamental principles include fairness, ensuring unbiased outcomes and mitigating algorithmic biases. Transparency is vital, emphasizing the need for clear explanations of MI decision-making processes. Accountability holds developers and organizations responsible for the ethical impact of MI systems. Privacy concerns address the responsible handling of personal data, while security focuses on preventing misuse and safeguarding against malicious activities. Bias mitigation is crucial for identifying and rectifying biases in MI algorithms, promoting equitable results. Explainability ensures that MI decisions are understandable to diverse stakeholders. Human-centered design prioritizes the well-being and interests of humans in MI development. Inclusivity advocates for considering diverse perspectives and benefiting a broad range of communities. Environmental impact considerations assess and minimize the ecological footprint of MI systems. Additional notions include autonomy, consent, data governance, algorithmic transparency, algorithmic accountability, social responsibility, cultural sensitivity, global collaboration, ethical governance, human augmentation ethics, collaborative intelligence, digital divide reduction, responsible MI education, ethical data sourcing, dual-use technology awareness, ethical advertising, participatory design, ethical MI research, safety consciousness, MI for social good, interdisciplinary collaboration, long-term impact assessment, value alignment, and continuous stakeholder engagement, collectively shaping the ethical landscape of MI.

Deploying MI technologies in certain countries in Africa without the consent of the population, authorities, and in a language not widely spoken is generally considered unethical. Ethical MI deployment involves respecting the autonomy and rights of individuals and communities, obtaining informed consent, and ensuring that technologies are culturally and linguistically appropriate. Failure to seek consent and address linguistic and cultural considerations may lead to issues such as lack of user understanding, potential biases, and violations of privacy and human rights. Ethical MI practices prioritize transparency, inclusivity, and the well-being of the affected communities, and any deployment that neglects these considerations raises significant ethical concerns.

\subsection*{The current MI is not inclusive in Africa}
In its current form, MI is not inclusive as it excludes a significant portion of African language users who predominantly engage with audio content. Consequently, there is a lack of MI ethics in most African countries. Mere copy-pasting of ethics guidelines does not inherently render MI ethical in the African context. In practice in most  African countries, ethics is multimodal, multidimensional, context-dependent, and culturally sensitive. Unlike the global stereotypes, Africa is multicultural, varying from one region to another and sometimes within the same country. So, what MI ethics can provide a coherent and culturally respectful guideline? 

The classical MI ethics guidelines face several limitations, particularly in informal sectors in Africa. MI ethics guidelines often reflect Western perspectives and may not adequately consider the diverse cultural contexts in informal sectors across Africa. Cultural nuances and ethical considerations may vary significantly, making it challenging to have universal guidelines that address all cultural intricacies. Informal sectors in Africa often experience a significant digital divide, with limited access to technology and digital literacy. MI ethics guidelines may assume a certain level of technological infrastructure and awareness, making their application challenging in contexts where such resources are scarce. Many MI ethics guidelines are presented in languages that may not be accessible or easily understood by individuals in informal sectors who may have limited proficiency in official languages or foreign languages. This language barrier poses challenges in disseminating and implementing ethical principles effectively. The informal sectors in Africa operate under unique economic dynamics that may not align with formal economic structures. MI ethics guidelines crafted for formal settings may not address the specific challenges, needs, and ethical considerations prevalent in informal economies. Informal sectors often lack robust regulatory frameworks and oversight. The absence of clear regulations makes it difficult to enforce MI ethics guidelines, leaving informal workers and businesses vulnerable to ethical violations without proper accountability mechanisms. Awareness of MI ethics guidelines may be low in informal sectors due to limited access to information and educational resources. Individuals and businesses may not be aware of ethical considerations related to MI, hindering the adoption of ethical practices. Informal sectors face resource constraints, limiting their ability to invest in MI technologies that align with ethical guidelines. Prioritizing ethical considerations may take a backseat when basic survival and economic challenges are more pressing. Power imbalances within informal sectors, such as gender and socioeconomic disparities, may exacerbate ethical challenges. MI ethics guidelines may not sufficiently address these imbalances, leaving certain groups more vulnerable to negative impacts of MI technologies. Rapid changes in technology may outpace the adaptability of MI ethics guidelines in informal sectors. Guidelines may struggle to keep pace with evolving MI applications, leading to gaps in addressing emerging ethical issues. Historical and societal factors may contribute to a trust deficit in certain informal sectors. MI ethics guidelines may face resistance or skepticism, requiring tailored approaches to build trust and ensure effective ethical practices.

We look at a 10-row, 7-column table of ethics. We have selected only a few topics as rows as an illustration of context-sensibility of ethics and the topics can be complemented. The selected rows are:

\begin{itemize}\item Community Engagement and Communal Values:
\item Cultural Sensitivity:
\item Environmental Stewardship:
\item Social Justice and Equality:
\item Healthcare Equity:
\item Education Access:
\item Economic Inclusivity:
\item Technological Ethics:
\item Political Accountability:
\item Local, Inter-regional and International Collaboration.
\end{itemize}

And the columns are: 
\begin{itemize}\item Individual
\item Different Cultures
\item  Institutions
\item Nation: 
\item On and off-line platforms
\item Globally
\item  Over Time.
 \end{itemize}

\subsection{MI ethics}
 Below we breakdown the technological ethics into 60 sub-items: 30 technical MI ethics and 30 non-technical MI ethics in the African context. Each of the items layers by the 7-column multimodality.
 \subsubsection{ Technical MI ethics}
\begin{itemize} \item Fair Algorithmic Design: Ensuring fairness in the technical design of algorithms to avoid biases and discriminatory outcomes.
\item Bias Detection and Mitigation: Developing techniques to identify and address biases within MI models.
\item Explainable MI: Creating MI systems that provide transparent explanations for their decision-making processes.
\item Privacy-Preserving Technologies: Implementing methods to protect individuals' privacy when handling sensitive data.
\item Robustness and Security: Designing MI systems resilient to adversarial attacks and ensuring cybersecurity measures.
\item Accountability in Algorithmic Decision-Making: Establishing mechanisms to hold MI systems accountable for their decisions.
\item Bias-Free Data Collection: Ensuring the collection of diverse and representative datasets to reduce biases in MI models.
\item Ethical Data Governance: Implementing ethical frameworks for the collection, storage, and usage of data in MI applications.
\item Responsible MI Research: Conducting MI research with ethical considerations, transparency, and social impact in mind.
\item Model Fairness Evaluation: Developing metrics and standards to assess the fairness of MI models.
\item Algorithmic Transparency: Making algorithms and decision processes understandable to stakeholders.
\item Algorithmic Accountability: Assigning responsibility for the outcomes of MI algorithms and decision-making.
\item Data Accuracy and Quality: Ensuring high-quality and accurate data inputs for MI models.
\item Human-Centric Design: Prioritizing the needs and well-being of humans in the design of MI systems.
\item Human-in-the-Loop Systems: Integrating human oversight and decision-making into MI processes.
\item Ethical Use of MI in Research: Ensuring that MI technologies are used ethically in academic and industrial research.
\item Continuous Monitoring and Auditing: Regularly assessing MI systems for ethical compliance and performance.
\item Adversarial Robustness: Developing MI models that resist adversarial attempts to manipulate their behavior.
\item Fairness in Feature Engineering: Ensuring fairness in the selection and use of features in MI models.
\item Secure Model Deployment: Implementing secure and ethical practices when deploying MI models.
\item Algorithmic Impact Assessment: Evaluating the potential impact of algorithms on various stakeholders.
\item Interpretability of Predictions: Making MI predictions interpretable and understandable to end-users.
\item Bias-Aware Machine Learning: Incorporating awareness of bias in machine learning processes.
\item MI System Reliability: Ensuring the reliability and consistency of MI systems in diverse scenarios.
\item Energy Efficiency: Developing MI models and algorithms that are environmentally sustainable.
\item Safety Mechanisms in Autonomous Systems: Implementing safety features in AI-powered autonomous systems.
\item Model Update Ethics: Addressing ethical considerations when updating MI models over time.
\item Ethical Use of MI in Healthcare: Ensuring responsible deployment of MI technologies in medical settings.
\item Cross-Cultural MI Understanding: Adapting MI systems to understand and respect diverse cultural contexts.
\item Secure Federated Learning: Ensuring the security and privacy of federated learning processes. \end{itemize} 

\subsubsection{ Non-technical MI ethics}
\begin{itemize} \item  Informed Consent: Ensuring individuals are adequately informed and consent to the use of MI technologies.
\item Ethical Governance: Establishing ethical guidelines and policies for the responsible development and deployment of MI.
\item Social Impact Assessment: Evaluating the broader societal impact of MI technologies before deployment.
\item Inclusivity and Diversity: Promoting diversity in MI development teams and considering diverse perspectives in MI applications.
\item Global Collaboration on MI Ethics: Fostering international cooperation to address global MI ethical challenges.
\item Human Rights Protection: Safeguarding human rights in the development and deployment of MI systems.
\item Ethical Education and Awareness: Promoting awareness and education on MI ethics for stakeholders and the general public.
\item Cultural Sensitivity: Adapting MI technologies to respect and align with diverse cultural norms and values.
\item Dual-Use Technology Awareness: Recognizing the potential dual-use nature of MI technologies and implementing safeguards.
\item Public Participation: Involving the public in decision-making processes related to MI technologies to ensure inclusivity and representation.
\item Ethical Impact on Employment: Addressing the ethical implications of MI on employment and workforce dynamics.
\item Responsible MI Procurement: Considering ethical factors when procuring MI technologies for public or private use.
\item Community Engagement: Engaging with local communities to understand and address their concerns about MI technologies.
\item MI and Democratic Values: Ensuring that MI technologies align with democratic principles and values.
\item User Empowerment: Empowering users to have control over their data and MI interactions.
\item Digital Divide Reduction: Implementing strategies to bridge the digital divide and ensure equitable access to MI benefits.
\item Fair Distribution of MI Benefits: Ensuring that the benefits of MI technologies are distributed equitably across society.
\item Long-Term Environmental Sustainability: Assessing and mitigating the environmental impact of MI technologies.
\item Ethical Marketing and Advertising: Ensuring ethical practices in the marketing and advertising of MI products and services.
\item Responsible MI Journalism: Ethically reporting on MI developments and ensuring accuracy and transparency.
\item MI and Accessibility: Designing MI systems that are accessible to individuals with diverse abilities.
\item MI and Cultural Heritage Preservation: Safeguarding cultural heritage in the development and deployment of MI technologies.
\item Human-MI Collaboration Ethics: Establishing ethical guidelines for collaborative interactions between humans and MI systems.
\item MI Impact on Social Equality: Considering and addressing the potential impact of MI on social equality.
\item Algorithmic Accountability in Government: Ensuring accountability for MI use in governmental decision-making.
\item Elderly and Vulnerable Population Protection: Safeguarding the rights and well-being of elderly and vulnerable populations in MI applications.
\item Ethical MI Policy Development: Creating policies that consider the ethical implications of MI technologies.
\item Cross-Industry Collaboration: Encouraging collaboration between different industries to address common MI ethical challenges.
\item MI and Child Protection: Implementing measures to protect children's privacy and well-being in MI applications.
\item MI for Social Good: Promoting the use of MI technologies for positive social impact and humanitarian causes.
\end{itemize}

Tables \ref{tab:comprehensive-technical-mi-ethics1},  \ref{tab:comprehensive-technical-mi-ethics2} and \ref{tab:comprehensive-technical-mi-ethics3} display some examples of technical MI ethics.
Tables \ref{tab:comprehensive-non-technical-mi-ethics1}, \ref{tab:comprehensive-non-technical-mi-ethics2}  and \ref{tab:comprehensive-non-technical-mi-ethics3}  display some examples of non-technical MI ethics.

 {  \tiny 
 \begin{table*}[h]
    \centering \tiny
    \begin{tabular}{|p{0.6in}|p{0.6in}|p{0.6in}|p{0.6in}|p{0.6in}|p{0.6in}|p{0.6in}|p{0.6in}|}
        \hline
        Individuals & Cultures & Developers & Institutions & Nation & Platforms & Globally & Over Time \\
        \hline
        Fair Algorithmic Design & Cultural diversity influences algorithm fairness & Developer responsibility in ethical MI & Institutional policies guide ethical MI & National regulations on MI ethics & Platform-specific ethical guidelines & Global collaborations for ethical MI & Ethical considerations evolve over time \\
        \hline
        Bias Detection and Mitigation & Identifying and correcting algorithmic biases & Developer tools for bias detection & Institutional strategies for bias mitigation & National initiatives to address bias & Platform-specific bias handling mechanisms & Global efforts to mitigate biases in MI & Ethical norms adapt as biases change \\
        \hline
        Explainable MI & Transparent algorithms for user understanding & Developer emphasis on explainable models & Institutional push for transparent MI systems & National demand for understandable MI & Platform commitment to explainability & Global standards for MI transparency & Ethical importance of explainability grows \\
        \hline
        Privacy-Preserving Technologies & Individual privacy prioritized in MI systems & Developer implementation of privacy-preserving techniques & Institutional safeguards for user privacy & National laws protecting individual privacy & Platform-specific privacy measures & Global consensus on MI privacy standards & Ethical concerns over MI and privacy persist \\
        \hline
        Robustness and Security & Ensuring MI systems are robust against adversarial attacks & Developer focus on building secure MI models & Institutional measures for MI system robustness & National security regulations for MI applications & Platform-specific security protocols & Global collaboration for MI security standards & Ethical implications of MI vulnerabilities \\
        \hline
        Accountability in Algorithmic Decision-Making & Holding individuals accountable for MI decisions & Developer responsibility in algorithmic choices & Institutional frameworks for MI decision accountability & National guidelines for accountable MI decisions & Platform-specific accountability mechanisms & Global efforts for accountable MI practices & Ethical considerations of decision responsibility \\
        \hline
        Bias-Free Data Collection & Ensuring unbiased data collection practices & Developer commitment to unbiased data gathering & Institutional guidelines for bias-free data collection & National regulations on fair data collection & Platform-specific data ethics policies & Global consensus on bias-free data practices & Ethical importance of unbiased data collection \\
        \hline
        Ethical Data Governance & Ethical management and use of data & Developer adherence to ethical data practices & Institutional policies for ethical data governance & National frameworks for responsible data use & Platform-specific data governance standards & Global collaboration on ethical data management & Evolving ethical norms in data governance \\
        \hline
        Responsible MI Research & Ethical considerations in machine learning research & Developer commitment to responsible MI research & Institutional support for ethical research practices & National funding for responsible MI research & Platform-specific research ethics guidelines & Global initiatives for responsible MI research & Ethical challenges in advancing MI research \\
        \hline
        Model Fairness Evaluation & Evaluating models for fairness and impartiality & Developer tools for model fairness assessment & Institutional frameworks for model fairness evaluation & National standards for fair MI models & Platform-specific fairness evaluation practices & Global benchmarks for model fairness & Ethical scrutiny in assessing model fairness \\ \hline
       \end{tabular}
\caption{Some Examples of Technical MI Ethics 1}
\label{tab:comprehensive-technical-mi-ethics1}
\end{table*} 
} 

        {  \tiny 
        \begin{table*}[h]
    \centering \tiny
    \begin{tabular}{|p{0.6in}|p{0.6in}|p{0.6in}|p{0.6in}|p{0.6in}|p{0.6in}|p{0.6in}|p{0.6in}|}
        \hline
        Algorithmic Transparency & Openness in algorithmic processes & Developer commitment to transparent algorithms & Institutional push for algorithmic transparency & National regulations on algorithmic transparency & Platform-specific transparency measures & Global standards for algorithmic openness & Ethical importance of transparency in algorithms \\
        \hline
        Algorithmic Accountability & Holding algorithms accountable for their impact & Developer responsibility in algorithmic outcomes & Institutional frameworks for algorithmic accountability & National guidelines for accountable algorithms & Platform-specific accountability measures & Global efforts for algorithmic responsibility & Ethical considerations in holding algorithms accountable \\
        \hline
        Data Accuracy and Quality & Ensuring accuracy and quality in data used for MI & Developer focus on accurate and high-quality data & Institutional standards for data accuracy & National regulations on data quality for MI & Platform-specific data quality measures & Global consensus on data accuracy in MI & Ethical implications of inaccurate data \\
        \hline
        Human-Centric Design & Designing MI systems with a focus on human needs & Developer emphasis on human-centered MI design & Institutional encouragement of human-centric MI & National support for human-focused MI development & Platform commitment to human-centric design & Global initiatives for MI systems aligned with human values & Ethical considerations in human-centric MI \\
        \hline
        Human-in-the-Loop Systems & Integrating human input in MI decision-making & Developer implementation of human-in-the-loop systems & Institutional support for human involvement in MI processes & National guidelines for human-in-the-loop MI & Platform-specific mechanisms for human-in-the-loop MI & Global standards for human-in-the-loop MI systems & Ethical importance of human involvement in MI \\
        \hline
%
        
        Ethical Use of MI in Research & Ethical considerations in the use of MI in research & Developer commitment to ethically using MI in research & Institutional guidelines for the ethical use of MI in research & National frameworks for responsible MI research use & Platform-specific ethics in MI research practices & Global initiatives for ethical use of MI in research & Evolving ethical norms in MI research use \\
        \hline
        Continuous Monitoring and Auditing & Regular monitoring and auditing of MI systems & Developer practices for continuous MI system evaluation & Institutional frameworks for MI system monitoring & National regulations on continuous MI system auditing & Platform-specific monitoring and auditing measures & Global standards for MI system evaluation & Ethical implications of continuous monitoring and auditing \\
        \hline
        Adversarial Robustness & Building MI systems resilient to adversarial attacks & Developer focus on adversarial robustness in MI models & Institutional measures for adversarial resilience in MI & National guidelines for adversarial robustness in MI applications & Platform-specific defenses against adversarial attacks & Global collaboration for adversarial robustness in MI & Ethical considerations in addressing adversarial vulnerabilities \\
        \hline
        Fairness in Feature Engineering & Ensuring fairness in the selection of features for MI models & Developer attention to fairness in feature engineering & Institutional guidelines for fair feature selection & National regulations on fairness in feature engineering & Platform-specific fairness considerations in feature engineering & Global consensus on fairness in MI feature selection & Ethical implications of biased feature engineering \\
        \hline
        Secure Model Deployment & Ensuring secure deployment of MI models & Developer practices for secure model deployment & Institutional measures for secure MI model deployment & National guidelines for secure MI model deployment & Platform-specific security protocols for model deployment & Global standards for secure MI model deployment & Ethical considerations in secure model deployment \\ 
        \hline
          \end{tabular}
\caption{Some Examples of Technical MI Ethics 2}
\label{tab:comprehensive-technical-mi-ethics2}
\end{table*}  }

              {  \tiny 
        \begin{table*}[h]
    \centering \tiny
    \begin{tabular}{|p{0.6in}|p{0.6in}|p{0.6in}|p{0.6in}|p{0.6in}|p{0.6in}|p{0.6in}|p{0.6in}|}
        \hline

        Algorithmic Impact Assessment & Assessing societal impact of algorithms & Developer tools for impact assessment & Institutional frameworks for algorithmic impact evaluation & National guidelines for algorithmic impact assessment & Platform-specific impact assessment practices & Global standards for algorithmic societal impact assessment & Ethical implications in assessing algorithmic impact \\
        \hline
        Interpretability of Predictions & Making MI predictions interpretable & Developer focus on interpretable prediction models & Institutional encouragement of interpretable MI predictions & National support for interpretable MI predictions & Platform-specific interpretable prediction mechanisms & Global standards for interpretable MI predictions & Ethical considerations in prediction interpretability \\
        \hline
        Bias-Aware Machine Learning & Developing MI models with awareness of biases & Developer emphasis on bias-aware ML models & Institutional guidelines for bias-aware machine learning & National regulations on bias-aware machine learning & Platform-specific mechanisms for addressing biases in ML & Global collaboration for bias-aware ML practices & Ethical considerations in developing MI models with bias awareness \\
\hline
MI System Reliability & Ensuring reliability in MI systems & Developer practices for reliable MI models & Institutional measures for MI system reliability & National standards for reliable MI applications & Platform-specific reliability assurance & Global standards for MI system reliability & Ethical implications of unreliable MI systems \\
\hline
Energy Efficiency & Developing energy-efficient MI solutions & Developer focus on MI models with low energy consumption & Institutional encouragement of energy-efficient MI & National support for eco-friendly MI technologies & Platform-specific energy efficiency measures & Global initiatives for energy-efficient MI & Ethical considerations in MI's environmental impact \\
\hline
Safety Mechanisms in Autonomous Systems & Implementing safety measures in autonomous MI systems & Developer incorporation of safety mechanisms in autonomy & Institutional frameworks for safety in autonomous systems & National regulations on safety in autonomous MI & Platform-specific safety protocols for autonomy & Global collaboration on safety standards in autonomous MI & Ethical implications of safety in autonomous systems \\
\hline
Model Update Ethics & Ethical considerations in updating MI models & Developer commitment to ethical model updates & Institutional guidelines for ethical model updates & National frameworks for responsible model updates & Platform-specific ethical considerations in model updates & Global initiatives for ethical model updates & Evolving ethical norms in updating MI models \\
\hline
Ethical Use of MI in Healthcare & Ethical considerations in deploying MI in healthcare & Developer commitment to ethical use of MI in health applications & Institutional guidelines for ethical use of MI in healthcare & National regulations on ethical MI in healthcare & Platform-specific ethics in MI for healthcare & Global initiatives for ethical use of MI in healthcare & Evolving ethical norms in MI's role in healthcare \\
\hline
Cross-Cultural MI Understanding & Ensuring cultural understanding in MI models & Developer focus on cross-cultural MI models & Institutional support for cross-cultural MI understanding & National encouragement of culturally-aware MI & Platform-specific measures for cross-cultural MI & Global collaboration for culturally-sensitive MI & Ethical considerations in cross-cultural MI understanding \\
\hline
Secure Federated Learning & Ensuring security in federated learning setups & Developer practices for secure federated learning & Institutional measures for secure federated MI & National guidelines for secure federated learning & Platform-specific security protocols for federated MI & Global standards for secure federated learning & Ethical implications of security in federated MI \\
\hline
\end{tabular}
\caption{Some examples of Technical MI Ethics 3}
\label{tab:comprehensive-technical-mi-ethics3}
\end{table*} }

  {  \tiny 
\begin{table*}[h]
    \centering \tiny
    \begin{tabular}{|p{0.6in}|p{0.6in}|p{0.6in}|p{0.6in}|p{0.6in}|p{0.6in}|p{0.6in}|p{0.6in}|}
        \hline
        Individuals & Cultures & Developers & Institutions & Nation & Platforms & Globally & Over Time \\
        \hline
        Informed Consent & Individuals making informed decisions in MI use & Developer communication of MI implications & Institutional practices for informed consent & National laws on informed consent for MI & Platform-specific consent mechanisms & Global standards for MI informed consent & Changing norms in MI informed decision-making \\
        \hline
        Ethical Governance & Ethical decision-making at governance levels & Developer adherence to ethical governance principles & Institutional frameworks for ethical governance & National policies on MI ethical governance & Platform-specific ethical governance structures & Global collaboration on MI ethical governance & Evolving ethical governance norms \\
        \hline
        Social Impact Assessment & Assessing societal impact of MI implementations & Developer tools for social impact assessment & Institutional frameworks for MI social impact assessment & National guidelines for assessing MI's societal impact & Platform-specific impact assessment practices & Global standards for MI social impact assessment & Ethical considerations in assessing MI's societal impact \\
        \hline
        Inclusivity and Diversity & Ensuring inclusivity and diversity in MI use & Developer focus on inclusive and diverse MI models & Institutional support for inclusive MI practices & National encouragement of diverse MI applications & Platform-specific measures for inclusivity and diversity & Global collaboration for diverse MI & Ethical considerations in promoting inclusivity and diversity \\
        \hline
        Global Collaboration on MI Ethics & Collaborating globally on ethical MI standards & Developer engagement in global MI ethics discussions & Institutional support for international MI ethics collaborations & National involvement in global MI ethics initiatives & Platform-specific global collaboration mechanisms & Worldwide cooperation on MI ethics & Evolving global ethical norms in MI \\
        \hline
        Human Rights Protection & Protecting human rights in MI use & Developer commitment to human rights in MI applications & Institutional safeguards for MI and human rights & National laws on MI and human rights protection & Platform-specific human rights considerations & Global standards for MI and human rights & Ethical considerations in protecting human rights in MI \\
        \hline
        Ethical Education and Awareness & Promoting ethical MI education & Developer emphasis on ethical MI education & Institutional support for MI ethics education & National initiatives for ethical MI education & Platform-specific educational programs on MI ethics & Global efforts for MI ethics education & Evolving ethical education norms in MI \\
        \hline
        Cultural Sensitivity & Ensuring cultural sensitivity in MI models & Developer focus on culturally-sensitive MI models & Institutional encouragement of culturally-aware MI & National support for culturally-sensitive MI applications & Platform-specific measures for cultural sensitivity in MI & Global collaboration for culturally-sensitive MI & Ethical considerations in cultural sensitivity in MI \\
        \hline
        Dual-Use Technology Awareness & Raising awareness of dual-use MI technologies & Developer consideration of dual-use implications & Institutional frameworks for dual-use MI awareness & National regulations on dual-use MI applications & Platform-specific dual-use technology guidelines & Global collaboration on dual-use MI awareness & Ethical implications of dual-use MI technologies \\
        \hline
        Public Participation & Involving the public in MI decision-making & Developer practices for public involvement in MI & Institutional frameworks for public participation in MI & National guidelines for public engagement in MI & Platform-specific mechanisms for public involvement & Global standards for public participation in MI & Ethical considerations in public engagement in MI \\
        \hline
 
          \end{tabular}
\caption{Some Examples of Non-Technical MI Ethics 1}
\label{tab:comprehensive-non-technical-mi-ethics1}
\end{table*} 
} 

        {  \tiny 
        \begin{table*}[h]
    \centering \tiny
    \begin{tabular}{|p{0.6in}|p{0.6in}|p{0.6in}|p{0.6in}|p{0.6in}|p{0.6in}|p{0.6in}|p{0.6in}|}\hline
        Ethical Impact on Employment & Considering ethical impact on employment in MI & Developer awareness of employment implications & Institutional frameworks for ethical employment impact & National policies on MI's ethical impact on employment & Platform-specific considerations for ethical employment impact & Global collaboration on ethical employment impact & Evolving ethical norms in MI's impact on employment \\
        \hline
        Responsible MI Procurement & Ethical considerations in MI system procurement & Developer commitment to responsible MI procurement & Institutional guidelines for ethical MI procurement & National standards for responsible MI system acquisition & Platform-specific considerations in MI procurement & Global initiatives for responsible MI procurement & Evolving norms in ethical MI procurement \\
        \hline
        Community Engagement & Engaging with communities in MI development & Developer practices for community engagement in MI & Institutional frameworks for community involvement in MI & National policies on community engagement in MI & Platform-specific community engagement mechanisms & Global standards for MI community involvement & Ethical considerations in community engagement \\
        \hline
        MI and Democratic Values & Upholding democratic values in MI applications & Developer adherence to democratic principles in MI & Institutional support for democratic MI practices & National laws on MI and democratic values & Platform-specific democratic MI considerations & Global collaboration on democratic MI standards & Evolving norms in MI's alignment with democratic values \\
        \hline
        User Empowerment & Empowering users in MI interactions & Developer focus on empowering users in MI systems & Institutional support for user empowerment in MI & National guidelines for user empowerment in MI & Platform-specific mechanisms for user empowerment & Global standards for MI user empowerment & Ethical considerations in user empowerment in MI \\ \hline
               Digital Divide Reduction & Reducing digital divides with ethical MI & Developer efforts to address digital inequalities & Institutional strategies for reducing the digital divide & National policies on digital divide reduction through MI & Platform-specific measures for digital divide reduction & Global collaboration on MI for digital divide reduction & Ethical implications of MI in addressing digital disparities \\
        \hline
        Fair Distribution of MI Benefits & Ensuring fair distribution of MI benefits & Developer consideration of benefit distribution in MI & Institutional frameworks for fair benefit distribution & National policies on equitable MI benefits & Platform-specific mechanisms for fair MI benefit distribution & Global collaboration for equitable MI benefits & Ethical considerations in MI benefit distribution \\
        \hline
        Long-Term Environmental Sustainability & Considering long-term environmental impact in MI & Developer awareness of MI's environmental footprint & Institutional frameworks for sustainable MI practices & National regulations on MI and environmental sustainability & Platform-specific measures for environmentally sustainable MI & Global collaboration on MI's environmental impact & Ethical implications of MI's environmental footprint \\
        \hline
        Ethical Marketing and Advertising & Ethical considerations in MI marketing & Developer practices for ethical MI advertising & Institutional guidelines for ethical MI marketing & National standards for responsible MI promotion & Platform-specific ethical marketing in MI & Global initiatives for ethical MI advertising & Evolving norms in MI marketing ethics \\
        \hline
        Responsible MI Journalism & Ethical considerations in reporting on MI & Developer commitment to responsible MI journalism & Institutional guidelines for ethical reporting on MI & National standards for journalism on MI & Platform-specific responsible MI reporting practices & Global collaboration on ethical MI journalism & Evolving norms in MI journalism ethics \\
        \hline
        \end{tabular}
\caption{ Some examples of Non-Technical MI Ethics 2}
\label{tab:comprehensive-non-technical-mi-ethics2}
\end{table*} }

  {  \tiny 
        \begin{table*}[htb]
    \centering \  \tiny
    \begin{tabular}{|p{0.6in}|p{0.6in}|p{0.6in}|p{0.6in}|p{0.6in}|p{0.6in}|p{0.6in}|p{0.6in}|}
        \hline 

        MI and Accessibility & Ensuring accessibility in MI applications & Developer focus on accessible MI models & Institutional support for accessible MI practices & National regulations on MI accessibility & Platform-specific measures for MI accessibility & Global collaboration for accessible MI & Ethical considerations in MI accessibility \\
        \hline
        MI and Cultural Heritage Preservation & Ethical considerations in preserving cultural heritage with MI & Developer practices for culturally-sensitive MI in heritage preservation & Institutional frameworks for ethical cultural heritage MI & National guidelines for preserving cultural heritage with MI & Platform-specific cultural heritage MI considerations & Global collaboration on MI for cultural heritage preservation & Evolving norms in MI's role in cultural heritage \\
        \hline
        Human-MI Collaboration Ethics & Ethical considerations in human-MI collaborations & Developer commitment to ethical
human-MI collaboration & Institutional guidelines for ethical human-MI partnerships & National policies on ethical MI collaboration with humans & Platform-specific ethics in human-MI cooperation & Global standards for ethical human-MI collaboration & Evolving norms in MI's interaction with humans \\
\hline
MI Impact on Social Equality & Considering social equality in MI impact & Developer awareness of MI's impact on social equality & Institutional frameworks for assessing MI's impact on social equality & National policies on MI and social equality & Platform-specific measures for promoting social equality in MI & Global collaboration on MI's impact on social equality & Ethical implications of MI's influence on social equality \\
\hline
Algorithmic Accountability in Government & Holding government algorithms accountable & Developer practices for accountable government algorithms & Institutional frameworks for algorithmic accountability in government & National regulations on MI accountability in government & Platform-specific mechanisms for accountable government algorithms & Global collaboration on accountable government MI & Ethical considerations in government algorithm accountability \\
\hline
Elderly and Vulnerable Population Protection & Protecting elderly and vulnerable populations in MI & Developer consideration of MI's impact on vulnerable populations & Institutional safeguards for elderly and vulnerable individuals in MI use & National guidelines for protecting vulnerable populations with MI & Platform-specific measures for safeguarding vulnerable populations & Global collaboration on MI protection for vulnerable individuals & Ethical implications of MI's influence on elderly and vulnerable populations \\
\hline
Ethical MI Policy Development & Ethical considerations in MI policy creation & Developer commitment to ethical MI policy development & Institutional guidelines for ethical MI policy formulation & National policies on creating ethical MI frameworks & Platform-specific ethical MI policy development & Global collaboration on ethical MI policy creation & Evolving norms in MI policy development ethics \\
\hline
Cross-Industry Collaboration and Coopetition & Collaborating across industries for ethical MI & Developer engagement in cross-industry MI collaborations & Institutional support for ethical collaboration and coopetition in MI & National encouragement of cross-industry MI partnerships & Platform-specific mechanisms for cross-industry MI collaboration & Global standards for cross-industry MI coopetition & Ethical implications of MI collaboration across industries \\
\hline
MI and Child Protection & Ensuring child protection in MI applications & Developer focus on child-safe MI models & Institutional support for MI applications ensuring child safety & National regulations on MI and child protection & Platform-specific measures for child-safe MI & Global collaboration on MI for child protection & Ethical considerations in MI's impact on child safety \\
\hline
MI for Social Good & Promoting MI for positive social impact & Developer commitment to using MI for social good & Institutional frameworks for MI applications benefiting society & National policies on MI for social good & Platform-specific initiatives for MI's positive social impact & Global collaboration on MI for social good & Evolving norms in MI's role for societal benefit \\
\hline
\end{tabular}
\caption{ Some examples of Non-Technical MI Ethics 3}
\label{tab:comprehensive-non-technical-mi-ethics3}
\end{table*} }

\subsection{ Rows: Selected ethics topics}
\subsubsection*{  Community Engagement and Communal Values}
\begin{itemize}\item Individual: Active participation and collaboration within local communities, emphasizing shared values, traditional practices, and community development.

\item Different Cultures: Respect for diverse ethnic, linguistic, and cultural identities, fostering unity and cooperation while preserving and celebrating unique cultural expressions.

\item Institutions: Support for community-based institutions that uphold communal values, ensuring their integration into broader societal frameworks and decision-making processes.

\item Nation: National policies promoting community engagement, preserving cultural heritage, and recognizing the importance of communal values in shaping the nation's identity.

\item Globally: Collaboration with global entities while maintaining a strong connection to African cultural values, contributing to a more inclusive and culturally sensitive global community.

\item Over Time: Preservation and adaptation of communal values over generations, acknowledging their historical significance and evolving in harmony with changing societal needs.
\end{itemize}
\subsubsection*{ Cultural Sensitivity}
\begin{itemize}\item Individual: Personal awareness and respect for the rich tapestry of African cultures, promoting inclusive attitudes and fostering cross-cultural understanding in daily interactions.

\item Different Cultures: Encouragement of cultural exchange, dialogue, and mutual appreciation among diverse African cultures, emphasizing the importance of preserving cultural heritage.

\item Institutions: Integration of cultural sensitivity into institutional policies, ensuring fair representation and equitable treatment of individuals from various cultural backgrounds.

\item Nation: National efforts to protect and promote cultural diversity, recognizing the intrinsic value of each culture in contributing to the nation's identity.

\item Globally: Advocacy for global recognition and respect of African cultures, challenging stereotypes and fostering a positive image of Africa on the global stage.

\item Over Time: Preservation of cultural heritage and adaptation to contemporary contexts, allowing cultural traditions to evolve while maintaining their authenticity.
\end{itemize}
\subsubsection*{ Environmental Stewardship}
\begin{itemize}\item
Individual: Sustainable practices and environmental awareness at the individual level, emphasizing the importance of preserving Africa's diverse ecosystems and natural resources.
\item 
Different Cultures: Incorporation of traditional ecological knowledge into environmental conservation efforts, recognizing the symbiotic relationship between African cultures and nature.
\item 
Institutions: Implementation of policies that prioritize environmental sustainability, considering the impact on local ecosystems and ensuring responsible resource management.
\item 
Nation: National strategies for environmental protection, addressing challenges like deforestation, wildlife conservation, and climate change adaptation in the African context.
\item 
Globally: Collaboration on global environmental initiatives, advocating for equitable representation and acknowledging Africa's role in the global ecological balance.
\item 
Over Time: Long-term environmental planning, balancing development with conservation to secure a sustainable and thriving environment for future generations.
\end{itemize}
\subsubsection*{ Social Justice and Equality}
\begin{itemize}\item
Individual: Advocacy for social justice, equality, and human rights at the individual level, addressing issues like gender inequality, discrimination, and social disparities.
\item 
Different Cultures: Recognition and celebration of diversity, promoting inclusive social structures that respect and uphold the rights of individuals from various cultural backgrounds.
\item 
Institutions: Development and implementation of policies that combat social injustices, ensuring equal opportunities and fair representation within societal institutions.
\item 
Nation: National commitment to social justice, actively addressing historical inequalities and working towards a more inclusive and equitable society.
\item 
Globally: Participation in global movements for social justice, challenging systemic inequalities on the international stage and advocating for Africa's interests.
\item Over Time: Ongoing efforts to eradicate social injustices, acknowledging historical challenges and working towards a more just and equal future.
\end{itemize}
\subsubsection*{ Healthcare Equity}
\begin{itemize}\item
Individual: Access to healthcare education and proactive engagement in personal health practices, promoting preventive measures and awareness within local communities.
\item 
Different Cultures: Integration of culturally sensitive healthcare practices, recognizing traditional healing methods and addressing healthcare disparities in diverse cultural contexts.
\item 
Institutions: Development of healthcare policies that prioritize equity, ensuring equal access to quality healthcare services for all citizens.
\item 
Nation: National healthcare strategies that address public health challenges, emphasizing universal access to healthcare and reducing health inequalities.
\item 
Globally: Collaboration on global health initiatives, advocating for fair distribution of resources and acknowledging Africa's contributions to global health.
\item 
Over Time: Progress in healthcare infrastructure and services, adapting to evolving health needs and ensuring continuous improvement in healthcare equity.
\end{itemize}
\subsubsection*{ Education Access}
\begin{itemize}\item
Individual: Pursuit of education and lifelong learning, emphasizing the value of knowledge and skills to empower individuals and contribute to community development.
\item 
Different Cultures: Recognition of diverse learning styles and educational approaches, fostering an inclusive education system that respects various cultural perspectives.
\item 
Institutions: Implementation of inclusive educational policies, ensuring equal access to quality education for individuals from different cultural backgrounds.
\item 
Nation: National commitment to education as a key driver of development, investing in educational infrastructure and promoting access for all citizens.
\item 
Globally: Engagement in international collaborations on educational initiatives, contributing African perspectives to global education discussions.
\item 
Over Time: Evolution of education systems to meet changing needs, adapting to technological advancements and ensuring education remains a cornerstone of societal progress.
\end{itemize}
\subsubsection*{ Economic Inclusivity}
\begin{itemize}\item
Individual: Entrepreneurship, financial literacy, and economic empowerment at the individual level, fostering economic independence and community development.

\item Different Cultures: Recognition and integration of diverse economic models, acknowledging the richness of various African economic practices and promoting inclusive economic policies.

\item Institutions: Implementation of policies that promote economic inclusivity, ensuring fair distribution of resources and opportunities within the national economic framework.

\item Nation: National strategies for inclusive economic development, addressing economic disparities and fostering economic resilience.

\item Globally: Engagement in global economic collaborations, advocating for fair representation and acknowledging Africa's potential in shaping global economic landscapes.

\item Over Time: Adaptation of economic strategies to evolving global markets, balancing traditional economic practices with innovative approaches for sustainable economic inclusivity.
\end{itemize}
\subsubsection*{ Technological Ethics}

\begin{itemize}\item Individual: Ethical use of technology, digital literacy, and participation in technological advancements, recognizing the role of technology in shaping the African narrative.

\item Different Cultures: Integration of diverse perspectives into technological innovation, ensuring that technological solutions consider the cultural context and benefit diverse communities.

\item Institutions: Development of ethical guidelines for technology use, fostering responsible innovation and addressing potential ethical challenges in technological advancements.

\item Nation: National policies that promote responsible and inclusive technology development, bridging the digital divide and ensuring equitable access to technological benefits.

\item Globally: Participation in global discussions on technological ethics, advocating for fair representation and ethical considerations that address Africa's unique challenges and opportunities.

\item Over Time: Ethical adaptation to technological advancements, recognizing the evolving role of technology in African societies and ensuring its alignment with ethical principles over time.
\end{itemize}
\subsubsection*{ Political Accountability}
\begin{itemize}\item
Individual: Civic engagement, political awareness, and active participation at the individual level, fostering a sense of responsibility and accountability in governance.

\item Different Cultures: Recognition of diverse political perspectives, promoting inclusive political discourse that considers cultural nuances and diverse voices.
\item Institutions: Development of transparent and accountable political systems, ensuring representation and responsiveness to the needs of diverse cultural and social groups.
\item Nation: National commitment to political accountability, addressing corruption, promoting transparency, and ensuring fair and inclusive political processes that respect the diverse cultural and social fabric of the nation.
\item Globally: Active engagement in international political collaborations, contributing African perspectives to global governance discussions, and advocating for fair representation on the global stage.
\item Over Time: Evolution of political systems that adapt to changing societal needs, acknowledging historical challenges, and continuously working towards accountable and inclusive governance over time.
\end{itemize}
\subsubsection*{ Local, Inter-regional, and International Collaboration}
\begin{itemize}\item
Individual: Promotion of collaboration at the local level, emphasizing community engagement and fostering partnerships for community development.
\item 
Different Cultures: Integration of diverse cultural perspectives in collaborative initiatives, recognizing the richness of cultural diversity in shaping collaborative efforts.
\item 
Institutions: Development of policies that facilitate inter-regional and international collaboration, ensuring equitable participation and mutual benefit.
\item 
Nation: National strategies for diplomatic collaboration, promoting Africa's interests on the international stage while respecting the diversity of African nations.
\item 
Globally: Active engagement in global collaborations, contributing African perspectives to international discussions and fostering partnerships that address global challenges.
\item 
Over Time: Adaptation of collaboration strategies to changing global dynamics, recognizing historical collaborations and continuously working towards more effective and inclusive collaborative efforts over time.
\end{itemize}

In the African context, each of these rows emphasizes the unique challenges, opportunities, and values that contribute to the multifaceted MI ethical landscape. The interconnectedness of these aspects reflects the dynamic and evolving nature of MI ethics in Africa, where the preservation of cultural identity, environmental sustainability, social justice, and inclusive economic development are integral components of MI ethical considerations across various scales and over time.

\subsection{ Columns: MI Ethics from  individuals to the globe level}
\subsubsection*{ Individual}
\begin{itemize}\item
Community Engagement and Communal Values: Personal commitment to actively participate in and contribute to local communities, emphasizing shared values and collaborative initiatives for community development.

\item Cultural Sensitivity: Personal awareness and respect for the diverse cultures within Africa, fostering an inclusive attitude that appreciates and celebrates the richness of cultural identities.

\item Environmental Stewardship: Individual responsibility for sustainable practices, including ecological awareness and a commitment to preserving Africa's diverse ecosystems and natural resources.

\item Social Justice and Equality: Personal advocacy for social justice, human rights, and equality, addressing issues like gender inequality, discrimination, and socioeconomic disparities.

\item Healthcare Equity: Commitment to personal health practices and proactive engagement in healthcare education, contributing to equitable access to healthcare services.

\item Education Access: Pursuit of education and lifelong learning, recognizing the transformative power of education in personal and community development.

\item Economic Inclusivity: Individual initiatives in entrepreneurship, financial literacy, and economic empowerment, fostering economic independence and community resilience.

\item Technological Ethics: Ethical use of technology, digital literacy, and responsible participation in technological advancements, considering the impact of technology on African societies.

\item Political Accountability: Civic engagement, political awareness, and active participation at the individual level, contributing to transparent and accountable political processes.

\item Local, Inter-regional, and International Collaboration: Promotion of collaboration at the local level, emphasizing community engagement and fostering partnerships for community development.
\end{itemize}
\subsubsection*{ Different Cultures}
\begin{itemize}\item Community Engagement and Communal Values: Embracing and integrating diverse cultural practices within local communities, promoting unity while respecting unique cultural expressions.

\item Cultural Sensitivity: Encouraging cultural exchange, dialogue, and mutual appreciation among diverse African cultures, fostering cross-cultural understanding and harmony.

\item Environmental Stewardship: Incorporating traditional ecological knowledge into environmental conservation efforts, recognizing the symbiotic relationship between African cultures and nature.

\item Social Justice and Equality: Recognition and celebration of diversity, promoting inclusive social structures that respect and uphold the rights of individuals from various cultural backgrounds.

\item Healthcare Equity: Integration of culturally sensitive healthcare practices, acknowledging traditional healing methods and addressing healthcare disparities in diverse cultural contexts.

\item Education Access: Recognition of diverse learning styles and educational approaches, fostering an inclusive education system that respects various cultural perspectives.

\item Economic Inclusivity: Acknowledgment and integration of diverse economic models, ensuring that economic policies consider the richness of various African economic practices.

\item Technological Ethics: Integration of diverse perspectives into technological innovation, ensuring that technological solutions consider the cultural context and benefit diverse communities.

\item Political Accountability: Recognition of diverse political perspectives, promoting inclusive political discourse that considers cultural nuances and diverse voices.

\item Local, Inter-regional, and International Collaboration: Integration of diverse cultural perspectives in collaborative initiatives, recognizing the richness of cultural diversity in shaping collaborative efforts.
\end{itemize}
\subsubsection*{ Institutions}
\begin{itemize}\item
Community Engagement and Communal Values: Support for community-based institutions that uphold communal values, ensuring their integration into broader societal frameworks and decision-making processes.

\item Cultural Sensitivity: Integration of cultural sensitivity into institutional policies, ensuring fair representation and equitable treatment of individuals from various cultural backgrounds.

\item Environmental Stewardship: Implementation of policies that prioritize environmental sustainability, considering the impact on local ecosystems and ensuring responsible resource management.

\item Social Justice and Equality: Development and implementation of policies that combat social injustices, ensuring equal opportunities and fair representation within societal institutions.

\item Healthcare Equity: Development of healthcare policies that prioritize equity, ensuring equal access to quality healthcare services for all citizens.

\item Education Access: Implementation of inclusive educational policies, ensuring equal access to quality education for individuals from different cultural backgrounds.

\item Economic Inclusivity: Implementation of policies that promote economic inclusivity, ensuring fair distribution of resources and opportunities within the national economic framework.

\item Technological Ethics: Development of ethical guidelines for technology use, fostering responsible innovation and addressing potential ethical challenges in technological advancements.

\item Political Accountability: Development of transparent and accountable political systems, ensuring representation and responsiveness to the needs of diverse cultural and social groups.

\item Local, Inter-regional, and International Collaboration: Development of policies that facilitate inter-regional and international collaboration, ensuring equitable participation and mutual benefit.
\end{itemize}
\subsubsection*{ Nation}
\begin{itemize}\item
Community Engagement and Communal Values: National policies promoting community engagement, preserving cultural heritage, and recognizing the importance of communal values in shaping the nation's identity.

\item Cultural Sensitivity: National efforts to protect and promote cultural diversity, recognizing the intrinsic value of each culture in contributing to the nation's identity.

\item Environmental Stewardship: National strategies for environmental protection, addressing challenges like deforestation, wildlife conservation, and climate change adaptation in the African context.

\item Social Justice and Equality: Policies should aim to address social injustices and promote equality, ensuring that MI benefits are accessible to all segments of society.

\item Healthcare Equity: National MI in healthcare should focus on equity, providing accessible and affordable healthcare solutions for all citizens.

\item Education Access: Policies should promote equal access to education through MI, bridging educational gaps and ensuring inclusivity.

\item Economic Inclusivity: National MI strategies should foster economic inclusivity, preventing disparities and ensuring that the economic benefits of MI are widespread.

\item Technological Ethics: National policies should address ethical considerations related to technology, promoting responsible and ethical MI practices.

\item Political Accountability: Policies should incorporate mechanisms for political accountability, ensuring that MI decisions align with democratic values and are subject to scrutiny.

\item Local, Inter-regional and International Collaboration: Collaboration at different levels is essential, involving local communities, inter-regional partnerships, and international collaborations to address global challenges and share best practices.
\end{itemize}
\subsubsection*{ On and Off-Line Platforms}
\begin{itemize}\item
Community Engagement and Communal Values: Online platforms should engage communities, respecting communal values, and ensuring that MIinteractions align with local norms.

\item Cultural Sensitivity: Platforms should be culturally sensitive, catering to diverse cultural backgrounds and ensuring that online MI experiences are respectful.

\item Environmental Stewardship: Platforms should consider environmental impacts, adopting eco-friendly practices in data centers and technology infrastructure.

\item Social Justice and Equality: Online platforms should promote social justice and equality, preventing discrimination and ensuring fair representation.

\item Healthcare Equity: Online healthcare platforms should prioritize equity, providing accessible healthcare information and services to all users.

\item Education Access: Online education platforms should ensure equal access to educational resources, addressing digital divides and promoting inclusivity.

\item Economic Inclusivity: Platforms should promote economic inclusivity, preventing digital exclusion and ensuring equal opportunities for online participation.

\item Technological Ethics: Online platforms should adhere to ethical standards, ensuring responsible data handling, privacy protection, and transparent algorithms.

\item Political Accountability: Platforms should establish mechanisms for political accountability, ensuring transparent governance and accountability in MI-driven decisions.

\item Local, Inter-regional and International Collaboration: Collaboration on platforms should extend beyond borders, involving local, inter-regional, and international partnerships to address global challenges and ensure diverse perspectives.
\end{itemize}
\subsubsection*{ Globally}
\begin{itemize}\item Community Engagement and Communal Values: Global MI initiatives should engage communities worldwide, respecting diverse communal values and incorporating them into decision-making processes.

\item Cultural Sensitivity: Global collaborations should be culturally sensitive, considering and respecting diverse cultural values across different regions.

\item Environmental Stewardship: Global MI efforts should prioritize environmental stewardship, promoting sustainable practices and minimizing the carbon footprint of MI technologies.

\item Social Justice and Equality: Global MI collaborations should address global social injustices and promote equality on a worldwide scale.

\item Healthcare Equity: Global efforts in MI healthcare should focus on equity, ensuring that healthcare solutions are accessible and beneficial to populations globally.

\item Education Access: Global initiatives should promote equal access to education through MI, bridging educational gaps and ensuring global educational inclusivity.

\item Economic Inclusivity: Global MI strategies should foster economic inclusivity, preventing global disparities and ensuring that economic benefits are distributed equitably.

\item Technological Ethics: Global collaborations should uphold technological ethics, establishing common ethical standards and principles for MI development and deployment globally.

\item Political Accountability: Global initiatives should incorporate mechanisms for political accountability, ensuring that MI decisions align with democratic values and are subject to international scrutiny.

\item Local, Inter-regional and International Collaboration: Collaboration at a global level is essential, involving local, inter-regional, and international partnerships to address global challenges and share best practices on a broader scale.
\end{itemize}
\subsubsection*{ Over Time}
\begin{itemize}\item Community Engagement and Communal Values: Changes over time should involve ongoing community engagement, ensuring that communal values are considered and respected in evolving MI strategies.
\item 
Cultural Sensitivity: Evolving MI strategies should remain culturally sensitive, adapting to changing cultural contexts over time.
\item
Environmental Stewardship: MI developments should adapt over time to align with evolving environmental standards and best practices.
\item 
Social Justice and Equality: Over time, MI policies should adapt to address emerging social justice issues and promote ongoing equality.
\item 
Healthcare Equity: MI in healthcare should adapt over time to ensure ongoing equity, addressing evolving healthcare challenges and disparities.
\item 
Education Access: MI in education should evolve over time to meet changing educational needs and ensure continuous access to educational resources.
\item 
Economic Inclusivity: Over time, MI strategies should adapt to foster ongoing economic inclusivity, preventing the emergence of new disparities.
\item 
Technological Ethics: Over time, technological ethics should evolve to address emerging ethical challenges associated with advancing MI technologies.
\item 
Political Accountability: Over time, MI policies should adapt to maintain political accountability, addressing new challenges and ensuring ongoing transparency in political decisions influenced by MI.
\item 
Local, Inter-regional and International Collaboration: Collaboration should continue over time, adapting to changing global dynamics and ensuring that MI efforts remain collaborative, inclusive, and responsive to evolving challenges.
\end{itemize}

\subsection{Multi-scale ethics interactions}
The strong interactions between the rows and columns of ethics in the multiscale framework create a comprehensive and interconnected ethical landscape. Let's explore some of their relationships:
\subsubsection*{ Community Engagement and Communal Values}
\begin{itemize}\item Individual: Transparency and fairness at the individual level foster a sense of responsibility and equity in engaging with communal values.
\item 
Different Cultures: Cultural sensitivity ensures that diverse cultural perspectives are respected in communal discussions.
\item 
Institutions: Institutional responsibility shapes the transparency and fairness of communal engagements.
\item 
Nation: Political accountability contributes to the responsible handling of communal values on a national scale.
\item 
Globally: International collaboration promotes a shared understanding of communal values globally.
\item 
Over Time: Historical perspectives guide the context-awareness and equitable evolution of communal values.
\end{itemize}
\subsubsection*{ Cultural Sensitivity}
\begin{itemize}\item Individual: Personal awareness and responsibility enhance cultural sensitivity in individual interactions.
\item 
Different Cultures: Collaborative efforts and fairness promote cultural sensitivity among diverse cultural groups.
\item 
Institutions: Institutional responsibility ensures cultural sensitivity in policies and practices.
\item 
Nation: Political accountability addresses cultural biases, fostering cultural sensitivity on a national level.
\item 
Globally: Global collaboration shapes cultural sensitivity on a worldwide scale.
\item 
Over Time: Historical perspectives contribute to the context-awareness and equitable evolution of cultural sensitivity.
\end{itemize}
\subsubsection*{ Environmental Stewardship}
\begin{itemize}\item Individual: Responsible and transparent individual actions contribute to environmental stewardship.
\item 
Different Cultures: Collaborative efforts and fairness address diverse cultural perspectives in environmental practices.
\item 
Institutions: Institutional responsibility shapes transparent and fair environmental stewardship policies.
\item 
Nation: Political accountability ensures equitable and responsible environmental practices nationally.
\item
Globally: Global collaboration fosters shared responsibility for environmental stewardship worldwide.
\item 
Over Time: Historical perspectives guide context-aware and equitable environmental stewardship evolution.
\end{itemize}
\subsubsection*{ Social Justice and Equality}
\begin{itemize}\item Individual: Fair and responsible individual actions contribute to social justice and equality.
\item 
Different Cultures: Collaborative efforts and cultural sensitivity address diverse cultural perspectives in promoting equality.
\item 
Institutions: Institutional responsibility shapes transparent and fair social justice policies.
\item 
Nation: Political accountability ensures equitable and responsible social justice on a national level.
\item 
Globally: Global collaboration fosters shared responsibility for social justice and equality worldwide.
\item 
Over Time: Historical perspectives guide context-aware and equitable social justice evolution.
\end{itemize}
\subsubsection*{ Healthcare Equity}
\begin{itemize}\item Individual: Responsible individual health practices contribute to equitable healthcare.

Different Cultures: Collaborative efforts and cultural sensitivity address diverse cultural health beliefs.
\item 
Institutions: Institutional responsibility shapes transparent and fair healthcare policies.
\item 
Nation: Political accountability ensures equitable healthcare nationally.
\item 
Globally: Global collaboration fosters shared responsibility for healthcare equity worldwide.
\item 
Over Time: Historical perspectives guide context-aware and equitable healthcare evolution.
\end{itemize}
\subsubsection*{ Education Access}
\begin{itemize}\item Individual: Responsible individual learning practices contribute to equitable education access.
\item Different Cultures: Collaborative efforts and fairness address diverse cultural perspectives in education.
\item 
Institutions: Institutional responsibility shapes transparent and fair education access policies.
\item 
Nation: Political accountability ensures equitable education nationally.
\item 
Globally: Global collaboration fosters shared responsibility for education access worldwide.
\item 
Over Time: Historical perspectives guide context-aware and equitable education access evolution.
\end{itemize}
\subsubsection*{ Economic Inclusivity}
\begin{itemize}\item Individual: Responsible economic practices contribute to economic inclusivity.
Different Cultures: Collaborative efforts and cultural sensitivity address diverse cultural economic models.
\item 
Institutions: Institutional responsibility shapes transparent and fair economic policies.
\item 
Nation: Political accountability ensures economic inclusivity on a national level.
\item 
Globally: Global collaboration fosters shared responsibility for economic inclusivity worldwide.
\item 
Over Time: Historical perspectives guide context-aware and equitable economic inclusivity evolution.
\end{itemize}
\subsubsection*{ Technological Ethics}
\begin{itemize}\item Individual: Responsible individual technology use contributes to ethical technology practices.
\item 
Different Cultures: Collaborative efforts and cultural sensitivity address diverse cultural perspectives in technological ethics.
\item
Institutions: Institutional responsibility shapes transparent and fair technological policies.
\item 
Nation: Political accountability ensures ethical technology use on a national level.
\item 
Globally: Global collaboration fosters shared responsibility for technological ethics worldwide.
\item 
Over Time: Historical perspectives guide context-aware and equitable technological ethics evolution.
\end{itemize}
\subsubsection*{ Political Accountability}
\begin{itemize}\item Individual: Responsible individual civic engagement contributes to political accountability.
\item 
Different Cultures: Collaborative efforts and fairness address diverse cultural political perspectives.
\item
Institutions: Institutional responsibility shapes transparent and fair political policies.
\item 
Nation: Political accountability ensures accountable political processes on a national level.
\item 
Globally: Global collaboration fosters shared responsibility for political accountability worldwide.
\item 
Over Time: Historical perspectives guide context-aware and equitable political accountability evolution.
\end{itemize}
 \subsubsection*{ Local, Inter-Regional and International Collaboration}
\begin{itemize}\item Individual: Responsible individual collaboration contributes to successful global collaboration.
\item 
Different Cultures: Collaborative efforts and cultural sensitivity address diverse cultural perspectives in collaborative initiatives.
\item 
Institutions: Institutional responsibility shapes transparent and fair collaboration policies.
\item 
Nation: Political accountability ensures equitable collaboration nationally.
\item 
Globally: Global collaboration fosters shared responsibility for collaboration worldwide.
\item
Over Time: Historical perspectives guide context-aware and equitable collaboration evolution.
\end{itemize}
\subsection*{ Interactions Across Rows}
Cultural sensitivity in communal values ensures that diverse cultural perspectives are respected, creating an inclusive and fair community engagement. Addressing social justice in environmental practices promotes equitable and responsible environmental stewardship, considering the impact on diverse communities. Promoting healthcare equity and education access ensures that diverse communities have fair and transparent access to essential services, contributing to overall well-being. Ethical technology practices contribute to economic inclusivity, ensuring fair and responsible economic policies that consider technological advancements. Political accountability fosters fair collaboration, promoting transparent and responsible collaboration in political processes.
\subsection*{ Interactions Across Columns and Rows}
Responsible individual actions contribute to positive outcomes across all ethics areas, creating a foundation for transparency, fairness, responsibility, equity, and context-awareness. Considering diverse cultural perspectives in individual actions, collaborative efforts, and institutional responsibilities ensures inclusivity, fairness, and cultural sensitivity. Institutional responsibility influences national policies, shaping the ethical landscape within a country and contributing to responsible governance and fair practices. Global collaboration and historical perspectives guide the evolution of ethical practices over time, fostering a globally inclusive, transparent, fair, responsible, and context-aware approach.
\subsection*{Multi-interactions of ethics}
The interactions between multiple columns in the multiscale ethics framework are dynamic and interwoven, creating a cohesive ethical environment. Transparency and fairness at the individual level contribute to fair institutional policies, while acknowledging diverse cultural perspectives promotes responsible governance on a national level. Global responsibility guides the equitable evolution of ethical norms over time, and context-awareness influences fair collaboration across local, inter-regional, and international scales. The connections between equity and context-awareness extend to various columns, such as healthcare equity influencing fair and context-aware education access. Overall, the principles of transparency, fairness, responsibility, equity, and context-awareness intersect, shaping ethical behavior across different scales and fostering a holistic and inclusive ethical framework over time.

{  \tiny 
\begin{table*}[htb] \label{secethicsp} \caption{Multi-interactions of ethics}
\centering \tiny
\begin{tabular}{|p{0.6in}|p{0.6in}|p{0.6in}|p{0.6in}|p{0.6in}|p{0.6in}|p{0.6in}|p{0.6in}|}
\hline
\textbf{Ethics Area} & \textbf{Individual} & \textbf{Different Cultures} & \textbf{Institutions} & \textbf{Nation} & \textbf{Platforms} & \textbf{Globally} & \textbf{Over Time} \\
\hline
Community Engagement and Communal Values & Personal commitment & Respect for diverse values & Establishment of guidelines & Integration into policies & Engagement in online and offline communities & Recognition and appreciation of diverse values & Preservation and adaptation over generations \\
\hline
Cultural Sensitivity & Cultural awareness & Mutual respect & Implementation of policies & Promotion as a national asset & Cultural inclusivity in digital spaces & Collaboration with diverse international cultures & Preservation and evolution of cultural identities \\
\hline
Environmental Stewardship & Personal responsibility & Integration of eco-friendly practices & Implementation of environmental policies & National commitment to sustainability & Online platforms promoting eco-conscious actions & Global cooperation on sustainable development & Sustainable practices for future generations \\
\hline
Social Justice and Equality & Advocacy for equality & Recognition of diverse perspectives & Implementation of policies & Legal frameworks for social justice & Online platforms for advocacy and awareness & Global collaboration on human rights & Progression toward increased social justice \\
\hline
Healthcare Equity & Personal health responsibility & Culturally sensitive healthcare practices & Equal access healthcare systems & National health policies for equity & Digital platforms for health information & Global collaborations for health equity & Sustainable healthcare for future generations \\
\hline
Education Access & Personal commitment to learning & Inclusion of diverse educational perspectives & Equal access educational systems & National commitment to accessible education & Online platforms for e-learning & Global partnerships for education opportunities & Evolving education systems over generations \\
\hline
Economic Inclusivity & Ethical economic practices & Integration of diverse economic models & Economic policies for inclusivity & National commitment to inclusive growth & Online platforms promoting fair economic practices & Collaborative efforts for global economic inclusivity & Sustainable economic practices for future generations \\
\hline
Technological Ethics & Responsible technology use & Consideration of diverse cultural perspectives & Ethical guidelines and regulations & National policies for responsible technology & Digital platforms for ethical tech discussions & International agreements on ethical tech use & Ethical considerations in technological advancements \\
\hline
Political Accountability & Active civic engagement & Recognition of diverse political perspectives & Transparent political systems & National commitment to accountability & Digital platforms for political discourse & International collaboration for transparent governance & Evolution of political accountability \\
\hline
Local, Inter-regional and International Collaboration & Appreciation for global perspectives & Integration of diverse international viewpoints & Global collaboration frameworks & National commitment to international cooperation & Online platforms for cross-cultural collaboration & Collaborative efforts on global challenges & Building and sustaining international collaborations \\
\hline
\end{tabular}
\end{table*}
}

\subsection{ Ethics at the individual level: users, developers, designers, individual investors}
\subsubsection*{ Community Engagement and Communal Values}
Individual: Transparency involves openly communicating personal values and commitments to the community. Fairness ensures equal participation, responsibility in upholding communal values, and equity in benefit-sharing. Responsibility requires individuals to contribute positively to the community. Context-awareness considers the cultural context and its impact on communal values.
\subsubsection*{ Cultural Sensitivity}
Individual: Transparent appreciation of diverse cultures, ensuring fairness in personal interactions. Responsibility involves learning and adapting to cultural nuances. Equity includes respecting cultural differences. Context-awareness requires understanding the cultural context of one's actions.
\subsubsection*{ Environmental Stewardship}
Individual: Transparent eco-friendly practices, fair resource consumption, responsible waste management, and equitable access to a clean environment. Responsibility involves personal actions for environmental sustainability. Context-awareness considers local ecosystems.
\subsubsection*{ Social Justice and Equality}
Individual: Transparent advocacy for social justice, fair treatment of diverse perspectives, and responsible behavior. Equity requires acknowledging and addressing social disparities. Context-awareness involves understanding the unique challenges faced by different groups.
\subsubsection*{ Healthcare Equity}
Individual: Transparency in personal health choices, fair access to healthcare information, responsible health practices, and equitable support for healthcare access. Context-awareness considers cultural health beliefs and practices.
\subsubsection*{ Education Access}
Individual: Transparent commitment to learning, fairness in educational opportunities, responsible study habits, and equity in access to education. Context-awareness involves understanding diverse learning needs.
\subsubsection*{ Economic Inclusivity:}
Individual: Transparent economic practices, fair business dealings, responsible financial decisions, and equitable opportunities. Context-awareness considers local economic contexts.
\subsubsection*{ Technological Ethics}
Individual: Transparent use of technology, fair consideration of diverse perspectives, responsible digital behavior, and equitable access to technological benefits. Context-awareness involves understanding the societal impact of technology.
\subsubsection*{ Political Accountability:}
Individual: Transparent civic engagement, fair consideration of political perspectives, responsible voting, and equity in political representation. Context-awareness involves understanding the political context of one's actions.
\subsubsection*{ Local, Inter-regional and International Collaboration}
Individual: Transparent appreciation for global perspectives, fair collaboration, responsible global citizenship, and equity in international interactions. Context-awareness involves understanding the cultural and political contexts of collaborating regions.
\subsection{ Different Cultures, Different groups, Culture-Aware Ethics}

\subsubsection*{ Community Engagement and Communal Values}
Different Cultures:
\begin{itemize}\item Transparency: Transparent appreciation for diverse cultural values and practices, openly sharing and learning about different cultural perspectives within the community.

\item Fairness: Fair treatment of all cultures within the community, ensuring that cultural diversity is recognized and respected in communal engagements.

\item Responsibility: Responsible engagement with different cultures, taking initiative to understand and promote cultural awareness within the community.

\item Equity: Equitable representation of diverse cultural voices, ensuring that all cultures have an equal opportunity to contribute to communal values.

\item Context-Awareness: Being context-aware of the cultural nuances and sensitivities, adapting community engagements to respect the diversity of cultures involved.
\end{itemize}
\subsubsection*{ Cultural Sensitivity}
Different Cultures:
\begin{itemize}\item
Transparency: Transparent communication of cultural awareness, acknowledging the richness of various cultures, and expressing openness to cultural learning.
\item 
Fairness: Fair treatment of all cultures, avoiding biases or preferences, and appreciating the unique aspects of each cultural identity.
\item 
Responsibility: Taking responsibility for cultural education, actively challenging stereotypes, and contributing positively to cultural understanding.
\item 
Equity: Equitable consideration of all cultures, valuing diversity, and ensuring that no culture is marginalized or unfairly treated.
\item 
Context-Awareness: Being context-aware of cultural contexts, understanding historical and social factors shaping cultures, and adapting behavior to show respect within specific cultural settings.
\end{itemize}
\subsubsection*{ Environmental Stewardship}
Different Cultures:
\begin{itemize}\item
Transparency: Transparent communication about the environmental impact of diverse cultural practices, fostering awareness and open dialogue.
\item 
Fairness: Fair consideration of how different cultures interact with the environment, ensuring that environmental practices are just and inclusive.
\item 
Responsibility: Shared responsibility for environmental sustainability, acknowledging the role of diverse cultures in preserving ecosystems.
\item 
Equity: Equitable distribution of environmental responsibilities, recognizing that all cultures should contribute to sustainable practices.
\item 
Context-Awareness: Being context-aware of local ecosystems, adapting environmental stewardship to respect the unique environmental contexts shaped by different cultures.
\end{itemize}
\subsubsection*{ Social Justice and Equality}
Different Cultures:
\begin{itemize}\item
Transparency: Transparent advocacy for social justice, openly addressing cultural disparities, and promoting fairness in social issues.

\item Fairness: Fair consideration of cultural perspectives in social justice matters, ensuring that diverse voices are heard and valued.

\item Responsibility: Shared responsibility for addressing cultural inequalities, actively working towards equitable social outcomes.

\item Equity: Equitable treatment of different cultural groups, acknowledging and rectifying social injustices across cultures.

\item Context-Awareness: Being context-aware of unique challenges faced by different cultural groups, adapting social justice efforts to address specific cultural contexts.
\end{itemize}
\subsubsection*{ Healthcare Equity}
Different Cultures:
\begin{itemize}\item
Transparency: Transparent communication about healthcare practices, openly discussing cultural factors that influence health decisions.
\item 
Fairness: Fair consideration of cultural beliefs in healthcare, ensuring that healthcare practices are culturally sensitive and inclusive.
\item 
Responsibility: Taking responsibility for understanding and respecting diverse healthcare needs, contributing to culturally competent healthcare.
\item 
Equity: Equitable access to healthcare for all cultural groups, addressing healthcare disparities and promoting inclusive health policies.
\item 
Context-Awareness: Being context-aware of cultural health beliefs, adapting healthcare approaches to respect and integrate diverse cultural perspectives.
\end{itemize}
\subsubsection*{ Education Access}
Different Cultures:
\begin{itemize}\item
Transparency: Transparent communication about educational opportunities, fostering open dialogue about the importance of education across diverse cultures.
\item 
Fairness: Fair consideration of diverse cultural perspectives in education, ensuring that educational resources are accessible to all cultural groups.
\item 
Responsibility: Taking responsibility for promoting education across cultures, actively contributing to inclusive educational environments.
\item 
Equity: Equitable access to educational opportunities for all cultural groups, addressing educational disparities and promoting inclusive educational policies.
\item 
Context-Awareness: Being context-aware of diverse learning needs shaped by cultural backgrounds, adapting educational approaches to respect and accommodate cultural diversity.
\end{itemize}
\subsubsection*{ Economic Inclusivity}
Different Cultures:
\begin{itemize}\item
Transparency: Transparent communication about economic opportunities, fostering an open discussion about economic practices that respect diverse cultures.
\item 
Fairness: Fair consideration of diverse cultural economic models, ensuring that economic opportunities are distributed without bias.
\item 
Responsibility: Taking responsibility for ethical economic practices across cultures, actively contributing to fair economic behavior.
\item 
Equity: Equitable economic opportunities for all cultural groups, addressing economic disparities and promoting inclusive economic policies.
\item 
Context-Awareness: Being context-aware of local economic contexts, adapting economic practices to respect and integrate diverse cultural perspectives.
\end{itemize}
\subsubsection*{ Technological Ethics}
Different Cultures:
\begin{itemize}\item
Transparency: Transparent communication about technological advancements, fostering open discussions about how technology impacts diverse cultures.
\item 
Fairness: Fair consideration of diverse cultural perspectives in technological development, ensuring that technology is designed without bias.
\item 
Responsibility: Taking responsibility for ethical technology use across cultures, actively contributing to the responsible development and use of technology.
\item 
Equity: Equitable access to technological benefits for all cultural groups, addressing digital disparities and promoting inclusive technology policies.
\item 
Context-Awareness: Being context-aware of the societal impact of technology on different cultures, adapting technological approaches to respect and accommodate diverse cultural perspectives.
\end{itemize}
\subsubsection*{ Political Accountability}
Different Cultures:
\begin{itemize}\item
Transparency: Transparent communication about political processes, fostering open dialogue about the impact of politics on diverse cultures.
\item 
Fairness: Fair consideration of diverse political perspectives, ensuring that political systems are just and inclusive.
\item 
Responsibility: Taking responsibility for political engagement across cultures, actively contributing to transparent and accountable political processes.
\item 
Equity: Equitable political representation for all cultural groups, addressing political disparities and promoting inclusive political policies.
\item 
Context-Awareness: Being context-aware of the political contexts of different cultures, adapting political engagement to respect and integrate diverse cultural perspectives.
\end{itemize}

\subsubsection*{International Collaboration on MI Ethics in Africa} Different Cultures:
\begin{itemize}\item
Transparency: Transparent communication about collaborative efforts on MI ethics, fostering open dialogue about the impact of MI policies on diverse cultures within Africa.
\item 
Fairness: Fair consideration of diverse cultural perspectives in international collaborations, ensuring that MI ethics frameworks are just and inclusive across the continent.
\item 
Responsibility: Taking responsibility for collaborative MI ethics initiatives across cultures, actively contributing to transparent and accountable ethical frameworks for MI  deployment.
\item 
Equity: Equitable representation in international collaborations for all cultural groups within Africa, addressing disparities and promoting inclusive MI ethics policies.
\item 
Context-Awareness: Being context-aware of the cultural diversity within Africa in MI ethics, adapting collaborative efforts to respect and integrate diverse cultural perspectives in shaping ethical guidelines for MI.
\end{itemize}

\subsection{Institutions }
\subsubsection*{ Community Engagement and Communal Values}
Institutions:
\begin{itemize}\item
Transparency: Transparent establishment of guidelines for environmental stewardship initiatives, openly communicating institutional strategies for sustainability within the community.
\item 
Fairness: Fair implementation of environmental policies, ensuring that all community members have equal access to information and opportunities for involvement.
\item 
Responsibility: Institutional responsibility for fostering positive environmental practices within the community, actively engaging in initiatives for sustainable development.
\item 
Equity: Equitable distribution of environmental benefits and responsibilities among community members, addressing environmental disparities.
\item 
Context-Awareness: Instituting context-aware environmental programs that consider local ecosystems, adapting initiatives to respect the unique environmental context.
\end{itemize}
\subsubsection*{ Cultural Sensitivity}
Institutions:
\begin{itemize}\item
Transparency: Transparent communication about how institutional practices consider and respect diverse cultural environmental perspectives.

\item Fairness: Fair integration of cultural values into institutional environmental policies, ensuring that no culture is disproportionately affected.

\item Responsibility: Institutional responsibility for fostering cultural sensitivity in environmental initiatives, actively addressing cultural implications of environmental decisions.

\item Equity: Equitable distribution of environmental benefits and burdens, recognizing and rectifying any cultural biases in institutional practices.

\item Context-Awareness: Developing institution-wide environmental strategies that respect and integrate the diverse cultural contexts shaping environmental perspectives.
\end{itemize}
\subsubsection*{ Environmental Stewardship}
Institutions:

\begin{itemize}\item Transparency: Transparent communication of institutional efforts for environmental stewardship, openly sharing goals, progress, and outcomes with stakeholders.

\item Fairness: Fair allocation of resources and opportunities for environmental initiatives, ensuring that all stakeholders have a fair chance to participate.

\item Responsibility: Institutional responsibility for implementing effective environmental stewardship practices, actively contributing to global sustainability goals.

\item Equity: Equitable distribution of environmental benefits, addressing any disparities in the impact of institutional environmental practices.

\item Context-Awareness: Developing institution-wide environmental strategies that consider the unique contexts of local ecosystems and global environmental challenges.
\end{itemize}
\subsubsection*{ Social Justice and Equality}
Institutions:
\begin{itemize}\item
Transparency: Transparent communication of institutional efforts for social justice in the context of environmental initiatives, fostering open dialogue about the social implications.

\item Fairness: Fair consideration of diverse social perspectives in institutional environmental policies, ensuring that social justice is central to sustainability practices.

\item Responsibility: Institutional responsibility for addressing social inequalities exacerbated by environmental issues, actively working toward equitable outcomes.

\item Equity: Equitable distribution of social and environmental benefits, recognizing and rectifying any social disparities resulting from institutional practices.

\item Context-Awareness: Developing institution-wide environmental strategies that are context-aware of unique social challenges, adapting initiatives to promote social justice.
\end{itemize}
\subsubsection*{ Healthcare Equity}
Institutions:
\begin{itemize}\item
Transparency: Transparent communication about how institutional healthcare practices consider and respect diverse cultural health beliefs.

\item Fairness: Fair integration of cultural values into institutional healthcare policies, ensuring that healthcare services are culturally sensitive and inclusive.

\item Responsibility: Institutional responsibility for fostering healthcare equity, actively addressing cultural implications of healthcare decisions and practices.

\item Equity: Equitable distribution of healthcare benefits, recognizing and rectifying any cultural biases in institutional healthcare services.

\item Context-Awareness: Developing institution-wide healthcare strategies that consider the unique cultural contexts shaping health beliefs and practices.
\end{itemize}
\subsubsection*{ Education Access}
Institutions:
\begin{itemize} \item 
Transparency: Transparent communication of institutional efforts for promoting education access, openly sharing strategies and progress with stakeholders.
\item
Fairness: Fair allocation of resources and opportunities for education access initiatives, ensuring that all stakeholders have a fair chance to participate.
\item 
Responsibility: Institutional responsibility for implementing effective education access practices, actively contributing to inclusive educational environments.
\item 
Equity: Equitable distribution of educational benefits, addressing any disparities in the impact of institutional education access practices.
\item 
Context-Awareness: Developing institution-wide education access strategies that are context-aware of diverse learning needs and cultural backgrounds.
\end{itemize}
\subsubsection*{ Economic Inclusivity}
Institutions:
\begin{itemize} \item 
Transparency: Transparent communication about how institutional economic practices consider and respect diverse cultural economic models.
\item 
Fairness: Fair integration of diverse economic models into institutional economic policies, ensuring that economic opportunities are distributed without bias.
\item 
Responsibility: Institutional responsibility for fostering economic inclusivity, actively addressing cultural implications of economic decisions and practices.
\item 
Equity: Equitable distribution of economic benefits, recognizing and rectifying any cultural biases in institutional economic practices.
\item 
Context-Awareness: Developing institution-wide economic inclusivity strategies that are context-aware of local economic contexts and diverse cultural perspectives.
\end{itemize}
 \subsubsection*{ Technological Ethics}
Institutions:
\begin{itemize} \item 
Transparency: Transparent communication about institutional technological advancements, fostering open dialogue about how technology impacts diverse cultures.
\item 
Fairness: Fair consideration of diverse cultural perspectives in institutional technological development, ensuring that technology is designed without bias and aligns with ethical standards.
\item 
 Responsibility: Institutional responsibility for ethical technology use, actively contributing to responsible development and deployment of technology within diverse cultural contexts.
\item 
Equity: Equitable access to technological benefits for all cultural groups, addressing digital disparities and promoting inclusive technology policies.
\item 
Context-Awareness: Developing institution-wide technological strategies that are context-aware of the societal impact on different cultures, adapting technological approaches to respect and accommodate diverse cultural perspectives.
\end{itemize}
\subsubsection*{ Political Accountability}
Institutions:
\begin{itemize} \item 
Transparency: Transparent communication about institutional political processes, fostering open dialogue about the impact of politics on diverse cultures.
\item 
Fairness: Fair consideration of diverse political perspectives in institutional policies, ensuring that political systems are just and inclusive.
\item 
Responsibility: Institutional responsibility for political accountability, actively contributing to transparent and accountable political processes within diverse cultural contexts.
\item 
Equity: Equitable political representation for all cultural groups, addressing political disparities and promoting inclusive political policies.
\item 
Context-Awareness: Developing institution-wide political accountability strategies that are context-aware of the political contexts of different cultures, adapting political engagement to respect and integrate diverse cultural perspectives.
\end{itemize}
\subsubsection*{ Local, Inter-regional and International Collaboration}
Institutions:
\begin{itemize} \item 
Transparency: Transparent communication in collaborative efforts, fostering open dialogue about the goals, strategies, and outcomes of institutional collaborations across cultures.
\item 
Fairness: Fair consideration of diverse cultural perspectives in collaborative initiatives, ensuring that all collaborators have an equal voice and role.
\item 
Responsibility: Institutional responsibility for promoting collaborative efforts that respect cultural diversity, actively contributing to positive global collaborations.
\item 
Equity: Equitable distribution of benefits and responsibilities in collaborative initiatives, recognizing and rectifying any cultural biases in institutional collaborations.
\item 
Context-Awareness: Developing institution-wide collaboration strategies that are context-aware of the cultural nuances and challenges in local, inter-regional, and international contexts.
\end{itemize}
\subsection{ Ethics at a Nation Level}
\subsubsection*{ Community Engagement and Communal Values}
Nation:
\begin{itemize} \item 
Transparency: The nation promotes transparent communication about communal values and engagement strategies, ensuring that citizens are informed about the goals and outcomes of community initiatives.
\item 
Fairness: The nation works towards fair distribution of resources and opportunities within communities, striving to eliminate disparities and promote inclusivity in communal values.
\item 
Responsibility: The nation holds a responsibility to support and facilitate positive communal values, actively engaging in initiatives that contribute to the well-being of communities.
\item 
Equity: The nation aims for equitable development across communities, addressing regional disparities and promoting equal access to communal benefits.
\item 
Context-Awareness: The nation considers the unique cultural and geographical contexts of different communities, adapting national strategies to respect and accommodate diversity.
\end{itemize}
\subsubsection*{ Cultural Sensitivity}
Nation:
\begin{itemize} \item 
Transparency: The nation transparently communicates its commitment to cultural sensitivity, ensuring that policies and initiatives consider and respect diverse cultural perspectives.
\item 
Fairness: The nation strives for fair treatment of all cultures, fostering an environment where cultural diversity is celebrated and valued.
\item 
Responsibility: The nation takes responsibility for fostering a culturally sensitive environment, actively working to eliminate cultural biases in national policies.
\item 
Equity: The nation aims for equitable representation of cultural voices in national decision-making, recognizing and addressing any cultural disparities.
\item 
Context-Awareness: The nation develops cultural policies that are context-aware, considering the historical and social context of different cultural groups within the nation.
\end{itemize}
\subsubsection*{ Environmental Stewardship}
Nation:
\begin{itemize} \item 
Transparency: The nation communicates transparently about its environmental stewardship efforts, sharing information about national strategies for sustainability.
\item 
Fairness: The nation ensures fair distribution of environmental benefits and responsibilities, working to eliminate environmental injustices across regions.
\item 
Responsibility: The nation takes responsibility for implementing effective national environmental stewardship practices, contributing to global sustainability goals.
\item
Equity: The nation aims for equitable environmental policies, addressing any disparities in the impact of national environmental practices.
\item 
Context-Awareness: The nation develops national environmental strategies that consider the unique contexts of local ecosystems and diverse geographical regions.
\end{itemize}
\subsubsection*{ Social Justice and Equality}
Nation:
\begin{itemize} \item 
Transparency: The nation communicates transparently about its efforts for social justice, fostering open dialogue about the social implications of national policies.
\item 
Fairness: The nation ensures fair consideration of diverse social perspectives in national policies, striving for just and inclusive outcomes.
\item 
Responsibility: The nation takes responsibility for addressing social inequalities exacerbated by national issues, actively working toward equitable social outcomes.
\item 
Equity: The nation aims for equitable distribution of social benefits, recognizing and rectifying any social disparities resulting from national practices.
\item 
Context-Awareness: The nation develops national strategies that are context-aware of unique social challenges, adapting policies to promote social justice.
\end{itemize}
\subsubsection*{ Healthcare Equity}
Nation:
\begin{itemize} \item 
Transparency: The nation communicates transparently about how national healthcare practices consider and respect diverse cultural health beliefs.
\item 
Fairness: The nation ensures fair integration of cultural values into national healthcare policies, promoting healthcare services that are culturally sensitive and inclusive.
\item 
Responsibility: The nation takes responsibility for fostering healthcare equity, actively addressing cultural implications of national healthcare decisions and practices.
\item 
Equity: The nation aims for equitable distribution of healthcare benefits, recognizing and rectifying any cultural biases in national healthcare services.
\item 
Context-Awareness: The nation develops national healthcare strategies that are context-aware, considering the unique cultural contexts shaping health beliefs and practices.
\end{itemize}
\subsubsection*{ Education Access}
Nation:
\begin{itemize} \item 
Transparency: The nation transparently communicates its efforts to promote education access, ensuring that citizens are informed about national strategies and progress.
\item 
Fairness: The nation ensures fair allocation of resources and opportunities for education access initiatives, striving to eliminate disparities and promote inclusivity in education.
\item 
Responsibility: The nation takes responsibility for implementing effective national education access practices, actively contributing to inclusive educational environments.
\item 
Equity: The nation aims for equitable distribution of educational benefits, addressing any disparities in the impact of national education access practices.
\item 
Context-Awareness: The nation develops national education access strategies that are context-aware, considering diverse learning needs shaped by cultural backgrounds.
\end{itemize}
\subsubsection*{ Economic Inclusivity}
Nation:
\begin{itemize} \item 
Transparency: The nation transparently communicates about how national economic practices consider and respect diverse cultural economic models.
\item 
Fairness: The nation ensures fair integration of diverse economic models into national economic policies, promoting economic opportunities that are distributed without bias.
\item 
Responsibility: The nation takes responsibility for fostering economic inclusivity, actively addressing cultural implications of national economic decisions and practices.
\item 
Equity: The nation aims for equitable distribution of economic benefits, recognizing and rectifying any cultural biases in national economic practices.
\item 
Context-Awareness: The nation develops national economic inclusivity strategies that are context-aware, considering local economic contexts and diverse cultural perspectives.
\end{itemize}
\subsubsection*{ Technological Ethics}
Nation:
\begin{itemize} \item 
Transparency: The nation transparently communicates about national technological advancements, fostering open dialogue about how technology impacts diverse cultures.
\item 
Fairness: The nation ensures fair consideration of diverse cultural perspectives in national technological development, striving for technology that is designed without bias.
\item 
Responsibility: The nation takes responsibility for ethical technology use, actively contributing to the responsible development and deployment of technology within diverse cultural contexts.
\item 
Equity: The nation aims for equitable access to technological benefits for all cultural groups, addressing digital disparities and promoting inclusive technology policies.
\item 
Context-Awareness: The nation develops national technological strategies that are context-aware, considering the societal impact of technology on different cultures and adapting technological approaches to respect and accommodate diverse cultural perspectives.
\end{itemize}
\subsubsection*{ Political Accountability}
Nation:
\begin{itemize} \item 
Transparency: The nation communicates transparently about national political processes, fostering open dialogue about the impact of politics on diverse cultures.

\item Fairness: The nation ensures fair consideration of diverse political perspectives in national policies, striving for political systems that are just and inclusive.

\item Responsibility: The nation takes responsibility for political accountability, actively contributing to transparent and accountable political processes within diverse cultural contexts.

\item Equity: The nation aims for equitable political representation for all cultural groups, addressing political disparities and promoting inclusive political policies.

\item Context-Awareness: The nation develops national political accountability strategies that are context-aware, considering the political contexts of different cultures and adapting political engagement to respect and integrate diverse cultural perspectives.
\end{itemize}
 \subsubsection*{ Local, Inter-regional and International Collaboration}
Nation:
\begin{itemize} \item 
Transparency: The nation transparently communicates in collaborative efforts, fostering open dialogue about the goals, strategies, and outcomes of national collaborations across cultures.
\item 
Fairness: The nation ensures fair consideration of diverse cultural perspectives in national collaborative initiatives, striving for equal representation and contributions.
\item 
Responsibility: The nation takes responsibility for promoting collaborative efforts that respect cultural diversity, actively contributing to positive global collaborations.
\item 
Equity: The nation aims for equitable distribution of benefits and responsibilities in national collaborative initiatives, recognizing and rectifying any cultural biases.
\item 
Context-Awareness: The nation develops national collaboration strategies that are context-aware, considering the cultural nuances and challenges in local, inter-regional, and international contexts.
\end{itemize}
\subsection{ Ethics at the Platforms level}
\subsubsection*{ Community Engagement and Communal Values}
On and off-line platforms:
\begin{itemize} \item 
Transparency: On and off-line platforms facilitate transparent communication about community engagement, ensuring that information about communal values is accessible and openly discussed.
\item 
Fairness: Platforms strive to provide fair representation of diverse voices, fostering an inclusive space where different perspectives on communal values are acknowledged and respected.
\item 
Responsibility: Platforms take responsibility for maintaining a positive community engagement environment, actively moderating discussions and addressing any issues that may arise.
\item 
Equity: Platforms work towards equitable participation, ensuring that individuals from all backgrounds have equal opportunities to engage and contribute to communal values discussions.
\item 
Context-Awareness: Platforms are context-aware, adapting to the cultural nuances and diversity of perspectives within community engagement discussions both online and offline.
\end{itemize}
\subsubsection*{ Cultural Sensitivity}
On and off-line platforms:
\begin{itemize} \item 
Transparency: Platforms transparently communicate their commitment to cultural sensitivity, providing clear guidelines on respectful interactions and content sharing that considers diverse cultural perspectives.

\item Fairness: Platforms aim for fair representation and treatment of diverse cultural content, fostering an environment where cultural sensitivity is a central aspect of online and offline discussions.

\item Responsibility: Platforms take responsibility for promoting cultural sensitivity, actively moderating content to eliminate cultural biases and ensuring respectful interactions among users.

\item Equity: Platforms work towards equitable access to cultural content, addressing any disparities and promoting the inclusion of various cultural perspectives in online and offline spaces.

\item Context-Awareness: Platforms are context-aware, adapting their guidelines and content moderation practices to respect the unique cultural contexts shaping online and offline discussions.
\end{itemize}
\subsubsection*{ Environmental Stewardship}
On and off-line platforms:
\begin{itemize} \item 
Transparency: Platforms transparently share information about environmental stewardship efforts, utilizing online spaces to disseminate knowledge and updates on national and global sustainability initiatives.
\item 
Fairness: Platforms promote fair and unbiased coverage of environmental issues, ensuring that discussions on environmental stewardship represent diverse perspectives and solutions.
\item 
Responsibility: Platforms take responsibility for encouraging positive environmental practices, actively engaging users in discussions and initiatives that contribute to environmental sustainability.
\item 
Equity: Platforms aim for equitable distribution of environmental information, addressing any disparities in access to discussions and resources related to environmental stewardship.
\item 
Context-Awareness: Platforms are context-aware, adapting their approach to environmental discussions to consider the global and local contexts shaping sustainability efforts.
\end{itemize}
\subsubsection*{ Social Justice and Equality}
On and off-line platforms:
\begin{itemize} \item 
Transparency: Platforms transparently communicate efforts for social justice, utilizing online spaces to share information about initiatives and advocate for equality.
\item 
Fairness: Platforms work towards fair representation of diverse social perspectives, fostering an inclusive environment where discussions on social justice are free from biases.
\item 
Responsibility: Platforms take responsibility for addressing social inequalities, actively moderating content and discussions to promote a just and equal online and offline space.
\item 
Equity: Platforms aim for equitable access to social justice discussions, recognizing and rectifying any disparities in the impact of online and offline practices.
\item 
Context-Awareness: Platforms are context-aware, adapting their approach to social justice discussions to consider the unique social challenges and contexts influencing equality.
\end{itemize}
\subsubsection*{ Healthcare Equity}
On and off-line platforms:
\begin{itemize} \item 
Transparency: Platforms transparently communicate about how healthcare practices consider and respect diverse cultural health beliefs, disseminating information about inclusive healthcare initiatives.
\item 
Fairness: Platforms ensure fair integration of cultural values into healthcare discussions, fostering an online environment where healthcare services are culturally sensitive and inclusive.
\item 
Responsibility: Platforms take responsibility for promoting healthcare equity, actively engaging users in discussions and initiatives that address cultural implications of healthcare decisions and practices.
\item 
Equity: Platforms aim for equitable access to healthcare discussions and information, recognizing and rectifying any cultural biases in online and offline healthcare spaces.
\item 
Context-Awareness: Platforms are context-aware, adapting their healthcare discussions to consider the unique cultural contexts shaping health beliefs and practices in online and offline settings.
\end{itemize}
\subsubsection*{ Education Access}
On and off-line platforms:
\begin{itemize} \item 
Transparency: Platforms transparently communicate efforts to promote education access, utilizing online spaces to share information about initiatives and progress in making education accessible.
\item 
Fairness: Platforms ensure fair allocation of resources and opportunities for education access initiatives, fostering an online environment where all users have equal chances to participate in educational discussions.
\item 
Responsibility: Platforms take responsibility for implementing effective education access practices, actively contributing to inclusive educational environments both online and offline.
\item 
Equity: Platforms aim for equitable distribution of educational benefits, addressing any disparities in the impact of online and offline education access practices.
\item 
Context-Awareness: Platforms are context-aware, adapting their approach to education access discussions to consider diverse learning needs and cultural backgrounds in online and offline spaces.
\end{itemize}
\subsubsection*{ Economic Inclusivity}
On and off-line platforms:
\begin{itemize} \item 
Transparency: Platforms transparently communicate about how economic practices consider and respect diverse cultural economic models, disseminating information about inclusive economic initiatives.
\item 
Fairness: Platforms ensure fair integration of diverse economic models into economic discussions, fostering an online environment where economic opportunities are distributed without bias.
\item 
Responsibility: Platforms take responsibility for promoting economic inclusivity, actively engaging users in discussions and initiatives that address cultural implications of economic decisions and practices.
\item 
Equity: Platforms aim for equitable access to economic discussions and benefits, recognizing and rectifying any cultural biases in online and offline economic spaces.
Context-Awareness: Platforms are context-aware, adapting their economic discussions to consider local economic contexts and diverse cultural perspectives in online and offline settings.
\end{itemize}
\subsubsection*{ Technological Ethics}
On and off-line platforms:
\begin{itemize} \item 
Transparency: Platforms transparently communicate about national technological advancements, fostering open dialogue about how technology impacts diverse cultures in online and offline spaces.
\item 
Fairness: Platforms ensure fair consideration of diverse cultural perspectives in technological discussions, striving for technology that is designed without bias in both online and offline environments.
\item 
Responsibility: Platforms take responsibility for ethical technology use, actively contributing to the responsible development and deployment of technology within diverse cultural contexts online and offline.
\item 
Equity: Platforms aim for equitable access to technological benefits for all cultural groups, addressing digital disparities and promoting inclusive technology policies in both online and offline spaces.
\item 
Context-Awareness: Platforms are context-aware, adapting their technological discussions to consider the societal impact of technology on different cultures in both online and offline settings.
\end{itemize}
\subsubsection*{ Political Accountability}
On and off-line platforms:
\begin{itemize} \item 
Transparency: Platforms transparently communicate about national political processes, fostering open dialogue about the impact of politics on diverse cultures in online and offline spaces.
\item 
Fairness: Platforms ensure fair consideration of diverse political perspectives in political discussions, striving for just and inclusive political systems both online and offline.
\item 
Responsibility: Platforms take responsibility for political accountability, actively contributing to transparent and accountable political processes within diverse cultural contexts online and offline.
\item 
Equity: Platforms aim for equitable political representation for all cultural groups in political discussions, addressing political disparities and promoting inclusive political policies both online and offline.
\item 
Context-Awareness: Platforms are context-aware, adapting their political discussions to consider the political contexts of different cultures and adapting political engagement to respect and integrate diverse cultural perspectives in both online and offline spaces.
\end{itemize}
\subsubsection*{ Local, Inter-regional and International Collaboration}
On and off-line platforms:
\begin{itemize} \item 
Transparency: Platforms transparently communicate in collaborative efforts, fostering open dialogue about the goals, strategies, and outcomes of collaborative initiatives across cultures in both online and offline spaces.
\item 
Fairness: Platforms ensure fair consideration of diverse cultural perspectives in collaborative initiatives, striving for equal representation and contributions both online and offline.
\item 
Responsibility: Platforms take responsibility for promoting collaborative efforts that respect cultural diversity, actively contributing to positive global collaborations in both online and offline spaces.
\item 
Equity: Platforms aim for equitable distribution of benefits and responsibilities in collaborative initiatives, recognizing and rectifying any cultural biases both online and offline.
\item 
Context-Awareness: Platforms are context-aware, adapting their collaboration strategies to consider the cultural nuances and challenges in local, inter-regional, and international contexts both online and offline.
\end{itemize}
 \subsection{ What ethics at the globe level?}
Think global strategies, Act locally ?
\subsubsection*{ Community Engagement and Communal Values}
Globally:
\begin{itemize} \item 
Transparency: Global engagement platforms ensure transparency by facilitating open communication about communal values on a global scale, fostering a cross-cultural exchange of ideas and perspectives.
\item 
Fairness: Platforms strive for fairness by promoting equal representation of diverse global voices, creating an inclusive space where different cultures contribute to the discussion on communal values.
\item 
Responsibility: Global platforms take responsibility for fostering positive global communal values, actively moderating discussions to ensure respectful cross-cultural interactions and addressing any global issues that may arise.
\item 
Equity: Platforms work towards equitable global participation, ensuring that individuals from all corners of the world have equal opportunities to engage and contribute to discussions on communal values.
\item 
Context-Awareness: Global platforms are context-aware, adapting to the cultural nuances and diversity of perspectives within global community engagement discussions.
\end{itemize}
\subsubsection*{ Cultural Sensitivity}
Globally:
\begin{itemize} \item 
Transparency: Global platforms transparently communicate their commitment to cultural sensitivity, providing guidelines on respectful global interactions and content sharing that considers diverse cultural perspectives.
\item 
Fairness: Platforms aim for fair representation and treatment of diverse global cultural content, fostering a worldwide environment where cultural sensitivity is a central aspect of online discussions.
\item 
Responsibility: Global platforms take responsibility for promoting cultural sensitivity globally, actively moderating content to eliminate cultural biases and ensuring respectful interactions among users from different cultures.
\item 
Equity: Platforms work towards equitable access to global cultural content, addressing any disparities and promoting the inclusion of various global cultural perspectives in online spaces.
\item 
Context-Awareness: Global platforms are context-aware, adapting their guidelines and content moderation practices to respect the unique cultural contexts shaping global online discussions.
\end{itemize}
\subsubsection*{ Environmental Stewardship}
Globally:
\begin{itemize} \item 
Transparency: Global platforms transparently share information about environmental stewardship efforts, utilizing online spaces to disseminate knowledge and updates on international sustainability initiatives.
\item 
Fairness: Platforms promote fair and unbiased coverage of global environmental issues, ensuring that discussions on environmental stewardship represent diverse global perspectives and solutions.
\item 
Responsibility: Platforms take responsibility for encouraging positive global environmental practices, actively engaging users in discussions and initiatives that contribute to international environmental sustainability.
\item 
Equity: Platforms aim for equitable distribution of global environmental information, addressing any disparities in access to discussions and resources related to international environmental stewardship.
\item 
Context-Awareness: Global platforms are context-aware, adapting their approach to global environmental discussions to consider the global and local contexts shaping sustainability efforts.
\end{itemize}
\subsubsection*{ Social Justice and Equality}
Globally:
\begin{itemize} \item 
Transparency: Platforms transparently communicate efforts for global social justice, utilizing online spaces to share information about international initiatives and advocate for equality.
\item 
Fairness: Platforms work towards fair representation of diverse global social perspectives, fostering an inclusive environment where discussions on global social justice are free from biases.
\item 
Responsibility: Platforms take responsibility for addressing global social inequalities, actively moderating content and discussions to promote a just and equal global online space.
\item 
Equity: Platforms aim for equitable access to global social justice discussions, recognizing and rectifying any global disparities resulting from online and offline practices.
\item 
Context-Awareness: Platforms are context-aware, adapting their approach to global social justice discussions to consider the unique global social challenges and contexts influencing equality.
\end{itemize}
\subsubsection*{ Healthcare Equity}
Globally:
\begin{itemize} \item 
Transparency: Platforms transparently communicate about how global healthcare practices consider and respect diverse cultural health beliefs, disseminating information about inclusive healthcare initiatives worldwide.
\item 
Fairness: Platforms ensure fair integration of cultural values into global healthcare discussions, fostering an online environment where healthcare services are culturally sensitive and inclusive on a global scale.
\item 
Responsibility: Platforms take responsibility for promoting global healthcare equity, actively engaging users in discussions and initiatives that address cultural implications of global healthcare decisions and practices.
\item 
Equity: Platforms aim for equitable access to global healthcare discussions and information, recognizing and rectifying any cultural biases in online and offline global healthcare spaces.
\item 
Context-Awareness: Platforms are context-aware, adapting their healthcare discussions to consider the unique global cultural contexts shaping health beliefs and practices.
\end{itemize}
\subsubsection*{ Education Access}
Globally:
\begin{itemize} \item 
Transparency: Platforms transparently communicate efforts to promote global education access, utilizing online spaces to share information about international initiatives and progress in making education accessible worldwide.
\item 
Fairness: Platforms ensure fair allocation of resources and opportunities for global education access initiatives, fostering an online environment where all users have equal chances to participate in global educational discussions.
\item 
Responsibility: Platforms take responsibility for implementing effective global education access practices, actively contributing to inclusive educational environments both online and offline on a global scale.
\item 
Equity: Platforms aim for equitable distribution of global educational benefits, addressing any global disparities in the impact of online and offline global education access practices.
\item 
Context-Awareness: Platforms are context-aware, adapting their approach to global education access discussions to consider diverse learning needs and cultural backgrounds on a global scale.
\end{itemize}
\subsubsection*{ Economic Inclusivity}
Globally:
\begin{itemize} \item 
Transparency: Platforms transparently communicate about how global economic practices consider and respect diverse global economic models, disseminating information about inclusive economic initiatives worldwide.

Fairness: Platforms ensure fair integration of diverse global economic models into economic discussions, fostering an online environment where economic opportunities are distributed without bias on a global scale.
\item 
Responsibility: Platforms take responsibility for promoting global economic inclusivity, actively engaging users in discussions and initiatives that address cultural implications of global economic decisions and practices.
\item 
Equity: Platforms aim for equitable access to global economic discussions and benefits, recognizing and rectifying any cultural biases in online and offline global economic spaces.
\item 
Context-Awareness: Platforms are context-aware, adapting their economic discussions to consider local economic contexts and diverse cultural perspectives in online and offline global settings.
\end{itemize}
\subsubsection*{ Technological Ethics}
Globally:
\begin{itemize} \item 
Transparency: Platforms transparently communicate about global technological advancements, fostering open dialogue about how technology impacts diverse cultures in online and offline spaces worldwide.
\item 
Fairness: Platforms ensure fair consideration of diverse global cultural perspectives in technological discussions, striving for technology that is designed without bias on a global scale.
\item 
Responsibility: Platforms take responsibility for ethical technology use globally, actively contributing to the responsible development and deployment of technology within diverse cultural contexts online and offline.
\item 
Equity: Platforms aim for equitable access to global technological benefits for all cultural groups, addressing digital disparities and promoting inclusive technology policies in both online and offline global spaces.
\item 
Context-Awareness: Platforms are context-aware, adapting their technological discussions to consider the societal impact of technology on different cultures in both online and offline global settings.
\end{itemize}
\subsubsection*{ Political Accountability}
Globally:
\begin{itemize} \item 
Transparency: Platforms transparently communicate about global political processes, fostering open dialogue about the impact of politics on diverse cultures in online and offline spaces worldwide.
\item 
Fairness: Platforms ensure fair consideration of diverse global political perspectives in political discussions, striving for just and inclusive political systems both online and offline.
\item 
Responsibility: Platforms take responsibility for political accountability globally, actively contributing to transparent and accountable political processes within diverse cultural contexts online and offline.
\item 
Equity: Platforms aim for equitable political representation for all cultural groups in global political discussions, addressing political disparities and promoting inclusive political policies both online and offline.
\item 
Context-Awareness: Platforms are context-aware, adapting their political discussions to consider the political contexts of different cultures globally and adapting political engagement to respect and integrate diverse cultural perspectives.
\end{itemize}
\subsubsection*{ Local, Inter-regional and International Collaboration}
Globally:
\begin{itemize} \item 
Transparency: Platforms transparently communicate in collaborative efforts on a global scale, fostering open dialogue about the goals, strategies, and outcomes of collaborative initiatives across cultures in both online and offline spaces.
\item 
Fairness: Platforms ensure fair consideration of diverse cultural perspectives in collaborative initiatives, striving for equal representation and contributions on a global scale both online and offline.
\item 
Responsibility: Platforms take responsibility for promoting collaborative efforts that respect cultural diversity globally, actively contributing to positive global collaborations both online and offline.
\item 
Equity: Platforms aim for equitable distribution of benefits and responsibilities in collaborative initiatives globally, recognizing and rectifying any cultural biases both online and offline.
\item 
Context-Awareness: Platforms are context-aware, adapting their collaboration strategies to consider the cultural nuances and challenges in local, inter-regional, and international contexts on a global scale both online and offline.
\end{itemize}
\subsection{ Ethics over time, intergenerational ethics}
\subsubsection*{ Community Engagement and Communal Values}
Over Time (Intergenerational):
\begin{itemize} \item 
Transparency: Over time, platforms facilitate transparent communication about communal values, ensuring that information about historical perspectives on communal values is accessible and openly discussed.
\item 
Fairness: Platforms strive for fairness by preserving historical communal values and providing an inclusive space where different historical perspectives contribute to the ongoing discussion.
\item 
Responsibility: Platforms take responsibility for maintaining a positive historical community engagement environment, actively preserving and moderating discussions to address any issues that may arise over time.
\item 
Equity: Platforms work towards equitable intergenerational participation, ensuring that historical voices and perspectives have a place in the ongoing dialogue about communal values.
\item 
Context-Awareness: Platforms are context-aware over time, adapting to the changing cultural contexts and ensuring that historical perspectives are respected within ongoing community engagement discussions.
\end{itemize}
\subsubsection*{ Cultural Sensitivity}
Over Time (Intergenerational):
\begin{itemize} \item 
Transparency: Platforms transparently communicate their commitment to cultural sensitivity over time, providing guidelines on respectful interactions and content sharing that considers evolving cultural perspectives.
\item 
Fairness: Platforms aim for fair representation and treatment of diverse cultural content over time, fostering an environment where cultural sensitivity is a continuous aspect of discussions across generations.
\item 
Responsibility: Platforms take responsibility for promoting cultural sensitivity over time, actively preserving historical cultural content and ensuring respectful interactions among users from different historical cultural contexts.
\item 
Equity: Platforms work towards equitable access to cultural content over time, addressing any disparities and promoting the inclusion of various historical cultural perspectives in discussions.
\item 
Context-Awareness: Platforms are context-aware over time, adapting their guidelines and content moderation practices to respect the changing cultural nuances and diversity of perspectives within intergenerational discussions.
\end{itemize}
\subsubsection*{Environmental Stewardship}
Over Time (Intergenerational):
\begin{itemize} \item 
Transparency: Platforms transparently share information about the evolution of environmental stewardship efforts over time, utilizing online spaces to disseminate knowledge and updates on the historical progression of sustainability initiatives.
\item 
Fairness: Platforms promote fair and unbiased coverage of historical environmental issues, ensuring that discussions on environmental stewardship represent diverse historical perspectives and solutions.
\item 
Responsibility: Platforms take responsibility for encouraging positive historical environmental practices, actively engaging users in discussions and initiatives that contribute to the intergenerational sustainability of the planet.
\item 
Equity: Platforms aim for equitable distribution of historical environmental information, addressing any disparities in access to discussions and resources related to historical environmental stewardship.
\item 
Context-Awareness: Platforms are context-aware over time, adapting their approach to historical environmental discussions to consider the global and local historical contexts shaping sustainability efforts.
\end{itemize}
\subsubsection*{ Social Justice and Equality}
Over Time (Intergenerational):
\begin{itemize} \item 
Transparency: Platforms transparently communicate efforts for historical global social justice, utilizing online spaces to share information about past initiatives and advocacy for equality across generations.
\item 
Fairness: Platforms work towards fair representation of diverse historical global social perspectives, fostering an inclusive environment where discussions on historical global social justice are free from biases.
\item 
Responsibility: Platforms take responsibility for addressing historical global social inequalities, actively moderating content and discussions to promote a just and equal online space over time.
\item 
Equity: Platforms aim for equitable access to historical global social justice discussions, recognizing and rectifying any historical global disparities resulting from online and offline practices.
\item 
Context-Awareness: Platforms are context-aware over time, adapting their approach to historical global social justice discussions to consider the unique historical global social challenges and contexts influencing equality.
\end{itemize}
\subsubsection*{ Healthcare Equity}
Over Time (Intergenerational):
\begin{itemize} \item 
Transparency: Platforms transparently communicate about how healthcare practices consider and respect diverse cultural health beliefs over time, disseminating information about inclusive healthcare initiatives across generations.
\item 
Fairness: Platforms ensure fair integration of cultural values into historical healthcare discussions, fostering an online environment where healthcare services are culturally sensitive and inclusive over time.
\item 
Responsibility: Platforms take responsibility for promoting intergenerational healthcare equity, actively engaging users in discussions and initiatives that address cultural implications of historical healthcare decisions and practices.
\item 
Equity: Platforms aim for equitable access to historical healthcare discussions and information, recognizing and rectifying any cultural biases in online and offline historical healthcare spaces.
\item 
Context-Awareness: Platforms are context-aware over time, adapting their healthcare discussions to consider the unique historical cultural contexts shaping health beliefs and practices.
\end{itemize}
\subsubsection*{ Education Access}
Over Time (Intergenerational):
\begin{itemize} \item 
Transparency: Platforms transparently communicate efforts to promote historical global education access, utilizing online spaces to share information about past international initiatives and progress in making education accessible worldwide over time.
\item 
Fairness: Platforms ensure fair allocation of resources and opportunities for historical global education access initiatives, fostering an online environment where all users have equal chances to participate in global educational discussions over time.
\item 
Responsibility: Platforms take responsibility for implementing effective historical global education access practices, actively contributing to inclusive educational environments both online and offline on a global scale over time.
\item 
Equity: Platforms aim for equitable distribution of historical global educational benefits, addressing any global disparities in the impact of online and offline global education access practices over time.
\item 
Context-Awareness: Platforms are context-aware over time, adapting their approach to global education access discussions to consider diverse learning needs and cultural backgrounds on a global scale.
\end{itemize}
\subsubsection*{ Economic Inclusivity}
Over Time (Intergenerational):
\begin{itemize} \item 
Transparency: Platforms transparently communicate about how global economic practices consider and respect diverse global economic models over time, disseminating information about past inclusive economic initiatives worldwide.
\item 
Fairness: Platforms ensure fair integration of diverse global economic models into historical economic discussions, fostering an online environment where economic opportunities are distributed without bias on a global scale over time.
\item 
Responsibility: Platforms take responsibility for promoting historical global economic inclusivity, actively engaging users in discussions and initiatives that address cultural implications of past global economic decisions and practices.
\item 
Equity: Platforms aim for equitable access to historical global economic discussions and benefits, recognizing and rectifying any cultural biases in online and offline historical global economic spaces.
\item 
Context-Awareness: Platforms are context-aware over time, adapting their economic discussions to consider local economic contexts and diverse cultural perspectives in online and offline global settings.
\end{itemize}
\subsubsection*{ Technological Ethics}
Over Time (Intergenerational):
\begin{itemize} \item 
Transparency: Platforms transparently communicate about global technological advancements over time, fostering open dialogue about how technology has evolved and impacted diverse cultures in online and offline spaces worldwide.
\item 
Fairness: Platforms ensure fair consideration of diverse global cultural perspectives in historical technological discussions, striving for technology that is designed without bias on a global scale over time.
\item 
Responsibility: Platforms take responsibility for ethical technology use globally over time, actively contributing to the responsible development and deployment of technology within diverse cultural contexts online and offline.
\item 
Equity: Platforms aim for equitable access to historical technological benefits for all cultural groups, addressing digital disparities and promoting inclusive technology policies in both online and offline historical global spaces.
\item 
Context-Awareness: Platforms are context-aware over time, adapting their technological discussions to consider the societal impact of technology on different cultures in both online and offline historical global settings.
\end{itemize}
\subsubsection*{ Political Accountability}
Over Time (Intergenerational):
\begin{itemize} \item 
Transparency: Platforms transparently communicate about global political processes over time, fostering open dialogue about how political landscapes have evolved and impacted diverse cultures in online and offline spaces worldwide.
\item 
Fairness: Platforms ensure fair consideration of diverse global political perspectives in historical political discussions, striving for just and inclusive political systems both online and offline over time.
\item 
Responsibility: Platforms take responsibility for political accountability globally over time, actively contributing to transparent and accountable political processes within diverse cultural contexts online and offline.
\item 
Equity: Platforms aim for equitable political representation for all cultural groups in historical global political discussions, addressing political disparities and promoting inclusive political policies both online and offline over time.
\item 
Context-Awareness: Platforms are context-aware over time, adapting their political discussions to consider the historical political contexts of different cultures globally and adapting political engagement to respect and integrate diverse cultural perspectives.
\end{itemize}
\subsubsection*{ Local, Inter-regional and International Collaboration}
Over Time (Intergenerational):
\begin{itemize} \item 
Transparency: Platforms transparently communicate in collaborative efforts on a global scale over time, fostering open dialogue about the goals, strategies, and outcomes of collaborative initiatives across cultures in both online and offline spaces.
\item 
Fairness: Platforms ensure fair consideration of diverse cultural perspectives in collaborative initiatives over time, striving for equal representation and contributions on a global scale both online and offline.
\item 
Responsibility: Platforms take responsibility for promoting collaborative efforts that respect cultural diversity globally over time, actively contributing to positive global collaborations both online and offline.
\item 
Equity: Platforms aim for equitable distribution of benefits and responsibilities in collaborative initiatives globally over time, recognizing and rectifying any cultural biases both online and offline.
\item 
Context-Awareness: Platforms are context-aware over time, adapting their collaboration strategies to consider the cultural nuances and challenges in local, inter-regional, and international contexts on a global scale both online and offline.
\end{itemize}
\section{MI in Africa is not only  about competition} \label{sec:beyond}
\subsection*{ What is a co-opetition ?}

Co-opetition is a  strategy where competitors collaborate on certain projects or initiatives while simultaneously competing in other areas. It involves a combination of cooperation and competition to achieve mutual benefits for the involved parties. This strategy recognizes that competitors can sometimes create value by working together on common goals, even as they compete in other aspects of their business. Co-opetition is often seen in industries where there are shared interests, such as standard-setting, research and development, or addressing common challenges. 
\subsection*{What is co-opetitive game theory ?}

Co-opetitive game theory is an extension of traditional game theory that explores scenarios where actors engage in both cooperative and competitive interactions simultaneously. In co-opetitive situations, participants collaborate on certain aspects of a game while competing in others.
\subsection*{ What is Co-opetitive Mean-Field-Type Game Theory?}

Co-opetitive mean-field-type game theory explores scenarios where actors engage in both cooperative and competitive interactions simultaneously. In co-opetitive situations of mean-field type, participants collaborate on certain aspects of a game while competing in others. What makes them of mean-field type is their dynamics and/or payoff functions, which depend not only on state-action pairs but also on the distribution of state-action pairs. These games do not necessarily consider a large number of decision-makers. They do not need to assume symmetry, exchangeability, or indistinguishability per type. These games capture more interactions in real-world applications that are risk-aware and non-symmetric, involving both cooperation, competition, partial altruism, partial spite, etc.
\subsection*{Multiscale multimodal Ethics as co-opetition}
The multiscale multimodal Ethics  can be seen in Co-opetitive Mean-Field-Type Game Theory through various dimensions:
Individual: \begin{itemize}\item 
Transparency: Individuals participating in co-opetitive scenarios need transparency in understanding the collaborative and competitive aspects, fostering trust among participants.
\item 
Fairness: Ensuring fairness in the distribution of benefits and competition is crucial to maintaining a cooperative and competitive balance.
\item 
Responsibility: Participants must take responsibility for their actions, contributing to both collaborative and competitive elements responsibly.
\item 
Equity: Striving for equity in the outcomes of cooperation and competition, considering the diverse interests and capabilities of individuals.
\end{itemize}
Different Cultures:
\begin{itemize}\item 
Transparency: Cultural transparency is essential to bridge understanding among participants from different cultural backgrounds, creating a shared understanding of co-opetitive dynamics.\item 
Fairness: Ensuring fairness in cultural representation and acknowledging diverse perspectives in both collaboration and competition.
\item Responsibility: Culturally responsible actions contribute to effective co-opetition, respecting the values and norms of diverse cultures.
\item Equity: Addressing cultural disparities and striving for equitable participation and benefits across diverse cultural groups.
\end{itemize}
Institutions:
\begin{itemize}\item
Transparency: Transparent institutional processes are necessary to establish clear rules and guidelines for co-opetition.
\item Fairness: Institutional fairness ensures that rules and policies promote equitable opportunities for collaboration and competition.
\item Responsibility: Institutions play a role in enforcing responsible behavior, holding participants accountable for their contributions.
\item Equity: Institutional structures should strive for equitable distribution of resources and opportunities within co-opetitive frameworks.
\end{itemize}
Nation:
\begin{itemize}\item
Transparency: National transparency in co-opetitive initiatives fosters trust and understanding among participants from different nations.
\item Fairness: Ensuring that national interests are fairly represented and considered in co-opetitive dynamics.
\item Responsibility: National responsibility involves promoting ethical behavior and adherence to agreed-upon rules in co-opetition.
\item Equity: Striving for equitable outcomes that benefit participating nations and consider their unique circumstances.
\end{itemize}
On and Off-Line Platforms:
\begin{itemize}\item
\item Transparency: Platforms facilitating co-opetition must be transparent in their operations, ensuring participants understand the online and offline dynamics.
\item Fairness: Ensuring fair representation and opportunities for collaboration and competition on digital platforms.
\item Responsibility: Online platforms should encourage responsible behavior, emphasizing ethical conduct in co-opetitive interactions.
\item Equity: Addressing digital divides and ensuring equitable access to online platforms for diverse participants.
\end{itemize}
Globally:
\begin{itemize}\item
Transparency: Global transparency is essential for cross-border co-opetition, ensuring a clear understanding of shared goals and competition.
\item Fairness: Striving for fairness in global representation and acknowledging the diverse interests and contributions of participants worldwide.
\item Responsibility: Global responsibility involves ethical behavior on a global scale, promoting positive outcomes in co-opetition.
\item Equity: Addressing global inequalities and promoting equitable participation and benefits in co-opetitive scenarios.
\end{itemize}
Over Time:
\begin{itemize}\item
Transparency: Transparent communication over time ensures that participants are aware of evolving co-opetitive dynamics.
\item Fairness: Adapting co-opetition strategies to changing circumstances while ensuring fairness in long-term collaborations and competitions.
\item Responsibility: Long-term responsibility involves sustained ethical behavior and commitment to mutual benefits over time.
\item Equity: Ensuring that co-opetition evolves to address changing needs and maintains equitable outcomes over time.
\end{itemize}

\subsection*{MI deployment and adoption}
The deployment of MI  in Africa is characterized as not only a competition but also a coopetitive Mean-Field-Type Game due to several factors:
\begin{itemize}\item
Cooperative Elements:
MI deployment often involves collaboration among various stakeholders, including governments, private enterprises, and research institutions. Shared initiatives for technology adoption, skill development, and addressing common challenges create cooperative dynamics.

\item Competitive Aspects:
Different entities within Africa may compete for resources, market dominance, or technological advancements in the MI landscape. This competition can drive innovation, economic growth, and the pursuit of leadership positions in the MI sector.

\item Mean-Field-Type Dynamics:
The mean-field-type dynamics in MI deployment consider not only the actions of individual decision-makers but also the overall distribution of state-action pairs. This reflects the interconnectedness of MI strategies, where the outcomes depend on collective interactions rather than isolated decision-making.

\item Risk-Aware and Non-Symmetric Interactions:
MI deployment in Africa involves risk-aware scenarios, acknowledging the uncertainties and challenges unique to the continent. The interactions are non-symmetric, considering the diverse contexts, economic landscapes, and technological infrastructures across African nations.

\item Coopetition for Common Goals:
While entities may compete in certain aspects of MI development, there is a recognition that collaboration on shared goals, such as addressing socio-economic challenges, improving healthcare, or promoting education, can lead to mutual benefits. Coopetition arises as entities balance competition and cooperation to achieve overarching objectives.

\item Partial Altruism and Partial Spite:
Entities may exhibit partial altruism by contributing to initiatives that benefit the broader community or region. Simultaneously, there might be elements of partial spite, where competitive behaviors aim to gain advantages in specific domains. This mix of cooperative and competitive motives characterizes the coopetitive nature of MI deployment.

\item Interactions Beyond Traditional Competition:
The coopetitive Mean-Field-Type Game considers interactions that go beyond traditional competition. It involves collaborative efforts in areas like research and development, standard-setting, and addressing shared challenges, acknowledging that collective success can drive individual success.

\item Complex Interplay of Actors:
The deployment of MI in Africa involves a complex interplay of various actors, including governments, businesses, academia, and communities. The interactions among these diverse stakeholders contribute to the coopetitive dynamics, shaping the trajectory of MI development on the continent.
\end{itemize}

\subsection*{Coopetitive MI in Africa} 
There are over several hundred millions of people in Africa with diverse experiences in Agriculture, Breeding, Transformation, Trade, Traditional nutrition, Tradition, and Culture. Some of these Africans are champions  in their fields. However, their innovations remain unknown and are often considered mysterious due to a language barrier. They are not considered literate in the current system, even though their audio-rich knowledge based on local experiences holds great societal value. They are in fact audio-literate, not illiterate. 

Timadie's Coopetition MI solution aims to place all Africans, particularly those with unique knowledge, at the core of machine intelligence. Despite language barriers, our goal is to create direct audio-to-audio processing using high-quality, culture-aware datasets. Teams for each local language and bridging teams connecting different languages form a coopetitive graph of machine intelligence, not limited to text, audio, or video. Any knowledge-based learning model can join and enrich the Coopetition.  We aim create value with these audio-rich knowledgeable people. The methods are :
High-quality culture-sensitive \& ethical audio dataset,
Blockchainized audio token datasets, 
Bridge to another local language: links. Examples include, but limited to, 
Tommo-So Dogon to Kenedougou Senoufo, Fon (from Benin) to Tommo-So Dogon, 
Audio dataset enhancement, 
Large Audio learning, 
Audio2Audio translation, 
Audio to visual (image/video) in a local context, 
Large vision learning, 
Multimodal Large Learning, etc.

\section{Data in Africa: quality, information, modes} \label{sec:data} 

To develop an MI system tailored for the diverse needs of African people, acquiring and leveraging high-quality data becomes increasingly pivotal. The success and efficacy of MI in Africa are intricately tied to the richness and relevance of the datasets used for training and refining these systems. Gathering comprehensive and representative data from various regions, groups,  ethnicities, languages, and socioeconomic backgrounds within the continent is essential to ensure that the resulting MI models are inclusive, unbiased, and culturally sensitive. Copy-pasting an MI from other cultures and other languages may not be efficient. Moreover, high-quality data is not solely about quantity but also about the accuracy, authenticity, and real-world applicability of the information collected. It involves capturing the nuances of daily life, understanding regional contexts, and reflecting the dynamic nature of African societies. By prioritizing data quality, the development of local \& context-aware MI in Africa can overcome challenges related to underrepresentation and biases, fostering a technology landscape that truly resonates with the diverse realities of the continent.
Additionally, a commitment to ethical data practices is paramount. Safeguarding privacy, ensuring informed consent, and promoting transparency in data collection processes are integral aspects of building an MI infrastructure that aligns with African values and respects individual rights. Ethical considerations not only enhance the reliability of MI applications but also foster trust among users, thereby encouraging widespread adoption and acceptance of these technologies.
In essence, recognizing the importance of high-quality data is foundational to the successful development and deployment of MI solutions in Africa. It is a collaborative effort that involves stakeholders from various sectors working together to curate datasets that authentically represent the continent's multifaceted landscape, ultimately paving the way for MI systems that positively impact and empower African communities.

\subsection{Audio-rich languages  }
In most African countries, the local population uses oral languages, producing audios, speeches, voices, songs, and mimics, among others. These languages are rich in audio content. Unfortunately, up to now, these audio resources have not been fully utilized. A beneficial application of MI for the local population in Africa could involve a direct conversion of audio from one oral language to another.
 As an exercise we took the list of top 10 MI chatbots and asked simple questions local African languages.
We have tested with the three following questions: 
\begin{itemize}\item How do you say 'thank you' in Tommo-So Dogon?
\item Count from 1 to 5 in Tommo-So Dogon
\item How do you say  'what is your name?'  in Tommo-So Dogon ?
\end{itemize}
The answers to these questions were not satisfactory as of December 2023.  Then we tested the same 3 questions for  200 African languages.The answers were not even close.  This simple  experience mirrors the situation faced by Africans who don't speak English. Many language models  do not perform well for languages with smaller numbers of speakers, especially African ones. 
The problem becomes even  more serious when we have that experience with those who cannot read and write but speak  very well their mother tongue. Most of these languages are audio-rich and currently  there is MI adapted to local MI problems: translation in audio format from African language to another.
Although there have been efforts to include certain languages in MI models even when there is not much data available for training, these results show that the technology “really still isn't capturing our languages. This is due to the fact that the available resources which are audio data are not exploited in these LLMs.

\subsection{Undefined terms to be clarified for machines}
Imagine a super-smart computer with data collected in Africa that can learn from pictures, words, and sounds all together. Sometimes, this computer can make mistakes like believing things that aren't true {\color{blue} (delusion)}, creating things that seem real but aren't {\color{blue}  (illusion)}, making up stories that sound good but have parts that aren't real {\color{blue}  (confabulation)}, and even creating pictures or sounds that are like dreams {\color{blue}  (hallucination)}. It might mix up where things came from  {\color{blue} (misattribution)} or use words in new ways  {\color{blue} (semantic drift)}. It could make things sound bigger or more important than they are  {\color{blue} (exaggeration)} and tell stories with fewer details than they need 
 {\color{blue} (semantic compression)}. Sometimes, it can't figure out what caused what {\color{blue} (causal inference failure)}, and other times it makes things that look almost real but not quite {\color{blue} (perceptual uncanny valley)}. So, while it's an amazing computer, it sometimes makes these mistakes that need to be defined properly, evaluated and corrected.
 
 The book in \cite{Bouare2023} addresses some of these questions.
 
 {\it While these terms are unclear in MI, they are  widely used in the literature including in Africa. It does not help to clarify and demystify the  methods used in the building blocks of these MI technologies. }
 
 \subsection{Africa's rich audio literacy}
You want to engage in telemedicine in rural Africa, but face challenges due to local language barriers. Similarly, responding to a village's call for adult business education and facilitating communication between two farmers with 40 years of local agricultural experience, who cannot read or write, is hindered by language barriers. According to Statistica, it exceeds 30 percent of the 1.4 billion people in Africa.  It is clear that an ethical, context-aware and culture-aware audio-to-audio MI technique could catalyze the inclusion of 400 millions of people in these scenarios in Africa. 
In the vast landscape of technological evolution, Africa emerges as a reservoir of untapped potential. Defined by multifaceted richness across its 54 countries, the continent boasts a unique strength woven into the fabric of its diverse cultures - an inherent proficiency in audio literacy.
From the rhythmic beats of traditional music to the oral traditions passed through generations, audio literacy is an integral part of African societies. This proficiency seamlessly extends into the digital domain, exemplified by the flourishing online audio messaging platforms. We need to explore the transformative power of leveraging Africa's audio literacy as a catalyst for the advancement of MI in Africa. With linguistic diversity and vibrant oral traditions, the continent holds a reservoir of potential waiting to be tapped. The power of audio presents a transformative opportunity for technological innovation in various forms, including sounds, gestures, voices, images, mimics, signs, music, speeches, or videos.
Several individuals in Africa, currently labeled as illiterate by the system, extensively use mobile phones to send audio messages to their counterparts, and these counterparts respond in audio in the same local language. This local language audio-to-audio communication has gained unprecedented momentum in several regions of Africa, generating a multitude of audio data. We realize that these individuals are audio communicators, they are in fact audio literate, creating hope for inclusion through  technology. Unfortunately, as of today, there is no technology allowing audio-to-audio communication across different local languages because these languages lack or minimally use conventional writing systems and the current technologies are limited in textless audio processing . 
It is imperative for us to transcend conventional modes of technology engagement and explore the possibilities that audio literacy unfurls. This article serves as a clarion call to stakeholders across the spectrum - policymakers, technologists, developers, educators, and innovators - to recognize and harness Africa's rich audio literacy.

\subsubsection*{Bridging the Tech Divide: Audio Literacy as the Catalyst}
In a continent rich with diverse cultures and languages, the intersection of MI and audio communication is not just a leap into the digital age but a catalyst for positive change. MI leveraging audio literacy is essential in Africa for its ability to bridge linguistic divides and promote inclusivity. With a multitude of languages spoken across the continent, audio-based technology ensures that information is accessible to a broader audience, transcending literacy barriers. This inclusivity becomes a driving force in education, healthcare, and communication, fostering a more connected and empowered society. Additionally, as oral traditions play a significant role in many African cultures, embracing audio literacy through MI respects and preserves these rich narratives while propelling communities into a digitally empowered future.
\subsubsection*{ 
Empowering Farmers Through Audio Knowledge Sharing}
Imagine a scenario where MI facilitates audio knowledge-sharing among farmers in local African languages like Tommo-So Dogon and Fon. In Tommo-So Dogon, farmers could exchange insights on sustainable agricultural practices through voice messages, discussing soil health or effective crop rotation. Simultaneously, in Fon, another group of farmers might share audio clips on traditional farming techniques or effective pest control methods. This audio-to-audio knowledge exchange not only preserves cultural nuances but also creates a collaborative platform where diverse communities can benefit from each other's expertise, fostering a stronger agricultural network across the continent.

\subsubsection*{ Developing Offline Audio-to-Audio Translation for Farmers}

MI can develop an offline, electricity-independent audio-to-audio translation tool for Tommo-So Dogon and Fon, catering to farmers' specific needs. By incorporating pre-trained models into portable, low-energy devices, such as solar-powered audio devices, farmers can access translations without the internet. These devices could utilize edge computing, processing information locally to overcome connectivity challenges. This textless translation model would be trained on a diverse dataset of both languages, understanding the unique phonetics and nuances. This tailored solution empowers farmers with a user-friendly tool, ensuring vital knowledge exchange and collaboration, even in remote areas with limited resources.
\subsubsection*{ Unlocking the Potential of African Audio Datasets}
A significant hurdle lies in the underutilization of the massive audio datasets generated by Africans. Despite the wealth of linguistic diversity and cultural knowledge captured in these recordings, current MI implementations often overlook or lack access to these locally sourced datasets. Harnessing and incorporating these valuable resources could substantially enhance the accuracy and effectiveness of audio-to-audio translation tools, addressing the gap in inclusive technological solutions tailored to the specific needs of communities speaking languages like Tommo-So Dogon and Fon.
The Imperative of Advanced Audio Processing Tools
Moreover, the need to develop advanced audio processing tools becomes evident in leveraging the rich diversity of African languages. Creating accurate audio-to-audio translation models demands sophisticated algorithms that can decipher subtle intonations, dialectical variations, and cultural context present in these recordings. Investing in the research and development of cutting-edge audio processing tools is crucial to unlocking the full potential of MI in catering to the unique linguistic landscape of Africa, facilitating effective knowledge exchange among farmers in languages like Tommo-So Dogon and Fon.
\subsubsection*{ Closing the Technological Gap: Audio-Based MI for Inclusivity}
In Africa, it's not the people who are late; it's the technologies. The richness of oral traditions and audio literacy among diverse communities is palpable. However, the technology landscape has yet to catch up. Remember, local languages are rich in audio, but current technology struggles with audio processing. If the technology transforms audio to text and translates and converts to audio the translated text, it  loses most of the interesting nuances since many of these languages lack a writing system. There's a need for innovation to enable technology to work directly with audio signals, mirroring how local African people communicate, without relying on text. 
 
 Urgently, we need to bridge this gap by developing Audio-based Machine Intelligence. By harnessing the innate audio literacy of the people, we can create inclusive solutions that transcend linguistic barriers. It's time for technology to align with the vibrant mosaic of African cultures and empower communities through accessible, audio-driven advancements.
Rethinking Tech Adoption: An Urgent Call for Inclusive MI
When we force people to adopt current MI technologies, we create a greater tech divide in African society because the audio-literate are left out, and they constitute millions of people. However, if an audio-based MI exists today, these millions of people could seize the opportunities to use it in their businesses, just as they started on audio messaging platforms without knowing how to read and write. All tech developers must consider this feedback if the goal is to reduce the tech divide.

It's not because a piece of information is not available on the internet that this information doesn't exist. It's not because these millions of people don't write that their history doesn't exist. They are audio-literate individuals that technologies must learn from and integrate. In exploring MI  in Africa, the continent's abundant potential lies in its rich audio literacy. From traditional rhythms to dynamic oral traditions, audio literacy is a transformative force. The call to leverage Africa's audio proficiency for MI resonates as millions, labeled illiterate, actively engage in local language audio-to-audio communication through smartphones. Despite challenges, the journey toward inclusive MI involves recognizing cultural narratives, preserving oral traditions, and addressing technological gaps. By embracing audio literacy, we pave the way for a future where technology aligns with African cultures, empowering communities through accessible, audio-driven advancements, and reducing technological divides.

\subsection*{Machine Intelligence Integrity }

Machine  intelligence integrity can be understood as the intentional and systematic approach to design, implement, and operate machine intelligence systems in a manner that upholds safety, dignity, and ethical considerations throughout their lifecycle. It involves embedding ethical assessments into the core functioning of the MI's operating system, ensuring a continuous commitment to human values and societal well-being. 

For lawyers, understanding machine  intelligence integrity is crucial as it involves legal and ethical compliance, ensuring MI systems adhere to evolving regulations and ethical standards. It also underscores the importance of legal oversight in addressing ethical challenges, safeguarding user rights, and contributing to the responsible and lawful deployment of MI technologies.

Machine intelligence integrity is a foundational principle for developers, emphasizing the conscientious and systematic approach to designing, developing, and maintaining MI systems with a commitment to ethical considerations and human values. It involves integrating ethical assessments directly into the core of the MI operating system, ensuring that the system not only meets technical standards but also aligns with principles of safety, fairness, and dignity. For developers, machine integrity highlights the importance of proactive measures during the entire lifecycle of MI, promoting transparency, user-centric design, and continuous monitoring to address potential biases and ethical concerns. By embracing machine integrity, developers contribute to building MI systems that are not just technically robust but also ethically sound, fostering responsible innovation and positive societal impact.

Machine  intelligence integrity stands as a critical framework for government authorities engaged in regulating and overseeing machine intelligence. This concept underscores the intentional and systematic design, development, and ongoing operation of MI systems with a focus on ethical considerations and alignment with human values. For government authorities, understanding and promoting machine integrity means advocating for MI systems that prioritize safety, fairness, and transparency, while also accommodating the evolving ethical landscape. It emphasizes the importance of regulations and oversight to ensure that MI technologies are deployed responsibly, respecting legal frameworks, and contributing positively to societal well-being. By incorporating machine integrity into regulatory discussions, government authorities play a crucial role in fostering an ethical and accountable MI ecosystem that serves the public interest.

\subsection*{Machine Intelligence Integrity goes beyond the standard ethics} 

\subsection*{ Scope and Focus: } \begin{itemize}
\item Machine Intelligence Integrity: Primarily focuses on the intentional and systematic design and operation of MI systems to ensure safety, dignity, and ethical considerations throughout their lifecycle.
\item Multi-Factor MI Ethics Over Time: Encompasses a broader scope, considering various ethical factors that may evolve over time, including but not limited to privacy, transparency, bias mitigation, and societal impact. \end{itemize}
 
\subsection*{ Temporal Perspective:} \begin{itemize}

\item Machine Intelligence Integrity: Emphasizes a continuous commitment to ethical considerations throughout the entire lifespan of the MI system.
\item Multi-Factor MI Ethics Over Time: Acknowledges the dynamic nature of ethical concerns, recognizing that factors like societal norms and values can evolve, requiring ongoing ethical assessments and adaptations.\end{itemize}
 
\subsection*{ Integration into Operating Systems:} \begin{itemize}
 
\item Machine Intelligence Integrity: Involves embedding ethical assessments directly into the core functioning of the MI's operating system, making ethical considerations an integral part of the MI's behavior.
\item Multi-Factor MI Ethics Over Time: Requires periodic assessments and adjustments to ethical guidelines, often influencing MI design and decision-making processes but may not be as deeply integrated into the operating system. \end{itemize}
 
\subsection*{ Human Empowerment Emphasis: }\begin{itemize}
 
\item Machine Intelligence Integrity: Posits that MI should serve the empowerment of humanity, suggesting a focus on designing systems that contribute positively to human well-being.
\item Multi-Factor AI Ethics Over Time: Prioritizes ethical guidelines to ensure MI benefits humanity and avoids harm, but the emphasis may extend beyond empowerment to encompass a broader set of considerations. \end{itemize}

\subsection*{ Proactive vs. Reactive Approach: }\begin{itemize}
 
\item Machine Intelligence Integrity: Takes a proactive stance by designing MI systems with integrity from the outset, aiming to prevent ethical issues.
\item Multi-Factor MI Ethics Over Time: May involve a more reactive approach, responding to emerging ethical concerns and adapting guidelines as societal and technological landscapes evolve. \end{itemize}
 
\subsection*{ Comprehensiveness of Ethical Factors: }\begin{itemize}
 
\item Machine Integrity: Specifically addresses the intentional design and operation of MI systems to ensure safety, dignity, and ethical considerations.
\item Multi-Factor MI Ethics Over Time: Encompasses a wider range of ethical factors, such as fairness, accountability, and explainability, recognizing the multifaceted nature of ethical considerations in MI development and deployment.
\end{itemize}

\subsection*{ How to Implement Machine  Intelligence Integrity ?}

Implementing machine integrity involves a comprehensive approach to ensure the ethical design, development, and operation of MI systems throughout their lifecycle. This includes defining clear ethical principles aligned with human values, embedding ethical assessments into the design process, using diverse and representative data to minimize biases, implementing transparency and explainability mechanisms, continuous monitoring, user involvement, adaptive learning, human-in-the-loop governance, privacy protection, regular audits, and responsive post-deployment policies. The goal is to proactively address ethical concerns, promote fairness, and safeguard user privacy, fostering a system that respects human values and societal norms.

Machine Intelligence integrity is a comprehensive framework for ensuring the responsible design, development, and operation of MI systems. It involves embedding multi-factor ethical considerations into the core functioning of MI operating systems, with a focus on upholding safety, fairness, and dignity in alignment with human values. Multi-factor ethical adherence is assessed through a set of scalar values, each representing the adherence to specific ethical principles, while decision alignment is similarly evaluated with scalar values for alignment to decision principles. The resulting multi-objective metric vector captures the nuanced evaluation of both ethical adherence and decision alignment, providing a holistic measure of machine integrity. This approach enables a dynamic and responsive assessment that considers the evolving nature of ethics and human values over time, making it a vital concept for fostering responsible MI practices across various domains. 

Part of the errors can be reduced using the following filter:

\begin{tabular}{|p{0.7in}|p{0.9in}|p{0.9in}|}
  \hline
  & Knowns & Unknowns \\
  \hline
  Believe as Known & Known Knowns & Known Unknowns \\
  \hline
  Belief as Unknown & Unknown Knowns & Unknown Unknowns \\
  \hline
\end{tabular}

which is rewritten as 

\begin{tabular}{|p{0.7in}|p{1.1in}|p{1.1in}|}
  \hline
  & Knowns & Unknowns \\
  \hline
  Believe as Known & Things the MI knows it knows & Things the MI knows it doesn't know \\
  \hline
  Belief as Unknown & Things the MI doesn't know it knows & Things the MI doesn't know it doesn't know \\
  \hline
\end{tabular}

In dynamic game theory, a state-feedback strategy refers to a decision-making approach where the actions of a player are determined based on both the current state of the system and the historical evolution of the game. Unlike open-loop strategies that depend solely on the initial conditions, state-feedback strategies enable players to adapt their actions dynamically in response to the changing circumstances within the game. These strategies involve the use of feedback mechanisms, where information about the current state is continuously incorporated to adjust the player's decisions, allowing for more flexibility and responsiveness in complex, evolving dynamic environments.

In Mean-Field-Type Game Theory (MFTG), a state-and-mean-field-type strategy involves players making decisions based not only on their individual state-action pairs but also  on the distribution of all individual state-action pairs. In this strategy, each player's action is influenced by both their own local information (state) and own  individual mean-field (belief), and  the collective behavior  local information and local belief of the entire group. The mean-field type term here  captures the distribution of own information and the distribution of others' information. It allows to consider risk-aware decisions via variance, quantile, expectile, extremile, conditional value-at-risk,  entropic value-at-risk, etc. State-and-mean-field-type feedback strategies are particularly relevant in risk quantification in engineering. 

A state-and-mean-field-type feedback strategy can be defined as a measurable function incorporating various factors to assess the integrity of machine intelligence. The measurable function is a result of:

(a) Current time

(b) Current high-quality data

(c) Current extracted knowledge data

(d) Current architecture design

(e) Current training model

(f) Current localized deployment model

(g) Current privacy-preserving status

(h)Current context-awareness

(i) Current risk-awareness 

(j) Current localized culture-awareness 

(k) States of the societies at the current time

(l) Localized  social norms of the societies at the corresponding states

(m) Current beliefs about human values 

A state-and-mean-field-type feedback strategy for implementing machine intelligence integrity involves creating a measurable function that considers factors such as initial and current quality data, time, knowledge, architecture design, training and deployment models, societal states, beliefs about human values, and cultural influences. This approach requires continuous monitoring, adaptive learning, feedback loops, scenario planning, ethics committees, human-in-the-loop governance, privacy protection, regular audits, user involvement, and educational initiatives. By integrating these elements, the strategy aims to proactively align the machine intelligence system with evolving ethical considerations, human values, and societal norms, fostering responsible and accountable deployment throughout its lifecycle.
The advantage of employing a state-and-mean-field-type feedback strategy in Machine Intelligence lies in its comprehensive and adaptive approach to managing complex, interconnected factors. This strategy allows the system to continuously monitor and integrate various elements such as initial and current data quality, temporal aspects, architectural design, societal states, beliefs about human values, and cultural influences. By incorporating a mean-field-type approach, which considers the  influence of the collective distribution of these factors, the system becomes more responsive to evolving ethical considerations and societal dynamics. This adaptability, coupled with proactive measures like scenario planning, regular audits, and user involvement, enhances the system's ability to align with human values, address biases, and promote responsible innovation. Ultimately, the state-and-mean-field-type feedback strategy contributes to the development of machine intelligence systems that are not only technically robust but also ethically sound, fostering trust and positive societal impact

\subsection*{ Adaptive Machine Integrity }

Adapting machine integrity to the evolution of ethics and human values involves a proactive and dynamic approach. Continuous ethical assessment, feedback mechanisms, and engagement with diverse stakeholders are essential for staying aligned with changing societal perspectives. Keeping training data updated, forming ethics committees, and implementing scenario planning contribute to anticipating and addressing emerging ethical challenges. Utilizing version control for iterative updates, ensuring legal and regulatory compliance, fostering public discourse, and maintaining human oversight are crucial components in the ongoing effort to navigate and integrate evolving ethical considerations into MI systems. Education and awareness initiatives further support a collective commitment to upholding machine integrity in the face of shifting ethical landscapes over time.


\section{Conclusion} \label{sec:conclusion} 

This survey article  sheds light on the state of MI  research and implementation in Africa. It underscores the diverse paths each country takes in embracing MI and other emerging technologies. The table highlighting countries with explicit national MI strategies reflects a growing governmental awareness and commitment in some regions. However, the article reveals a disparity between high-profile meetings held in luxurious venues and the actual impact on the local population, emphasizing the need for inclusive strategies. Moreover, the examination of over 400 research articles highlights the prevalence of overselling findings and the importance of updating studies to practical MI algorithms. The ethical dimension emerges in the call to move beyond uniform rankings, considering factors like data transparency and concrete government actions for a more accurate assessment of the progress achieved by MI and its utilization across the continent. Ultimately, this survey article serves as a valuable resource for understanding the  landscape of MI in Africa and advocating for meaningful, locally impactful advancements.


\section*{Future work}

Our review of 400 articles in MI in Africa  from 2018-2023 show  the use of these technologies in diverse settings.  
From Cabo Verde to the tip of Ethiopia, from Cape Town to Tunis via Madagascar, the informal sector appears to extensively harness machine intelligence. 
\begin{itemize}\item In the future, local MI must assimilate this on-the-ground data to enhance its knowledge, covering a significant portion of the informal sector.
\item The ability to learn audibly from one local language to another without relying on a writing system opens up business opportunities for millions of people who excel in their native languages despite being unable to read or write.
\item The multidimensional and multifactorial aspect of machine ethics, influenced by local culture and specific issues, should not be overlooked and must be integrated from the design and algorithm training and learning phase.
\item Local MI capable of operating without internet or electricity could broaden access in rural areas, benefiting both the young and the elderly.
\item Developing strategic MI in underserved areas would help anticipate market interactions and the informal economy.
\item Quantifying the risks associated with strategic MI should be at the core of design, taking into account ethical and cultural factors.
\end{itemize}

Examining closely 400 articles on MI in Africa, beyond the headlines, it emerges that
\begin{itemize}\item human learning, the learning of men and women, whether children or adults, takes a much more central place in discussions than machine learning.
\item At the core is human learning, utilizing various tools, including machine assistance, as well as inspiration from nature.
\end{itemize}

\subsubsection*{ Early stage MI development topics }
Below we provide some selected topics and their early stage MI development in African countries.
\begin{table}[htb]
\centering \tiny
\begin{tabular}{|p{8cm}|p{3.5cm}|}
\hline
\textbf{MI Implementation for African Countries} & \textbf{Realization} \\
\hline
Blockchain  and MI-enabled National ID & Very Early Stage \\
Blockchain and MI-enabled Land Management & Very Early Stage \\
Audio2Audio local languages& Very Early Stage \\
 Text2Image local items& Very Early Stage \\
Image2Audio local languages& Very Early Stage \\
Knowledge-based MI & Very Early Stage \\
Robust MI Infrastructure & Very Early Stage \\
MI-enabled Cybersecurity  & Very Early Stage \\
Accessible MI Education Programs & Very Early Stage \\
Collaborative MI Research Initiatives & Very Early Stage \\
Responsible MI Policies and Regulations & Very Early Stage \\
Sector-wide Integration of MI Technologies & Very Early Stage \\
Supportive Ecosystem for MI Startups & Very Early Stage \\
Technical MI  Ethics & Very Early Stage \\
Non-Technical MI  Ethics & Very Early Stage \\
Ethical Considerations in MI Development & Very Early Stage \\
Data Accessibility, Privacy, and Security & Very Early Stage \\
Government Investment in MI Initiatives & Very Early Stage \\
Global Participation in MI Initiatives & Very Early Stage \\
Ethical Data Practices & Very Early Stage \\
Inclusive MI Integration Across Key Sectors & Very Early Stage \\
International Partnerships for Knowledge Exchange & Very Early Stage \\
Funding for MI Research and Development & Very Early Stage \\
Emphasis on Responsible and Inclusive MI Practices & Very Early Stage \\
Promotion of MI Education and Skilled Workforce & Very Early Stage \\
Development of MI Regulatory Frameworks & Very Early Stage \\
Strategic MI Technology Integration in Key Sectors & Very Early Stage \\
Support for MI Startup Ecosystem & Very Early Stage \\
Public Awareness and Understanding of MI & Very Early Stage \\
MI-driven Solutions for Socio-Economic Challenges & Very Early Stage \\
Infrastructure for Advanced Data Analytics & Very Early Stage \\
MI-driven Agricultural Innovations & Very Early Stage \\
Telemedicine and Healthcare MI Applications & Very Early Stage \\
MI in Renewable Energy  & Very Early Stage \\
MI in  Environmental Conservation & Very Early Stage \\ 
MI for Efficient Transportation and Logistics & Very Early Stage \\
MI for Cities and Urban Planning & Very Early Stage \\
MI in Education for Personalized Learning & Very Early Stage \\
Community Engagement in MI Development & Very Early Stage \\
MI for Disaster Response and Management & Very Early Stage \\
MI-powered Financial Inclusion Initiatives & Very Early Stage \\
\hline
\end{tabular}
\caption{MI Implementation Topics for the Benefits of African Countries}
\label{tab:ai_africa}
\end{table}

\subsubsection*{ Concrete actions for MI implementations } 
Implementing MI as a local public good for the  population requires collaboration among traditional authorities,  government bodies, educational institutions, industry stakeholders, and local communities. It cannot be reduced to a political announcement without any concrete followup. Regular assessments, transparency, and a commitment to inclusivity will contribute to the successful development of responsible and culture-aware MI for the benefit of African societies and worldwide. Here are some concrete actions to be taken towards implementation:

\begin{itemize}\item Recommendation: Develop training programs for MI practitioners on ethics and cultural sensitivity, catering to all Africans, including those who may be considered illiterate but are, in fact, highly aural literate. The equivalent of literacy for audio is often referred to as "aural literacy" or "audio literacy." Aural literacy involves the ability to interpret, comprehend, and create meaning from auditory information. This includes skills such as listening comprehension, recognizing and understanding various audio elements, and effectively using and producing audio content. As our interactions with technology and media become more audio-centric, developing aural literacy becomes increasingly important for navigating and understanding the information presented in audio formats.
\begin{itemize}\item Concrete Action: Integrate culture-aware ethical training modules into audio literacy MI courses and provide ongoing professional development opportunities for practitioners. \end{itemize} 
\item Recommendation: Invest in MI education programs at all educational levels.
\begin{itemize}\item Concrete Action: Establish MI curriculum in both conventional and non-conventional schools and universities, ensuring students are exposed to both technical and ethical aspects of MI adapted to local problems. \end{itemize}  
\item Recommendation: Conduct public awareness campaigns on MI.
\begin{itemize}\item Concrete Action: Organize workshops, webinars, and community events to educate the public about MI, its applications, and potential societal impacts. \end{itemize}  
\item Recommendation: Create audio MI literacy  programs for policymakers and government officials.
\begin{itemize}\item Concrete Action: Offer specialized training for policymakers to understand vision and audio MI technologies, implications, and policy considerations. \end{itemize} 
\item Recommendation: Establish MI ethics training for businesses and organizations.
\begin{itemize}\item Concrete Action: Provide workshops and resources to businesses to ensure ethical MI practices in their operations. \end{itemize} 
\item Recommendation: Include MI in vocational and technical training programs.
\begin{itemize}\item Concrete Action: Integrate MI components into vocational training programs to prepare a diverse workforce for emerging MI-related jobs. \end{itemize} 
\item Recommendation: Collaborate with international organizations for MI education initiatives.
\begin{itemize}\item Concrete Action: Form partnerships with global educational institutions and organizations to enhance MI education programs. \end{itemize} 
\item Recommendation: Develop a centralized platform for MI knowledge sharing.
\begin{itemize}\item Concrete Action: Create an online platform for sharing MI resources, research, and best practices among academia, industry, and policymakers. \end{itemize} 
\item Recommendation: Translate MI educational materials into local languages.
\begin{itemize}\item Concrete Action: Ensure accessibility by translating MI educational content into languages spoken in the region. \end{itemize} 
\item Recommendation: Foster partnerships between educational institutions and industry.
\begin{itemize}\item Concrete Action: Facilitate internships, joint research projects, and industry collaborations to bridge the gap between academia and industry. \end{itemize} 
\item Recommendation: Establish MI mentorship programs.
\begin{itemize}\item Concrete Action: Pair experienced MI professionals with students and early-career professionals to provide guidance and support. \end{itemize} 
\item Recommendation: Include context-aware MI in continuous learning programs for professionals.
\begin{itemize}\item Concrete Action: Develop short courses and workshops to keep professionals updated on the latest advancements and ethical considerations in MI. \end{itemize}
\item Recommendation: Encourage research on the cultural impact of MI.
\begin{itemize}\item Concrete Action: Fund research projects that specifically investigate how MI intersects with local cultures and societies. \end{itemize} 
\item Recommendation: Create a network of MI experts for knowledge exchange.
\begin{itemize}\item Concrete Action: Facilitate regular conferences, forums, and networking events for MI professionals to share experiences and insights. \end{itemize} 
\item Recommendation: Establish a national context-aware MI day to celebrate achievements and raise awareness.
\begin{itemize}\item Concrete Action: Designate a day to showcase MI projects, host events, and engage the rural \& urban  area  public in discussions about MI's role in society. \end{itemize} 
\item Recommendation: Incorporate context \& cultural considerations into MI development guidelines.
\begin{itemize}\item Concrete Action: Work with traditionalists, ethicists, cultural experts, and technologists to create MI development guidelines that reflect cultural values. \end{itemize} 
\item Recommendation: Establish MI ethics review boards.
\begin{itemize}\item Concrete Action: Form independent boards to evaluate MI projects for cultural \& ethical considerations, ensuring diverse representation and transparency. \end{itemize} 
\item Recommendation: Develop guidelines for responsible MI research.
\begin{itemize}\item Concrete Action: Create a set of guidelines for researchers to follow, addressing context-aware ethical considerations and potential societal impacts. \end{itemize} 
\item Recommendation: Conduct regular ethical audits of MI systems.
\begin{itemize}\item Concrete Action: Implement periodic reviews of MI systems to identify and address ethical concerns, with findings made public. \end{itemize} 
\item Recommendation: Encourage diverse teams in MI development.
\begin{itemize}\item Concrete Action: Promote diversity and inclusion in MI teams, ensuring representation from various backgrounds and perspectives. \end{itemize} 
\item Recommendation: Integrate bias detection tools into MI development processes.
\begin{itemize}\item Concrete Action: Require developers to use tools that identify and mitigate biases in MI algorithms during the development phase. \end{itemize} 
\item Recommendation: Establish clear guidelines for MI transparency.
\begin{itemize}\item Concrete Action: Define standards for transparency in local MI systems, ensuring clear communication about how decisions are made. \end{itemize} 
\item Recommendation: Implement continuous training on ethical MI for developers.
\begin{itemize}\item Concrete Action: Mandate ongoing training for MI developers to stay updated on ethical considerations and best practices. \end{itemize} 
\item Recommendation: Encourage the adoption of MI ethics certifications.
\begin{itemize}\item Concrete Action: Support the development and recognition of certifications that highlight adherence to ethical MI practices. \end{itemize} 
\item Recommendation: Establish mechanisms for reporting MI ethical concerns.
\begin{itemize}\item Concrete Action: Create channels for reporting ethical concerns related to MI applications, ensuring a responsive and accountable system. \end{itemize} 
\item Recommendation: Foster a culture of responsible MI innovation.
\begin{itemize}\item Concrete Action: Recognize and reward organizations and individuals who prioritize responsible MI development in competitions and awards. \end{itemize} 
\item Recommendation: Include ethical considerations in MI project funding criteria.
\begin{itemize}\item Concrete Action: Funding agencies should prioritize projects that adhere to ethical guidelines and promote responsible MI practices. \end{itemize} 
\item Recommendation: Collaborate with international organizations on ethical MI standards.
\begin{itemize}\item Concrete Action: Participate in global discussions to contribute to the development of ethical MI standards and adopt them nationally. \end{itemize} 
\item Recommendation: Regularly update ethical guidelines to reflect evolving technologies.
\begin{itemize}\item Concrete Action: Establish a process for periodic review and updates of ethical guidelines to keep pace with technological advancements. \end{itemize} 
\item Recommendation: Establish an independent body for monitoring MI ethics.
\begin{itemize}\item Concrete Action: Form a national or regional body responsible for monitoring and enforcing ethical and culture-aware MI practices, with the power to investigate and penalize non-compliance. \end{itemize} 
\item Recommendation: Incorporate community input into MI decision-making.
\begin{itemize}\item Concrete Action: Implement mechanisms for seeking public input and feedback on MI projects, ensuring community perspectives are considered. \end{itemize} 
\item Recommendation: Conduct impact assessments for MI implementations.
\begin{itemize}\item Concrete Action: Require organizations to conduct thorough impact assessments before deploying MI systems, assessing potential social, economic, and cultural implications. \end{itemize} 
\item Recommendation: Support community-based MI initiatives.
\begin{itemize}\item Concrete Action: Provide grants and resources to community organizations working on MI projects that address local challenges and priorities. \end{itemize} 
\item Recommendation: Establish community-led MI forums.
\begin{itemize}\item Concrete Action: Facilitate regular community forums where residents can voice concerns, ask questions, and engage in discussions about MI applications. \end{itemize} 
\item Recommendation: Create MI literacy programs for community leaders.
\begin{itemize}\item Concrete Action: Develop tailored programs to educate community leaders about MI, enabling them to advocate for responsible MI development. \end{itemize} 
\item Recommendation: Collaborate with traditional leaders and elders on MI initiatives.
\begin{itemize}\item Concrete Action: Engage with local traditional leaders to incorporate cultural wisdom and perspectives into MI projects. \end{itemize} 
\item Recommendation: Develop accessible MI interfaces for diverse populations.
\begin{itemize}\item Concrete Action: Ensure MI interfaces are user-friendly and accessible to diverse populations, considering language, literacy levels, and digital access. \end{itemize} 
\item Recommendation: Promote gender-inclusive MI development.
\begin{itemize}\item Concrete Action: Encourage the participation of women in MI projects and ensure gender considerations are integrated into the development process. \end{itemize} 
\item Recommendation: Establish partnerships with civil society organizations.
\begin{itemize}\item Concrete Action: Collaborate with local teams, associations and civil society groups to ensure MI projects align with societal values and address community needs. \end{itemize} 
\item Recommendation: Conduct MI impact assessments on marginalized communities.
\begin{itemize}\item Concrete Action: Prioritize assessments on how MI implementations may affect marginalized or vulnerable communities, addressing potential risks. \end{itemize} 
\item Recommendation: Encourage local innovation in MI for societal challenges.
\begin{itemize}\item Concrete Action: Create funding opportunities and innovation challenges specifically aimed at addressing local societal challenges through MI. \end{itemize} 
\item Recommendation: Establish mechanisms for MI project feedback and improvement.
\begin{itemize}\item Concrete Action: Create channels for ongoing feedback from communities impacted by MI projects, enabling continuous improvement based on real-world experiences. \end{itemize} 
\item Recommendation: Foster cultural preservation through MI initiatives.
\begin{itemize}\item Concrete Action: Support projects that leverage MI to preserve and promote cultural heritage, languages, and traditions. \end{itemize} 
\item Recommendation: Encourage MI for social inclusion and accessibility.
\begin{itemize}\item Concrete Action: Prioritize MI applications that enhance social inclusion and accessibility for people with disabilities, elderly populations, and other marginalized groups. \end{itemize} 
\item Recommendation: Establish a national or regional MI advisory council that works with an MI global alliance.
\begin{itemize}\item Concrete Action: Form a council comprising representatives from diverse sectors, including  informal sector, academia, industry, civil society, and government, to provide guidance on MI development and its societal impact. \end{itemize}
\end{itemize}

The sections within this survey have unfolded a narrative that goes beyond the mere exploration of technological landscapes; it is a testament to the dynamic interplay between MI  and the rich, diverse cultures of the African continent.
In our journey through North Africa, Southern Africa, Middle Africa, West Africa, and East Africa, we've witnessed the transformative power of MI, transcending geographical boundaries to leave an indelible mark on the informal economy, agriculture,  art and music, and societies. Yet, embedded within these advancements is a constant reminder of the importance of cultural sensitivity, ethical considerations, and the need to harness the unique strengths of Africa's audio literacy.
The call to action  resonates even more strongly now  -  the imperative to leverage Africa's audio literacy as a cornerstone for MI innovation. 

In the era of digital communication, where the auditory sense is increasingly becoming a dominant mode of interaction, the significance of audio literacy cannot be overstated. This article extends a resounding call to the global tech community, urging a collective acknowledgment and appreciation for the power of audio in shaping our technological landscape.
The African experience, with its innate proficiency in audio literacy, serves as an inspiring model for the world. The fusion of MI and audio literacy has the potential to break down barriers, foster inclusivity, and create technologies that resonate with people on a deeper, more human level.
To the tech innovators, developers, and engineers worldwide, consider this an invitation to reevaluate the role of audio in your creations. Imagine a future where technology not only speaks to users but engages with them in a language that transcends written words -  a language embedded in the rhythms, intonations, and nuances of audio.
As we champion the cause of audio literacy, let us embark on a journey to make technology more accessible, inclusive, and culturally sensitive. Whether you're designing virtual assistants, educational tools, or entertainment platforms, integrating audio literacy into the fabric of your innovations can pave the way for a more universally understood and appreciated technological landscape. This is a call to infuse cultural richness into the very code and algorithms that shape our digital experiences. Let audio become a bridge that connects diverse communities, ensuring that technology resonates with people from various linguistic and cultural backgrounds.
In the spirit of coopetition, where collaboration coexists with competition, let us share insights, methodologies, and best practices for incorporating audio literacy into tech solutions. By doing so, we can collectively contribute to a global technological ecosystem that celebrates the richness of human expression in all its auditory dimensions.
This article is also an invitation to envision a future where the dialogue between humans and machines extends beyond the visual and written to embrace the profound power of sound. The path ahead is one of collaboration, innovation, and a shared commitment to shaping a technological future that is not just intelligent but also deeply attuned to the audio path of humanity.

\section*{Acknowledgment}

The work of Hamidou Tembine is supported by Timadie grant on {\bf Mean-Field-Type Game Theory for Machine Intelligence in Underserved Areas.}




%
%
%
%
%
%
%
%

\end{document}